\documentclass[pdflatex,sn-mathphys-num]{sn-jnl}

\usepackage{graphicx}%
\usepackage{multirow}%
\usepackage{amsmath,amssymb,amsfonts}%
\usepackage{amsthm}%
\usepackage{mathrsfs}%
\usepackage[title]{appendix}%
\usepackage{xcolor}%
\usepackage{textcomp}%
\usepackage{manyfoot}%
\usepackage{booktabs}%
\usepackage{algorithm}%
\usepackage{algorithmicx}%
\usepackage{algpseudocode}%
\usepackage{listings}%

\usepackage{rotating}
\usepackage{cuted}
\usepackage{stfloats}
\usepackage{adjustbox}
\usepackage{colortbl}
\usepackage{float}
\usepackage{soul}
\usepackage{newtxtext,newtxmath}
\usepackage{amsmath}
\usepackage[absolute,overlay]{textpos}
\usepackage{subcaption}

\textblockorigin{0cm}{0cm}
\usepackage[percent]{overpic}

\usepackage{subcaption} 
\usepackage{threeparttable} 

\usepackage{color}

\theoremstyle{thmstyleone}%
\theoremstyle{thmstyletwo}%

\theoremstyle{thmstylethree}%

\usepackage{placeins}
\usepackage{capt-of}
\colorlet{teal}{black}
\begin{document}

\title[Article Title]{\textcolor{black}{Demographic inference in exoplanet chemistry is shaped by retrieval assumptions}}


\author*[1,2]{\fnm{Dwaipayan} \sur{Dubey}}\email{dwaipayan.dubey25@gmail.com,ddubey@usm.lmu.de}

\author[3]{\fnm{Karan} \sur{Molaverdikhani}}

\author[1,2]{\fnm{Barbara} \sur{Ercolano}}
\author[1,2]{\fnm{Surya } \sur{Shivaprasad}}
\author[1,2]{\fnm{Fabian} \sur{Grübel}}
\author[1,2]{\fnm{Rosa} \sur{Arenales-Lope}}

\affil*[1]{\orgdiv{Universitäts-Sternwarte, Fakultät für Physik}, \orgname{Ludwig-Maximilians-Universität München}, \orgaddress{\street{Scheinerstr. 1}, \city{ D-81679 München}, \country{Germany}}}
\affil[2]{\orgdiv{Exzellenzcluster `Origins’}, \orgaddress{\street{ Boltzmannstr. 2}, \city{D-85748 Garching}, \country{Germany}}}
\affil[3]{\orgdiv{Max-Planck-Institut für extraterrestrische Physik}, \orgaddress{\street{ Gießenbachstr. 1}, \city{D-85748 Garching}, \country{Germany}}}


\abstract{Understanding the chemical diversity of exoplanet atmospheres is crucial for linking atmospheric composition to planetary formation history. \textcolor{teal}{Atmospheric C/O can provide clues about where and how planets form, particularly through the accretion of gas and solids across disk snowlines. Since planetary migration can change orbital distances and hence equilibrium temperatures ($T_{\rm eq}$), investigating C/O across different planetary populations may provide insights into their formation and evolution.} As observational precision improves with the James Webb Space Telescope (JWST), the limits of standard atmospheric retrieval methods that assume altitude-independent molecular abundances have become apparent. \textcolor{black}{We present atmospheric retrievals for 20 transiting planets with published JWST transmission spectra, including complementary HST spectra for targets where such data are available, spanning gas giants, warm Neptunes, and temperate/warm sub-Neptunes, to test how population-level chemical inferences depend on retrieval assumptions.} \textcolor{teal}{Free-chemistry retrievals show a visually non-monotonic C/O-$T_{\rm eq}$ distribution, but this apparent pattern is not statistically robust and is sensitive to sample composition.} When molecular abundances are constrained by a grid of disequilibrium models, the inferred C/O distribution becomes more restricted, showing that demographic chemical trends can depend strongly on the adopted retrieval parameterization. Several hot Jupiters converge to low C/O ($\leq$ 0.05), whereas colder, cloudy planets retain broad posteriors that reflect limited information content of their spectra. Our results suggest that population-level chemical trends inferred from free-chemistry retrievals should be interpreted with caution unless sample selection, residual structure, prior sensitivity, and chemical assumptions are explicitly tested.}




\maketitle

\section{Introduction}\label{sec1}

Atmospheric elemental abundances in giant exoplanets provide one of the few direct probes of their formation histories and interior compositions \citep{oberg2011effects,madhusudhan2014toward,mordasini2016imprint,teske2024star}. In our own Solar System, Jupiter’s supersolar enrichments in carbon, sulfur, nitrogen, and noble gases indicate that heavy elements were delivered by solid planetesimals rather than by gas alone \citep{owen1999low,atreya2005coupled,mousis2019jupiter}. This link between atmospheric composition and formation has motivated the use of transmission and emission spectra to trace accretion history in exoplanets \citep{oberg2011effects,madhusudhan2016exoplanetary,cridland2020connecting,teske2024star}.

Among the various elemental ratios, the carbon-to-oxygen ratio (C/O) has been the most widely investigated. Early models showed that the C/O of a planet’s envelope depends on where it forms relative to the major volatile snowlines of $\mathrm{H_2O}$, $\mathrm{CO_2}$, and CO, which divide carbon and oxygen between gas and solids \citep{oberg2011effects,cridland2020connecting,oberg2019jupiter}. \textcolor{black}{\textcolor{black}{In disk-chemistry models, gas inside the $\mathrm{H_2O}$ snowline has approximately stellar or slightly super-stellar C/O.} A subsolar atmospheric C/O can therefore indicate enrichment by oxygen-rich solids rather than direct accretion of oxygen-rich gas. Beyond the $\mathrm{H_2O}$ snowline but inside the CO snowline, water ice is removed from the gas phase. This can result in a higher gas-phase C/O, so planets accreting a substantial fraction of their envelopes from this region may also reach higher atmospheric C/O \citep{oberg2011effects,madhusudhan2014toward,mordasini2016imprint}.} \textcolor{black}{Retrieval studies have reported a wide range of atmospheric C/O estimates, from higher or super-solar C/O in systems such as WASP-12 b, WASP-33 b, WASP-103 b, WASP-19 b, HD 149026 b, WASP-69 b, WASP-76 b, and WASP-121 b \citep{madhusudhan2011high,haynes2015spectroscopic,kreidberg2018global,bean2023high,schlawin2024multiple,weiner2024metallicity,smith2024roasting,saha2025highly} to near-solar or intermediate C/O in planets such as WASP-80 b, WASP-77 A b, and WASP-18 b \citep{wiser2025precise,august2023confirmation,brogi2023roasting}, and lower or subsolar C/O in planets such as HD 209458 b, HD 189733 b, WASP-107 b, and WASP-39 b \citep{xue2024jwst,fu2024hydrogen,welbanks2024high,alderson2023early}.} \textcolor{black}{However, C/O alone is not sufficient to trace planet formation pathways, as planets accrete both solids and gas in proportions that vary with time, disk chemistry, and dynamical history \citep{mordasini2018planetary,cridland2020connecting,eistrup2018molecular,crossfield2023volatile}. Robust formation interpretation therefore requires additional context beyond C/O alone, including host-normalized abundance ratios, volatile and refractory elemental ratios, disk chemistry, and migration history.}

Sulfur has recently emerged as a complementary tracer of solid accretion \citep{moses2011disequilibrium,zahnle2016photolytic,laas2019modeling,crossfield2023volatile}. In protoplanetary disks, sulfur is largely stored in refractory or moderately volatile compounds (e.g. FeS, MgS, organosulfur ices), unlike the highly volatile carriers of carbon and oxygen. As a result, the volatile-to-sulfur ratios, such as C/S and O/S, encode the balance between solid and gaseous heavy-element delivery \citep{laas2019modeling,keyte2024spatially}. Pebble-accretion models predict sulfur depletion and super-stellar C/S and O/S, whereas planetesimal-dominated accretion yields near-stellar values \citep{mordasini2016imprint,crossfield2023volatile}. Until recently, this idea could not be tested because sulfur-bearing molecules were not detected in exoplanet spectra. The James Webb Space Telescope (JWST) has now made sulfur chemistry observable in several systems \citep{welbanks2024high,beatty2024sulfur,fu2025statistical,fu2024hydrogen}, enabling first, albeit tentative, demographic explorations of C/S and O/S.

Early demographic studies based on the Hubble Space Telescope's Wide Field Camera (HST/WFC3) data pioneered the population approach but also exposed important limitations. The limited 1.1–1.7 $\mathrm{\mu m}$ band provided access to water vapour and cloud properties but not to key carbon and sulfur-bearing species such as CO, $\mathrm{CO_2}$, $\mathrm{CH_4}$, or $\mathrm{SO_2}$ \citep{barstow2020outstanding,lueber2024information,zhou2025hst}. In addition, non-uniform data reduction, noise treatment, and retrieval assumptions led to \textcolor{black}{different} abundance estimates \citep{mugnai2024comparing,barstow2022retrieval}. Atmospheric retrievals that assume thermochemical-equilibrium struggle in atmospheres where photochemistry and vertical mixing drive strong deviations from equilibrium scenarios \citep{moses2011disequilibrium,molaverdikhani2019cold1,tsai2017vulcan}. \textcolor{black}{Bardet et al. \citep{bardet25} recently re-analyzed ten hot-Jupiter atmospheres with the TauREx-FRECKLL disequilibrium-chemistry retrieval framework. They showed that including kinetic chemistry can significantly change the inferred metallicity and C/O compared to simpler chemistry assumptions, while the retrieved constraints remain sensitive to priors, spectral resolution, and data-reduction choices.}

JWST observations now allow the simultaneous detection of multiple species, yielding more informed constraints on C/O, metallicity, and sulfur content for planetary atmospheres \citep{madhusudhan2023carbon,grant2023jwst,powell2024sulfur,bean2023high,welbanks2024high}. \textcolor{teal}{Recent studies have compiled atmospheric composition measurements across exoplanet populations, providing a basis for comparing elemental ratios inferred from different observations and retrieval frameworks \citep{lothringer2025library}. Individual analyses of ultra-hot Jupiters, including WASP-19 b and WASP-178 b, have also reported elevated atmospheric C/O values \citep{saha2025highly,saha2025metal}. Since atmospheric C/O can depend on gas and solid accretion across disk snowlines and subsequent migration, its variation across planets with different equilibrium temperatures is worth investigating, although $T_{\rm eq}$ itself does not uniquely trace the location of planet formation.} \textcolor{black}{Building on this progress, we present a population-level analysis of elemental ratios (C/O, C/S, O/S) for 20 transiting planets selected from the set of targets with published JWST transmission spectra; for planets with suitable archival HST observations, we include those spectra together with the JWST data (see Supplementary Table \ref{tab:exoplanet_properties})}. The sample spans gas giants, warm Neptunes, and temperate/warm sub-Neptunes. We therefore treat demographic trends across the full sample with caution and explicitly test whether the results are sensitive to the inclusion of the smaller planets. We first perform free-chemistry retrievals using petitRADTRANS \citep{molliere2019petitradtrans,molliere2020retrieving,2024JOSS....9.5875N} to infer molecular abundances without enforcing chemical equilibrium, and then compare these results with retrievals in which molecular abundances are tied to a grid of chemical-kinetics models computed with VULCAN \citep{tsai2017vulcan,tsai2021comparative}. \textcolor{teal}{We therefore explore the C/O-$T_{\rm eq}$ distribution as a test case for assessing the sensitivity of population-level chemical inferences to retrieval assumptions, without assuming any underlying temperature-dependent relation.}

\textcolor{black}{In Section \ref{sec:results}, we describe the C/O behavior inferred from the free-chemistry retrievals, test the robustness of the apparent C/O-$T_{\rm eq}$ structure, examine residual patterns in the spectra, and compare the results with a chemically informed grid-retrieval framework.} Section \ref{sec:conclusion} discusses the implications for demographic studies and for using elemental ratios as tracers of planet formation.

\section{Results}
\label{sec:results}

\subsection{C/O ratio from free-chemistry retrievals and sensitivity to sample selection}
\label{result:free_chemistry}

C/O ratio is one of the most widely used indicators that can reveal planet formation environments. Most studies during the HST era and the early phase of JWST focused on individual targets to address this question by investigating chemical properties of their atmospheres. However, investigations have moved beyond the single-planet domain to explore population-level trends in recent times \citep{fu2025statistical,bardet25,lothringer2025library}. These studies raise an important question: how does atmospheric composition relate to a planet's mass, radius, and equilibrium temperature? \textcolor{teal}{Recent analyses of ultra-hot Jupiters, including WASP-19 b and WASP-178 b, have reported high atmospheric C/O values \citep{saha2025highly,saha2025metal}, motivating further comparisons across planets with different irradiation levels. The C/O-$T_{\rm eq}$ distribution considered here could reflect differences in gas and solid accretion histories and subsequent migration, although present-day $T_{\rm eq}$ does not uniquely trace where a planet formed. Irradiation-dependent chemistry, condensation, and retrieval assumptions may also influence the inferred C/O values.} \textcolor{black}{For this purpose, we performed ``free-chemistry'' retrievals for all planets in our sample, using the model setup described in Section \ref{sec:free_chemistry}.} For each target we retrieved molecular abundances, temperature structure, and cloud parameters, and then computed elemental ratios, in particular C/O, from the posterior distributions of the molecular abundances.

\begin{figure}
\centering
	\includegraphics[width=0.8\columnwidth]{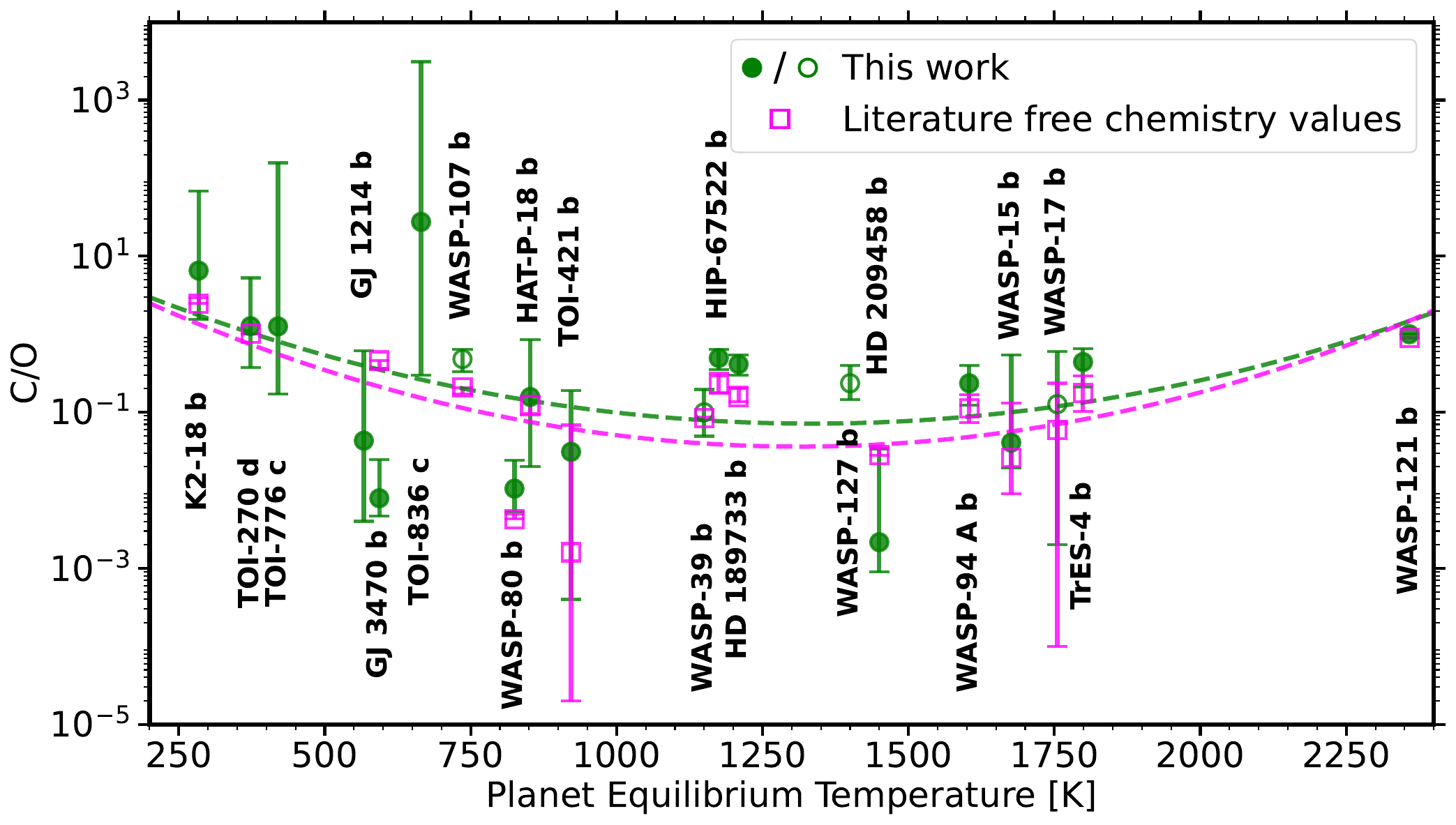}
  \caption{\textcolor{teal}{\textbf{Atmospheric C/O ratios from free-chemistry retrievals as a function of planetary equilibrium temperature.} Our retrieval results are shown as green circles, with solid circles representing planets with reduced $\chi^2_{\rm r} \leq 1.5$ and open circles representing planets with $\chi^2_{\rm r} > 1.5$. The points represent posterior median C/O values, with vertical error bars indicating the 16$\mathrm{^{th}}$--84$\mathrm{^{th}}$ percentile posterior intervals. The sample contains $n=20$ planets. Literature C/O estimates for 16 planets are shown as open magenta squares, calculated using the same elemental C/O definition. Dashed curves are shown only as visual guides. GJ 1214 b, TOI-836 c, TOI-776 c, and WASP-127 b lack literature free-chemistry C/O estimates. The discrepancy for GJ 3470 b is discussed in Supplementary Note \ref{sec:appendixA_free_planets}.}}
   \label{fig:co_vs_Tpl_free}
\end{figure}

\textcolor{black}{Figure \ref{fig:co_vs_Tpl_free} shows the retrieved C/O ratios as a function of $T_{\rm eq}$, together with literature estimates placed on a consistent elemental definition. For our retrievals, C/O was computed from the posterior molecular abundance samples by converting mass fractions to mole fractions and then summing the number of C and O atoms carried by each molecule. For literature comparisons, full molecular posterior samples were generally not available. We therefore computed approximate values from the published median molecular abundances and their reported 1$\sigma$ bounds using the same approach. \textcolor{teal}{The free-chemistry retrievals produce an apparent curved C/O-$T_{\rm eq}$ trend, although the statistical analysis does not support a significant correlation.} For the full 20-planet sample, the Spearman correlation between $\log_{10}({\rm C/O})$ and $T_{\rm eq}$ is not significant (Spearman rank correlation coefficient $\rho=-0.2452$, $p$-value=0.2974). The correlation also remains non-significant when the temperate/warm sub-Neptunes GJ 1214 b, K2-18 b, TOI-270 d, TOI-836 c, TOI-776 c, and TOI-421 b are excluded ($\rho=0.2882$, $p=0.3176$), when WASP-121 b alone is excluded ($\rho=-0.3335$, $p=0.1629$), and when both the sub-Neptune subset and WASP-121 b are excluded ($\rho=0.1100$, $p=0.7204$). Because $T_{\rm eq}$, planet mass ($M_{\rm p}$), and planet radius ($R_{\rm p}$) are strongly correlated in our chosen sample, we also tested whether C/O shows an independent dependence on these quantities. A multivariate regression of $\log_{10}({\rm C/O})$ against $T_{\rm eq}$, $\log M_{\rm p}$, and $\log R_{\rm p}$ does not identify any of them as a statistically significant independent predictor of C/O. We therefore do not interpret the C/O-$T_{\rm eq}$ plane as evidence for an isolated temperature-driven demographic relation. Instead, we use it as a diagnostic of retrieval-model sensitivity and sample composition.}

\begin{figure}
\centering
	\includegraphics[width=0.8\columnwidth]{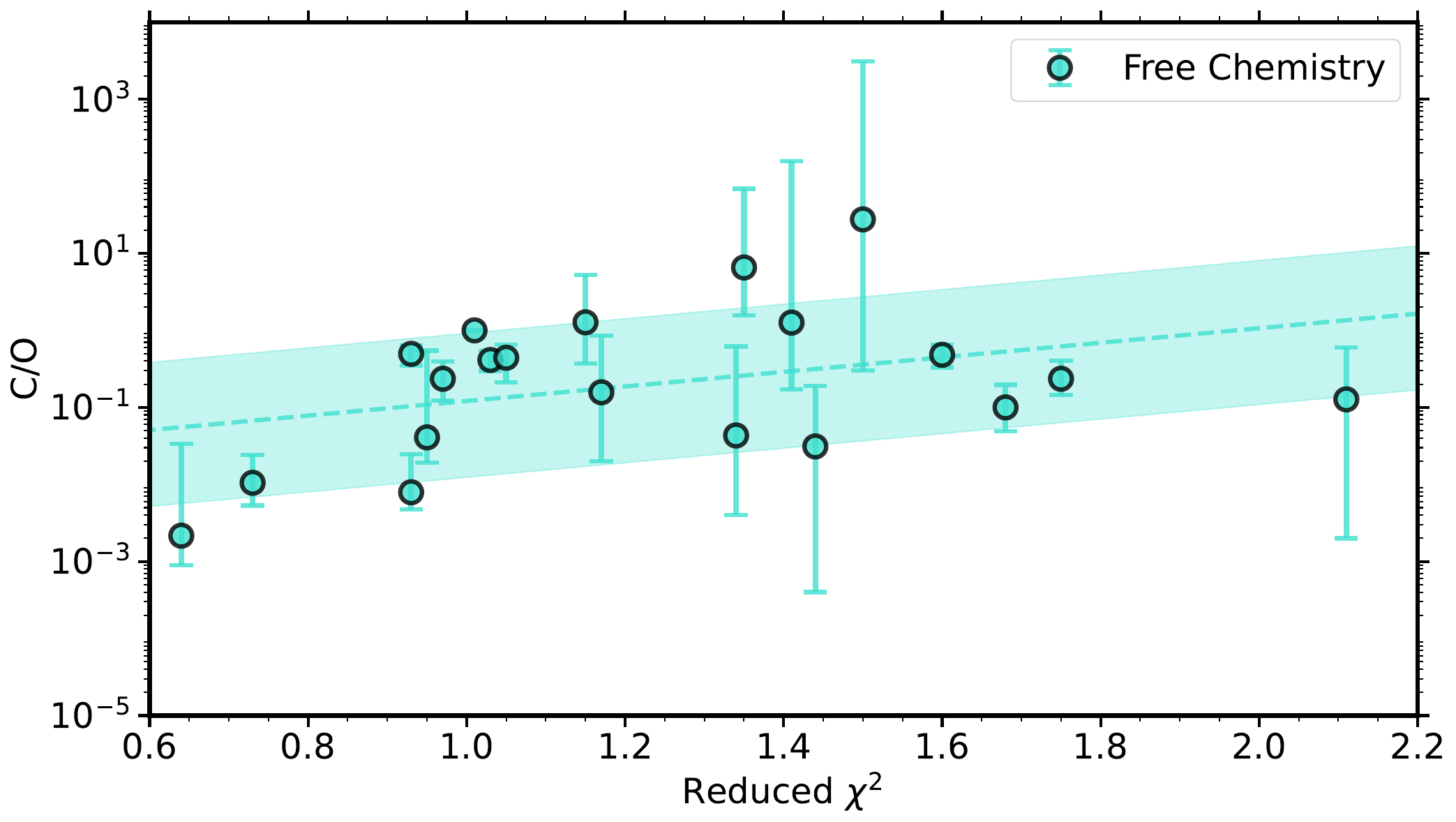}
  \caption{\textcolor{teal}{\textbf{Relationship between atmospheric C/O and reduced $\chi^2$ for free-chemistry retrievals.} The points represent posterior median C/O values from $n=20$ planets, with vertical error bars indicating the 16$\mathrm{^{th}}$--84$\mathrm{^{th}}$ percentile posterior intervals. The dashed line shows a linear fit, with the shaded area indicating its 68\% confidence interval. The correlation was evaluated using a two-sided Spearman rank correlation test ($\rho=0.3220$, $p=0.16$). The reported $p$-value was not adjusted for multiple comparisons.}}
  \label{fig:co_vs_chi2_free}
\end{figure}

\textcolor{black}{However, the free-chemistry framework also yields extremely wide and sometimes nonphysical C/O posteriors, spanning from the order of 10$^{-3}$ for planets such as HD 209458 b and GJ 3470 b to values $>$10 for TOI-836 c, K2-18 b, and TOI-776 c. Such broad or extreme posteriors, especially for smaller or cloudier planets, should not be interpreted as precise abundance constraints without considering the posterior width, prior range, and spectral information content. They indicate weakly constrained or prior-sensitive cases rather than robust data-driven C/O measurements. To assess whether the apparent C/O-$T_{\rm eq}$ relation reflects genuine physics and chemistry or a retrieval artifact, we examined how C/O varies with the approximate fit-quality indicator, reduced chi-squared ($\chi^2_{\rm r}$). We use $\chi^2_{\rm r}$ only descriptively, because the effective number of degrees of freedom is not rigorously defined for nonlinear retrieval models with priors and parameter degeneracies \citep{andrae2010and}.}

\begin{figure}
\centering
	\includegraphics[width=0.9\columnwidth]{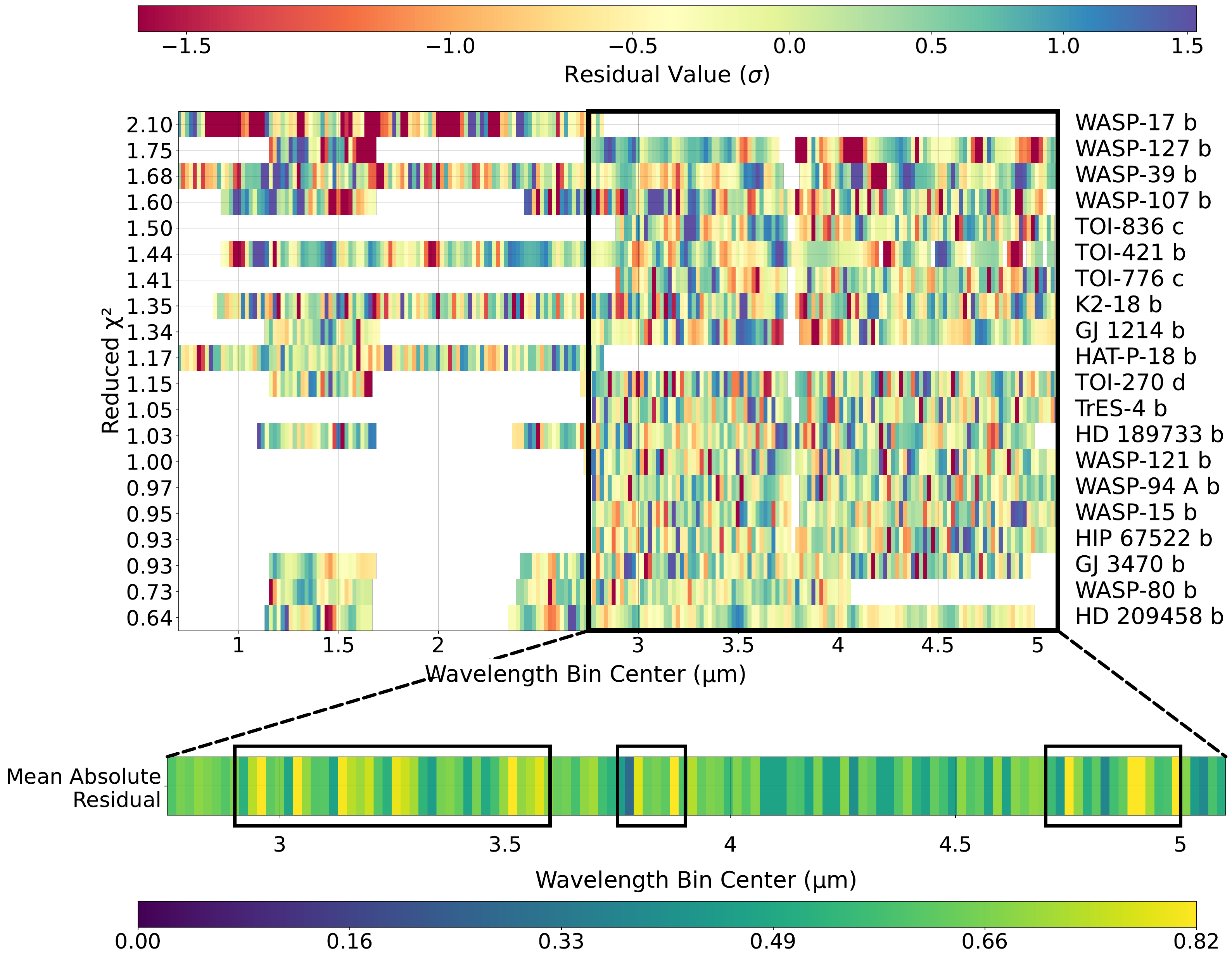}
  \caption{\textbf{Population-level spectral residuals from free-chemistry retrievals.} (Upper panel) Normalized residuals for all planets from the free-chemistry retrievals, arranged vertically by increasing reduced $\chi^2$ from bottom to top. The color scale shows $(D_{\rm obs}-D_{\rm model})/\sigma_{\rm obs}$, where $D_{\rm obs}$ is the observed transit depth, $D_{\rm model}$ is the best-fit model transit depth, and $\sigma_{\rm obs}$ is the observational uncertainty. The residuals are dimensionless and expressed in units of $\sigma$. (Lower panel) Mean absolute normalized residuals over all planets for the 2.75--5.10 $\mu$m range. \textcolor{teal}{The black-outlined boxes mark the wavelength intervals discussed in the text: 2.9--3.6 $\mu$m, 3.75--3.9 $\mu$m, and 4.7--5.0 $\mu$m. The corresponding mean absolute residuals are 0.660$\sigma$, 0.609$\sigma$, and 0.674$\sigma$, respectively, compared with a common mean of 0.594$\sigma$ for the 60 wavelength bins outside all three boxes. All three intervals have higher mean absolute residuals than this common reference, with the smallest difference in the 3.75--3.9 $\mu$m interval.} \textcolor{teal}{The statistical comparisons between these intervals and the common outside region, along with their limitations, are discussed in Section~\ref{result:free_chemistry}.}}
   \label{fig:residuals}
\end{figure}

\begin{figure}
\centering
	\includegraphics[width=0.9\columnwidth]{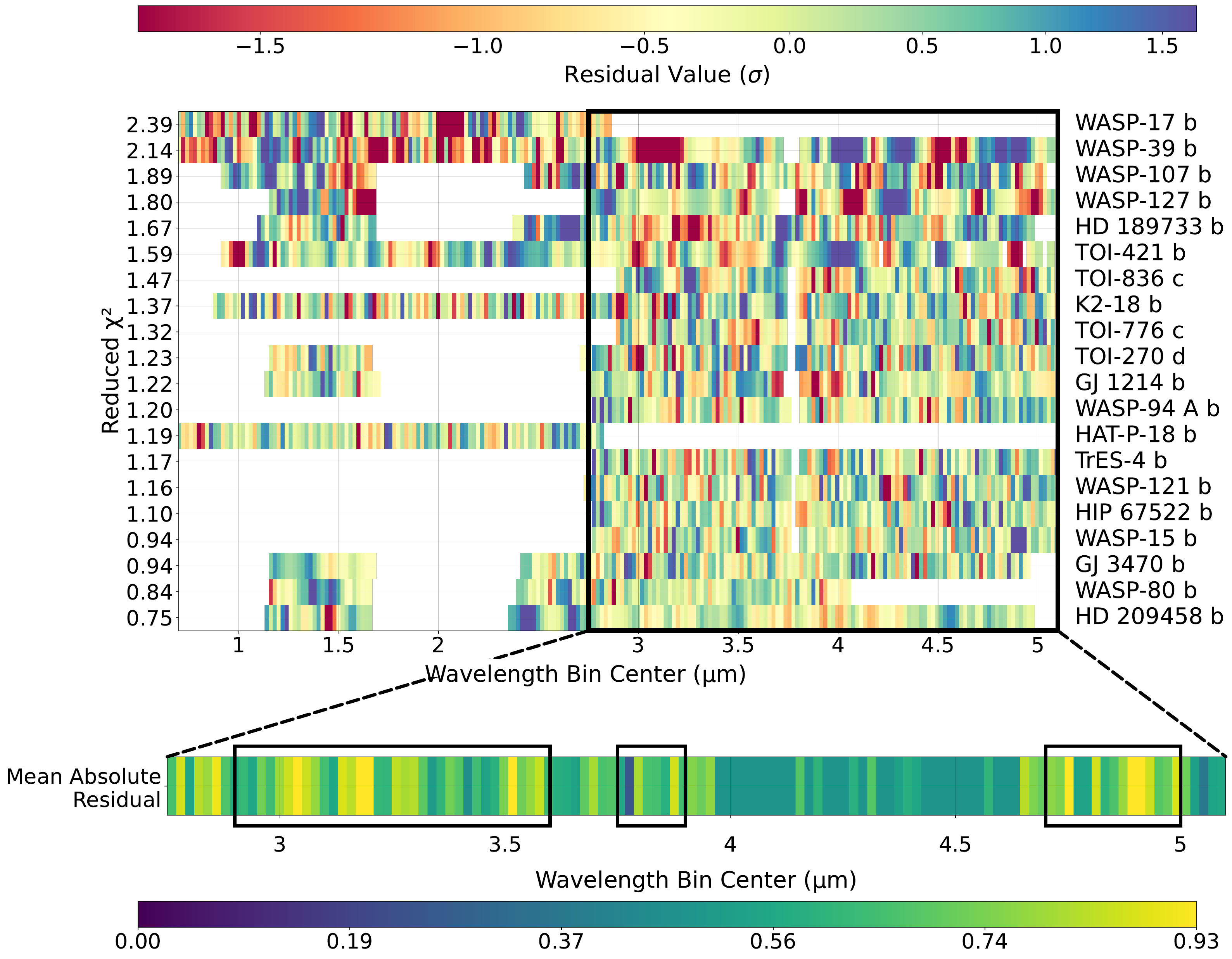}
  \caption{\textbf{Population-level spectral residuals from VULCAN disequilibrium-grid retrievals.} \textcolor{black}{Same residual construction as in Figure \ref{fig:residuals}, but using the best-fit spectra from the chemically constrained VULCAN grid-chemistry retrievals. Planets are ordered by their grid-retrieval reduced $\chi^2$. The mid-infrared residual structures identified in the free-chemistry retrievals are not uniformly removed by the disequilibrium-grid retrievals. This reflects the reduced flexibility of the grid-chemistry framework, in which molecular abundances are linked through pressure-dependent chemical profiles rather than varied independently.}}
  \label{fig:vulcan_residuals}
\end{figure}

Figure \ref{fig:co_vs_chi2_free} shows $\mathrm{\chi^2_r}$ versus C/O for the free-chemistry retrievals. We find a weak positive monotonic trend between C/O and $\chi^2_{\mathrm{r}}$ (Spearman rank correlation coefficient $\rho$ = 0.3220, $p$-value = 0.16): planets with poorer fits tend to have higher inferred C/O. \textcolor{black}{Since this correlation is not significant at the 5\% level, and since $\chi^2_{\rm r}$ is only an approximate fit-quality indicator for nonlinear retrievals, we do not treat this relation as proof of a statistically established bias. Instead, we use it only as a qualitative indication that some high-C/O solutions may arise when the free-chemistry model has difficulty matching the data.}

To localize the spectral regions responsible for this behavior, we constructed a population-level map of residuals between the best-fit free-chemistry models and the observed spectra. For each planet, residuals were interpolated onto a common 0.02 $\mu$m grid, and only points within 0.04 $\mu$m of original measurements were retained. The resulting two-dimensional residual matrix (as shown in Figure \ref{fig:residuals} upper panel) reveals localized regions of model-data mismatch across several planets. \textcolor{teal}{The lower panel of Figure \ref{fig:residuals} shows the mean absolute normalized residuals over 2.75--5.10 $\mu$m. We calculated the mean residual within each highlighted interval and compared it with a common outside mean obtained by excluding all three boxed regions. The four remaining wavelength intervals, 2.75--2.90, 3.60--3.75, 3.90--4.70, and 5.00--5.10 $\mu$m, have mean absolute residuals of 0.644$\sigma$, 0.624$\sigma$, 0.587$\sigma$, and 0.526$\sigma$, respectively. Together, these 60 wavelength bins give a common outside mean of 0.594$\sigma$ for the displayed lower-panel values.} The highlighted 2.9--3.6 $\mu$m region contains strong opacity contributions from H$_2$O, the CH$_4$ band near 3.3 $\mu$m, and additional carbon-bearing species such as C$_2$H$_2$ and other hydrocarbons. \textcolor{teal}{Its mean absolute residual is 0.660$\sigma$, approximately 1.11 times the common outside mean. The 4.7--5.0 $\mu$m region, which overlaps with the CO fundamental band and nearby opacity wings, has a mean absolute residual of 0.674$\sigma$, approximately 1.13 times the outside mean. The 3.75--3.9 $\mu$m interval has a mean absolute residual of 0.609$\sigma$, approximately 1.03 times the outside mean, although this value is slightly lower than that of the adjacent 3.60--3.75 $\mu$m interval.} This residual structure suggests that opacity overlap, continuum placement, and simplified chemistry or temperature structure may contribute to the apparent demographic behavior inferred from free-chemistry retrievals.

\textcolor{teal}{To assess these differences statistically, we performed one-sided Mann--Whitney U tests by comparing each highlighted interval with the same common outside region. The resulting $p$-values are 0.0011, 0.40, and 0.19 for the 2.9--3.6, 3.75--3.9, and 4.7--5.0 $\mu$m intervals, respectively. The corresponding bootstrap 95\% confidence intervals for the differences in mean absolute residuals are $[0.024,0.108]\sigma$, $[-0.105,0.124]\sigma$, and $[-0.017,0.187]\sigma$. Only the 2.9--3.6 $\mu$m interval shows a statistically significant difference under these tests. However, the statistical significance may be affected by correlations between adjacent wavelength bins, which are treated as independent in this analysis. Splitting the 2.9--3.6 $\mu$m interval into 2.9--3.25 and 3.25--3.6 $\mu$m gives nearly identical mean absolute residuals of 0.660$\sigma$ in both regions, indicating that neither sub-interval alone accounts for the observed difference.} 

\textcolor{black}{To test whether these residual structures are reduced when the abundances are constrained by chemical kinetics, we repeated the same residual matrix calculation for the VULCAN disequilibrium-grid retrievals (Figure \ref{fig:vulcan_residuals}). The mid-infrared residual structure is not uniformly removed in the grid-chemistry retrievals. Instead, the residuals remain comparable, and in some wavelength regions become slightly larger, because the disequilibrium-grid retrievals impose chemically linked, pressure-dependent abundance profiles and therefore have less freedom to adjust individual molecular abundances independently. Over the 2.75-5.10 $\mu$m window, the mean absolute residual is slightly higher for the grid-chemistry retrievals than for the free-chemistry retrievals. The 4.7-5.0 $\mu$m interval also remains mildly elevated relative to the surrounding wavelengths. We therefore interpret the residual comparison as evidence that the chemically constrained retrievals change the abundance interpretation, but do not necessarily remove all local spectral mismatches.}

\textcolor{black}{Taken together, the C/O-$T_{\rm eq}$ sensitivity tests, the weak C/O-$\chi^2_{\rm r}$ diagnostic, and the residual maps show that the apparent free-chemistry demographic structure is not robust enough to be interpreted as a confirmed formation trend. Instead, the C/O-$T_{\rm eq}$ plane provides a useful test for retrieval assumptions. Addressing these systematic effects will be essential before drawing robust conclusions about atmospheric elemental ratios and their implications for planet formation.}


\subsection{Physical interpretation of C/O ratio from grid-chemistry retrieval}
\label{result:grid_chemistry}

\textcolor{black}{To test how sensitive the inferred population-level C/O distribution is to chemical parameterization, we reanalyzed the full sample with grid-chemistry retrievals that tie the molecular abundances to pre-computed VULCAN disequilibrium models (see Section \ref{sec:grid_chemistry} for details).} The chemical grid spans different C/O, metallicity [M/H], equilibrium temperature, and Eddy diffusion coefficient ($\mathrm{K_{zz}}$), and is interpolated during the retrieval to provide self-consistent vertical abundance profiles for all major species.

\begin{figure}
\centering
	\includegraphics[width=0.9\columnwidth]{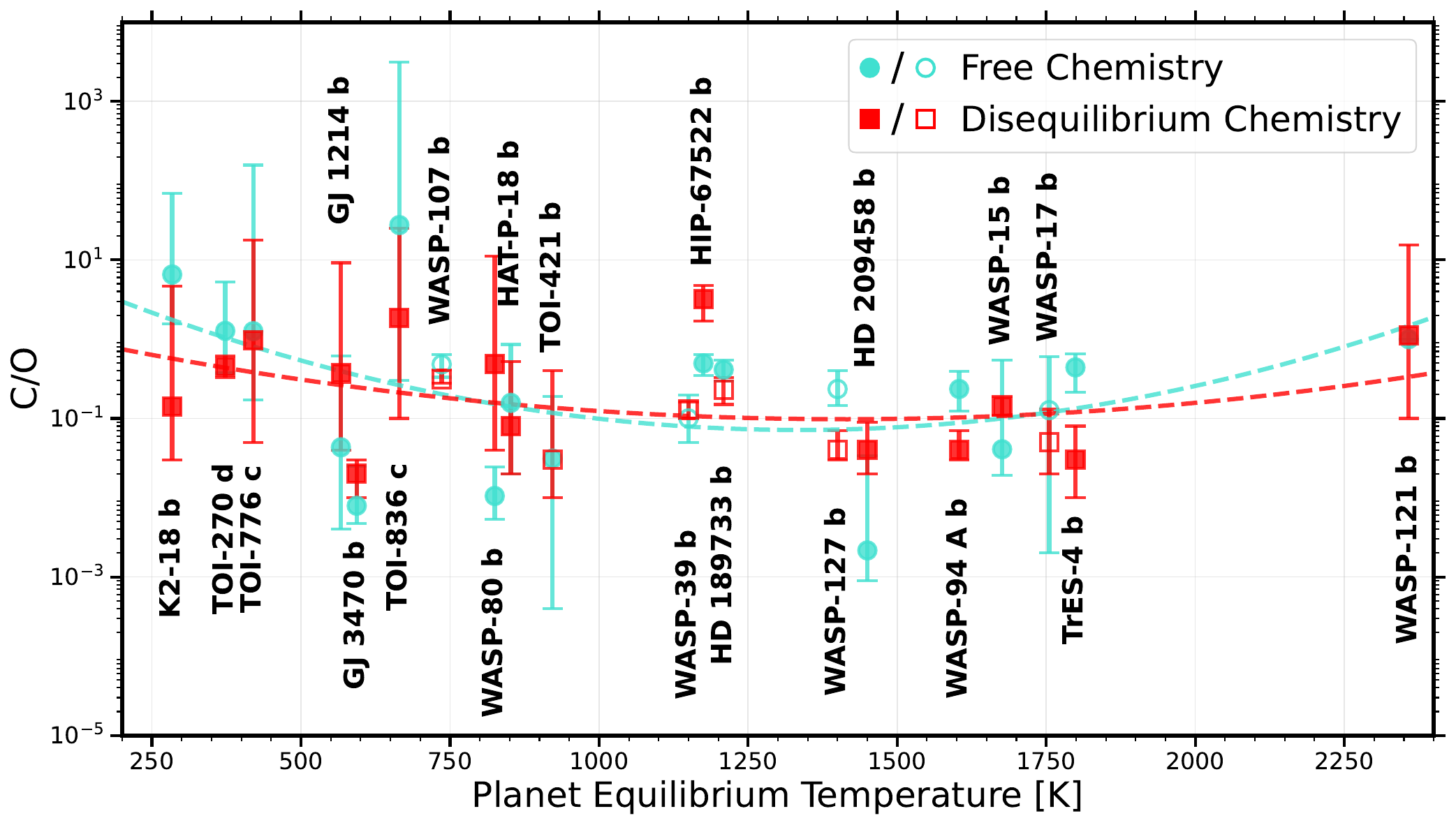}
  \caption{\textcolor{teal}{\textbf{Comparison of atmospheric C/O ratios across planetary equilibrium temperatures for free-chemistry and grid-chemistry retrievals.} Free-chemistry results are shown as turquoise circles, while grid-chemistry results are shown as red squares. For each retrieval framework, solid symbols represent planets with reduced $\chi^2_{\rm r} \leq 1.5$, while open symbols represent planets with $\chi^2_{\rm r} > 1.5$. The comparison includes $n=20$ planets, each analyzed using both retrieval frameworks. The points represent posterior median C/O values, with vertical error bars indicating the corresponding 16$\mathrm{^{th}}$--84$\mathrm{^{th}}$ percentile posterior intervals. Dashed curves are shown only as visual guides.}}
   \label{fig:free_grid_chem_a}
\end{figure}

\begin{figure}
\centering
	\includegraphics[width=0.9\columnwidth]{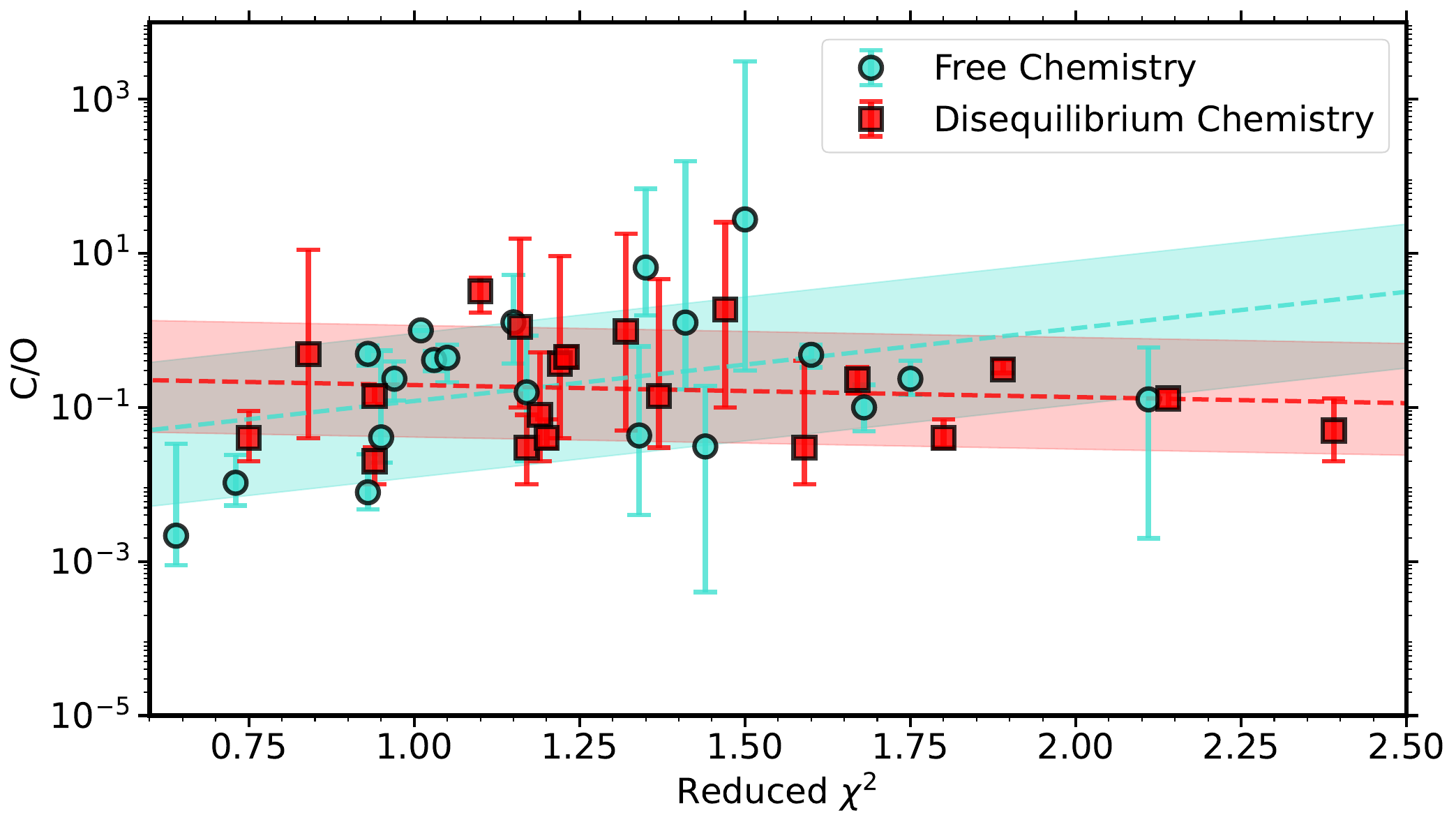}
  \caption{\textcolor{teal}{\textbf{Relationship between atmospheric C/O and reduced $\chi^2$ across retrieval frameworks.} Free-chemistry results are shown as turquoise circles, while grid-chemistry results are shown as red squares. The comparison includes $n=20$ planets, each analyzed using both retrieval frameworks. The points represent posterior median C/O values, with vertical error bars indicating the corresponding 16$\mathrm{^{th}}$--84$\mathrm{^{th}}$ percentile posterior intervals. The dotted lines show linear fits, with shaded bands indicating their 68\% confidence intervals. Two-sided Spearman rank correlation tests give $\rho=0.3220$ ($p=0.16$) for free chemistry and $\rho=-0.026$ ($p=0.91$) for grid chemistry. The reported $p$-values are unadjusted for these two framework-specific comparisons.}}
   \label{fig:free_grid_chem_b}
\end{figure}

\textcolor{black}{\textcolor{black}{On the population level, the chemically constrained retrievals yield a more restricted range of C/O values than the free-chemistry retrievals. However, the apparent free-chemistry C/O-$T_{\rm eq}$ structure is not statistically required by the data. To quantify this, we compared constant, linear, and quadratic descriptions of $\log_{10}({\rm C/O})$ as a function of $T_{\rm eq}$ using the Bayesian Information Criterion (BIC). For the full free-chemistry sample, the constant model has the lowest BIC (${\rm BIC}_{\rm constant}=1.3276$, ${\rm BIC}_{\rm linear}=3.6575$, ${\rm BIC}_{\rm quadratic}=2.3936$), and the same conclusion holds for the sub-samples excluding the temperate/warm sub-Neptunes and ultra hot-Jupiter WASP-121 b. We therefore treat the dashed curves in Figure \ref{fig:free_grid_chem_a} as visual guides rather than as statistically preferred demographic relations. The corresponding comparison with fit quality is shown in Figure \ref{fig:free_grid_chem_b}. While the free-chemistry results exhibit a weak positive relation between C/O and $\chi^2_{\rm r}$, the grid-chemistry retrievals show no monotonic relationship ($\rho=-0.026$, $p=0.91$). Since the free-chemistry C/O-$\chi^2_{\rm r}$ relation is not statistically significant, we interpret this as retrieval-model sensitivity rather than as proof that chemical self-consistency alone removes a statistically established bias.}}

\textcolor{black}{We quantified the relative support for the two retrieval frameworks using Bayesian evidences computed by MultiNest. For each planet we calculate the Bayes factor (K) (see Equation \ref{eq:bayes} from Section \ref{sec:grid_chemistry}) and interpret it following \cite{kass1995bayes}. Because the free-chemistry and grid-chemistry retrievals have different parameterizations, we interpret these Bayes factors as relative model-comparison diagnostics under the adopted priors, not as proof that either framework uniquely recovers the true atmospheric composition. As summarized in Table \ref{tab:bayes_factor}, 3 out of 20 planets show strong Bayesian preference ($K>10$) for the grid-chemistry model, while 2 systems show moderate preference ($3.2<K<10$) and another 3 show weak preference ($1<K<3.2$). For the remaining 12 planets, the free-chemistry model is favored ($K<1$) under the adopted priors, plausibly because its independent abundance parameters provide greater flexibility to absorb residual model-data mismatch.}
\begin{table}
\caption{\textbf{Bayesian evidence and atmospheric C/O ratios across retrieval frameworks.} For each planet, we report the Bayesian evidences ($\ln Z_{\mathrm{grid}}$ and $\ln Z_{\mathrm{free}}$) computed with MultiNest, along with the corresponding Bayes factor ($K$). Following \citet{kass1995bayes}, $K>10$ indicates strong preference for the grid-chemistry model, $3.2<K<10$ indicates moderate preference, and $1<K<3.2$ indicates weak preference. Values of $K<1$ favor the free-chemistry model. These values quantify the relative statistical support for the two retrieval frameworks for each planet.}
\label{tab:bayes_factor}%
\begin{tabular}{@{}lccccc@{}}

\toprule
& \multicolumn{2}{@{}c@{}}{Bayesian Evidence (ln Z)} & Bayes Factor (K) & \multicolumn{2}{@{}c@{}}{C/O}     \\
\cmidrule{2-3}\cmidrule{5-6}
Planet & Grid Retrieval & Free Retrieval   & $\mathrm{K} \ \text{=} \ \mathrm{e^{\left(ln \ Z_{grid} \ -\  ln \ Z_{free}\right)}}$ & Free Retrieval & Grid Retrieval\\
\midrule
HAT-P-18 b  & 6963.8 $\pm$ 0.0 & 6960.2 $\pm$ 0.1 & 36.60 $\pm$ 3.66 & 	0.157$\mathrm{^{\text{+} 0.701}_{- 0.137}}$ & 0.08$\mathrm{^{\text{+} 0.44}_{- 0.06}}$\\
\noalign{\smallskip}
WASP-17 b   & \textcolor{black}{1624.6} $\pm$ \textcolor{black}{0.1} & \textcolor{black}{1657.9} $\pm$ \textcolor{black}{0.0} & \textcolor{black}{3.46} $\times$ \textcolor{black}{$\mathrm{10^{-15}}$} $\pm$ \textcolor{black}{3.46} $\times$ \textcolor{black}{$\mathrm{10^{-16}}$} & \textcolor{black}{0.127$^{+0.476}_{-0.125}$} & 0.05$\mathrm{^{\text{+}0.08 }_{- 0.03}}$\\
\noalign{\smallskip}
GJ 3470 b   & 1537.9 $\pm$ 0.1 & 1536.8 $\pm$ 0.0 & 3.00 $\pm$ 0.30 & 0.008$\mathrm{^{\text{+} 0.017}_{- 0.003}}$ & 0.02$\mathrm{^{\text{+} 0.01}_{- 0.01}}$\\
\noalign{\smallskip}
GJ 1214 b   & 760.2 $\pm$ 0.6  & 758.2  $\pm$ 0.9 & 7.39 $\pm$ 7.99 & 0.004$\mathrm{^{\text{+}0.575 }_{- 0.004}}$ & 0.37$\mathrm{^{\text{+}8.83 }_{- 0.33}}$\\
\noalign{\smallskip}
K2-18 b     & 2971.3 $\pm$ 0.4 & 2969.5 $\pm$ 0.1  & 6.05 $\pm$ 2.49 & 6.540$\mathrm{^{\text{+} 62.2}_{- 4.99}}$ & 0.14$\mathrm{^{\text{+}4.49 }_{- 0.11}}$\\
\noalign{\smallskip}
TOI-270 d   & 1571.5 $\pm$ 0.0 & 1584.3 $\pm$ 0.0 & 2.76 $\times$ $\mathrm{10^{-6}}$ $\pm$ 0.00 & 1.270$\mathrm{^{\text{+} 4.000}_{- 0.897}}$ & 0.45$\mathrm{^{\text{+}0.16 }_{- 0.12}}$\\
\noalign{\smallskip}
HD 189733 b & 1404.2 $\pm$ 0.3 & 1454.3 $\pm$ 0.0 & 1.74 $\times$ $\mathrm{10^{-22}}$ $\pm$ 5.22 $\times$ $\mathrm{10^{-23}}$  & 0.412$\mathrm{^{\text{+} 0.127}_{- 0.114}}$ & 0.23$\mathrm{^{\text{+}0.10 }_{- 0.08}}$\\
\noalign{\smallskip}
HD 209458 b & 816.4 $\pm$ 0.1 &  820.8 $\pm$ 0.1 & 0.0123 $\pm$ 0.0017 & 0.002$\mathrm{^{\text{+} 0.003}_{- 0.001}}$ & 0.04$\mathrm{^{\text{+}0.05 }_{- 0.02}}$\\
\noalign{\smallskip}
HIP 67522 b & 1555.5 $\pm$ 0.0 & 1586.3 $\pm$ 0.0 & 4.20 $\times$ $\mathrm{10^{-14}}$ $\pm$ 0.00 & 0.495$\mathrm{^{\text{+} 0.147}_{- 0.145}}$ & 3.22$\mathrm{^{\text{+} 1.56}_{- 1.53}}$\\
\noalign{\smallskip}
WASP-80 b   & 990.6 $\pm$ 0.0  &   994.4 $\pm$ 0.2 & 0.0224 $\pm$ 0.0045 & 0.011$\mathrm{^{\text{+} 0.014}_{- 0.005}}$ & 0.49$\mathrm{^{\text{+} 10.65}_{-0.45 }}$\\
\noalign{\smallskip}
WASP-39 b  & 1494.8 $\pm$ 0.1  & 1602.6 $\pm$ 0.0 & 1.52 $\times$ $\mathrm{10^{-47}}$ $\pm$  1.52 $\times$ $\mathrm{10^{-48}}$  & 0.100$\mathrm{^{\text{+}0.096 }_{- 0.051}}$ & 0.13$\mathrm{^{\text{+}0.03 }_{- 0.03}}$\\
\noalign{\smallskip}
WASP-107 b  & 1538.6 $\pm$ 0.0 & 1537.7 $\pm$ 0.6 & 2.46 $\pm$ 1.48 & 0.481$\mathrm{^{\text{+} 0.162}_{- 0.150}}$ & 0.31$\mathrm{^{\text{+}0.10 }_{-0.03 }}$\\
\noalign{\smallskip}
WASP-15 b   & 1825.8 $\pm$ 0.1 &  1825.2 $\pm$ 0.5 & 1.82 $\pm$ 0.93 & 0.041$\mathrm{^{\text{+}0.503 }_{-0.021 }}$ & 0.14$\mathrm{^{\text{+} 0.05}_{- 0.03}}$\\
\noalign{\smallskip}
TOI-836 c   & 902.8 $\pm$ 0.2 & 905.4  $\pm$ 0.4 & 0.07 $\pm$ 0.03 & 27.5$\mathrm{^{\text{+}3.08\times 10^3}_{- 27.2}}$ & 1.86$\mathrm{^{\text{+}23.39 }_{- 1.76}}$\\
\noalign{\smallskip}
TOI-776 c   & 956.2 $\pm$ 0.4 & 953.8   $\pm$ 0.5 & 11.02 $\pm$ 7.06 & 1.26$\mathrm{^{\text{+} 1.56\times 10^2}_{- 1.09}}$ & 0.97$\mathrm{^{\text{+} 16.84}_{-0.92 }}$ \\
\noalign{\smallskip}
TrES-4 b    & 1719.2 $\pm$ 0.1 & 1731.4  $\pm$ 0.1 & 5.03 $\times$ $\mathrm{10^{-6}}$ $\pm$  7.11 $\times$ $\mathrm{10^{-7}}$ & 0.440$\mathrm{^{\text{+} 0.215}_{- 0.228}}$ & 0.03$\mathrm{^{\text{+}0.05 }_{-0.02 }}$\\
\noalign{\smallskip}
TOI-421 b   & 578.8 $\pm$ 0.1 & 576.3  $\pm$ 0.1 & 12.18 $\pm$ 1.72 & 0.031$\mathrm{^{\text{+} 0.158}_{- 0.030}}$ & 0.03$\mathrm{^{\text{+}0.37 }_{- 0.02}}$\\
\noalign{\smallskip}
WASP-94 A b & 1883.6 $\pm$ 0.0 & 1921.1 $\pm$ 0.0 & 5.17 $\times$ $\mathrm{10^{-17}}$ $\pm$ 0.00 & 0.235$\mathrm{^{\text{+} 0.160}_{-0.112 }}$ & 0.04$\mathrm{^{\text{+}0.03 }_{- 0.01}}$\\
\noalign{\smallskip}
WASP-121 b  & \textcolor{black}{2435.5 $\pm$ 0.0} & \textcolor{black}{2459.6 $\pm$ 0.1} & 3.42 $\times$ $\mathrm{10^{-11}}$ $\pm$ 0.34 $\times$ $\mathrm{10^{-11}}$ & \textcolor{black}{0.999$\mathrm{^{\text{+} 4.60\times 10^{-4}}_{- 1.01 \times 10^{-3}}}$} & \textcolor{black}{1.11$\mathrm{^{\text{+}14.42 }_{-1.04 }}$}\\
\noalign{\smallskip}
WASP-127 b  & 524.2 $\pm$ 0.0 & 537.5 $\pm$ 0.0 & 1.67 $\times$ $\mathrm{10^{-6}}$ $\pm$ 0.00 & 0.235$\mathrm{^{\text{+} 0.165}_{-0.090 }}$ & 0.04$\mathrm{^{\text{+} 0.03}_{- 0.03}}$\\
\botrule
\end{tabular}
\end{table}


Overall, the grid-chemistry analysis shows that the inferred population-level C/O distribution depends strongly on the chemical parameterization adopted in the retrieval. The chemically constrained framework restricts the explored abundance combinations to those allowed by the kinetic grid and yields a narrower range of C/O values for many planets.


\subsection{Implications for clouds and sulfur chemistry}
\label{result:clouds_sulfur}

\begin{figure*}[htbp]
    \centering
    \includegraphics[width=0.9\textwidth]{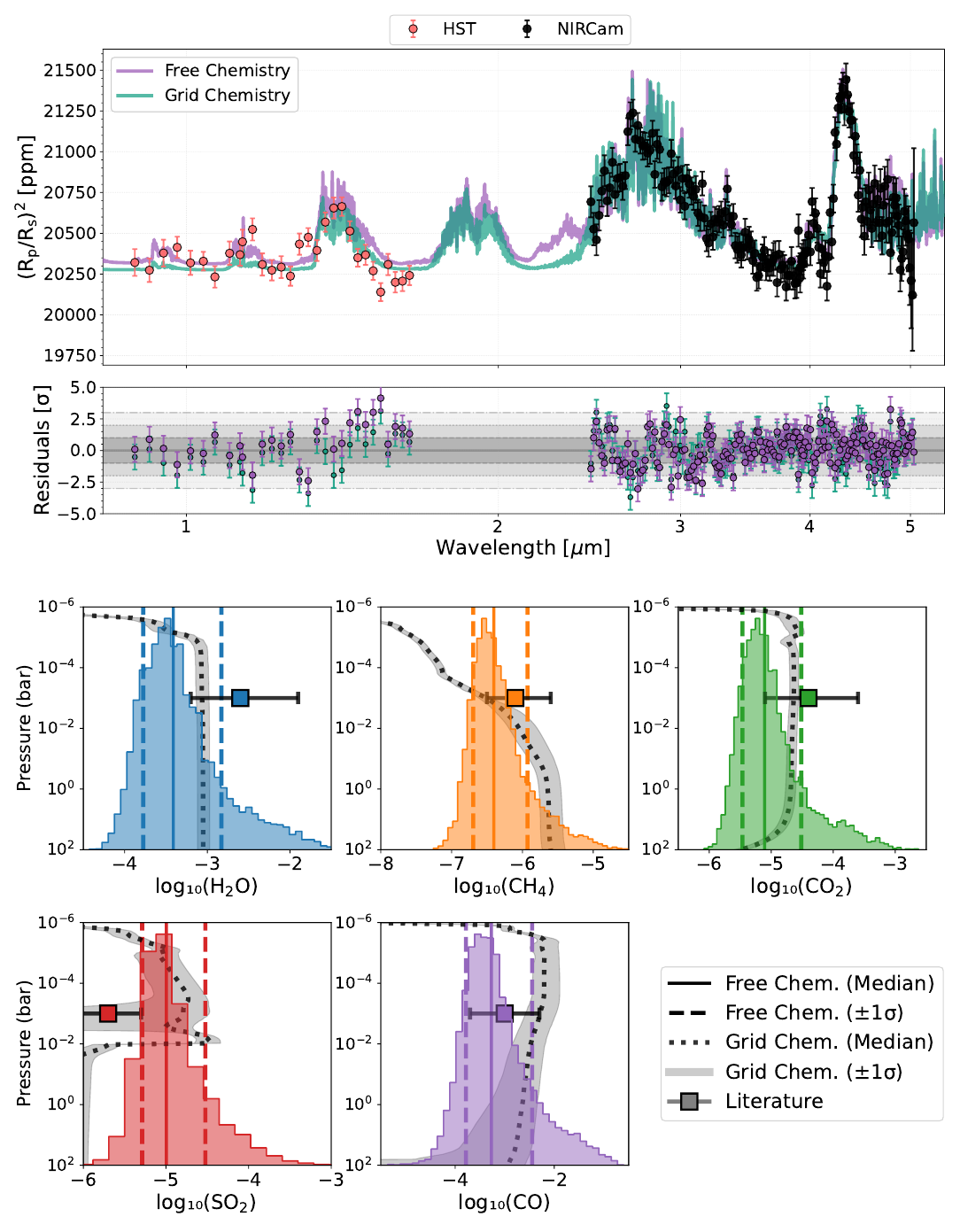}
    \caption{\textbf{Comparison of free-chemistry and grid-chemistry retrievals for WASP-107 b.} (Upper Panel) Best-fit model and residual comparison for WASP-107 b from free-chemistry (purple) and grid-chemistry (teal) retrieval approach. The HST and NIRCam data from the observation are shown in red and black respectively. (Lower Panel) Retrieved atmospheric composition of WASP-107 b from its transmission spectrum. The colored posterior distributions represent the inferred mass mixing ratios of the detected molecular species obtained from our free-chemistry retrieval, with the median values (solid colored line) and associated 68\% confidence intervals (dashed colored lines). For comparison, the dotted black profiles correspond to the predictions from the 1D-RCPE grid-based retrieval, showing their median and 68\% confidence ranges. Overall, the chemically self-consistent 1D-RCPE abundances agree with the free-chemistry estimates to within 1$\sigma$ for all major species. Colored square shaped markers mark the free-chemistry abundances reported by \protect\cite{welbanks2024high} and are shown here for reference.}
    \label{fig:W107b_best_fit}
\end{figure*}

\begin{figure*}[htbp]
    \centering
    \includegraphics[
        width=\textwidth,
        keepaspectratio
    ]{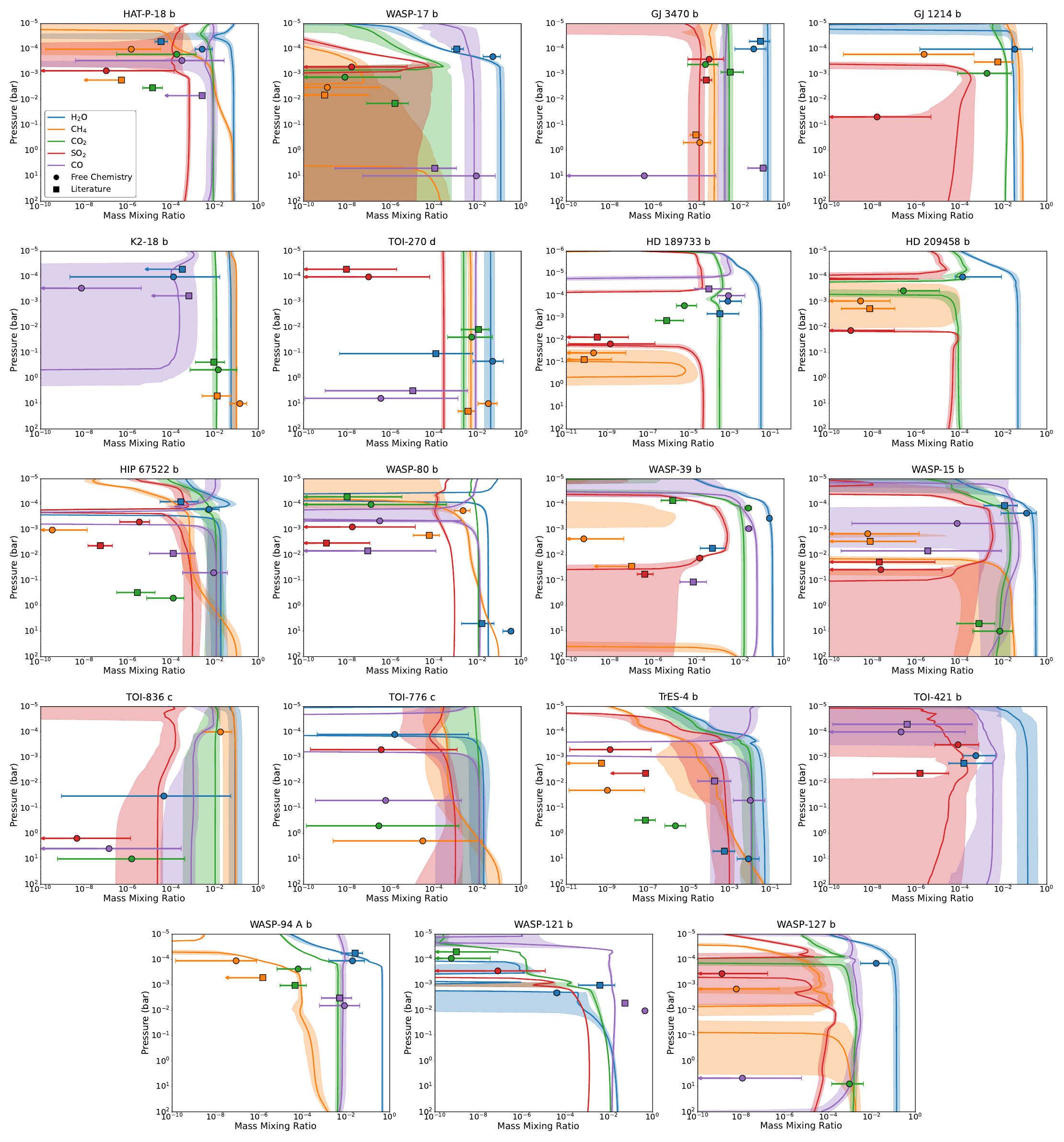}
    \caption{\textbf{Comparison of molecular abundance profiles from free-chemistry and VULCAN grid-chemistry retrievals.} Colored circular points with 1$\sigma$ error bars show the abundances retrieved in our free-chemistry analysis. These are compared with the corresponding abundance profiles predicted by the best-fit VULCAN self-consistent grid model, plotted as solid lines in matching colors. The shaded envelopes around each profile indicate the range of grid-model abundances within the 1$\sigma$ confidence interval of the grid fit. Colored square symbols denote the free-chemistry abundances from the literature, tabulated in Supplementary Table \ref{tab:molecule_abundance} for reference. We only focus on $\mathrm{H_2O}$, $\mathrm{CH_4}$, $\mathrm{CO_2}$,$\mathrm{SO_2}$ and CO in this plot, as these species produce the strongest spectral signatures and are the most robustly detected in free-retrieval analyses across the planets. \textcolor{black}{The pressure positions of the free-chemistry markers are chosen only for visual clarity and do not represent retrieved pressure levels, since the free-chemistry retrievals assume vertically constant abundances.}}
    \label{fig:chemisrty_extended}
\end{figure*}

Beyond C/O, the dual retrieval framework provides insight into cloud structures and reference pressure at the population level.  In free-chemistry retrievals, these parameters primarily adjust the spectral continuum because abundances do not vary with altitude. In the grid-chemistry retrievals, by contrast, these pressures are coupled to chemically varying profiles, so shifting them alters which parts of the vertical structure contribute most strongly to the transmission spectrum. As a result, the two frameworks often favor cloud and reference pressures that differ by several scale heights, even when they yield similar fit quality. This systematic difference suggests that free-chemistry models can compensate for missing or simplified chemistry by adjusting continuum parameters, whereas chemically informed retrievals attribute the same spectral features to changes in vertical composition. \textcolor{black}{This behavior is illustrated by WASP-107 b, for which the free- and grid-chemistry fits give similar spectra but the grid retrieval narrows the abundance posteriors (Figure \ref{fig:W107b_best_fit} and Supplementary FIgures \ref{fig:W107b_corner_free}-\ref{fig:W107b_corner_vulcan}), and by GJ 3470 b and WASP-39 b, where the grid retrieval maps SO$_2$-bearing spectra onto pressure-dependent abundance profiles rather than vertically constant abundances (Figure \ref{fig:chemisrty_extended}). Similar comparisons between flexible free retrievals and chemically consistent or microphysical grid models have been reported for WASP-107 b, GJ 3470 b, WASP-80 b, GJ 1214 b, and GJ 3090 b \citep{welbanks2024high,beatty2024sulfur,wiser2025precise,ohno2025possible,ahrer2025escaping}.}

The sulfur chemistry further illustrates the impact of a physically motivated prior. For planets with robust $\mathrm{SO_2}$ detections in JWST spectra (e.g. WASP-39 b, WASP-107 b, and GJ 3470 b), the grid-chemistry retrievals recover $\mathrm{SO_2}$ detections that are broadly consistent with the free-chemistry results and with dedicated photochemical studies, while placing tighter constraints on the corresponding C/O and [M/H]. At the same time, the grid-based models do not support some of the more tentative sulfur-bearing identifications suggested by free-chemistry retrievals (such as $\mathrm{H_2S}$ in HD 189733 b by \citet{fu2024hydrogen} and $\mathrm{CS_2}$ in TOI-270 d by \citet{holmberg2024possible}) which indicates that these features may be sensitive to degeneracies in the flexible parameterization. A planet-by-planet comparison of vertical mixing-ratio profiles for $\mathrm{H_2O}$, $\mathrm{CH_4}$, CO, $\mathrm{CO_2}$, and $\mathrm{SO_2}$ is provided in Figure \ref{fig:chemisrty_extended}.

\clearpage

\section{Discussion}
\label{sec:conclusion}

\textcolor{black}{This work presents a population-level retrieval analysis of atmospheric elemental ratios for a diverse sample of transiting planets selected from published JWST transmission observations, with HST spectra incorporated for targets where archival data are available. The sample spans gas giants, warm Neptunes, and temperate/warm sub-Neptunes.} Our aim is to test how sensitive population-level chemical inferences are to retrieval assumptions, sample composition, and chemical parameterization. The increasing precision of JWST spectra has made these questions timely, but it has also exposed limitations in atmospheric models and opacity data \citep{barstow2020outstanding,Rotman2025,thorngren2025bayesian,niraula2025detection}. The temperate/warm sub-Neptune part of the sample requires particular attention. For these planets, the H$_2$/He-rich interpretation is more model-dependent than for gas giants, and clouds, higher mean molecular weight compositions, weak molecular features, or limited wavelength coverage can strongly affect retrieved elemental ratios. Chemically consistent or disequilibrium modeling is therefore expected to be especially important in this cooler and smaller part of the sample, although such modeling is not exclusive to sub-Neptunes.

\textcolor{teal}{Our free-chemistry retrievals show a visually non-monotonic C/O-$T_{\rm eq}$ correlation, but this apparent behavior is not statistically significant in our sample.} However, we find a Spearman rank correlation coefficient $\rho$ = 0.3220 ($p$-value = 0.16) between C/O and $\mathrm{\chi^2_r}$: planets with poorer model fits tend to show higher inferred C/O values. \textcolor{teal}{The residual maps show higher mean absolute residuals in all three highlighted intervals (2.9--3.6, 3.75--3.9, and 4.7--5.0 $\mu$m) compared with the common outside region. However, only the 2.9--3.6 $\mu$m interval shows a statistically significant difference in the exploratory bin-level tests. These intervals overlap with H$_2$O, CH$_4$, CO, C$_2$H$_2$, and other opacity features, where opacity overlap and model simplifications can affect the inferred abundances.} Together, these findings show that residual systematics and incomplete opacity treatments, coupled with the flexibility of free-chemistry models, can make the exoplanet demographics fragile. \textcolor{black}{As a result, it can produce extreme or weakly constrained C/O values that may give the visual impression of demographic structure, even when the statistical support for a temperature-dependent trend is weak.} \textcolor{black}{For cloudy or low-information spectra, we therefore interpret C/O medians only together with their posterior widths, since broad posteriors can remain prior-sensitive.}

\textcolor{black}{The grid-chemistry retrievals provide a chemically informed comparison case. By restricting the explored compositions to those that arise from the adopted chemical-kinetics grid, the chemically constrained approach yields a different and often narrower population-level C/O distribution. This narrowing is expected, because the grid-chemistry retrieval has less chemical flexibility than the free-chemistry retrieval. However, tighter posteriors should not automatically be interpreted as more accurate constraints. If the grid is missing relevant chemical pathways or thermal-structure effects, the resulting solution can remain biased even if it is chemically plausible within the grid.} \textcolor{black}{An equilibrium-based chemically consistent treatment may therefore be sufficient for some hotter planets without strong photochemical tracers, while the kinetic framework becomes important for cooler planets and for species affected by vertical mixing or photochemistry, such as CH$_4$, NH$_3$, HCN, H$_2$S, and SO$_2$.} \textcolor{black}{Because Bayesian evidence depends on the adopted priors and parameterization, we interpret the Bayes factors as model-comparison diagnostics rather than as absolute evidence that one framework is uniquely correct.} For 3 out of 20 planets we find strong Bayesian preference (Bayes factor (K) $>$ 10) for the grid-chemistry model, while 2 show moderate preference and another 3 show weak preference. For the remaining 12 planets, the free-chemistry model is favored under the adopted priors.

\textcolor{black}{In the grid framework, C/O values become less coupled to fit quality, and uncertainties shrink for many planets, especially at higher temperatures. Several systems converge toward low C/O ($\lesssim$ 0.05). Such low C/O values are lower than the bulk of the population in existing compilation studies, which tend to find C/O $\geq$ 0.1 for most hot Jupiters. This may indicate that our sample, which is biased towards planets with strong water and sulfur features, probes a chemically distinct population, or that residual systematics still affect the carbon budget. If robust, low C/O values would disfavor formation scenarios in which those planets accreted predominantly carbon-rich gas beyond the $\mathrm{CO_2}$ or CO snowlines, and would instead be more consistent with mixed accretion of solids and gas with substantial delivery of oxygen-rich material, similar to enrichment scenarios proposed for Jupiter. We emphasize that this comparison is made relative to the adopted solar abundance reference, not relative to individual host-star C/O values. A host-normalized comparison of planetary C/O, C/S, and O/S ratios will require homogeneous stellar C, O, and S abundance measurements for the full sample, obtained through dedicated stellar abundance analysis. By contrast, cooler planets with featureless spectra retain broad error bars, reflecting the limited information content of their data. This behavior emphasizes that chemically informed retrievals provide a useful comparison by restricting the allowed abundance combinations, but their results must still be interpreted within the assumptions and limitations of the adopted chemical grid.}

We also explored the behavior of C/S and O/S ratios derived from the free-chemistry retrievals (see Supplementary Figure \ref{fig:CS_OS_Tpl}). Our free-chemistry retrievals suggest possible temperature-dependent patterns in C/S and O/S, but posterior uncertainties are broad and strongly affected by the same 2.9-5 $\mu$m residuals that bias C/O. Robust demographic inferences on C/S and O/S will therefore require retrieval methods that address the biases, ideally by combining physically informed chemistry.

\textcolor{black}{Overall, this demographic analysis shows that population-level trends inferred from free-chemistry retrievals must be interpreted with caution. Statistical flexibility, incomplete opacity treatments, and simplified atmospheric assumptions can together generate apparent correlations that may not reflect underlying physical processes.} \textcolor{black}{Chemically informed retrievals provide an important comparison by restricting the allowed abundance combinations, but they should also be interpreted within the assumptions of the adopted chemical grid.} Throughout this work we adopt a uniform diagonal noise model based on published uncertainties (see Supplementary Information Section \ref{sec:noise-likelihood}). Under this same likelihood, free-chemistry and grid-chemistry retrievals produce qualitatively different demographic trends. \textcolor{black}{This shows that chemical assumptions and model flexibility can substantially affect inferred demographic behavior under the adopted likelihood.} More sophisticated correlated-noise models, such as Gaussian-process likelihoods \citep{gibson2012gaussian,Rotman2025}, will be essential for future investigations but lie beyond the scope of the present study.


\section{Methodology}
\label{sec:methods}

\subsection{Sample Selection}
\label{sec:Data}

\textcolor{black}{\textcolor{black}{Our study focuses on 20 exoplanets observed in transmission for which published reduced JWST spectra are available. For targets that also have suitable published HST transmission spectra, we include the HST data together with the JWST spectra to extend the wavelength coverage. No planet in the sample is analyzed using HST data alone.} The selection criteria were: (i) at least one published transmission spectrum with informative near-infrared coverage is available, (ii) the system parameters required for retrievals, including planetary mass, planetary radius, equilibrium temperature, stellar radius, and orbital/stellar parameters, are available in the literature with reported uncertainties, and (iii) the planet has commonly been interpreted in the literature using an H$_2$/He-rich atmospheric retrieval framework. Planets with only emission spectra, or without sufficient near-infrared transmission coverage to constrain the main molecular absorbers considered here, were excluded.} \textcolor{black}{These planets span equilibrium temperatures of roughly 300-2500 K, radii from sub-Neptunes ($\sim$0.18$R_{\rm J}$) to inflated hot-Jupiters ($\sim$1.9$R_{\rm J}$) with masses from $\sim$0.014$M_{\rm J}$ to $>$1.2$M_{\rm J}$ (\textcolor{black}{the distribution of the planets in the mass, radius, and temperature domain is shown in Supplementary Fig. \ref{fig:population})}. The final sample includes gas giants, warm Neptunes, and temperate/warm sub-Neptunes. Because the H$_2$/He-rich assumption is usually not common for the smaller planets, we treat the temperate/warm sub-Neptunes as a separate subset in the demographic analysis.}

\textcolor{black}{We did not uniformly re-reduce the raw time-series data. Instead, we homogenize the atmospheric retrieval step by applying a common retrieval framework to the published spectra listed in Supplementary Table \ref{tab:exoplanet_properties}. The input spectra may therefore retain differences in light-curve extraction, systematics correction, stellar-contamination treatment, wavelength binning, and inter-mode offset assumptions. This limitation is one reason why we interpret the C/O-$T_{\rm eq}$ relation as a test of retrieval-model sensitivity rather than as a statistically significant demographic trend. Additional details on the sample-selection criteria, transmission-only choice, sub-Neptune caveats, and data-reduction heterogeneity are given in Supplementary Information Section \ref{sec:appendixA_sample}.}
\subsection{Atmospheric Retrieval Setup}
\label{sec:retreival_setup}

\subsubsection{Free Chemistry Retrieval}
\label{sec:free_chemistry}

We interpreted the transmission spectra of the planets using one-dimensional (1D) atmospheric models that capture a wide range of physical and chemical scenarios. \textcolor{black}{To perform these systematic retrieval calculations, we used the open-source petitRADTRANS retrieval framework \footnote{\href{https://petitradtrans.readthedocs.io/en/latest/}{https://petitradtrans.readthedocs.io/en/latest/}} (version 3.2.0). The framework computes the forward transmission spectra and interfaces with  MultiNest nested-sampling algorithm \citep{feroz2008multimodal, feroz2009multinest, 2019OJAp....2E..10F}, implemented through the Python wrapper, PyMultiNest \citep{buchner2014x}, with 1000 live points. Each atmosphere was treated under the plane-parallel assumption and discretized into 100 logarithmically spaced pressure layers from $10^2$ to $10^{-9}$ bar.} \textcolor{black}{All free-chemistry retrievals used the default petitRADTRANS sampling efficiency of 0.8. Atmospheric characterization was carried out within a Bayesian inference framework using flexible free-chemistry forward models, in which molecular abundances, temperature-pressure (T-P) structure, and cloud or haze properties are constrained directly from the observed spectra without imposing radiative-convective or thermochemical equilibrium assumptions.}

\textcolor{black}{To maintain comparability across the sample, we used a homogeneous core retrieval framework rather than an identical parameterization for every planet. All retrievals share the same radiative-transfer code, nested-sampling framework, pressure grid, H$_2$/He-dominated atmosphere, likelihood structure, molecular abundance priors, fixed-gravity baseline, and gray cloud-deck baseline. Target-specific extensions were introduced only when motivated by wavelength coverage, retrieval statistics, or previous dedicated analyses in the literature. These include treatment of gravity as free parameter for targets where this was favored by Bayesian evidence, additional cloud/scattering treatments for spectra with optical or continuum leverage, H$^{-}$ opacity for ultra-hot planets, and the SiO$_2$ cloud treatment for WASP-17 b. The molecular species lists follow the same baseline-plus-extension logic: H$_2$O, CH$_4$, CO, and CO$_2$ form the core C and O-bearing set, while additional species such as HCN, NH$_3$, C$_2$H$_2$, CS$_2$, H$_2$S, and SO$_2$ were included following the relevant planet-specific detection literature. The resulting setup is therefore not strictly identical for every planet, but it uses a common retrieval core with documented and physically motivated target-specific extensions.} The stellar radius for each system was fixed to its value reported in the literature, whereas parameters such as the planetary radius, equilibrium temperature, reference pressure, instrumental offsets (for cases combining multiple JWST or HST modes), cloud properties, and molecular abundances were treated as free parameters. The adopted cloud treatments varied depending on the spectral coverage available for each planet. \textcolor{black}{Most planets used a gray cloud deck \citep{kreidberg2015detection}, while spectra with short-wavelength leverage were tested with an additional power-law scattering component. WASP-17 b was treated separately with the \citet{ackerman2001precipitating} cloud model and crystalline SiO$_2$ Mie scattering, following \citet{louie2025jwst}. The cloud and scattering prescriptions, including the WASP-17 b-specific treatment, are described in Supplementary Information Section \ref{sec:appendixA_clouds}}.

We tested different treatments for gravity and T-P structures during the free-retrieval. Since the MultiNest algorithm directly computes the Bayesian evidence (ln Z) during sampling, the difference in ln Z between two competing models was used as a metric for model comparison \citep{kass1995bayes, tsiaras2016new, grubel2025detectability, arenales2025polycyclic}. In cases where the reduced-$\mathrm{\chi^2}$ ($\mathrm{\chi^2_r}$) and Bayesian evidence resulted into contradiction, the model with the higher ln Z was always given the priority to determine the best-fit solution. The surface gravity was investigated both as a fixed and as a free parameter in our retrievals, and the preferred model was selected based on the criteria mentioned above. The treatment of surface gravity showed mixed behavior: while most targets favor constant gravity treatment, HIP 67522 b, TOI-776 c, and TrES-4 b require gravity as a free parameter to achieve better Bayesian evidence (see Supplementary Table \ref{tab:gravity_comparison_overall}). We further investigated two prescriptions for the T–P profile: a simple isothermal model and the parametric profile of \citet{guillot2010radiative}, which permits non-isothermal gradients within the atmosphere. Among the isothermal and Guillot T-P profile, the isothermal profile provided higher Bayesian evidence for all planets and was therefore chosen for the analysis.

For all retrievals in this study, we assumed a $\mathrm{H_2}$–He dominated atmosphere in transmission geometry. It is consistent with the composition expected for $\mathrm{H_2}$-rich gas giant and sub-Neptune exoplanets. The molecular and atomic species were assigned vertically constant mass mixing ratios, and the remaining atmospheric mass fraction was distributed between molecular $\mathrm{H_2}$ (76.6\%) and helium (23.4\%). To maintain physical consistency, the likelihood function was set to zero whenever the sum of the retrieved mixing ratios exceeded 1 to prevent nonphysical solutions during sampling. To evaluate the sensitivity of our retrievals to the assumed $\mathrm{H_2}$–He composition, we performed a set of trial retrievals in which the $\mathrm{H_2}$/He ratio was treated as a free parameter. These tests were designed to examine whether variations in this ratio could influence the inferred molecular abundances, temperature structure, or overall model likelihood. In practice, allowing the $\mathrm{H_2}$/He ratio to vary did not lead to any statistically significant improvement in either the Bayesian evidence or the $\mathrm{\chi^2_r}$. It indicates that the available data do not meaningfully constrain this ratio within the explored wavelength range and signal-to-noise levels. Moreover, introducing $\mathrm{H_2}$/He as an additional free parameter substantially increases the dimensionality of the parameter space and prolongs convergence time, without improving the fit quality. For these reasons, and to maintain uniformity across all targets, we adopted a fixed $\mathrm{H_2}$-He composition in the final retrievals. This assumption is both computationally efficient and physically justified, as the $\mathrm{H_2}$/He ratio is expected to show only minor variations among $\mathrm{H_2}$-dominated planetary atmospheres.

\textcolor{black}{The molecular and atomic line lists used in this study were taken from the ExoMol \citep{tennyson2016exomol, chubb2021exomolop}, HITEMP \citep{rothman2014status}, and HITRAN \citep{gordon2016hitran2016} databases. For all free-chemistry and grid-chemistry retrievals, we used the petitRADTRANS correlated-$k$ opacity tables at a nominal resolving power of $R=1000$. The model spectra were then compared to the published transmission spectra at the corresponding observational wavelength bins. This opacity resolution is appropriate for the low- to moderate-resolution transmission spectra used in this population-level analysis, while avoiding the substantially higher computational cost of line-by-line opacity calculations.} We include the following line opacity species in our study: $\mathrm{H_2O}$, $\mathrm{CH_4}$, CO, $\mathrm{CO_2}$, $\mathrm{NH_3}$, HCN, $\mathrm{H_2S}$, $\mathrm{SO_2}$, $\mathrm{C_2H_2}$, $\mathrm{CS_2}$, SCO, $\mathrm{N_2O}$, AlO, SiO, SiH, Na, K (references are summarized in Supplementary Table \ref{tab:linelist}). In addition to molecular and atomic opacities, we also include Rayleigh scattering by $\mathrm{H_2}$ and He. We also include the continuum opacity arising from collision-induced absorption (CIA) between $\mathrm{H_2}$–$\mathrm{H_2}$ and $\mathrm{H_2}$–He pairs. These opacity sources contribute significantly to the overall continuum in $\mathrm{H_2}$-dominated atmospheres and are essential for reproducing the observed baseline levels in transmission spectra.

For WASP-121 b and WASP-17 b, we account for $\mathrm{H^{-}}$ bound–free absorption, which becomes significant in hot, ionized atmospheres. The inclusion of $\mathrm{H^{-}}$ in these two planets is motivated from \cite{gapp2025wasp} and \cite{louie2025jwst}, respectively. \textcolor{black}{$\mathrm{H^{-}}$ opacity mainly affects the continuum level against which molecular features are measured, and can therefore influence the retrieved reference radius/pressure and gas abundances for these two planets.} In these models, the bulk composition includes atomic H (2\%), molecular $\mathrm{H_2}$ (74\%), and He (4\%), along with free electrons represented by a fixed mass fraction of $\mathrm{10^{-10}}$. The full set of retrieval parameters and their prior ranges for all planets are summarized in Supplementary Table \ref{tab:priors}, and the list of molecular species included for each planet is provided in Supplementary Table \ref{tab:molecule_priors}. For all gas-phase molecules, the mass fractions were assigned log-uniform priors between 10$^{-12}$ and 10$^{-0.3}$. For the $\mathrm{SiO_2}$ and $\mathrm{H^{-}}$, we considered a log-uniform mass fraction prior of [10$^{-11}$, 10$^0$] and [10$^{-14.6}$, 10$^{0.4}$], respectively. This particular choice is motivated from \citet{louie2025jwst}.

\textcolor{black}{For a small number of targets, we excluded overlapping or tensioned spectral segments to avoid redundant wavelength coverage or unresolved multi-instrument offsets. The target-specific choices for WASP-39 b and WASP-107 b are described in Supplementary Information Section \ref{sec:appendixA_sample}.} \textcolor{black}{All free-chemistry and grid-chemistry retrievals used the same independent Gaussian likelihood based on the published 1$\sigma$ uncertainties, without additional jitter, diagonal error inflation, or wavelength-correlated noise terms. Reduced $\chi^2_{\rm r}$ values are reported only as descriptive fit-quality indicators, because the effective number of degrees of freedom is not rigorously defined for nonlinear retrievals with priors and parameter degeneracies. The full likelihood expression, covariance assumption, and limitations of this noise treatment are given in Supplementary Information Section \ref{sec:noise-likelihood}.}

\subsubsection{Grid Chemistry Retrieval}
\label{sec:grid_chemistry}

While free-chemistry retrievals provide more flexibility in fitting observed spectra, they offer limited physical insight into the underlying chemical and dynamical processes in a planetary atmosphere. This raises an important question: what do these retrieved abundances actually tell us about atmospheric chemistry? \textcolor{black}{Since the launch of JWST, an increasing number of studies have begun to interpret atmospheric observations using chemically informed models, including analyses of photochemistry, vertical mixing, sulfur chemistry, cloud and haze processes, and chemical pathways in individual planets (e.g. \cite{coulombe2023broadband,madhusudhan2023carbon,fournier2024near,holmberg2024possible,welbanks2024high,beatty2024sulfur,ohno2025possible,wiser2025precise,ahrer2025escaping,fu2025overcast}). These studies demonstrate the value of connecting retrieved abundances to atmospheric processes, but most such applications remain target-specific. In this work, we extend this perspective to a population-level comparison by explicitly incorporating chemical-kinetics grids across our sample.} We revisit the grid-chemistry retrievals for planets previously analyzed in the literature and extend this framework to several additional targets that are being investigated for the first time.

Atmospheric chemistry can generally be studied under two distinct paradigms: thermochemical equilibrium and chemical disequilibrium. In equilibrium models, the chemical composition is derived by minimizing the Gibbs free energy of the system for given temperature–pressure conditions. In contrast, disequilibrium models solve a coupled set of continuity equations for each species, accounting for both chemical kinetics and transport processes, until a steady state is achieved \citep{tsai2017vulcan,molaverdikhani2019cold1}. Because many planet atmospheres exhibit strong vertical mixing, photochemistry, and thermal gradients, we adopt the disequilibrium approach in this study, as it provides a more realistic representation of their atmospheric composition.

To model the atmospheric chemistry of our targets, we simulate a grid of the chemical kinetics models using VULCAN\footnote{\href{https://github.com/exoclime/VULCAN}{https://github.com/exoclime/VULCAN}} , which solves for the one-dimensional radiative–convective–photochemical equilibrium (RCPE). Each simulation begins with equilibrium abundance profiles generated by FastChem\footnote{\href{https://github.com/exoclime/FastChem}{https://github.com/exoclime/FastChem}} \citep{stock2018fastchem} based on the elemental abundances and T-P profiles specific to each planet.  \textcolor{black}{These starting conditions are not derived from the free-chemistry retrievals. Instead, for each VULCAN grid point, the elemental abundances are set by the chosen C/O and metallicity values, and the T-P structure is taken to be one of the planet-specific isothermal profiles listed in Supplementary Table \ref{tab:vulcan_temp_grids}. FastChem is used only to generate the initial thermochemical-equilibrium abundance profiles, which are then evolved by VULCAN under the adopted vertical-mixing and photochemical network.}\textcolor{black}{Thermochemical equilibrium is therefore a limiting case of the kinetic calculation: where chemical timescales are short compared with vertical-mixing and photochemical timescales, the final kinetic abundances remain close to the equilibrium solution. Deviations from equilibrium arise where transport, photolysis, or slow reaction pathways become important. We therefore do not interpret the kinetic grid as implying that disequilibrium chemistry is required for every planet; rather, it provides a common chemically consistent framework that reduces to equilibrium-like behavior where appropriate while allowing disequilibrium effects in cooler or photochemically active atmospheric regions.} These initial equilibrium chemical profiles are then evolved kinetically until the system reaches a steady state. The model accounts for some of the very important physical processes, including molecular diffusion, vertical mixing (parameterized by the Eddy diffusion), and photochemistry driven by incident stellar radiation. We adopted stellar spectra from the VULCAN repository or from MUSCLES \citep{france2016muscles,youngblood2016muscles,loyd2016muscles}, using proxy spectra of similar effective temperature when the host-star UV spectrum was not available. The MUSCLES spectra were rescaled to the stellar surface using the host-star radius and distance from Exo.MAST. The planet-specific UV fields are listed in Supplementary Information Section \ref{sec:appendixB_uv_fields}. \textcolor{black}{For the photochemical and kinetic network, we used the `\texttt{S-N-C-H-O full photo network}' available in the VULCAN repository for most planets. For WASP-39 b, WASP-107 b, and GJ 3470 b, we instead used the specific `\texttt{S-N-C-H-O photo network}' from \citet{tsai2023photochemically}. This choice was motivated by the sulfur-chemistry literature following the first JWST detection of SO$_2$ in WASP-39 b, where this network has been used to interpret SO$_2$-bearing atmospheres. We also tested the `\texttt{S-N-C-H-O full photo network}' for these three planets. The major molecular abundances were generally similar between the two networks, but the full network gave a poorer representation of the SO$_2$ feature in our tests. We therefore adopted the specific `\texttt{S-N-C-H-O photo network}'for the SO$_2$-relevant cases. This choice was motivated by the published SO$_2$ analyses and by our network comparison.}

This process is computationally expensive, as predicting the exact chemistry model that best fits the data requires generating thousands of converged models suitable for Bayesian inference (for example, see \cite{welbanks2024high,beatty2024sulfur,barat2025metal}). The problem becomes even more challenging when we aim to study multiple planets simultaneously at a population level. Producing such an extensive, fine model grid for 20 planets is extremely time-consuming. To reduce this computational complexity, we constructed a coarse grid of disequilibrium models based on four primary parameters:  C/O ratio (C/O = [0.01, 0.1, 0.55, 1, 10, 100]), metallicity (-2 $\leq$ log$\mathrm{_{10}}$[M/H] $\leq $ 2 in steps of 1), planet equilibrium temperature ($\mathrm{T_{eq}}$) (4 different values, see Supplementary Table \ref{tab:vulcan_temp_grids}) and $\mathrm{K_{zz}}$ (\textcolor{black}{6 $\leq$ log$\mathrm{_{10}}$$\mathrm{K_{zz}}$ $\leq $ 10 in steps of 1}). \textcolor{black}{For GJ 3470 b, WASP-107 b, and WASP-39 b, where SO$_2$ has been reported, we generated a larger grid of 900 disequilibrium models using a finer C/O grid, C/O = [0.01, 0.1, 0.2, 0.3, 0.4, 0.55, 1, 10, 100], while keeping the same grid spacing in metallicity, $T_{\rm eq}$, and $K_{zz}$. This finer C/O sampling was used because the SO$_2$ abundance is sensitive to the coupled C/O-sulfur chemistry and because these planets are the cases where sulfur photochemistry is most directly constrained by the observations.} \textcolor{black}{The coarse C/O grid, chosen for other planets, is intended to sample broad oxygen-rich, near-solar, and carbon-rich chemical regimes, rather than to provide finely resolved C/O constraints for each individual planet. Interpolation across this coarse grid may affect individual C/O medians and posterior widths, so the grid-retrieval C/O values should be interpreted as chemically informed comparison constraints rather than high-resolution abundance measurements.} Each simulation was performed on a vertical grid of 160 logarithmically spaced layers spanning $10^2$ to $10^{-7}$ bar. Additional details on the temperature-grid approximation, elemental-abundance prescription, and convergence statistics are given in Supplementary Information Section \ref{sec:appendixB_grid_details}.

Once the initial grid of models was generated, we implemented the Radial Basis Function (RBF) interpolation scheme \citep{buhmann2000radial,ghosh2001overview} to create a finer grid of models between the existing grid points. These finer grids have the interpolated vertical mixing ratio profiles for all molecular species considered in the retrievals. 

\textcolor{black}{The interpolated chemical profiles were then used in the grid-chemistry retrievals within the same petitRADTRANS retrieval framework. The grid-chemistry retrievals used the same nested-sampling settings as the free-chemistry retrievals: 1000 live points and the default petitRADTRANS sampling efficiency = 0.8.} \textcolor{black}{To test sensitivity to the sampling efficiency, we reran five representative planets, GJ 3470 b, WASP-80 b, HD 209458 b, WASP-15 b, and WASP-121 b, using a lower MultiNest sampling efficiency of 0.3 for both the free-chemistry and disequilibrium-grid retrievals, while keeping all other retrieval settings fixed. The free-chemistry posterior constraints and Bayesian evidence estimates were stable across this subset, and most VULCAN retrievals also remained consistent with the fiducial runs. \textcolor{teal}{The largest evidence difference occurred for the VULCAN retrieval of WASP-15 b, for which $\ln Z$ changed from $1825.8\pm0.1$ in the fiducial run to $1822.7\pm0.0$ at a sampling efficiency of 0.3. The evidence values for all five representative planets and both retrieval frameworks are provided in Supplementary Table \ref{tab:efficiency_lnz_comparison}.}} \textcolor{black}{At each sampling step, the abundance profiles were drawn from the interpolated VULCAN grid and passed to petitRADTRANS to compute the corresponding transmission spectrum. Thus, VULCAN provides the chemical abundance profiles, while petitRADTRANS performs the radiative-transfer calculation and retrieval sampling through its PyMultiNest interface.} During these retrievals, the abundances of all gas species were sampled directly from the interpolated model grid. For species not included in the kinetic network, such as Na, K, SiO, SiH, AlO, $\mathrm{SiO_2}$, and $\mathrm{H^{-}}$, their abundances were treated as free parameters with the same prior ranges used in the free-chemistry retrievals. In grid retrieval, the atmospheric pressure range was fixed between $\mathrm{10^2}$ and $\mathrm{10^{-7}}$ bar, uniformly discretized into 100 layers. Only the chemical abundance profiles were drawn from the precomputed models during the Bayesian inference. \textcolor{black}{The priors used in the chemically informed disequilibrium retrievals are listed in Supplementary Table \ref{tab:priors_chem}. These include the additional grid-retrieval parameters C/O, metallicity, $T_{\rm eq}$, and $K_{zz}$. The remaining parameters, including reference pressure, cloud properties, instrumental offsets, gravity treatment, and stellar radius, follow the final setup adopted in the free-chemistry retrievals unless stated otherwise.}

\textcolor{black}{The axes shown in the corner plots correspond to the actual prior ranges used in the retrievals. The reference pressure, $P_{\rm ref}$, retrieves the reference atmospheric pressure associated with the transmission radius. We adopt a broad prior, $\log_{10}(P_{\rm ref}/{\rm bar})=\mathcal{U}[-5,5]$, because the relevant transmission photosphere is expected to lie within this pressure interval.}

Finally, to compare the free-chemistry and grid-chemistry frameworks for each planet, we evaluate the Bayesian evidences computed with MultiNest and report the corresponding Bayes factors,

\begin{equation}
    \mathrm{K} \ \text{=} \ \mathrm{e^{\left(ln \ Z_{grid} \ -\  ln \ Z_{free}\right)}}
\label{eq:bayes}
\end{equation}

\textcolor{black}{following the convention of \citet{kass1995bayes}. Under this interpretation, K$>$10 (equivalently, lnK$>$2.3) indicates strong preference for the grid-chemistry model, $3.2<K<10$ indicates moderate preference, and $1<K<3.2$ indicates weak preference. Values of $K<1$ favor the free-chemistry model, with the strength of this preference determined by $1/K$. Because the free-chemistry and grid-chemistry retrievals have different parameterizations, these Bayes factors are interpreted as relative model support under the adopted priors, not as absolute proof that one parameterization is uniquely correct.} \textcolor{black}{These Bayes factors should also be interpreted with the usual numerical caveats associated with nested-sampling evidence estimates, since individual grid-chemistry cases can be sensitive to sampler settings such as the MultiNest sampling efficiency.}



\backmatter




\section*{Funding Statement}

D.D. discloses support for the research of this work from Excellence Cluster ORIGINS, which is funded by the Deutsche Forschungsgemeinschaft (DFG, German Research Foundation) under Germany’s Excellence Strategy – EXC-2094–390783311 (\href{http://www.universe-cluster.de/}{http://www.universe-cluster.de/}).

\section*{Acknowledgements}
This research benefited from resources available through the NASA Exoplanet Archive, operated by the California Institute of Technology for National Aeronautics and Space Administration’s Exoplanet Exploration Program. We would also like to thank the anonymous referees for constructive comments that helped improve the manuscript.

\section*{Author Contributions}
D.D., K.M., and B.E. jointly developed the scientific framework and analysis plan. D.D. carried out the modeling, data analysis, and visualization. The manuscript was written by D.D. and revised critically by K.M., B.E., S.S., R.A.-L., and F.G. S.S. assisted with the construction of the chemical model grids and the analysis of WASP-39 b and WASP-107 b. All authors contributed to the interpretation of the results and to the refinement of the final manuscript.



\section*{Data Availability}

The JWST and HST transmission spectra analyzed in this study were obtained from previously published observations, with their corresponding sources mentioned in the Supplementary Information. Additional data supporting the findings of this study are available from the corresponding author upon reasonable request.


\clearpage
\bibstyle{sn-mathphys-num}
\bibliography{sn-bibliography}


\clearpage
\clearpage
\begin{center}
    \section*{Supplementary Information}
\end{center}
\label{sec:supplementary}


\renewcommand{\thesection}{1.1}
\section{Planet sample and planet-by-planet comparisons from free-chemistry retrievals}
\label{sec:appendixA}
\subsection{Sample and data-selection details}
\label{sec:appendixA_sample}

\textcolor{black}{The sample was selected primarily by the availability of published reduced transmission spectra suitable for a homogeneous retrieval comparison. We did not require continuous wavelength coverage from 0.8-5 $\mu$m for every planet. Instead, we required at least one published transmission spectrum with sufficient near-infrared information to constrain the main molecular absorbers considered in this work. Targets with only emission spectra were excluded because emission spectra probe the dayside atmosphere and require additional assumptions about the temperature-pressure structure, heat redistribution, emission geometry, and possible longitudinal inhomogeneity. A joint transmission-emission analysis would therefore be a separate extension of the present population-level comparison.}

\textcolor{black}{For the system-parameter criterion, we did not apply a strict numerical uncertainty threshold. The requirement was that the parameters needed to run the retrievals were available in the literature with reported uncertainties, including planetary mass, planetary radius, equilibrium temperature, stellar radius, and the orbital/stellar parameters needed for the transmission calculation. Likewise, the H$_2$/He-rich criterion was not based on a strict mass limit, density limit, or position in the mass-radius diagram. It was based on whether the planet had previously been interpreted in the literature using an H$_2$/He-rich atmospheric retrieval framework.} \textcolor{black}{This assumption is less well established for the temperate/warm sub-Neptunes than for gas giants. For example, K2-18 b and TOI-270 d have been interpreted with H$_2$-rich retrieval frameworks \citep{madhusudhan2023carbon,holmberg2024possible}, whereas GJ 1214 b is consistent with a metal-rich atmosphere with a thick reflective haze and also with a hydrogen-rich but metallicity-enhanced atmosphere \citep{kempton2023reflective}. We treat the smaller planets as a cautionary subset and explicitly test whether the inferred C/O-$T_{\rm eq}$ structure depends on their inclusion.}

\textcolor{black}{The input spectra were compiled from different published reductions, as listed in Supplementary Table \ref{tab:exoplanet_properties}. They may differ in light-curve extraction, systematics correction, stellar-contamination treatment, wavelength binning, and instrument-offset assumptions. Our analysis homogenizes the atmospheric retrieval step, but not the original raw-data reduction step. When multiple instruments or observing modes were combined, relative offset parameters were included where required to account for constant baseline differences between datasets. These offsets cannot remove all wavelength-dependent reduction-level systematics, which is why the population-level C/O-$T_{\rm eq}$ relation is interpreted as a diagnostic of retrieval-model sensitivity and sample composition rather than as a confirmed demographic trend.}

\textcolor{black}{For a small number of targets, we made target-specific choices to exclude overlapping spectral segments from the final population-level retrievals. For WASP-39 b, we excluded the HST/WFC3 and JWST/NIRCam F322W2 spectra and did not include the NIRSpec PRISM spectrum. The HST WFC3 spectrum overlaps substantially with JWST/NIRISS SOSS, while NIRCam/F322W2 largely overlaps with the combined NIRISS/SOSS and NIRSpec/G395H wavelength range. Although NIRCam/F322W2 and NIRSpec/G395H show a known mode-to-mode offset in their overlap \citep{carter2024benchmark}, G395H was retained because it extends to the chemically important 5 $\mu$m region. Hence, we used NIRISS SOSS, NIRSpec G395H, and MIRI for the final WASP-39 b retrieval. For WASP-107 b, we omit the MIRI measurements from the final population-level retrieval. In initial tests, the MIRI segment introduced tension with the HST+NIRCam spectrum, mainly because part of the MIRI spectrum lies at lower transit depths relative to the HST+NIRCam continuum. Including MIRI increased the reduced $\chi^2$ from 1.60 for HST+NIRCam to 2.59 for HST+NIRCam+MIRI. The MIRI segmen is excluded to avoid propagating this unresolved multi-instrument tension into the retrieved abundances.}

\textcolor{black}{Compared to earlier HST-based surveys, which hinted at trends in cloudiness, water abundance, or atmospheric metallicity (e.g. \cite{sing2016continuum,welbanks2019mass,bardet25}), our analysis will benefit from the broader spectral coverage and sensitivity of JWST. Recent studies have begun to explore population-level patterns using these new datasets, but typically with fewer planets and atmospheric scale height-based trends \citep{fu2025statistical,davenport2025toi}. Similar studies were done before as well for the $\mathrm{H_2O}$ band observed with Hubble \citep{gao2021aerosols}. On the other hand, our sample allows us to test the sensitivity of inferred elemental ratios across a wider parameter space and across different planet classes. In doing so, we build on earlier work while extending demographic analyses into a multi-molecule, multi-planet regime.}

\subsection{Cloud and scattering prescriptions}
\label{sec:appendixA_clouds}

For the majority of targets, we assumed a gray cloud deck located at a specific pressure level (for example see \cite{kreidberg2015detection}). In this approach, the atmosphere becomes completely opaque below the defined cloud deck by assigning very high opacities to the deeper layers. \textcolor{black}{For WASP-17 b, we implemented the \citet{ackerman2001precipitating} microphysical cloud model, following the recent study by \citet{louie2025jwst}. In this scheme, the lognormal particle-size distribution width ($\mathrm{\sigma_{lnorm}}$), sedimentation factor ($\mathrm{f_{sed}}$), and eddy diffusion coefficient ($\mathrm{K_{zz}}$) were treated as free parameters. We also included crystalline SiO$_2$ as a Mie-scattering condensate species for WASP-17 b, again following \citet{louie2025jwst}. This component is parameterized by the mean particle radius ($\mathrm{r_m}$) and cloud-base pressure, which define the vertical distribution and optical depth of the condensate layer. The cloud-base pressure was drawn from a log-uniform prior spanning $10^{-6}$ to $10^{3}$ bar, while the mean particle radius was fixed at $\mathrm{r_m}$ = 0.06 $\mu$m.} In addition, for planets observed down to near-visible wavelengths ($\approx$ 0.8 $\mathrm{\mu}$m) with JWST/NIRISS (HAT-P-18 b, K2-18 b, WASP-39 b, and TOI-421 b), we additionally incorporated a power-law scattering model to capture the wavelength-dependent opacity in the optical region. The power-law cloud opacity is expressed as:

\begin{equation}
    \mathrm{\kappa} \ \text{=} \ \mathrm{\kappa_0} \times \left( \frac{\mathrm{\lambda}}{\mathrm{\lambda_0}} \right)^{\mathrm{\gamma_{scat}}}
\end{equation}

\noindent where both $\mathrm{\kappa_0}$ and $\mathrm{\gamma_{scat}}$ are free parameters and $\lambda_0$ is fixed at 350 nm. This parameterization produces a continuum slope in the transmission spectrum which is representative of small-particle scattering.

For TOI-421 b, however, the power-law scattering prescription did not improve the fit quality or Bayesian evidence, and we adopt a gray cloud deck for this planet in the final analysis. \textcolor{black}{For WASP-17 b, we also did not include an additional power-law scattering prescription because this planet was already treated with the Ackerman and Marley cloud model and crystalline SiO$_2$ Mie-scattering cloud prescription described above.} \textcolor{black}{This target-specific treatment can affect the retrieved continuum level, cloud pressure/particle parameters, and gas abundances for WASP-17 b, because aerosol opacity can partially trade off with molecular opacity in transmission spectra. That is why we do not compare the WASP-17 b cloud parameters one-to-one with the gray-cloud parameters used for most other planets. Instead, the SiO$_2$ cloud treatment is used to avoid forcing a gray or power-law cloud description onto a planet for which dedicated analyses have already reported condensate cloud opacity.} 

\subsection{Planet-by-planet comparisons from free-chemistry retrievals}
\label{sec:appendixA_free_planets}

GJ 3470 b stands out as an important example in this study. For GJ 3470 b, our retrieved C/O ratio is much lower than that inferred by \citet{beatty2024sulfur} because their analysis assumes a CO abundance several orders of magnitude higher. We adopt the published transmission spectrum from \citet{beatty2024sulfur} without re-reducing the raw data. 

\textcolor{black}{The discrepancy arises from differences in retrieval assumptions and parameterization, rather than from new spectral retrievals. In our free-chemistry retrieval, C/O is not directly sampled but is computed from the retrieved molecular abundances, for which gas-phase species were assigned log-uniform mass-fraction priors from $10^{-12}$ to $10^{-0.3}$. For the chemically constrained grid retrieval of GJ 3470 b, we used C/O = [0.01, 0.1, 0.2, 0.3, 0.4, 0.55, 1, 10, 100]. In contrast, the C/O value quoted by \citet{beatty2024sulfur} was obtained from their 1D-RCPE grid retrieval, which used a more restricted C/O grid spanning $0.01 \leq \mathrm{C/O} \leq 0.6$. It may also be affected by differences in cloud and scattering parameterization, since the continuum treatment for GJ 3470 b is relevant in previous analyses and can influence the inferred molecular abundances.} This underscores how strongly C/O can depend on retrieval frameworks even for the same underlying dataset. Constraining CO abundance from transmission spectra is challenging, as its bands are weak and often blended with noise. Different retrieval frameworks have reported similarly inconsistent CO abundances even for well-studied systems such as WASP-107 b \citep{welbanks2024high}.

\textcolor{black}{For TOI-270 d, the retrieved CS$_2$ abundance is substantially higher than the literature value ($\log_{10}{\rm CS_2}=-0.81^{+0.32}_{-0.50}$ compared with $-2.59^{+0.67}_{-0.95}$), and we treat it as a model-sensitive free-retrieval result rather than a robust abundance measurement. For WASP-39 b and WASP-80 b, some absolute free-retrieved molecular abundances are higher than previous estimates, indicating sensitivity to continuum placement, reference pressure, cloud opacity, and overlapping molecular features. For WASP-15 b, the retrieved H$_2$O abundance is higher than the literature median but overlaps within the quoted uncertainties. For WASP-121 b, the high CO abundance is qualitatively plausible for an ultra-hot Jupiter, but its precise value is model-sensitive because of thermal dissociation, H$^{-}$ opacity, non-isothermal structure, and day-night heterogeneity. We interpret these individual free-chemistry abundances cautiously and do not use any single high mass-mixing ratio as a demographic anchor. The main C/O interpretation is instead based on the comparison between the free-chemistry and chemically consistent retrievals.}

For GJ 1214 b, TOI-776 c, and TOI-836 c, the presence of high-altitude clouds is suspected to mute the spectral features. It causes weak abundance constraints for all molecules and hence, poor C/O constraint with larger uncertainties. \textcolor{black}{Localized model-data mismatches remain for a few individual targets. For WASP-17 b, the mismatch near $\sim$1 $\mu$m is most likely associated with the short wavelength continuum and the adopted cloud/scattering treatment. For WASP-39 b, the short-wavelength residuals may reflect instrumental systematics, inter-mode offsets, and saturation-affected wavelength regions \citep{carter2024benchmark}. For TOI-836 c, the mismatch near 3.25 $\mu$m lies in a region where CH$_4$, C$_2$H$_2$, and other hydrocarbon-bearing opacities overlap. So, we treat the retrieved CH$_4$ abundances for TOI-836 c as weakly constrained abundance measurements.}

\renewcommand{\thesection}{1.2}
\section{Grid-Chemistry Retrievals}

\subsection{Stellar UV fields used in the VULCAN grids}
\label{sec:appendixB_uv_fields}

For the VULCAN chemical-kinetics grids, the photochemistry is driven by the incident stellar UV field. Because our targets orbit stars with diverse spectral types and effective temperatures, each receives a distinct UV flux environment that strongly influences the resulting chemistry. To reflect this, we applied distinct stellar flux profiles from the VULCAN repository corresponding to each planet’s host star effective temperature - (1) HAT-P-18 b: `\texttt{sflux-epseri}', (2) GJ 3470 b: `\texttt{sflux-GJ436}', (3) GJ 1214 b : `\texttt{sflux-GJ1214}', (4) K2-18 b: `\texttt{sflux-GJ436}', (5) TOI-270 d: `\texttt{sflux-GJ436}', (6) HD 189733 b: `\texttt{sflux-HD189-Moses11}', (7) HD 209458 b: `\texttt{Gueymard-solar}', (8) HIP 67522 b: `\texttt{Gueymard-solar}', (9) TOI-776 c: `\texttt{sflux-GJ436}', (10) TOI-421 b: `\texttt{sflux-epseri}', (11) WASP-127 b: `\texttt{Gueymard-solar}', (12) WASP-39 b: `\texttt{Gueymard-solar}'. For the remaining planets, we used stellar spectra from the MUSCLES observations \citep{france2016muscles, youngblood2016muscles,loyd2016muscles} - (1) WASP-17 b: `\texttt{WASP-17}', (2) WASP-80 b: `\texttt{HD 85512}', (3) WASP-107 b: `\texttt{HD 85512}', (4) WASP-15 b: `\texttt{HD 149026}', (5) TOI-836 c: `\texttt{WASP-43}', (6) TrES-4 b: `\texttt{HD 149026}', (7) WASP-94 A b: `\texttt{HD 149026}', (8) WASP-121 b: `\texttt{WASP-17}'. When the UV flux for a specific host star was unavailable, we used a proxy spectrum from a star of comparable effective temperature. The MUSCLES spectra represent the measured fluxes on Earth. So we rescaled them to the stellar surface using the host star’s radius and distance from Earth, as listed in the \href{https://exo.mast.stsci.edu/}{Exo.MAST} database. This approach is consistent with previous studies: for instance, \citet{welbanks2024high} used the MUSCLES spectrum of HD 85512 for WASP-107 b, while \citet{beatty2024sulfur} adopted the GJ 176 spectrum for GJ 3470 b (we instead used GJ 436 as the proxy).


\subsection{Disequilibrium-grid construction}
\label{sec:appendixB_grid_details}

For temperature, the best practice is to compute the T–P profile using self-consistent radiative–convective equilibrium models and then feed that profile into the kinetics simulation \citep{molaverdikhani2019cold1,tsai2021comparative,dubey2024comparative,dubey2025quantified,welbanks2024high,barat2025metal}. Because, in reality, any change in molecular abundances affects the opacity, which in turn alters the atmospheric T–P structure \citep{drummond2016effects}. Fully self-consistent models account for this feedback, but given our large sample of twenty planets, such computations are impractical. In this work, we consider a simplified approach by using four different isothermal temperature levels for each planet. This simplified treatment is consistent with the approach used in the initial VULCAN benchmark study for HD 189733 b and HD 209458 b \citep{tsai2017vulcan}, which likewise implemented isothermal profiles to assess the dependence of chemical abundances on temperature. Because the equilibrium temperatures of our sample vary widely, the choice of the $\mathrm{T_{eq}}$ grid also differs from planet to planet. \textcolor{black}{For the coldest sub-Neptunes, K2 18 b and TOI-270 d, the temperature grid reaches the lower limit of 400 K. We treat their grid-retrieval C/O values as potentially sensitive to the cold-edge temperature grid and condensation physics.} We also need several star and planet parameters to run the simulations. The corresponding star and planet radius, planet surface gravity, star–planet distance, and the isothermal temperature grid are summarized in Supplementary Table \ref{tab:vulcan_temp_grids}.

\textcolor{black}{The elemental abundances used in the VULCAN grids are specified by the chemical-grid parameters and not by the free-chemistry retrieval posteriors.} To vary the C/O ratio, we follow the approach outlined in \citep{molliere2015model,woitke2018equilibrium,molaverdikhani2019cold2, dubey2023polycyclic,dubey2024comparative,dubey2025quantified}, where we adjust the oxygen elemental abundance and keep the carbon elemental abundance constant. For metallicity, $\log_{10}$[M/H] = 0 corresponds to the solar value, and the abundances of C, O, N, and S are scaled with metallicity, while H and He remain fixed.

For each planet, we computed a grid of 600 chemical models (except for GJ 3470 b, WASP-107 b, and WASP-39 b, for which we had 900 models because of more C/O ratio values) by varying four key parameters (equilibrium temperature, C/O, metallciity, and eddy diffusion coefficient). More than 80\% of the total models converged for all targets. The few non-convergent cases occurred mainly in the 400 K branch of the grid for hotter planets, where VULCAN is less suited for low-temperature conditions.

\subsection{Representative disequilibrium-grid case: WASP-107 b}

WASP-107 b provides a clear illustration of how free- and grid-chemistry retrievals compare when both models provide good fits. The free-chemistry abundances are consistent with those reported by \citet{welbanks2024high}. The grid-chemistry retrieval yields nearly identical values but with reduced uncertainties, demonstrating that a physically motivated chemical prior can tighten constraints while remaining consistent with the free-chemistry constraints. A detailed comparison (best-fit spectra comparison, abundance posteriors, and corner plots from both retrievals) particularly for WASP-107 b is shown in Figure \ref{fig:W107b_best_fit} and Supplementary Figures \ref{fig:W107b_corner_free} and \ref{fig:W107b_corner_vulcan}, respectively; analogous figures for the full sample are provided in Supplementary Figures \ref{fig:best_fit_corner_HAT-P-18b}–\ref{fig:best_fit_corner_WASP-127b}).

\subsection{Agreement between free and disequilibrium-grid chemical profiles}

Vertical mixing-ratio profiles for $\mathrm{H_2O}$, $\mathrm{CH_4}$, CO, $\mathrm{CO_2}$, and $\mathrm{SO_2}$ from the grid-chemistry retrievals are shown in Figure \ref{fig:chemisrty_extended}, together with our retrieved free-chemistry values and literature values. These species dominate the observed spectral signatures and contribute more in determining the elemental C, O, and S budgets.

As stated earlier, for GJ 3470 b, our free-chemistry retrieval retrieves CO at a much lower abundance than what was inferred by \citet{beatty2024sulfur}. However, the grid-chemistry retrieval also yields a lower CO abundance. The chemical profiles for GJ 3470 b from Figure \ref{fig:chemisrty_extended} indicates that our free-chemistry uncertainties for CO are broad enough for the grid-based value to fall within 1$\sigma$. Our free-chemistry retrieval produces C/O = 0.008$^{\mathrm{\text{+}0.017}}_{\mathrm{-0.003}}$, while the grid-chemistry retrieval yields C/O = 0.02$^{\mathrm{\text{+}0.01}}_{\mathrm{-0.01}}$. These values are statistically consistent with each other. \textcolor{black}{Most planets in our sample show good agreement in the CO abundance between the free- and grid-chemistry retrievals, except WASP-121 b. For this ultra-hot Jupiter, the free-chemistry retrieval yields a higher CO abundance than the grid-based result. This is qualitatively consistent with \citep{gapp2025wasp}, who show that CO dominates the opacity of WASP-121 b at wavelengths longer than $\sim$4.3 $\mu$m. However, they also emphasize that thermal dissociation and day-night heterogeneity can bias the inferred CO, H$_2$O, and C/O values.}

 For $\mathrm{H_2O}$, the grid-based profiles generally agree with the free-chemistry results, except for WASP-80 b and TOI-421 b where the free-retrieval values diverge from chemically expected abundances by single order. $\mathrm{CH_4}$ shows discrepancies for HIP 67522 b, GJ 1214 b, and TrES-4 b. The case of GJ 1214 b remains difficult due to high-altitude clouds that mute the spectral features. The strong 4.2 $\mathrm{\mu}$m $\mathrm{CO_2}$ band detected in many recent JWST spectra is reproduced by both retrieval approaches for most planets. However, the predicted abundance from both cases differ slightly for the planets HIP 67522 b and HD 189733 b. 


\subsection{\textcolor{black}{Comparison with equilibrium-chemistry retrievals}}
\label{sec:appendixB_equilibrium_validation}

\textcolor{black}{To test whether the computational approximations used in the population-level grid-chemistry retrievals could meaningfully affect the inferred C/O values, we performed an additional equilibrium-chemistry retrieval test for two representative planets, TOI-270 d and WASP-15 b. These planets were selected because they probe different regions of the sample in equilibrium temperature: TOI-270 d represents the lower-temperature sub-Neptune regime, whereas WASP-15 b represents a hotter gas-giant case. This comparison therefore provides a targeted check of whether the free-versus-grid C/O differences are robust across substantially different thermal regimes.}

\textcolor{black}{The equilibrium-chemistry retrievals use the petitRADTRANS pre-calculated equilibrium-chemistry framework, in which the molecular abundances are interpolated as a function of pressure, temperature, metallicity, and C/O. Unlike the disequilibrium grid used in the main analysis, this equilibrium setup does not include the full photochemical network. Instead, it provides an independent chemically constrained retrieval framework with a parametric Guillot temperature profile \citep{guillot2010radiative}. We used the Guillot profile rather than an isothermal profile because equilibrium chemistry is evaluated layer by layer and depends explicitly on the local temperature structure. An isothermal profile would suppress the temperature-driven vertical variation in the equilibrium abundances, whereas the Guillot profile provides a simple non-isothermal structure that allows the chemical profiles to vary more realistically with altitude. The priors and retrieved posterior constraints for these equilibrium-chemistry retrievals are listed in Supplementary Table \ref{tab:eq_validation_priors}. Both retrievals were run with MultiNest using 1000 live points, consistent with the sampling setup used for the main retrieval analysis. The molecular opacity species included in these equilibrium-chemistry calculations are H$_2$O, CH$_4$, CO$_2$, CO, HCN, NH$_3$, H$_2$S, and C$_2$H$_2$. Other species used in the disequilibrium grid and also, in free-chemistry retrievals, such as SO$_2$, OCS, CS$_2$, N$_2$O, and PH$_3$, were not included in these retrievals because they are not part of the pre-calculated equilibrium-chemistry grid of petitRADTRANS.}

\textcolor{black}{The resulting best-fit spectra and equilibrium-chemistry posterior distributions are shown in Supplementary Figures \ref{fig:TOI270d_eq_validation} and \ref{fig:WASP15b_eq_validation}. In each case, the best-fit spectrum is compared against the standard free-chemistry retrieval and the disequilibrium chemistry-grid retrieval. The purpose of this comparison is to check whether equilibrium chemistry constrained retrieval framework aligns with the disequilibrium chemistry retrieval results or leads to a qualitatively different C/O inference.}

\textcolor{black}{For TOI-270 d, the free-chemistry retrieval gives C/O = $1.270^{+4.000}_{-0.897}$, while the standard disequilibrium-grid retrieval gives C/O = $0.45^{+0.16}_{-0.12}$. The equilibrium-chemistry retrieval gives C/O = $0.26^{+0.28}_{-0.14}$. Although the median value from the equilibrium-chemistry retrieval is lower than the disequilibrium-grid median, the two chemically constrained estimates overlap within their 1$\sigma$ uncertainties. The equilibrium-chemistry value also overlaps with the broad free-chemistry posterior. Thus, for this lower-temperature planet TOI-270 d, the equilibrium-chemistry comparison does not indicate that the disequilibrium chemistry C/O constraint is driven primarily by the interpolation procedure or by the simplified thermal treatment.}

\textcolor{black}{As a second comparison case, we repeated the same equilibrium-chemistry test for WASP-15 b, which belongs to the hotter gas-giant regime of our sample. This choice allows us to check whether the comparison between the disequilibrium chemistry retrieval and the equilibrium-chemistry retrieval also holds outside the low-temperature sub-Neptune regime represented by TOI-270 d. For WASP-15 b, the free-chemistry retrieval gives C/O = $0.041^{+0.503}_{-0.021}$. The disequilibrium chemistry approach retrieves C/O = $0.14^{+0.05}_{-0.03}$, while the equilibrium-chemistry retrieval gives C/O = $0.27^{+0.22}_{-0.13}$. Although the median C/O values differ, the 1$\sigma$ interval from the equilibrium-chemistry retrieval overlaps with both the free-chemistry and disequilibrium-grid constraints.}
\renewcommand{\thesection}{1.3}
\section{Noise model and likelihood}
\label{sec:noise-likelihood}

For all retrievals in this work, we assume independent Gaussian measurement errors and construct a diagonal covariance matrix using the published 1$\sigma$ uncertainties for each wavelength bin. Let $d_i$ denote the observed transit depth in the $i$-th spectral channel, $m_i(\boldsymbol{\theta})$ the corresponding model prediction for a given parameter vector $\boldsymbol{\theta}$, and $\sigma_{i,\mathrm{pub}}$ the reported 1$\sigma$ uncertainty. We adopt
\begin{equation}
    \sigma_{i,\mathrm{eff}} \;=\; \sigma_{i,\mathrm{pub}}
\end{equation}
for all data points, i.e. we do not introduce any additional jitter or error inflation terms in the likelihood. The effective covariance matrix is therefore purely diagonal,
\begin{equation}
    \mathbf{C}_{ij} \;=\; \sigma_{i,\mathrm{eff}}^2 \,\delta_{ij},
\end{equation}
with no off-diagonal terms to account for correlated noise between spectral channels or between instruments/modes. When combining multiple datasets (e.g. different JWST or HST modes), we assume that the noise between modes is uncorrelated, so that the full covariance matrix is block-diagonal with the above form for each mode.

Under these assumptions, the log-likelihood for a given model spectrum $m_i(\boldsymbol{\theta})$ is
\begin{equation}
    \ln \mathcal{L}(\boldsymbol{\theta}) \;=\;
    -\frac{1}{2}\sum_{i}
    \left[
        \frac{\left(d_i - m_i(\boldsymbol{\theta})\right)^2}{\sigma_{i,\mathrm{eff}}^2}
        + \ln\left(2\pi \sigma_{i,\mathrm{eff}}^2\right)
    \right],
    \label{eq:loglike}
\end{equation}
which is the standard Gaussian likelihood adopted in the nested-sampling retrievals.

For diagnostic purposes, we compute the reduced chi-squared, $\chi^2_{\mathrm{r}}$, using the same effective uncertainties,
\begin{equation}
    \chi^2_{\mathrm{r}} \;=\;
    \frac{1}{\nu}\sum_i
    \frac{\left(d_i - m_i(\boldsymbol{\theta}_{\mathrm{bf}})\right)^2}{\sigma_{i,\mathrm{eff}}^2},
    \label{eq:chisq_red}
\end{equation}
\textcolor{black}{where $\boldsymbol{\theta}_{\mathrm{bf}}$ denotes the maximum-posterior (or maximum-likelihood) parameter set. We report $\chi^2_{\rm r}$ using the nominal choice of degrees of freedom $\nu = N_{\mathrm{data}} - N_{\mathrm{free}}$, where $N_{\mathrm{data}}$ is the total number of data points and $N_{\mathrm{free}}$ is the number of free parameters in the retrieval. However, for nonlinear retrieval models with priors and parameter degeneracies, this nominal degree of freedom estimate is only approximate \citep{andrae2010and}. We therefore use $\chi^2_{\rm r}$ only as a descriptive fit-quality indicator.}

\textcolor{black}{We emphasize that this likelihood does not model wavelength-correlated residual structure, for example from instrument systematics or forward-model incompleteness. It also does not include a diagonal error-inflation or jitter term that would rescale the published uncertainties \citep{line2015uniform}. Such a term could broaden posterior distributions and reduce overconfident abundance constraints when the published uncertainties underestimate the true scatter. However, it would not by itself impose chemically consistent abundance relationships, and therefore cannot replace chemically informed retrievals. More complete treatments, including diagonal error inflation and correlated-noise models such as Gaussian-process likelihoods, should be explored in future population-level retrieval studies.}

\clearpage
\renewcommand{\figurename}{Supplementary Fig.}
\renewcommand{\thefigure}{1}
\begin{figure}[h!]
\centering
	\includegraphics[width=\columnwidth]{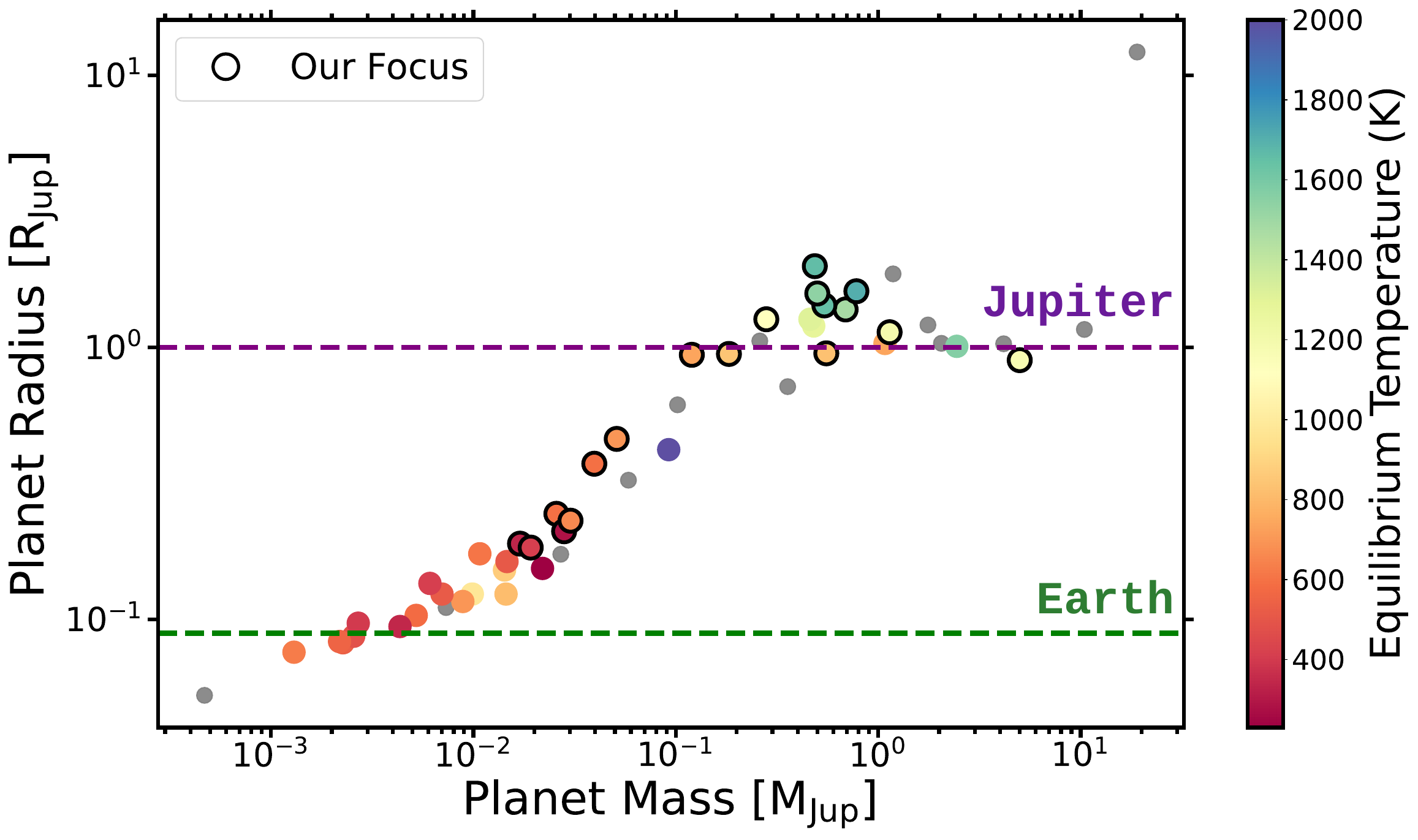}
  \caption{Mass-radius distribution of exoplanets observed with JWST. Gray points represent all planets observed with JWST (in transit, eclipse, or both), while colored points mark those observed specifically during transit. Planets highlighted with both color and a black outline denote the tailored sample used in this study. The sample selection is based on three criteria: (1) planets belonging to the Jupiter- or Neptune-class, (2) JWST observations during transit, and (3) publicly released, reduced spectra accessible to the community. The green and purple dashed line represents the Earth's and Jupiter's radius, respectively.}
   \label{fig:population}
\end{figure}

\renewcommand{\tablename}{Supplementary Table}
\renewcommand{\thetable}{1}
\begin{sidewaystable}
\caption{Exoplanet samples used in this study. For each planet, we list the system parameters (radius, mass, and equilibrium temperature) together with the literature references from which they were taken. We also provide the references for all Hubble and JWST transmission spectra used in this work, along with the corresponding JWST observing modes and program IDs. The mass of HIP 67522 b is taken from the \href{https://exo.mast.stsci.edu/}{Exo.MAST} database.}
\label{tab:exoplanet_properties}
\begin{tabular*}{\textheight}{@{\extracolsep\fill}lcccccccc}
\toprule%
Planet & Radius  & Mass  & $\mathrm{T_{eq}}$ & Parameter & \textit{Hubble}   & \textit{JWST} & \textit{JWST}   &  Program \\
 &  ($\mathrm{R_J}$) &  ($\mathrm{M_J}$) &  (K) & Ref. & Spectrum Ref. & Mode & Spectrum Ref.   & ID \\
\midrule
HAT-P-18 b       & 0.995 $\pm$ 0.052 & 0.197 $\pm$ 0.013 & 852 $\pm$ 28     & \cite{hartman2010hat}  &   \cite{tsiaras2018population}   & NIRISS  & \cite{fu2022water} & ERO 2734\\
\noalign{\smallskip}
WASP-17 b        & 1.87 $\pm$ 0.24 & 0.780 $\pm$ 0.23 & 1755 $\pm$ 28 & \cite{2017AJ....153..136S}, \cite{2012MNRAS.426.1338S}    &  \cite{saba2022transmission}     &  NIRISS \text{+} MIRI   &   \cite{louie2025jwst}  &  GTO 1353\\
\noalign{\smallskip}
GJ 3470 b        & 0.408 $\pm$ 0.016 & 0.0437 $\pm$ 0.0047 & 593.5 $\pm$ 96.5  & \cite{2016MNRAS.463.2574A}    &  \cite{beatty2024sulfur}   &   NIRCam   &   \cite{beatty2024sulfur}  & GTO 1185 \\
\noalign{\smallskip}
GJ 1214 b        & 0.2438 $\pm$ 0.0028 & 0.0265 $\pm$ 0.0011 & 567 $\pm$ 7.6    &   \cite{2024ApJ...963L..37M}   & \cite{berta2012flat}     &  NIRSpec   &    \cite{schlawin2024possible}  &  GTO 1185\\
\noalign{\smallskip}
        &  &  &      &      &      &  MIRI    &  \cite{kempton2023reflective}   &  GO 1803\\
\noalign{\smallskip}
K2-18 b          & 0.211 $\pm$ 0.020 & 0.0281 $\pm$ 0.00519 & 284 $\pm$  15   &  \cite{2018AJ....155..257S}   &  \cite{benneke2019water}    & NIRISS     & \cite{madhusudhan2023carbon}    &  GO 2722\\
\noalign{\smallskip}
        &  &  &      &      &      &  NIRSpec   &   \cite{madhusudhan2023carbon}  &   GO 2722\\
\noalign{\smallskip}
                &  &  &      &      &      &  MIRI   &  \cite{2025ApJ...983L..40M}   &   GO 2722\\
\noalign{\smallskip}
TOI-270 d        & 0.178 $\pm$ 0.004 & 0.0132 $\pm$ 0.0005 & 383  $\pm$ 9     &  \cite{2022MNRAS.510.5464K}     &  \cite{mikal2023hubble}    &  NIRSpec    &  \cite{holmberg2024possible}   &   GO 4098\\
\noalign{\smallskip}
HD 189733 b      & 1.119 $\pm$ 0.038 & 1.166 $\pm$  0.050 & 1209 $\pm$ 11  &  \cite{2019PASP..131k5003A}     &   \cite{sing2016continuum}   &  NIRCam    &    \cite{fu2024hydrogen} &  GO 1633\\
\noalign{\smallskip}
HD 209458 b      & 1.4  & 0.690   & 1450   &  \cite{salz2015high}    &  \cite{tsiaras2016new}    &  NIRCam    &   \cite{xue2024jwst}  &  GTO 1274\\
\noalign{\smallskip}
HIP 67522 b      & 0.891 $\pm$ 0.020 & 0.225  & 1175 $\pm$ 12.5     &  \cite{2024ApJ...973L..30B}     &   --   &  NIRSpec    &  \cite{thao2024featherweight}   &  GO 2498\\
\noalign{\smallskip}
WASP-39 b      & 1.279 $\pm$ 0.040 & 0.281 $\pm$ 0.032  & 1166 $\pm$ 14     & \cite{mancini2018gaps}    & --    &  NIRISS     &    \cite{carter2024benchmark} & ERS 1366 \\
\noalign{\smallskip}
        &  &  &      &      &      &  NIRSpec   &   \cite{carter2024benchmark}   &   ERS 1366\\
\noalign{\smallskip}
        &  &  &      &      &      &  MIRI   &   \cite{powell2024sulfur}  &   DD-2783\\
\noalign{\smallskip}
WASP-80 b        & 0.999 $\pm$ 0.0305 & 0.538 $\pm$ 0.0355   & 825 $\pm$ 19      &  \cite{2015MNRAS.450.2279T}    & \cite{wong2022hubble}     &   NIRCam    &  \cite{bell2023methane}   &  GTO 1185\\
\noalign{\smallskip}
WASP-107 b       & 0.924 $\pm$ 0.022 & 0.119 $\pm$ 0.014   & 736 $\pm$ 17   &   \cite{2017MNRAS.469.1622M}    & \cite{kreidberg2018water}     &  NIRCam     &  \cite{welbanks2024high}   &  GTO 1185\\
\noalign{\smallskip}
WASP-15 b        & 1.408 $\pm$ 0.046 & 0.592 $\pm$ 0.019  & 1676 $\pm$ 29  &   \cite{2013MNRAS.434.1300S}    &  --    & NIRSpec      &  \cite{kirk2025bowie}   &  GO 3838\\
\noalign{\smallskip}
TOI-836 c        & 0.231 $\pm$ 0.0079 & 0.0302 $\pm$ 0.0082  & 665 $\pm$ 27      &   \cite{2023MNRAS.520.3649H}    & --     &  NIRSpec     &   \cite{wallack2024jwst}  &  GO 2512\\
\noalign{\smallskip}
TOI-776 c        & 0.1826 $\pm$ 0.0071 & 0.022 $\pm$ 0.008  & 420  $\pm$ 10     &     \cite{fridlund2024planets}   & --     &  NIRSpec   &    \cite{teske2025jwst}   &  GO 2512\\
\noalign{\smallskip}
TrES-4 b         & 1.706 $\pm$ 0.056 & 0.917 $\pm$ 0.070  & 1778 $\pm$ 22  &      \cite{turner2016ground}    &  --  & NIRSpec     &   \cite{meech2025bowie}  &  GO 3838\\
\noalign{\smallskip}
TOI-421 b        & 0.236 $\pm$ 0.007 & 0.021 $\pm$ 0.002 & 922  $\pm$ 14     &     \cite{krenn2024characterisation}    &  --   &  NIRISS \text{+} NIRSpec\footnotemark[1]   &  \cite{davenport2025toi}   &  GO
1935\\
\noalign{\smallskip}
WASP-94 A b       & 1.72 $\pm$ 0.055  & 0.452 $\pm$ 0.033  & 1604 $\pm$ 23.5  &     \cite{neveu2014wasp}   &  --    &  NIRSpec    &   \cite{ahrer2025tracing}  &  GO 3154\\
\noalign{\smallskip}
WASP-121 b   & 1.865 $\pm$ 0.044 & 1.183 $\pm$ 0.063  & 2358 $\pm$ 52 &    \cite{2016MNRAS.458.4025D}    &   --    & NIRSpec    &  \cite{gapp2025wasp}   &  GO 1729\\
\noalign{\smallskip}
WASP-127 b  & 1.37 $\pm$ 0.04 & 0.18 $\pm$ 0.02 & 1400 $\pm$ 24 &     \cite{lam2017dense}    &  \cite{skaf2020ares}   &   NIRSpec   &   \cite{fu2025statistical}   & GO 2437 \\
\noalign{\smallskip}
\botrule
\end{tabular*}
\footnotetext[1]{\textit{Note}: TOI-421 b is observed in NIRSpec G395M mode. Other NIRSpec observations refer to G395H mode.}
\footnotetext{\textit{Note}: $\mathrm{R_J}$ is the radius of the Jupiter and $\mathrm{M_J}$ is the mass of the Jupiter. $\mathrm{T_{eq}}$ represents the equilibrium temperature of the planet.}
\end{sidewaystable}

\renewcommand{\tablename}{Supplementary Table}
\renewcommand{\thetable}{2}
\begin{sidewaystable*}
\rotatebox{0}{
           \begin{minipage}{\textwidth}
\centering
\caption{Retrieved molecular abundances and C/O from the free-chemistry analysis, compared with previous published results. For each planet, the first row (``This Work”) lists the median and 1$\sigma$ intervals of the mass fractions in log scale for the main investigated species. The second row gives the corresponding constraints reported in the literature, using the references indicated in the second column. A ``--” entry denotes species that were not included or not reported in the original study, while ``$<$” values indicate reported upper limits. These abundances form the basis for the C/O ratios estimation for the literature case, shown in Figure \ref{fig:co_vs_Tpl_free}.}
\label{tab:molecule_abundance}
\resizebox{\textwidth}{!}{
\begin{tabular}{cccccccccccccccc}
        \hline
        \hline
        \noalign{\smallskip}
       Name & Reference & log$\mathrm{_{10}}$($\mathrm{H_2O}$) & log$\mathrm{_{10}}$($\mathrm{CH_4}$) & log$\mathrm{_{10}}$($\mathrm{CO}$) & log$\mathrm{_{10}}$($\mathrm{CO_2}$) &log$\mathrm{_{10}}$($\mathrm{SiO}$) & log$\mathrm{_{10}}$($\mathrm{HCN}$) & log$\mathrm{_{10}}$($\mathrm{H_2S}$) & log$\mathrm{_{10}}$($\mathrm{SO_2}$) &  log$\mathrm{_{10}}$($\mathrm{C_2H_2}$) & log$\mathrm{_{10}}$($\mathrm{CS_2}$) & log$\mathrm{_{10}}$($\mathrm{SCO}$) & log$\mathrm{_{10}}$($\mathrm{N_2O}$)   & log$\mathrm{_{10}}$($\mathrm{AlO}$) & C/O\footnotemark[1]\\
        \noalign{\smallskip}
        \hline
        \noalign{\smallskip}
        HAT-P-18 b & This Work & -2.58 $\mathrm{^{\text{\text{+}}0.47}_{-0.33}}$ & -5.84$\mathrm{^{\text{\text{+}} 1.35}_{-3.93}}$ & -3.51$\mathrm{^{\text{\text{+}} 1.94}_{-4.89}}$ & -3.75$\mathrm{^{\text{\text{+}} 0.89}_{-2.76}}$ & -- & -8.45$\mathrm{^{\text{\text{+}} 2.25}_{-2.19}}$ & -8.07$\mathrm{^{\text{\text{+}} 2.54}_{-2.47}}$ & -6.99$\mathrm{^{\text{\text{+}} 3.13}_{-3.17}}$ & -- & -- & -- & -- & --  & 0.157$\mathrm{^{\text{+} 0.701}_{- 0.137}}$\\ 
        \noalign{\smallskip}
        & \cite{fournier2024near} & -4.47$\mathrm{^{\text{+} 0.30}_{- 0.28}}$ & $<$ -6.29 & $<$ -2.58 & -4.86$\mathrm{^{\text{+} 0.45}_{- 0.45}}$ & -- & $<$ -5.51 & -- & -- & -- & -- & -- & -- & -- & 0.125$\mathrm{^{\text{+} 0.0351}_{- 0.0331}}$ \\
         \noalign{\smallskip}
        \hline
        \noalign{\smallskip}
        WASP-17 b & This Work & \textcolor{black}{-1.34$\mathrm{^{\text{+} 0.37}_{-0.47}}$} &\textcolor{black}{ -8.19$\mathrm{^{\text{+} 2.43}_{-2.45}}$} & \textcolor{black}{-1.97$\mathrm{^{\text{+} 0.75}_{-4.93}}$} & \textcolor{black}{-8.10$\mathrm{^{\text{+} 2.48}_{-2.53}}$} & -- & -- & \textcolor{black}{-7.78$\mathrm{^{\text{+} 2.76}_{-2.69}}$} & \textcolor{black}{-7.71$\mathrm{^{\text{+} 2.33}_{-2.11}}$} & -- & -- & -- & -- & \textcolor{black}{-6.36$\mathrm{^{\text{+} 0.50}_{- 0.67}}$}   & 0.127$^{+0.476}_{-0.125}$\\ 
        \noalign{\smallskip}
        & \cite{louie2025jwst} & -2.96$\mathrm{^{\text{+} 0.31}_{- 0.24}}$ & -9.03$\mathrm{^{\text{+} 2.00}_{- 1.93}}$ & -3.97$\mathrm{^{\text{+} 1.00}_{- 4.59}}$ & -5.80$\mathrm{^{\text{+} 0.62}_{- 1.31}}$ & -- & -- & -- & -- & -- & -- & -- & --& -- & 0.0596$\mathrm{^{\text{+} 0.176}_{- 0.0589}}$\\
         \noalign{\smallskip}
        \hline
        \noalign{\smallskip}
        GJ 3470 b & This Work & -1.39$\mathrm{^{\text{+} 0.57}_{-0.87}}$ & -3.87$\mathrm{^{\text{+} 0.50}_{-0.76}}$ & -6.42$\mathrm{^{\text{+} 3.29}_{-3.58}}$ & -3.61$\mathrm{^{\text{+} 0.58}_{-0.82}}$ & -- & -8.37$\mathrm{^{\text{+} 2.31}_{-2.36}}$ & -- & -3.44$\mathrm{^{\text{+} 0.66}_{-0.98}}$ & -- & -- & -- & -- & -- & 0.008$\mathrm{^{\text{+} 0.017}_{- 0.003}}$ \\
        \noalign{\smallskip}
        & \cite{beatty2024sulfur} & -1.08$\mathrm{^{\text{+} 0.43}_{- 0.52}}$ & -4.05$\mathrm{^{\text{+} 0.25}_{- 0.27}}$ & -0.96$\mathrm{^{\text{+} 0.18}_{- 0.70}}$ & -2.47$\mathrm{^{\text{+} 0.61}_{- 0.43}}$ & -- & -8.14$\mathrm{^{\text{+} 2.55}_{- 2.46}}$ & -- & -3.57$\mathrm{^{\text{+} 0.26}_{- 0.25}}$ & -- & -- & -- & -- & --  & 0.460$\mathrm{^{\text{+} 0.000}_{- 0.096}}$  \\
         \noalign{\smallskip}
        \hline
        \noalign{\smallskip}
        GJ 1214 b & This Work & -1.47$\mathrm{^{\text{+} 0.81}_{-4.37}}$ & -5.64$\mathrm{^{\text{+}2.30}_{-3.71}}$ & -6.63$\mathrm{^{\text{+} 3.28}_{-3.23}}$ & -2.75$\mathrm{^{\text{+} 1.12}_{-1.35}}$ & -- & -7.28$\mathrm{^{\text{+} 2.99}_{-2.78}}$ & -6.53$\mathrm{^{\text{+} 3.17}_{-3.19}}$ & -7.80$\mathrm{^{\text{+} 2.48}_{-2.45}}$ & -7.96$\mathrm{^{\text{+} 2.50}_{-2.46}}$ & -7.22$\mathrm{^{\text{+} 3.04}_{-2.84}}$ & -7.93$\mathrm{^{\text{+} 2.50}_{-2.43}}$ & -- & -- & 0.004$\mathrm{^{\text{+}0.575 }_{- 0.004}}$ \\ 
         \noalign{\smallskip}
        & \cite{schlawin2024possible} & -- & -2.25$\mathrm{^{\text{+} 0.72}_{- 1.08}}$ & -- & -- & -- & -- & -- & -- & -- & -- & -- & -- & -- & -- \\
         \noalign{\smallskip}
        \hline
        \noalign{\smallskip}
        K2-18 b & This Work & -3.90$\mathrm{^{\text{+} 2.13}_{-4.76}}$ & -0.85$\mathrm{^{\text{+} 0.32}_{-0.46}}$ & -8.13$\mathrm{^{\text{+} 2.75}_{-2.56}}$ & -1.84$\mathrm{^{\text{+} 0.86}_{-1.30}}$ & -- & -7.31$\mathrm{^{\text{+} 2.79}_{-2.95}}$ & -- & -- & -- & -7.66$\mathrm{^{\text{+} 2.85}_{-2.77}}$ & -7.51$\mathrm{^{\text{+} 2.39}_{-2.91}}$ & -5.60$\mathrm{^{\text{+} 1.51}_{-3.14}}$ & -- & 6.540$\mathrm{^{\text{+} 62.2}_{- 4.99}}$\\ 
        \noalign{\smallskip}
        & \cite{madhusudhan2023carbon} & $<$ -3.49 & -1.89$\mathrm{^{\text{+} 0.63}_{- 0.70}}$ & $<$ -3.19 & -2.05$\mathrm{^{\text{+} 0.50}_{- 0.84}}$ & -- & $<$ -3.21 & -- & -- & -- & -- & -- & -- & -- & 2.48$\mathrm{^{\text{+} 0.457}_{- 0.000}}$ \\
         \noalign{\smallskip}
        \hline
        \noalign{\smallskip}
        TOI-270 d & This Work & -1.31$\mathrm{^{\text{+} 0.49}_{-0.91}}$ & -1.51$\mathrm{^{\text{+} 0.41}_{-0.48}}$ & -6.45$\mathrm{^{\text{+} 3.55}_{-3.49}}$ & -2.27$\mathrm{^{\text{+} 0.95}_{-1.11}}$ & -- & -8.50$\mathrm{^{\text{+} 2.33}_{-2.21}}$ & -2.27$\mathrm{^{\text{+} 1.52}_{-5.66}}$ & -7.01$\mathrm{^{\text{+} 2.81}_{-3.11}}$ & -8.87$\mathrm{^{\text{+} 2.12}_{-2.01}}$ & -0.81$\mathrm{^{\text{+} 0.32}_{-0.50}}$ & -4.54$\mathrm{^{\text{+} 0.75}_{-1.15}}$ & -7.72$\mathrm{^{\text{+} 2.47}_{-2.67}}$ &  -- & 1.270$\mathrm{^{\text{+} 4.000}_{- 0.897}}$\\
        \noalign{\smallskip}
        & \cite{holmberg2024possible} & -3.92$\mathrm{^{\text{+} 1.70}_{- 4.43}}$ & -2.44$\mathrm{^{\text{+} 0.34}_{- 0.46}}$ & -4.98$\mathrm{^{\text{+} 2.52}_{- 4.03}}$ & -1.96$\mathrm{^{\text{+} 0.49 }_{- 0.79}}$ & -- & -8.84$\mathrm{^{\text{+} 2.13}_{- 1.94}}$ & -6.16$\mathrm{^{\text{+} 3.25}_{- 3.57}}$ & -8.03$\mathrm{^{\text{+} 2.31}_{- 2.35}}$ & -9.31$\mathrm{^{\text{+} 1.73}_{- 1.65}}$ & -2.59$\mathrm{^{\text{+} 0.67}_{- 0.95}}$ & -8.42$\mathrm{^{\text{+} 2.06}_{- 2.15}}$ & -8.80$\mathrm{^{\text{+} 2.07}_{- 1.94}}$ & -- & 1.010$\mathrm{^{\text{+} 0.000}_{- 0.000}}$ \\
         \noalign{\smallskip}
        \hline
        \noalign{\smallskip}
        HD 189733 b & This Work & -3.09$\mathrm{^{\text{+} 0.67}_{-0.42}}$ & -9.67$\mathrm{^{\text{+} 1.58}_{-1.55}}$ & -3.06$\mathrm{^{\text{+} 0.81}_{-0.54}}$ & -5.22$\mathrm{^{\text{+} 0.60}_{-0.39}}$ & -- & -7.93$\mathrm{^{\text{+} 1.81}_{-2.65}}$ & -3.53$\mathrm{^{\text{+} 0.47}_{-0.32}}$ & -8.85$\mathrm{^{\text{+} 2.19}_{-2.04}}$ & -6.96$\mathrm{^{\text{+} 0.87}_{-3.19}}$ & -- & -- & -- & -- & 0.412$\mathrm{^{\text{+} 0.127}_{- 0.114}}$\\ 
        \noalign{\smallskip}
        & \cite{fu2024hydrogen} & -3.48$\mathrm{^{\text{+} 0.93}_{- 0.63}}$ & -10.13$\mathrm{^{\text{+} 1.35}_{- 1.15}}$ & -4.02$\mathrm{^{\text{+} 1.08}_{- 0.70}}$ & -6.09$\mathrm{^{\text{+} 0.83}_{- 0.55}}$ & -- & -8.49$\mathrm{^{\text{+} 1.55}_{- 2.20}}$ & -4.39$\mathrm{^{\text{+} 0.68}_{- 0.47}}$ & -9.49$\mathrm{^{\text{+} 1.53}_{- 1.58}}$ & -7.55$\mathrm{^{\text{+} 0.68}_{- 2.15}}$ & -- & -- & -- & -- & 0.157$\mathrm{^{\text{+} 0.051}_{- 0.020}}$ \\
         \noalign{\smallskip}
        \hline
        \noalign{\smallskip}
        HD 209458 b & This Work & -3.87$\mathrm{^{\text{+} 1.79}_{-0.34}}$ & -8.56$\mathrm{^{\text{+} 1.36}_{-2.14}}$ & -7.04$\mathrm{^{\text{+} 2.35}_{-3.13}}$ & -6.59$\mathrm{^{\text{+} 1.66}_{-0.24}}$ & -- & -9.23$\mathrm{^{\text{+} 1.83}_{-1.81}}$ & -8.53$\mathrm{^{\text{+} 2.39}_{-2.27}}$ & -9.01$\mathrm{^{\text{+} 2.00}_{-1.92}}$ & -9.32$\mathrm{^{\text{+} 1.79}_{-1.72}}$ & -- & -- & -- &  -- & 0.002$\mathrm{^{\text{+} 0.003}_{- 0.001}}$\\ 
        \noalign{\smallskip}
        & \cite{bachmann2025} &  -2.80$\mathrm{^{\text{+} 0.42}_{- 0.41}}$ & -9.42$\mathrm{^{\text{+} 2.24}_{- 2.26}}$ &  -4.15$\mathrm{^{\text{+} 0.51}_{- 0.30}}$ &  -6.16$\mathrm{^{\text{+} 0.38}_{- 0.39}}$ & -- & -9.74$\mathrm{^{\text{+} 2.32}_{- 2.21}}$ & -- & -- & -8.17$\mathrm{^{\text{+} 1.09 }_{- 1.22}}$ & -- & -- & -- & -- & 0.028$\mathrm{^{\text{+} 0.006}_{- 0.000}}$ \\
         \noalign{\smallskip}
        \hline
        \noalign{\smallskip}
        HIP 67522 b & This Work & -2.27$\mathrm{^{\text{+} 0.47}_{-1.15}}$ & -9.47$\mathrm{^{\text{+} 1.61}_{-1.55}}$ & -2.05$\mathrm{^{\text{+} 0.63}_{-1.43}}$ & -3.91$\mathrm{^{\text{+} 0.49}_{-1.22}}$ & -- & -- & -3.35$\mathrm{^{\text{+} 0.46}_{-1.06}}$  & -5.47$\mathrm{^{\text{+} 0.47}_{-0.90}}$ & -- & -- & -- & -- &  -- & 0.495$\mathrm{^{\text{+} 0.147}_{- 0.145}}$\\ 
        \noalign{\smallskip}
        & \cite{thao2024featherweight} & -3.56$\mathrm{^{\text{+} 0.81}_{- 0.96}}$ & -- & -3.91$\mathrm{^{\text{+} 1.01}_{- 1.09}}$ & -5.56$\mathrm{^{\text{+} 0.81}_{- 0.95}}$ & -- & -- & -4.91$\mathrm{^{\text{+} 0.65}_{- 0.78}}$ & -7.26$\mathrm{^{\text{+} 0.55}_{- 0.57}}$ & -- & -- & -- & -- & -- & 0.225$\mathrm{^{\text{+} 0.089}_{- 0.047}}$ \\
         \noalign{\smallskip}
        \hline
        \noalign{\smallskip}
        WASP-80 b & This Work & -0.47$\mathrm{^{\text{+} 0.12}_{-0.37}}$ & -2.67$\mathrm{^{\text{+} 0.32}_{-0.40}}$ & -6.50$\mathrm{^{\text{+} 3.74}_{-3.44}}$ & -6.90$\mathrm{^{\text{+} 3.47}_{-3.32}}$ & -- & -- & --  & -7.76$\mathrm{^{\text{+} 2.90}_{-2.66}}$ & -- & -- & -- & -- & -- & 0.011$\mathrm{^{\text{+} 0.014}_{- 0.005}}$\\ 
        \noalign{\smallskip}
        & \cite{bell2023methane} & -1.80$\mathrm{^{\text{+}0.55}_{-0.94}}$ & -4.22$\mathrm{^{\text{+}0.47}_{-0.74}}$ & -7.05$\mathrm{^{\text{+}3.13}_{-2.96}}$ & -8.00$\mathrm{^{\text{+}2.52}_{-2.48}}$ & -- & -- & -- & -8.95$\mathrm{^{\text{+}2.00}_{-1.88}}$ & -- & -- & -- & -- & -- & 0.0043$\mathrm{^{\text{+} 0.0007}_{- 0.0000}}$ \\
         \noalign{\smallskip}
        \hline
        \noalign{\smallskip}
        WASP-107 b & This Work & -3.41$\mathrm{^{\text{+} 0.58}_{-0.36}}$ & -6.40$\mathrm{^{\text{+} 0.48}_{-0.29}}$ & -3.27$\mathrm{^{\text{+} 0.82}_{-0.51}}$ & -5.11$\mathrm{^{\text{+} 0.59}_{-0.36}}$ & -- & -- & -8.47$\mathrm{^{\text{+} 2.65}_{-2.40}}$  & -5.00$\mathrm{^{\text{+} 0.47}_{-0.29}}$ & -- & -- & -- & -- &  -- & 0.481$\mathrm{^{\text{+} 0.162}_{- 0.150}}$\\ 
        \noalign{\smallskip}
        & \cite{welbanks2024high} & -2.60$\mathrm{^{\text{+}0.7}_{-0.6}}$ & -6.10$\mathrm{^{\text{+}0.5}_{-0.4}}$ & -3.00$\mathrm{^{\text{+}0.7}_{-0.7}}$ & -4.40$\mathrm{^{\text{+}0.8}_{-0.7}}$ & -- & -- & -8.60$\mathrm{^{\text{+}2.1}_{-2.2}}$ & -5.70$\mathrm{^{\text{+}0.4}_{-0.4}}$ & -- & -- & -- & -- & -- & 0.207$\mathrm{^{\text{+} 0.0007}_{- 0.0349}}$ \\
         \noalign{\smallskip}
        \hline
        \noalign{\smallskip}
         WASP-15 b & This Work & -0.93$\mathrm{^{\text{+} 0.46}_{-1.18}}$ & -8.23$\mathrm{^{\text{+} 2.37}_{-2.25}}$ & -4.12$\mathrm{^{\text{+} 2.77}_{-4.84}}$ & -2.16$\mathrm{^{\text{+} 0.58}_{-1.24}}$ & -- & -7.89$\mathrm{^{\text{+} 2.84}_{-2.53}}$ & -6.91$\mathrm{^{\text{+} 2.89}_{-3.02}}$ & -7.63$\mathrm{^{\text{+} 2.82}_{-2.69}}$ & -8.02$\mathrm{^{\text{+} 2.52}_{-2.43}}$ & -- & -5.36$\mathrm{^{\text{+} 1.60}_{-3.94}}$ & -- & -- & 0.041$\mathrm{^{\text{+}0.503 }_{-0.021 }}$\\ 
         \noalign{\smallskip}
          & \cite{kirk2025bowie} & -1.94$\mathrm{^{\text{+}0.59}_{-0.95}}$ & -8.12$\mathrm{^{\text{+}2.10}_{-2.33}}$ & -5.47$\mathrm{^{\text{+} 3.39}_{- 3.99}}$ & -3.12$\mathrm{^{\text{+} 0.73}_{- 1.02}}$ & -- & -- & -7.50$\mathrm{^{\text{+} 3.08}_{- 2.56}}$ & -7.70$\mathrm{^{\text{+} 2.56}_{- 2.61}}$ & -- & -- & -6.00$\mathrm{^{\text{+} 1.31}_{- 3.29}}$ & -- & -- & 0.026$\mathrm{^{\text{+} 0.106}_{- 0.017}}$ \\
         \noalign{\smallskip}
        \hline
        \noalign{\smallskip}
        TOI-836 c & This Work & -4.34$\mathrm{^{\text{+} 3.08}_{-4.71}}$ & -1.74$\mathrm{^{\text{+} 0.53}_{-0.84}}$ & -6.86$\mathrm{^{\text{+} 3.32}_{-3.14}}$ & -5.82$\mathrm{^{\text{+} 2.44}_{-3.41}}$ & -- & -5.01$\mathrm{^{\text{+} 2.90}_{-4.10}}$ & -6.65$\mathrm{^{\text{+} 3.22}_{-3.28}}$ & -8.34$\mathrm{^{\text{+} 2.47}_{-2.24}}$ & -2.84$\mathrm{^{\text{+} 1.10}_{-3.02}}$ & -7.59$\mathrm{^{\text{+} 2.78}_{-2.58}}$ & -8.08$\mathrm{^{\text{+} 2.36}_{-2.39}}$ & -- & -- & 27.5$\mathrm{^{\text{+}3.08\times 10^3}_{- 27.2}}$\\ 
        \noalign{\smallskip}
       & -- & -- & -- & -- & -- & -- & -- & -- & -- & -- & -- & -- & -- & -- & -- \\
         \noalign{\smallskip}
        \hline
        \noalign{\smallskip}
        TOI-776 c & This Work & -5.81$\mathrm{^{\text{+} 3.39}_{-3.57}}$ & -4.52$\mathrm{^{\text{+} 2.74}_{-4.12}}$ & -6.23$\mathrm{^{\text{+} 3.49}_{-3.23}}$ & -6.54$\mathrm{^{\text{+} 3.69}_{-3.33}}$ & -- & -6.84$\mathrm{^{\text{+} 3.37}_{-2.97}}$ & -6.36$\mathrm{^{\text{+} 3.31}_{-3.32}}$ & -6.43$\mathrm{^{\text{+} 3.49}_{-3.26}}$ & -6.87$\mathrm{^{\text{+} 3.28}_{-3.05}}$ & -5.92$\mathrm{^{\text{+} 3.21}_{-3.36}}$ & -6.84$\mathrm{^{\text{+} 3.25}_{-2.91}}$ & -- & -- & 1.26$\mathrm{^{\text{+} 1.56\times 10^2}_{- 1.09}}$\\ 
       \noalign{\smallskip}
        & -- & -- & -- & -- & -- & -- & -- & -- & -- & -- & -- & -- & -- & -- & -- \\
         \noalign{\smallskip}
        \hline
        \noalign{\smallskip}
        TrES-4 b & This Work & -2.08$\mathrm{^{\text{+} 0.53}_{-0.56}}$ & -8.99$\mathrm{^{\text{+} 1.81}_{-1.89}}$ & -1.99$\mathrm{^{\text{+} 0.71}_{-0.83}}$ & -5.66$\mathrm{^{\text{+} 0.50}_{-0.52}}$ & -- & -8.87$\mathrm{^{\text{+} 1.95}_{-1.94}}$ & -6.11$\mathrm{^{\text{+} 2.35}_{-3.72}}$  & -8.86$\mathrm{^{\text{+} 2.01}_{-1.98}}$ & -- & -- & -- & -- & -- & 0.440$\mathrm{^{\text{+} 0.215}_{- 0.228}}$\\ 
        \noalign{\smallskip}
        & \cite{meech2025bowie} & -3.26$\mathrm{^{\text{+} 0.51}_{- 0.56}}$ & $<$ -9.28 & -3.74$\mathrm{^{\text{+} 0.80}_{- 0.83}}$ & -7.13$\mathrm{^{\text{+} 0.49}_{- 0.52}}$ & -- & $<$ -7.14 & $<$ -4.57 & $<$ -7.12 & -- & -- & -- & --  & -- & 0.176$\mathrm{^{\text{+} 0.118}_{- 0.073}}$\\
        \noalign{\smallskip}
        \hline
        \noalign{\smallskip}
        TOI-421 b & This Work & -3.26$\mathrm{^{\text{+} 0.90}_{-0.59}}$ & -9.05$\mathrm{^{\text{+} 2.07}_{-1.87}}$ & -6.70$\mathrm{^{\text{+} 2.95}_{-3.32}}$ & -9.17$\mathrm{^{\text{+} 2.00}_{-1.80}}$ & -- & -7.86$\mathrm{^{\text{+} 2.64}_{-2.78}}$ & -7.16$\mathrm{^{\text{+} 2.84}_{-3.02}}$  & -4.08$\mathrm{^{\text{+} 0.96}_{-1.07}}$ & -- & -- & -- & -- & -- & 0.031$\mathrm{^{\text{+} 0.158}_{- 0.030}}$\\ 
        \noalign{\smallskip}
        & \cite{davenport2025toi} & -3.81$\mathrm{^{\text{+} 1.32}_{- 0.70}}$ & -8.39$\mathrm{^{\text{+} 1.91}_{- 2.18}}$ & -6.42$\mathrm{^{\text{+} 2.99}_{- 3.41}}$ & -9.14$\mathrm{^{\text{+} 2.02}_{- 1.71}}$ & -- & -- & -- & -5.83$\mathrm{^{\text{+} 1.32}_{- 2.16}}$ & -- & -- & -- & --  & -- & 0.002$\mathrm{^{\text{+} 0.0669}_{- 0.0016}}$\\
        
        \noalign{\smallskip}
        \hline
        \hline
\end{tabular}}
\end{minipage}}
\end{sidewaystable*}

\renewcommand{\tablename}{Supplementary Table}
\renewcommand{\thetable}{2 Continued}
\begin{sidewaystable*}
\vspace{-10cm}
\rotatebox{0}{
           \begin{minipage}{\textwidth}
\centering
\caption{}
\resizebox{\textwidth}{!}{
\begin{tabular}{cccccccccccccccc}
        \hline
        \hline
        \noalign{\smallskip}
      Name & Reference & log$\mathrm{_{10}}$($\mathrm{H_2O}$) & log$\mathrm{_{10}}$($\mathrm{CH_4}$) & log$\mathrm{_{10}}$($\mathrm{CO}$) & log$\mathrm{_{10}}$($\mathrm{CO_2}$) &log$\mathrm{_{10}}$($\mathrm{SiO}$) & log$\mathrm{_{10}}$($\mathrm{HCN}$) & log$\mathrm{_{10}}$($\mathrm{H_2S}$) & log$\mathrm{_{10}}$($\mathrm{SO_2}$) &  log$\mathrm{_{10}}$($\mathrm{C_2H_2}$) & log$\mathrm{_{10}}$($\mathrm{CS_2}$) & log$\mathrm{_{10}}$($\mathrm{SCO}$) & log$\mathrm{_{10}}$($\mathrm{N_2O}$)   & log$\mathrm{_{10}}$($\mathrm{AlO}$) & C/O\footnotemark[1]\\
        \noalign{\smallskip}
        \hline
        \noalign{\smallskip}
        WASP-94 A b & This Work & -1.71$\mathrm{^{\text{+} 0.55}_{-1.09}}$ & -7.06$\mathrm{^{\text{+} 0.94}_{-2.78}}$ & -2.08$\mathrm{^{\text{+} 0.70}_{-1.15}}$ & -4.21$\mathrm{^{\text{+} 0.58}_{-0.97}}$ & -- & -8.86$\mathrm{^{\text{+} 2.09}_{-2.04}}$ & -3.18$\mathrm{^{\text{+} 0.45}_{-1.25}}$  & -9.17$\mathrm{^{\text{+} 1.99}_{-1.89}}$ & -9.13$\mathrm{^{\text{+} 1.84}_{-1.86}}$ & -- & -- & -- & -- & 0.235$\mathrm{^{\text{+} 0.160}_{-0.112 }}$\\ 
        \noalign{\smallskip}
        & \cite{ahrer2025tracing} & -1.59$\mathrm{^{\text{+} 0.35}_{- 0.64}}$ & $<$ -5.83 & -2.30$\mathrm{^{\text{+} 0.55}_{- 0.84}}$ & -4.35$\mathrm{^{\text{+} 0.52}_{- 0.69}}$ & -- & $<$ -5.65 & -3.31$\mathrm{^{\text{+} 0.33}_{- 0.80}}$ & -- & $<$ -6.23 & -- & -- & --  & -- & 0.112$\mathrm{^{\text{+} 0.0545}_{- 0.0381}}$\\
        \noalign{\smallskip}
        \hline
        \noalign{\smallskip}
        WASP-39 b & This Work & -0.67$\mathrm{^{\text{+} 0.10}_{-0.11}}$ & -9.20$\mathrm{^{\text{+} 1.84}_{-1.76}}$ & -1.62$\mathrm{^{\text{+} 0.16}_{-0.17}}$ & -1.64$\mathrm{^{\text{+} 0.16}_{-0.17}}$ & -- & -8.96$\mathrm{^{\text{+} 2.06}_{-1.95}}$ & -6.99$\mathrm{^{\text{+} 3.07}_{-3.11}}$  & -3.87$\mathrm{^{\text{+} 0.14}_{-0.15}}$ & -- & -- & -- & -- & --  & 0.100$\mathrm{^{\text{+}0.096 }_{- 0.051}}$\\ 
        \noalign{\smallskip}
        & \cite{constantinou2023early} & -3.29$\mathrm{^{\text{+} 0.59}_{- 0.56}}$ & $\sim$ -7.00 & -4.17$\mathrm{^{\text{+} 0.61}_{- 0.61}}$ & -5.11$\mathrm{^{\text{+} 0.63}_{- 0.54}}$ & -- & $\sim$ -6.00 & -4.11$\mathrm{^{\text{+} 0.49}_{- 0.46}}$ & -6.40$\mathrm{^{\text{+} 0.39}_{-0.35}}$ & -- & -- & -- & --  & -- & 0.084$\mathrm{^{\text{+} 0.0028}_{- 0.0039}}$\\
        \noalign{\smallskip}
        \hline
        \noalign{\smallskip}
        WASP-121 b & This Work & \textcolor{black}{-4.40$\mathrm{^{\text{+} 0.13}_{-0.15}}$} & \textcolor{black}{-9.19$\mathrm{^{\text{+} 1.91}_{-1.75}}$} & \textcolor{black}{-0.35$\mathrm{^{\text{+} 0.04}_{-0.09}}$} &\textcolor{black}{ -9.30$\mathrm{^{\text{+} 1.73}_{-1.74}}$} & \textcolor{black}{-3.50$\mathrm{^{\text{+} 0.49}_{-2.44}}$} & -- & --  & \textcolor{black}{-7.65$\mathrm{^{\text{+} 2.53}_{-2.74}}$} & -- & -- & -- & -- & --& \textcolor{black}{0.999$\mathrm{^{\text{+} 4.60\times 10^{-4}}_{- 1.01 \times 10^{-3}}}$}\\ 
        \noalign{\smallskip}
        & \cite{gapp2025wasp} & -2.42$\mathrm{^{\text{+} 0.69}_{- 0.99}}$ & -- & -1.27$\mathrm{^{\text{+} 0.11}_{- 0.13}}$ & -9.02$\mathrm{^{\text{+} 1.92}_{- 1.54}}$ & -3.51$\mathrm{^{\text{+} 0.39}_{- 0.47}}$ & -- & -- & -- & -- & -- & -- & --  & -- & 0.898$\mathrm{^{\text{+} 0.000}_{- 0.000}}$\\
        \noalign{\smallskip}
        \hline
        \noalign{\smallskip}
        WASP-127 b & This Work & -1.80$\mathrm{^{\text{+} 0.58}_{-0.73}}$ & -8.23$\mathrm{^{\text{+} 1.96}_{-2.36}}$ & -7.95$\mathrm{^{\text{+} 2.72}_{-2.51}}$ & -3.03$\mathrm{^{\text{+} 0.64}_{-0.82}}$ & -- & -8.00$\mathrm{^{\text{+} 2.39}_{-2.55}}$ & -8.41$\mathrm{^{\text{+} 2.49}_{-2.30}}$  & -8.89$\mathrm{^{\text{+} 2.10}_{-1.95}}$ & -7.24$\mathrm{^{\text{+} 1.85}_{-2.97}}$ & -8.38$\mathrm{^{\text{+} 2.52}_{-2.33}}$ & -9.61$\mathrm{^{\text{+} 1.71}_{-1.51}}$ & -- & -- & 0.235$\mathrm{^{\text{+} 0.165}_{-0.090 }}$\\ 
        \noalign{\smallskip}
        & -- & -- & -- & -- & -- & -- & -- & -- & -- & -- & -- & -- & --  & -- & --\\

        \noalign{\smallskip}
        \hline
        \hline
\end{tabular}}
\end{minipage}}
\footnotetext[1]{\textit{Note}: \textcolor{black}{For our free-chemistry retrievals, C/O was computed from the posterior molecular-abundance samples by converting mass fractions to mole fractions and summing the number of C and O atoms carried by each molecule. For literature comparisons, full posterior samples were generally not available. We therefore recomputed approximate values from the published median molecular abundances and reported 1$\sigma$ bounds using the same  approach.}}
\end{sidewaystable*}

\renewcommand{\figurename}{Supplementary Fig.}
\renewcommand{\thefigure}{2}
\pagebreak
\begin{figure}
    \centering  
                \begin{minipage}[b]{\textwidth}
            \includegraphics[width=0.9\textwidth]{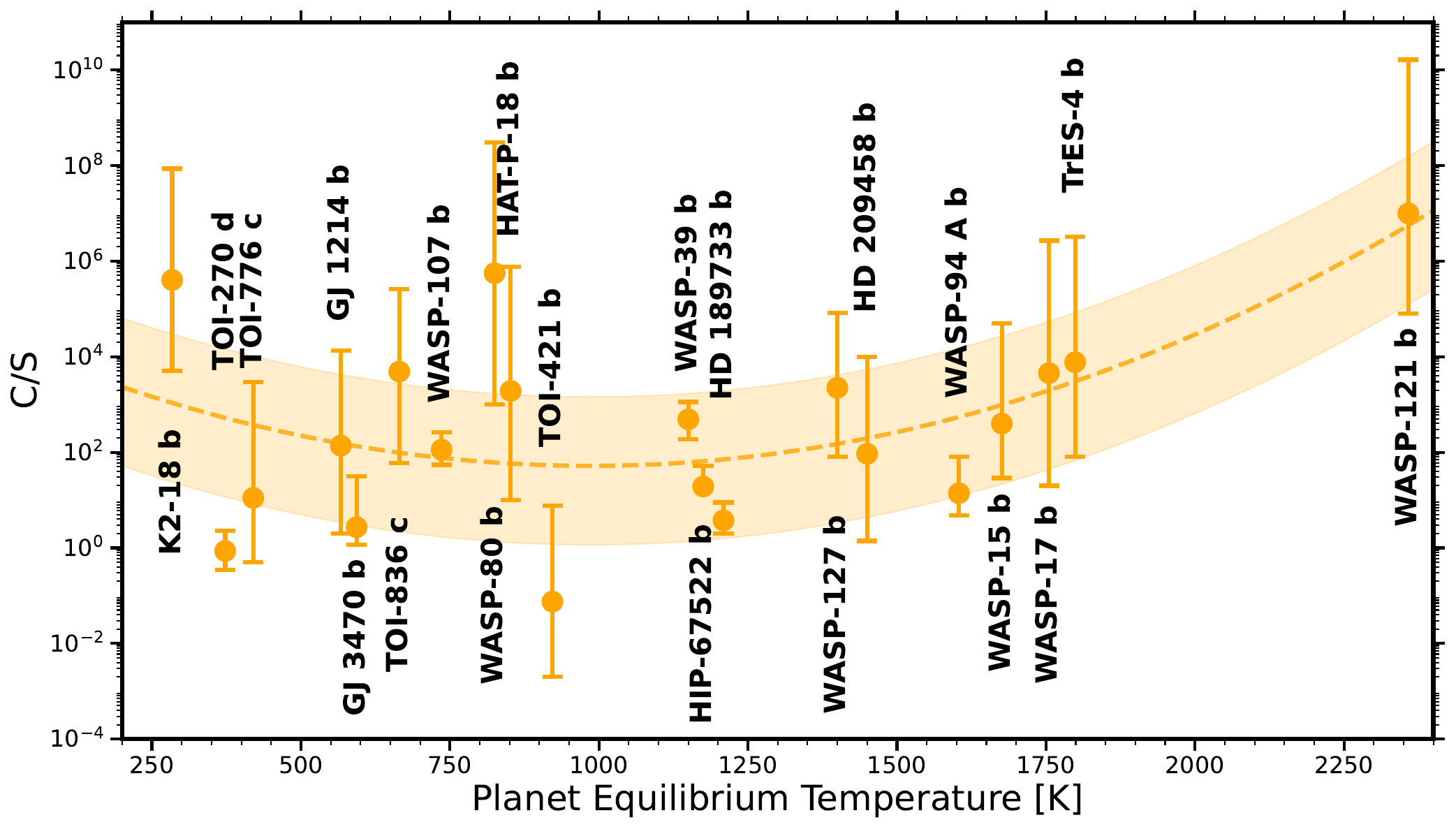}
        \end{minipage}
        \begin{minipage}[b]{\textwidth}
            \includegraphics[width=0.9\textwidth]{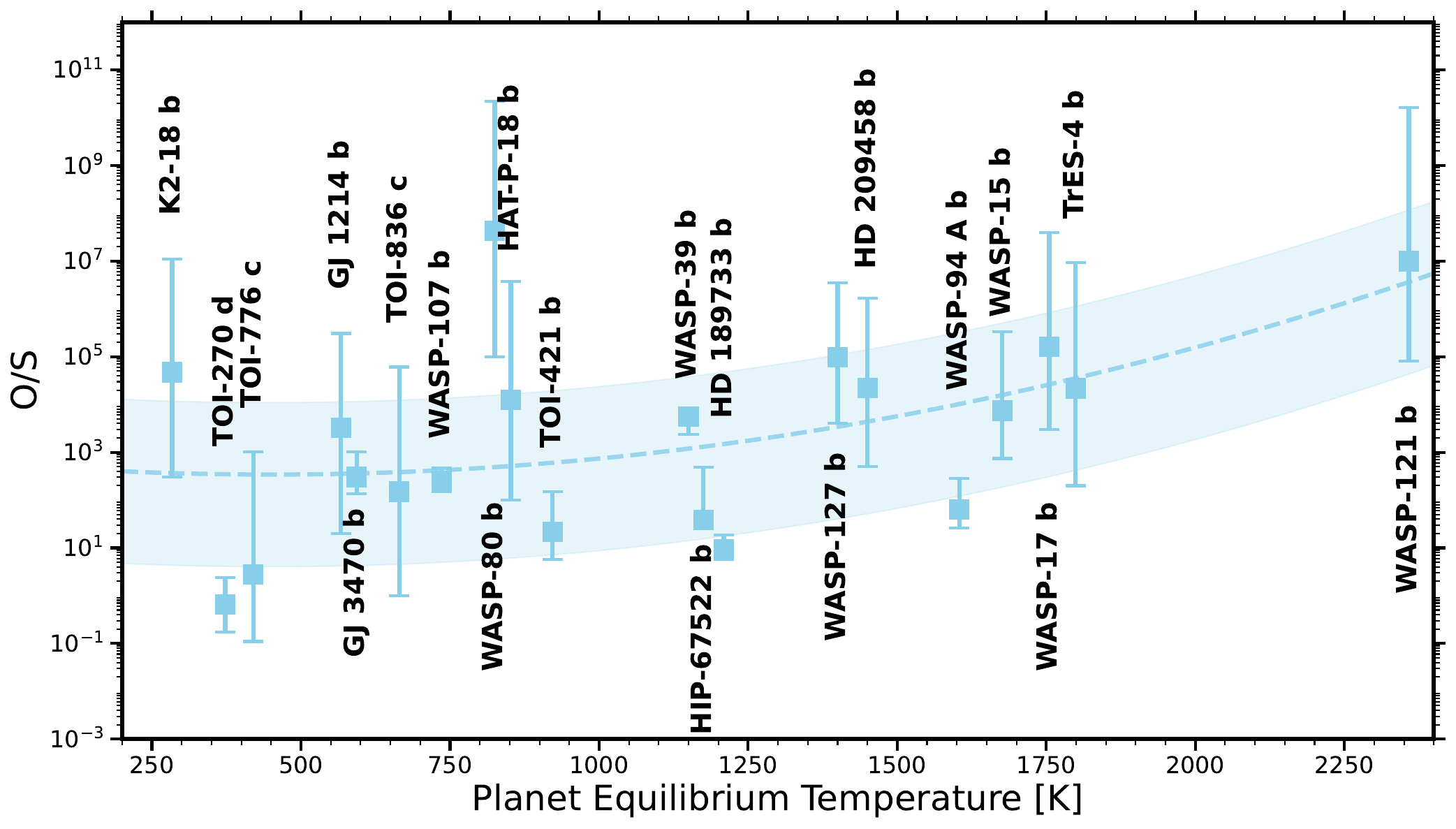}
        \end{minipage}
\caption{(Upper Panel) Atmospheric C/S ratio as a function of planetary equilibrium temperature for the free-chemistry retrievals. Here, the solid \textcolor{black}{orange} squares represent the median of the C/S value with 1$\sigma$ uncertainty bounds, estimated from the posterior distribution of C/S. \textcolor{black}{The dashed line shows a second-order polynomial fit to the data points, with the shaded region indicating the 68\% bound. The fit is shown only as a visual guide and is not interpreted as a statistically robust picture of exoplanet demographics. To compare the apparent $T_{\rm eq}$ trends across the three elemental ratios, we computed Spearman rank correlations using the median retrieved values for the full 20-planet free-chemistry sample. None of the three relations is significant at the 5\% level: $T_{\rm eq}$ vs. C/O gives $\rho=-0.25$, $p=0.30$; $T_{\rm eq}$ vs. C/S gives $\rho=0.29$, $p=0.21$; and $T_{\rm eq}$ vs. O/S gives $\rho=0.41$, $p=0.07$. The corresponding quadratic fits have $R^2$ =0.22, 0.35, and 0.29, with F-test $p$-values of 0.13, 0.025, and 0.057, respectively.} (Lower Panel) Similar figure for O/S. No robust trend is observed for both C/S and O/S with equilibrium temperature. Broad uncertainties and residual-driven degeneracies currently prevent firm demographic conclusions for C/S and O/S.}
\label{fig:CS_OS_Tpl}
\end{figure}

\pagebreak
\renewcommand{\figurename}{Supplementary Fig.}
\renewcommand{\thefigure}{3}
\begin{figure}[h!]
\centering
	\includegraphics[width=\textwidth]{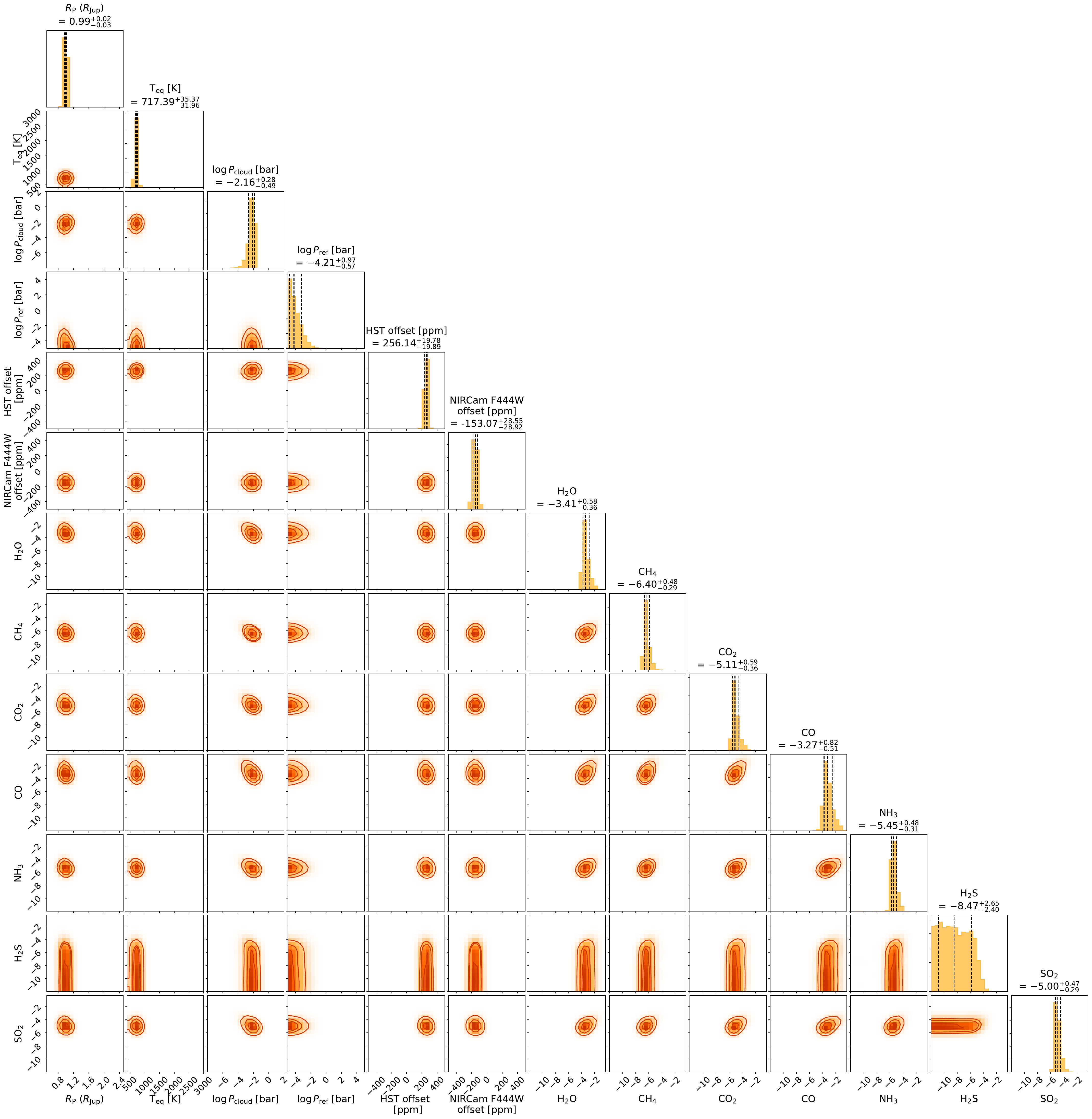}
  \caption{Corner plot of the free retrieval case for WASP-107 b. Posterior distributions are shown for all free parameters.}
   \label{fig:W107b_corner_free}
\end{figure}

\pagebreak
\renewcommand{\figurename}{Supplementary Fig.}
\renewcommand{\thefigure}{4}
\begin{figure}[h!]
\centering
	\includegraphics[width=\textwidth]{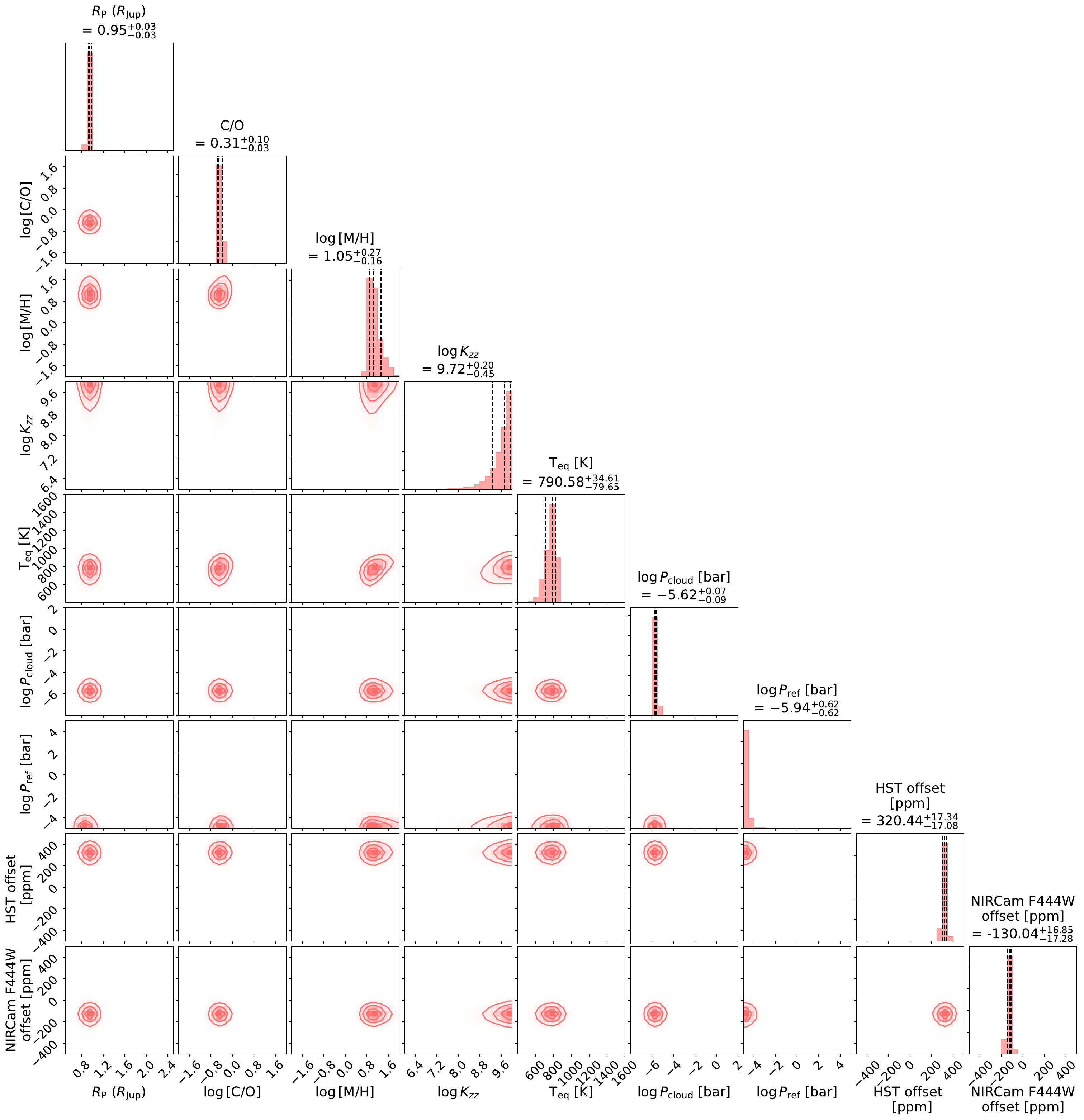}
  \caption{Corner plot of the grid retrieval case for WASP-107 b. Posterior distributions are shown for all free parameters.}
   \label{fig:W107b_corner_vulcan}
\end{figure}

\renewcommand{\figurename}{Supplementary Fig.}
\renewcommand{\thefigure}{5}
\begin{figure*}
\centering
\begin{overpic}[width=\textwidth]{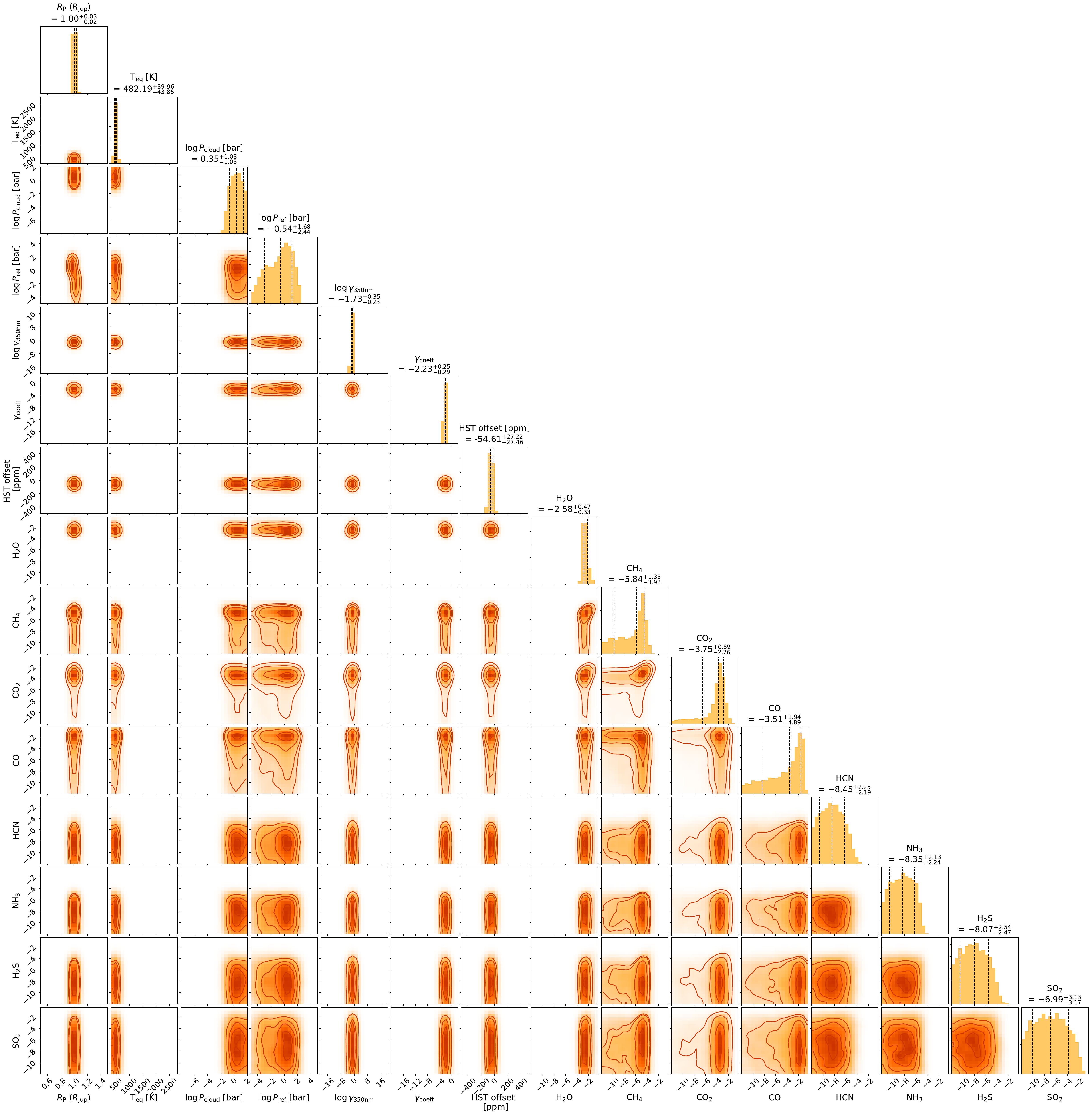}
  \put(45,70){\includegraphics[width=0.55\textwidth]{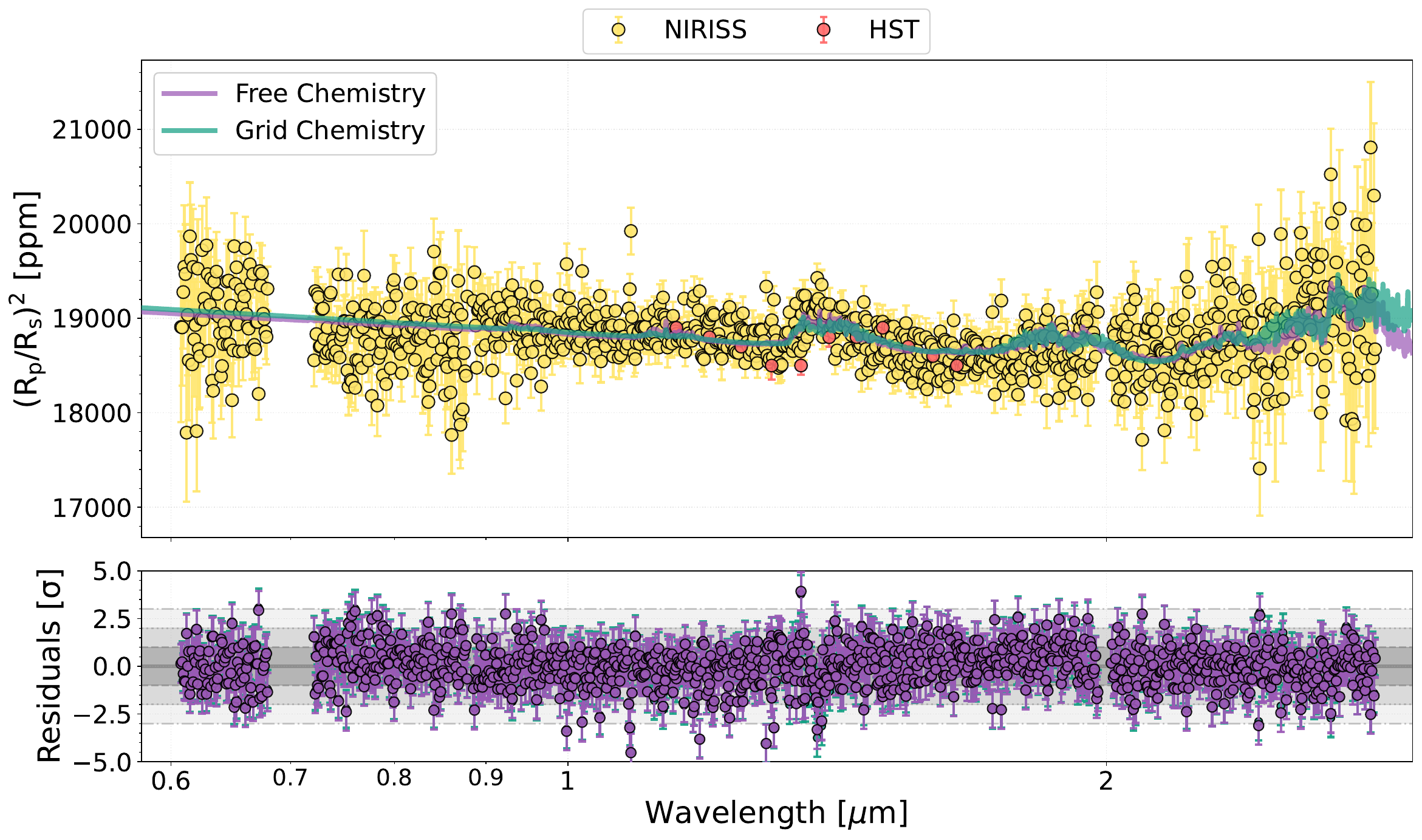}}
\end{overpic}
\caption{Best fit model comparison for free and grid chemistry retrievals for HAT-P-18 b. The corner plot presented here explains the posterior distribution of the free-parameters from the free-retrieval study of HAT-P-18 b.}
\label{fig:best_fit_corner_HAT-P-18b}
\end{figure*}

\renewcommand{\figurename}{Supplementary Fig.}
\renewcommand{\thefigure}{6}
\begin{figure*}
\centering
	\includegraphics[width=\textwidth]{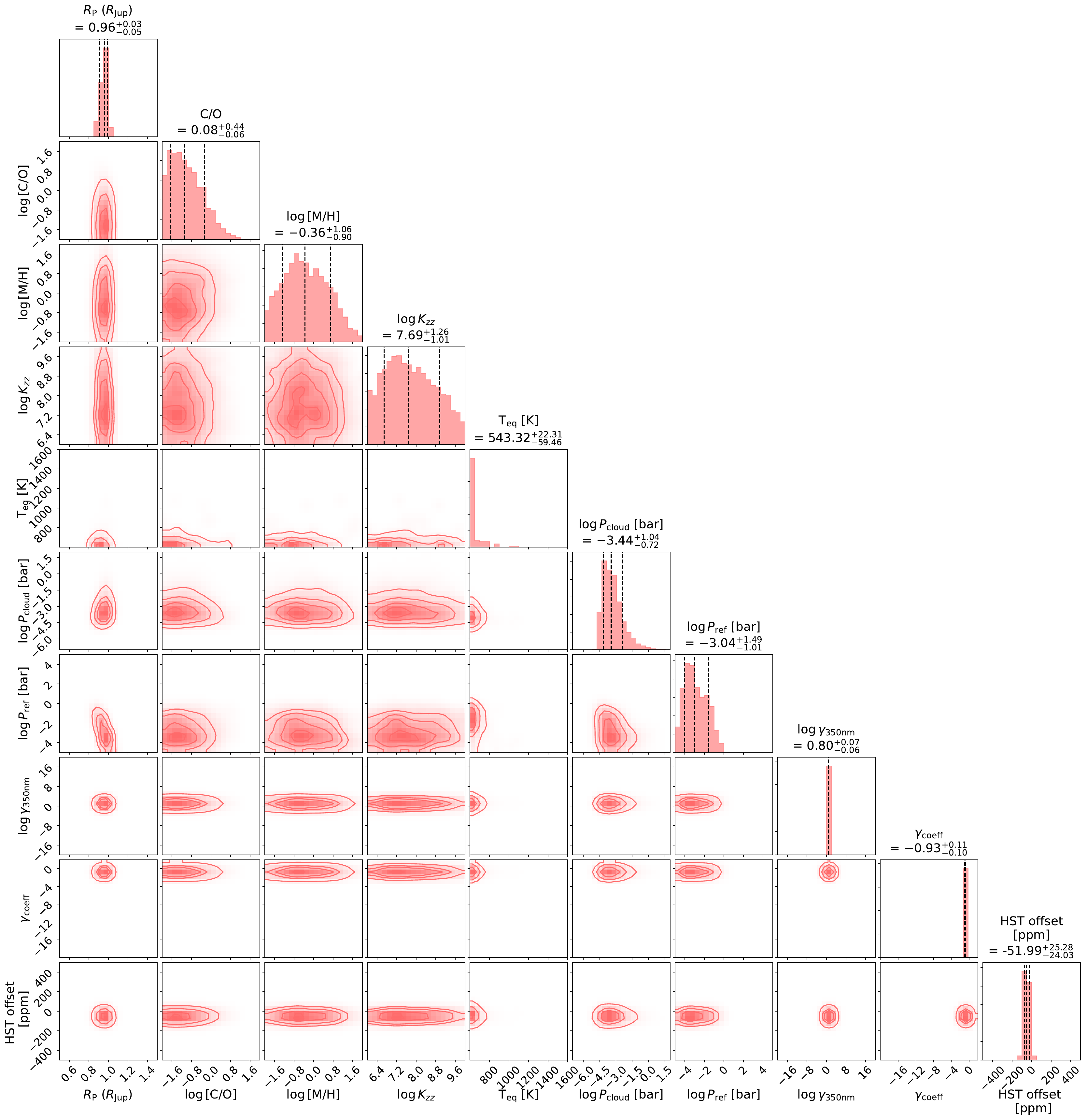}
  \caption{Corner plot of the grid retrieval case for HAT-P-18 b. Posterior distributions are shown for all free parameters.}
   \label{fig:corner_hat-p-18b_vulcan}
\end{figure*}

\renewcommand{\figurename}{Supplementary Fig.}
\renewcommand{\thefigure}{7}
\begin{figure*}
\centering
\begin{overpic}[width=\textwidth]{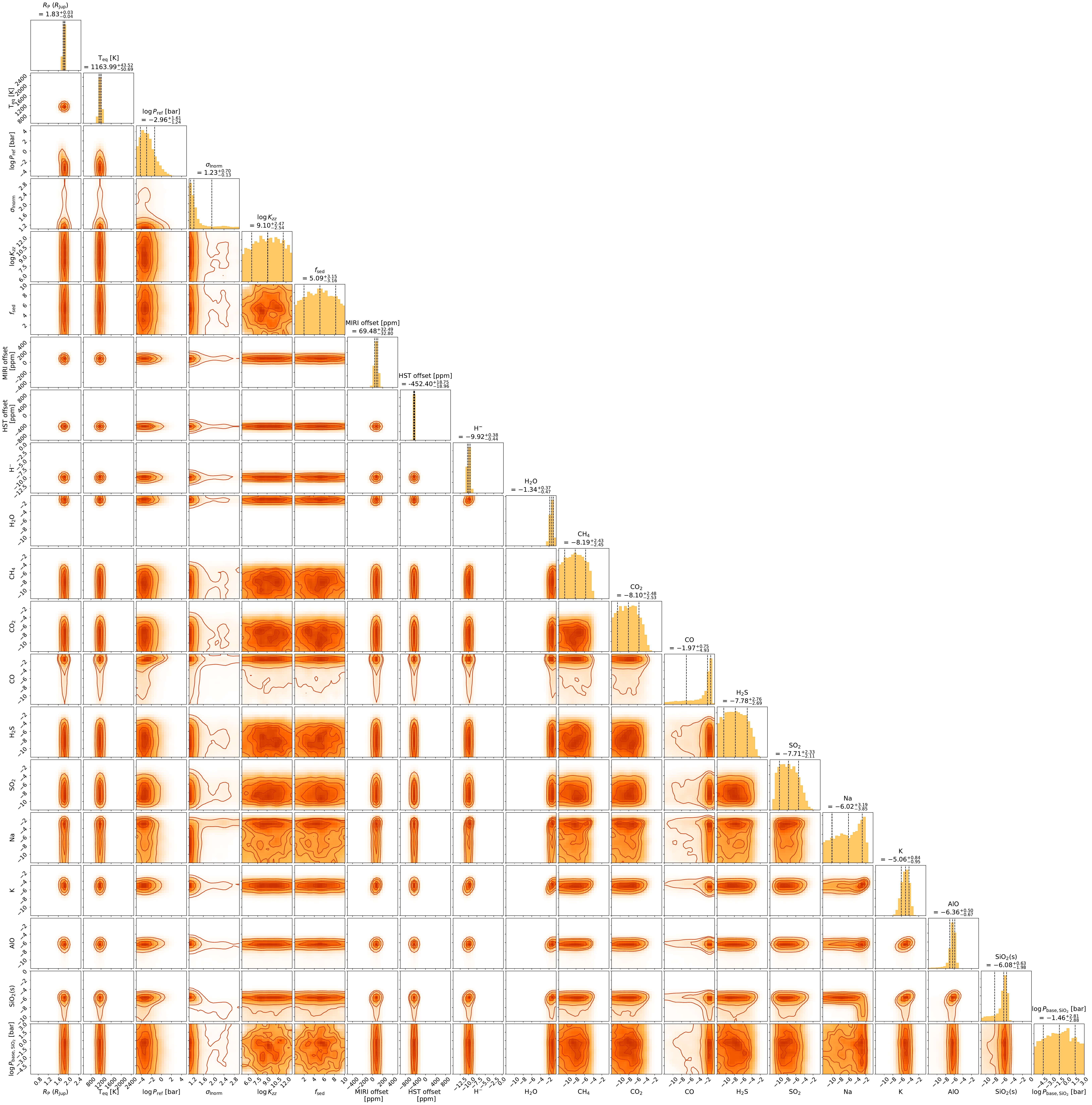}
  \put(45,70){\includegraphics[width=0.55\textwidth]{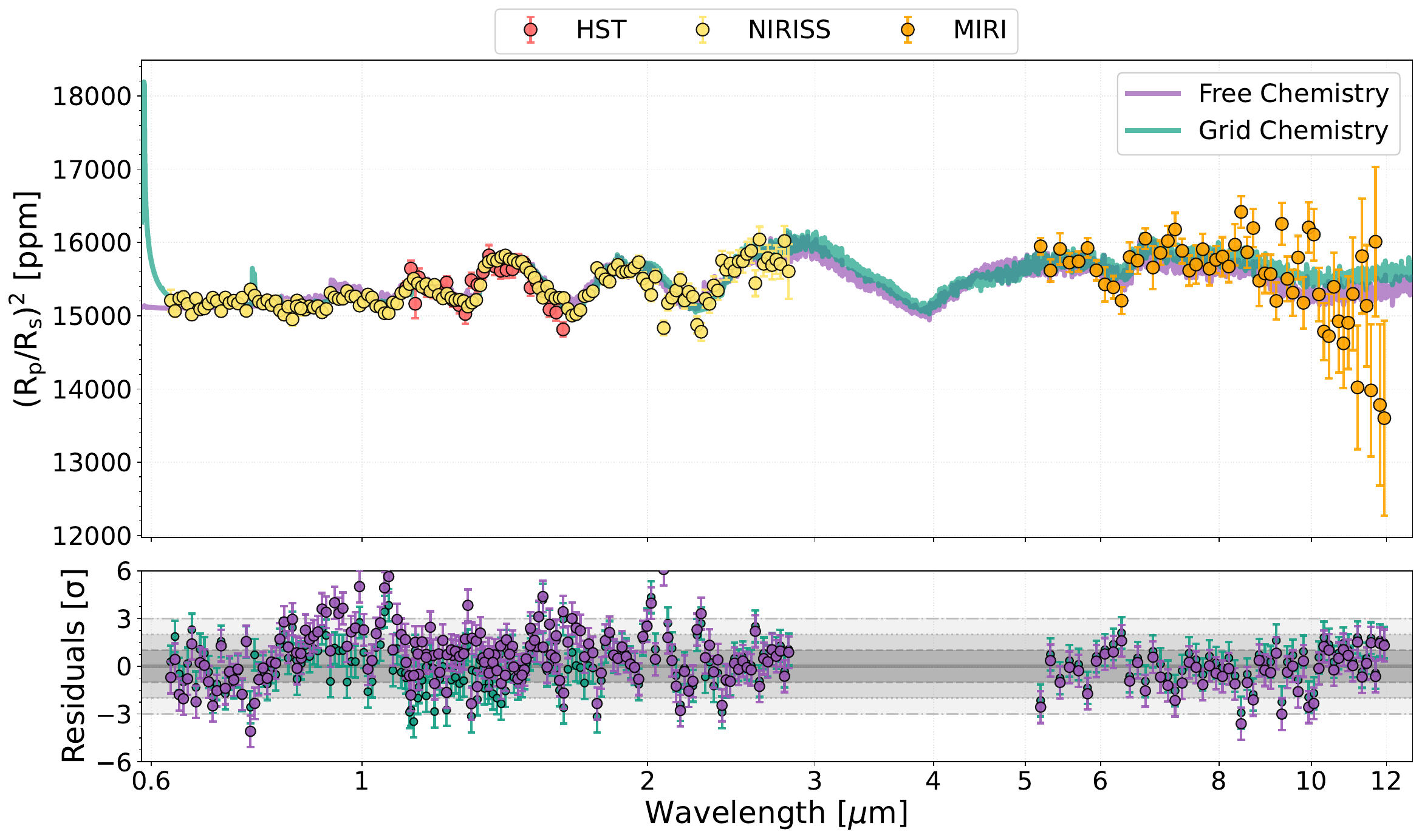}}
\end{overpic}
\caption{Similar to Figure \ref{fig:best_fit_corner_HAT-P-18b} for WASP-17 b.}
\end{figure*}

\renewcommand{\figurename}{Supplementary Fig.}
\renewcommand{\thefigure}{8}
\begin{figure*}
\centering
	\includegraphics[width=\textwidth]{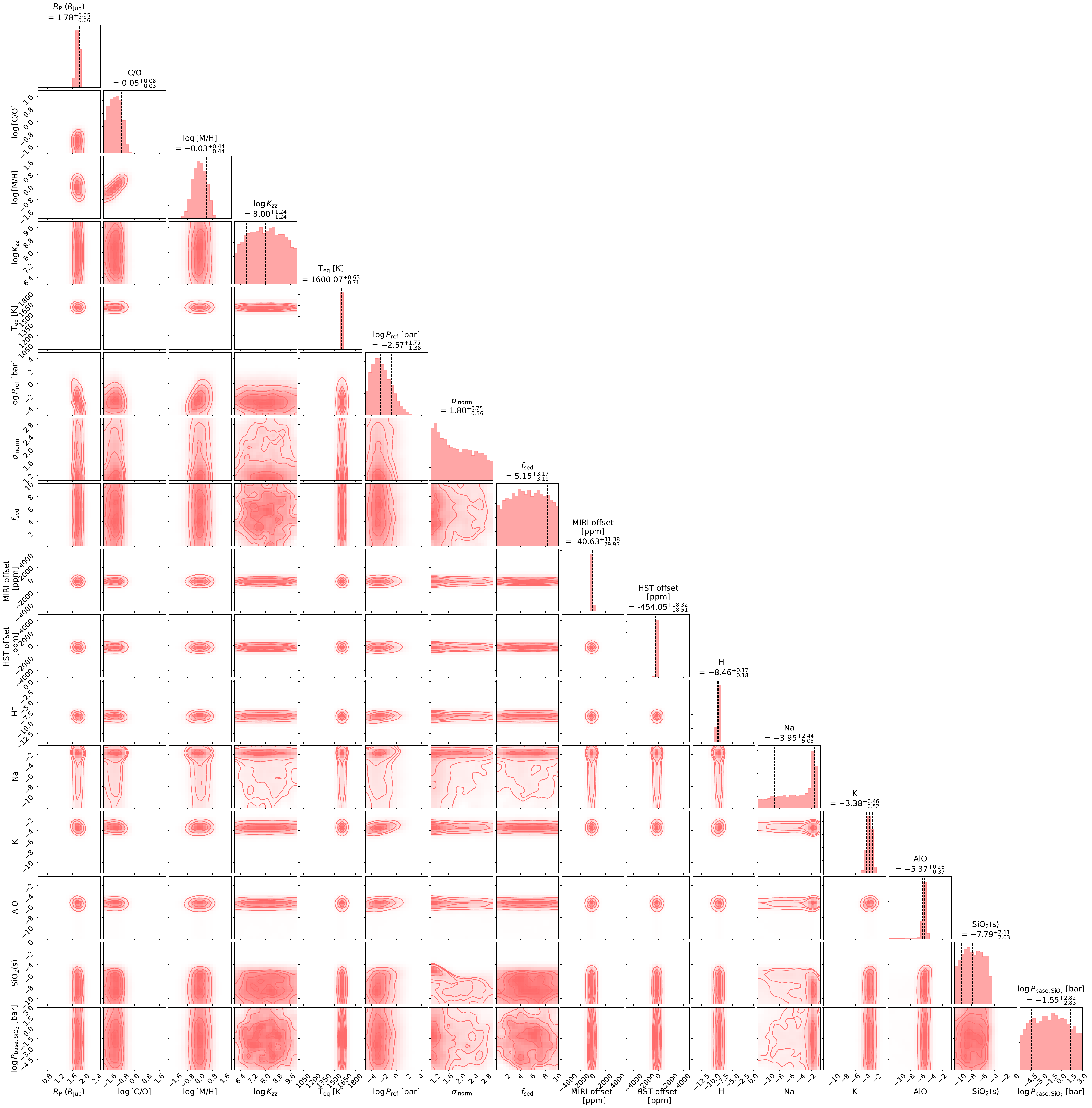}
  \caption{Similar to Figure \ref{fig:corner_hat-p-18b_vulcan} for WASP-17 b.}
\end{figure*}


\renewcommand{\thefigure}{9}
\begin{figure*}
\centering
\begin{overpic}[width=\textwidth]{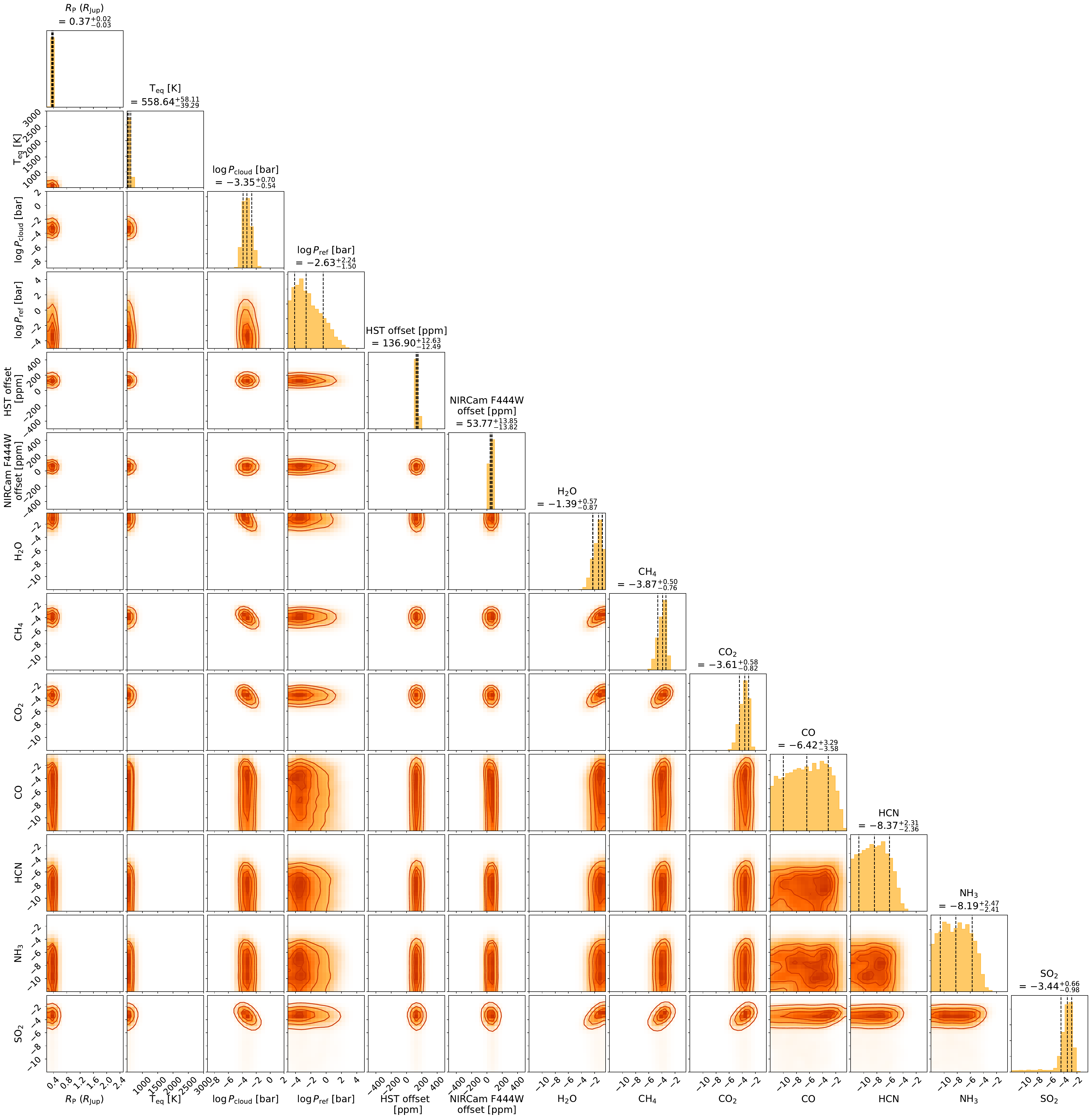}
  \put(45,70){\includegraphics[width=0.55\textwidth]{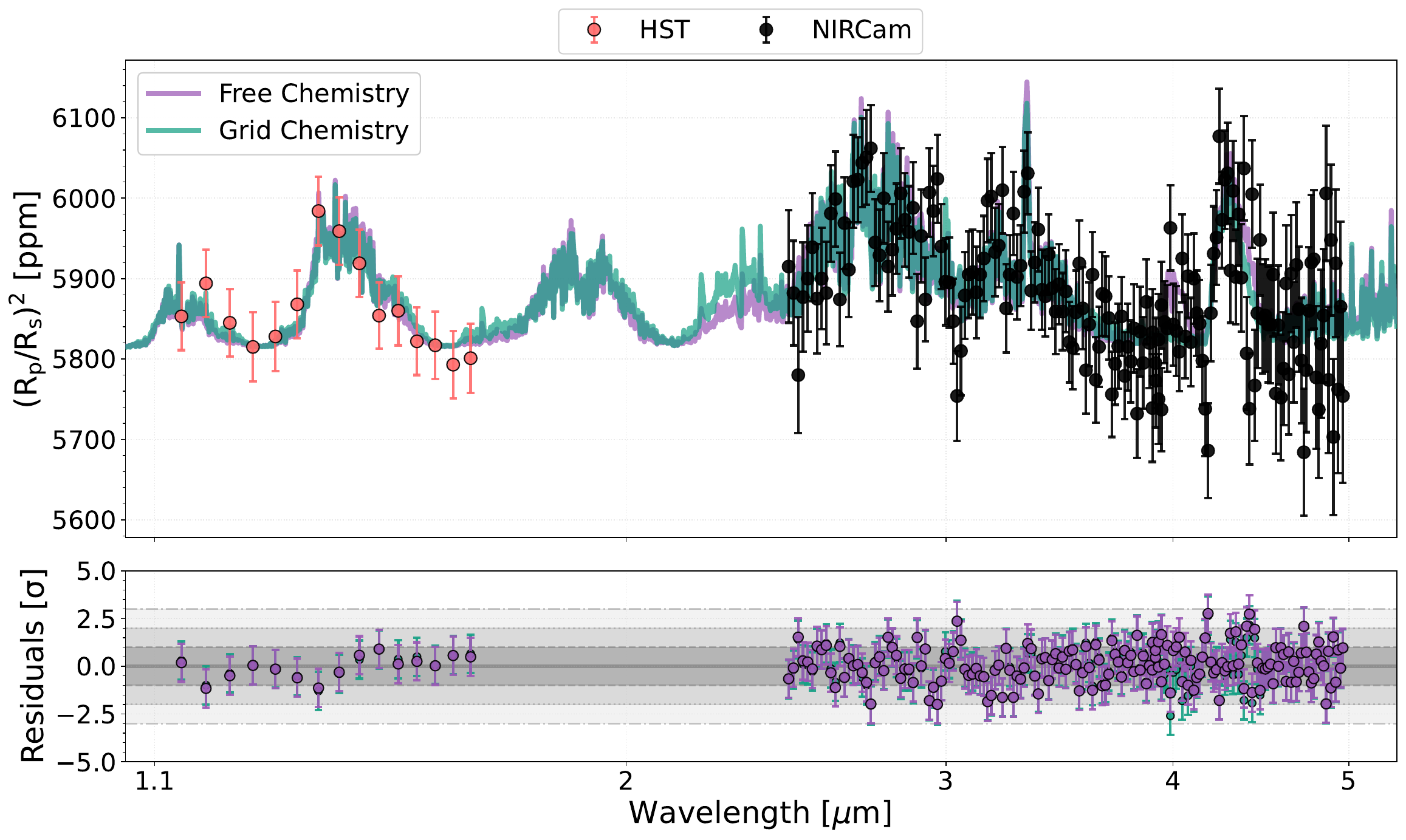}}
\end{overpic}
\caption{Similar to Figure \ref{fig:best_fit_corner_HAT-P-18b} for GJ 3470 b.}
\end{figure*}

\renewcommand{\thefigure}{10}
\begin{figure*}
\centering
	\includegraphics[width=\textwidth]{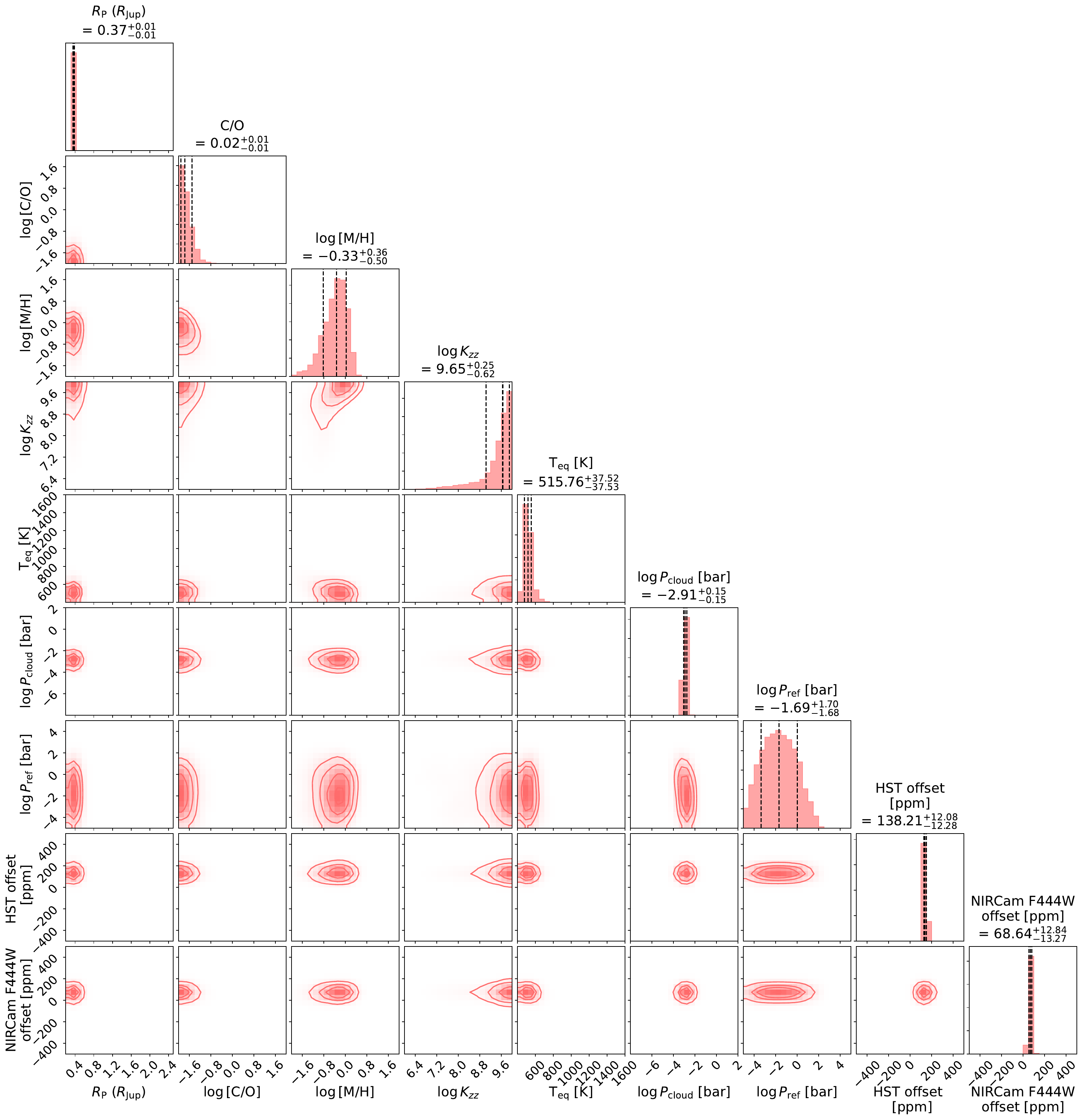}
  \caption{Similar to Figure \ref{fig:corner_hat-p-18b_vulcan} for GJ 3470 b.}
\end{figure*}


\renewcommand{\thefigure}{11}
\begin{figure*}
\centering
\begin{overpic}[width=\textwidth]{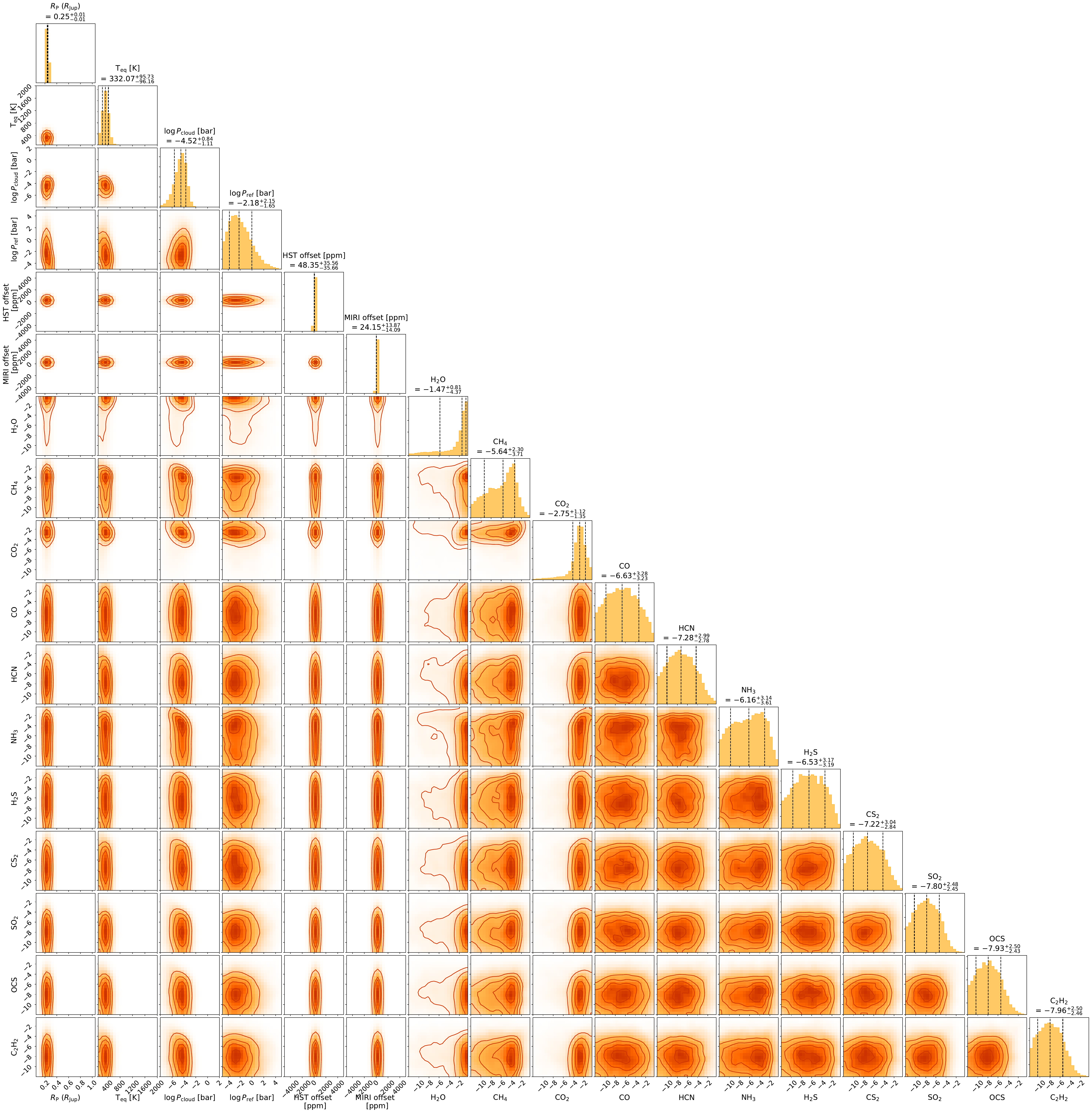}
  \put(45,70){\includegraphics[width=0.55\textwidth]{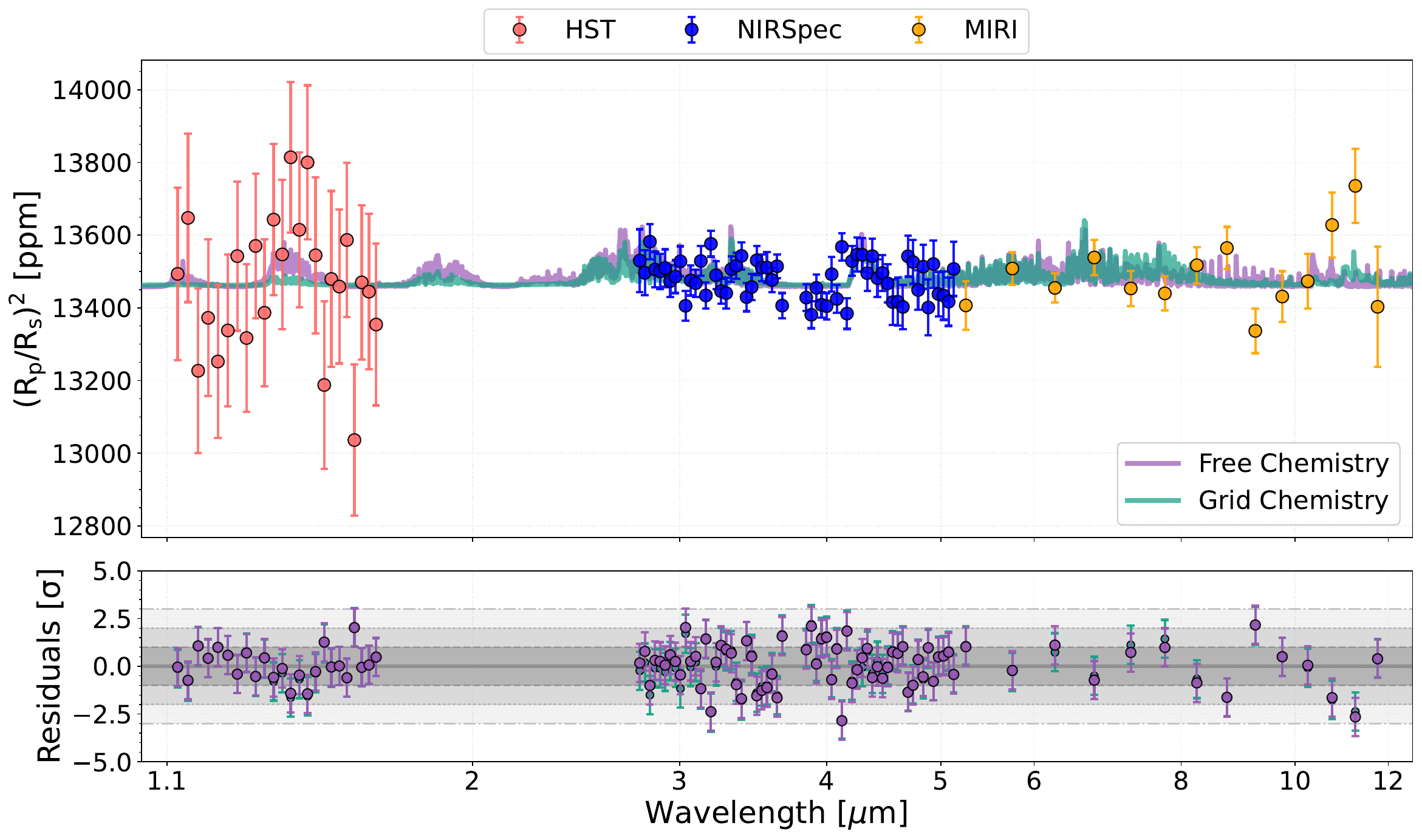}}
\end{overpic}
\caption{Similar to Figure \ref{fig:best_fit_corner_HAT-P-18b} for GJ 1214 b.}
\end{figure*}

\renewcommand{\thefigure}{12}
\begin{figure*}
\centering
	\includegraphics[width=\textwidth]{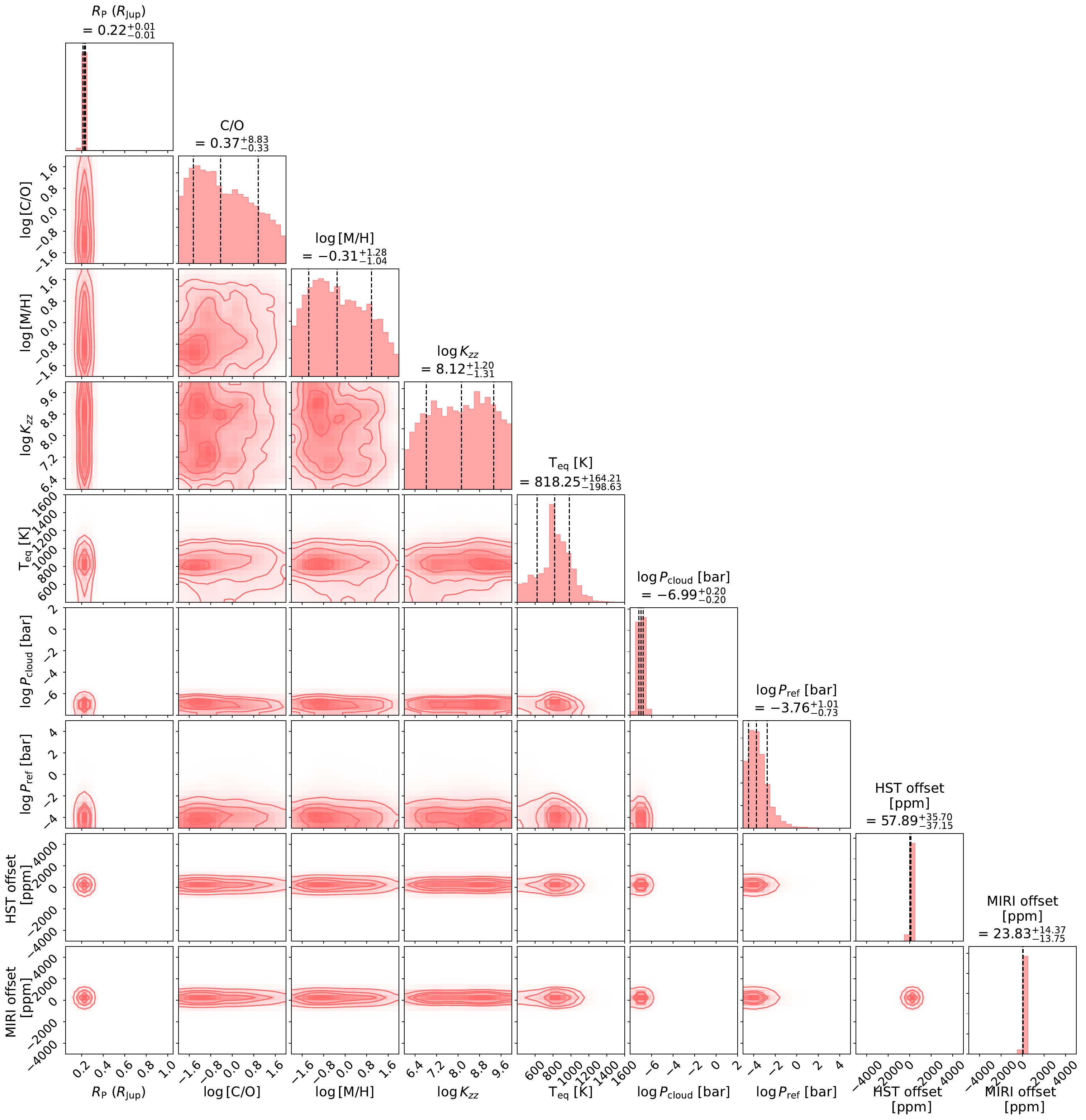}
  \caption{Similar to Figure \ref{fig:corner_hat-p-18b_vulcan} for GJ 1214 b.}
   \label{fig:}
\end{figure*}


\renewcommand{\thefigure}{13}
\begin{figure*}
\centering
\begin{overpic}[width=\textwidth]{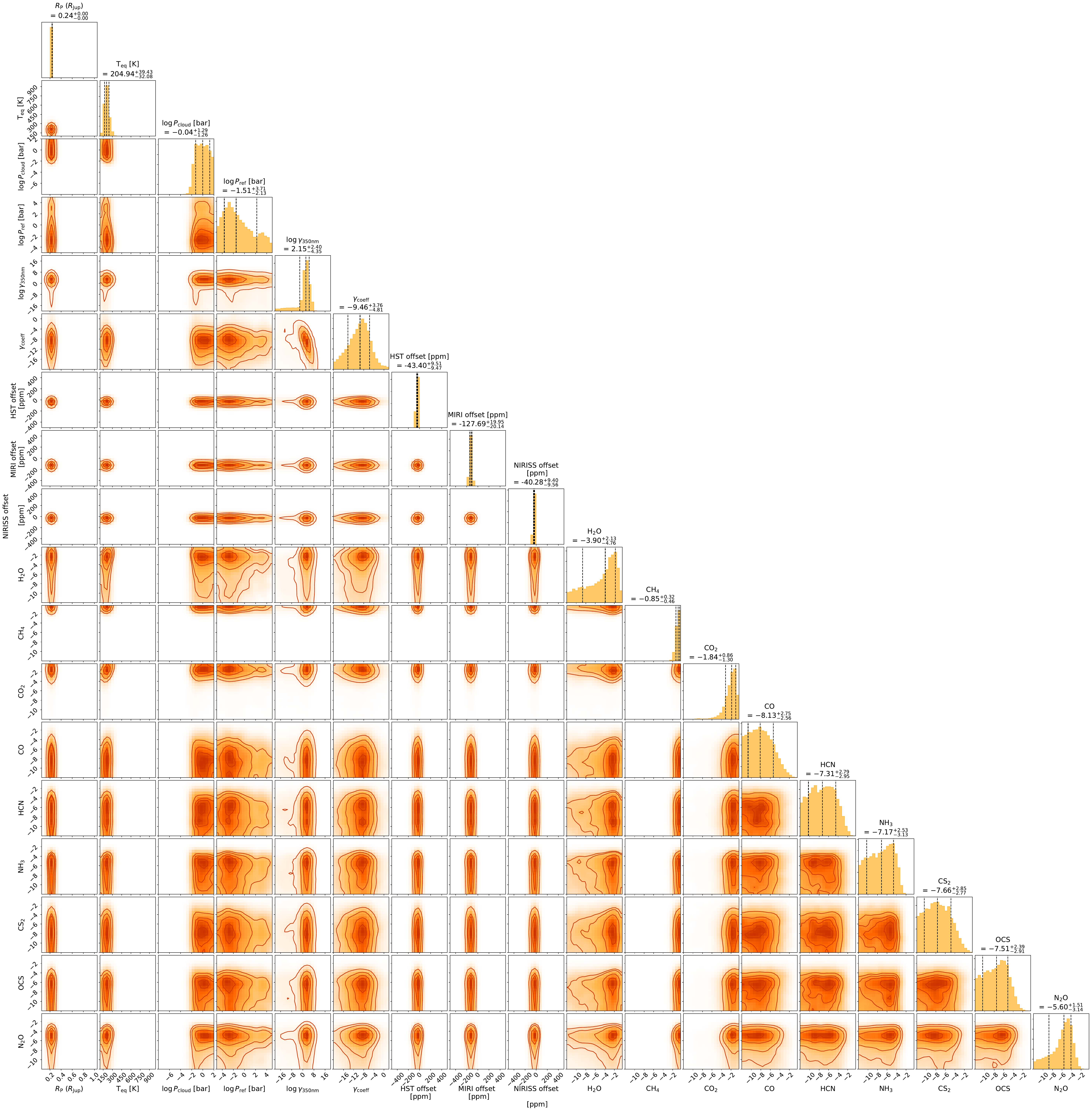}
  \put(45,70){\includegraphics[width=0.55\textwidth]{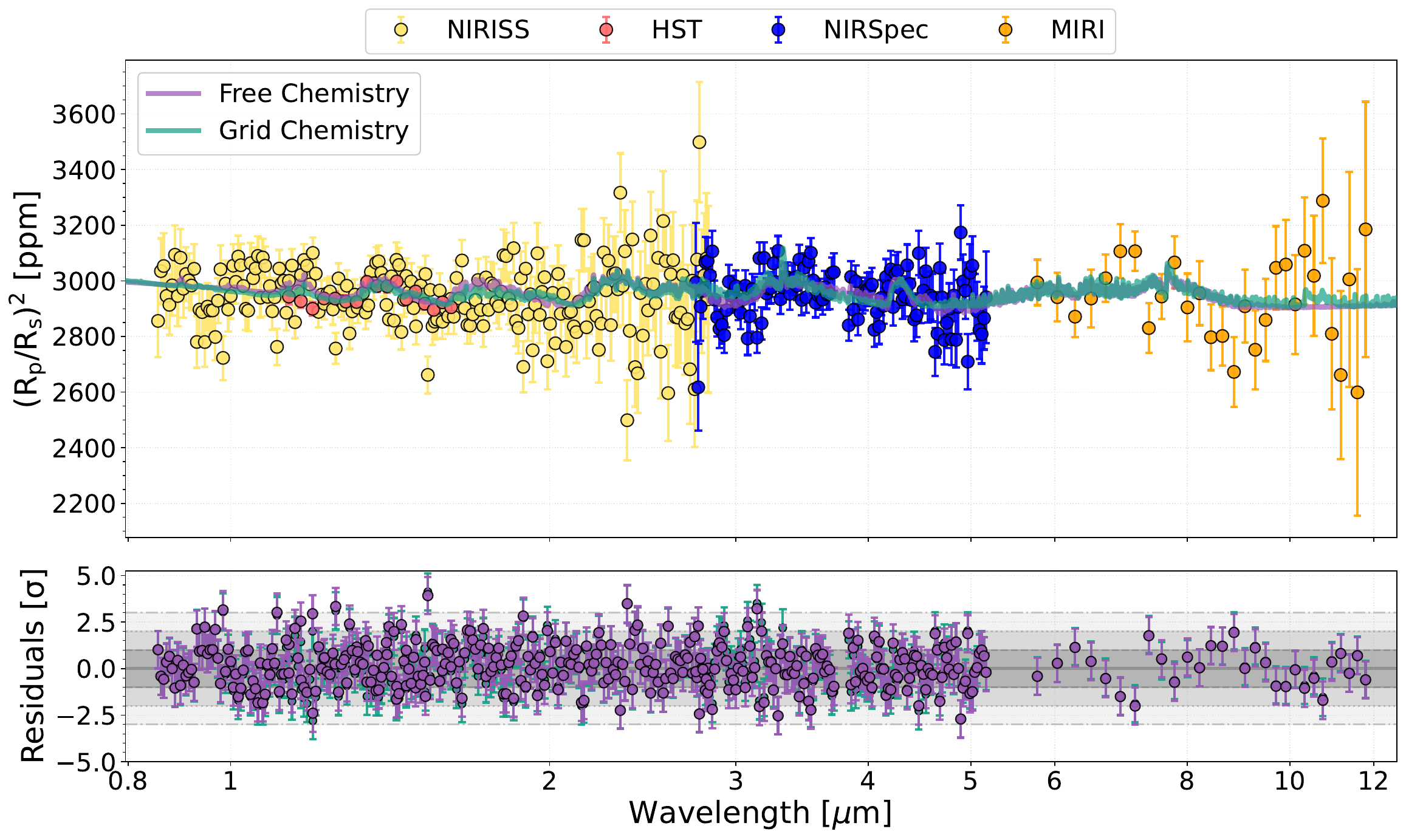}}
\end{overpic}
\caption{Similar to Figure \ref{fig:best_fit_corner_HAT-P-18b} for K2-18 b.}
\end{figure*}

\renewcommand{\thefigure}{14}
\begin{figure*}
\centering
	\includegraphics[width=\textwidth]{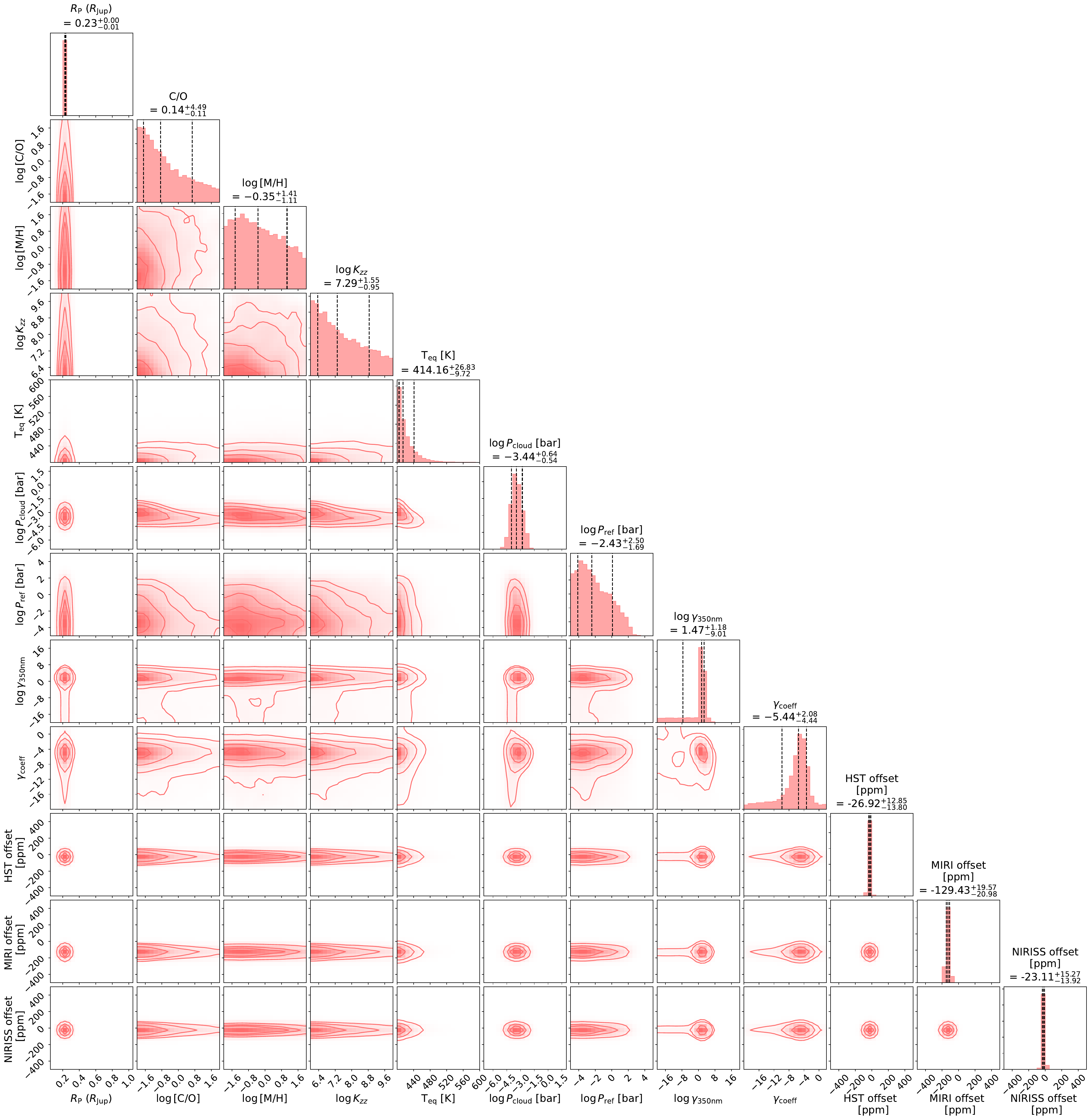}
  \caption{Similar to Figure \ref{fig:corner_hat-p-18b_vulcan} for K2-18 b.}
   \label{fig:}
\end{figure*}


\renewcommand{\thefigure}{15}
\begin{figure*}
\centering
\begin{overpic}[width=\textwidth]{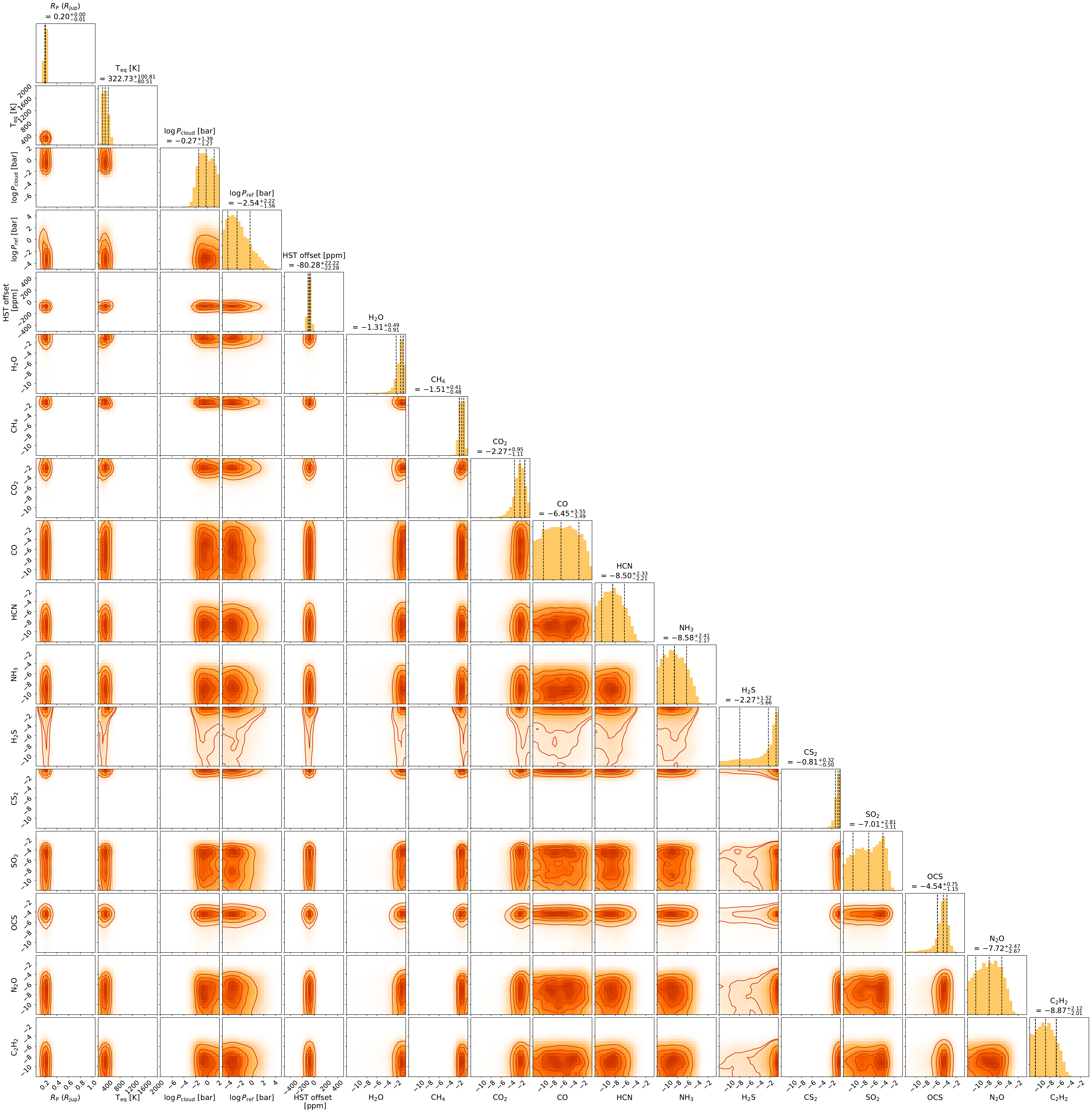}
  \put(45,70){\includegraphics[width=0.55\textwidth]{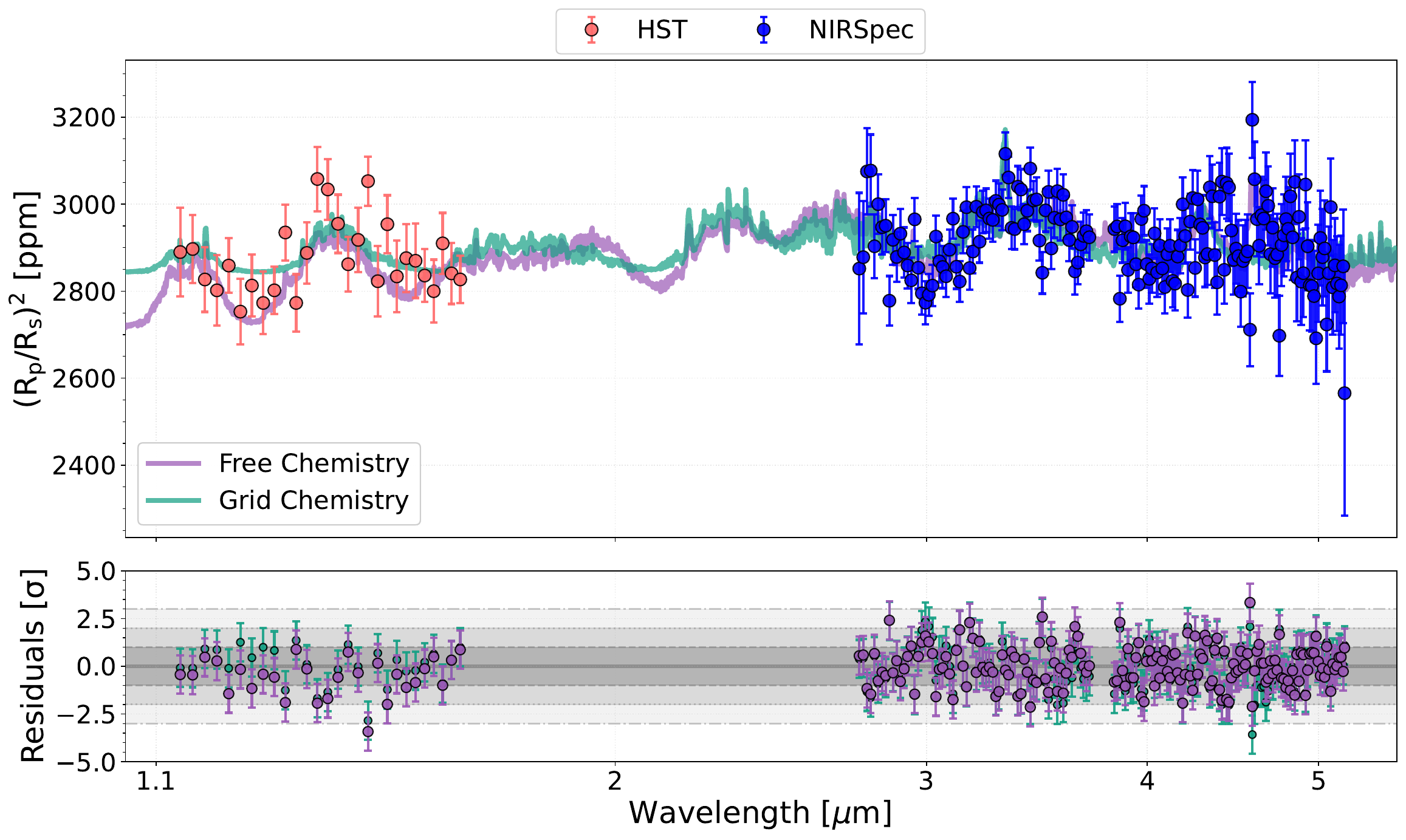}}
\end{overpic}
\caption{Similar to Figure \ref{fig:best_fit_corner_HAT-P-18b} for TOI-270d.}
\end{figure*}

\renewcommand{\thefigure}{16}
\begin{figure*}
\centering
	\includegraphics[width=\textwidth]{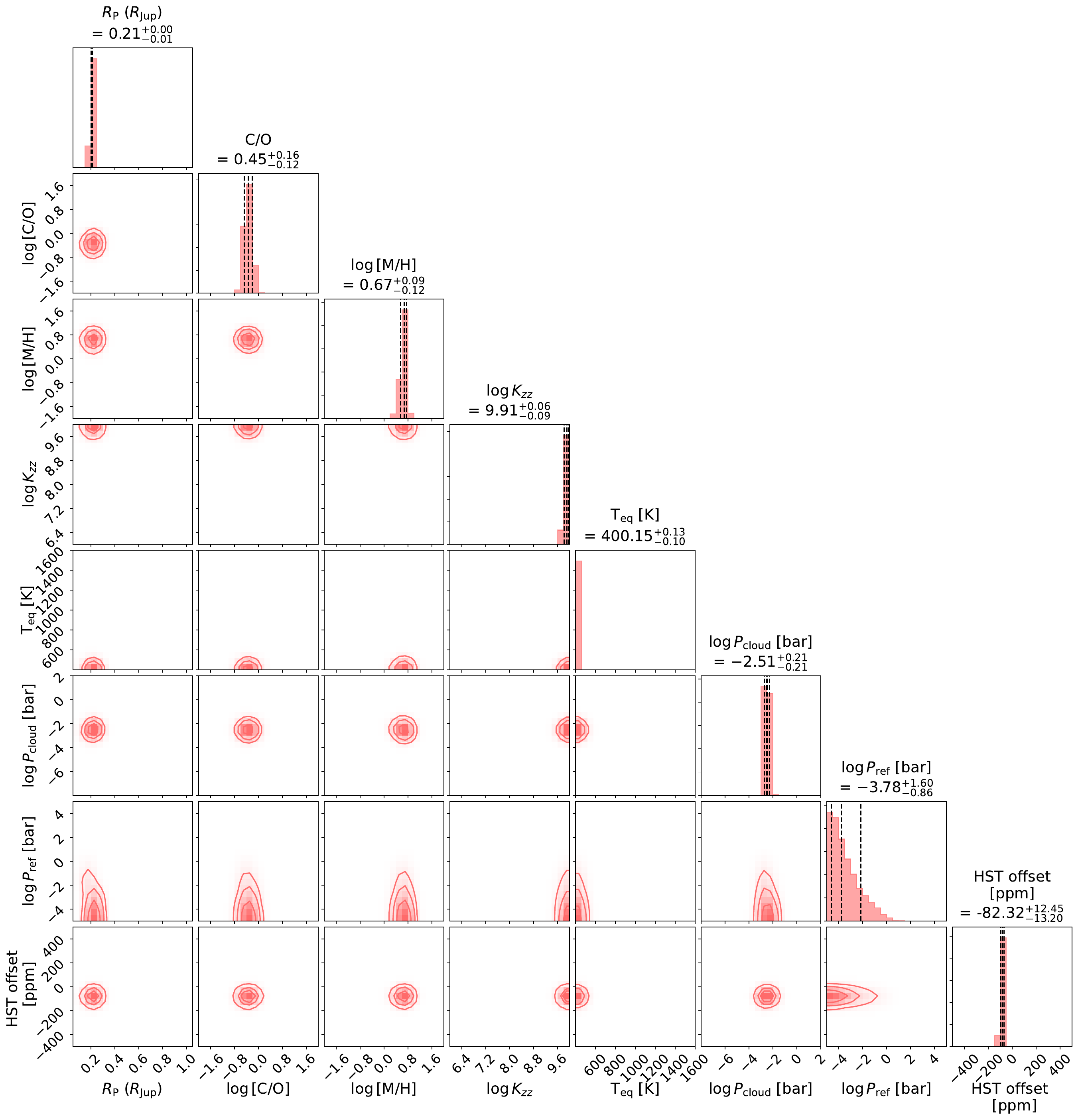}
  \caption{Similar to Figure \ref{fig:corner_hat-p-18b_vulcan} for TOI-270 d.}
   \label{fig:}
\end{figure*}

\renewcommand{\thefigure}{17}
\begin{figure*}
\centering
\begin{overpic}[width=\textwidth]{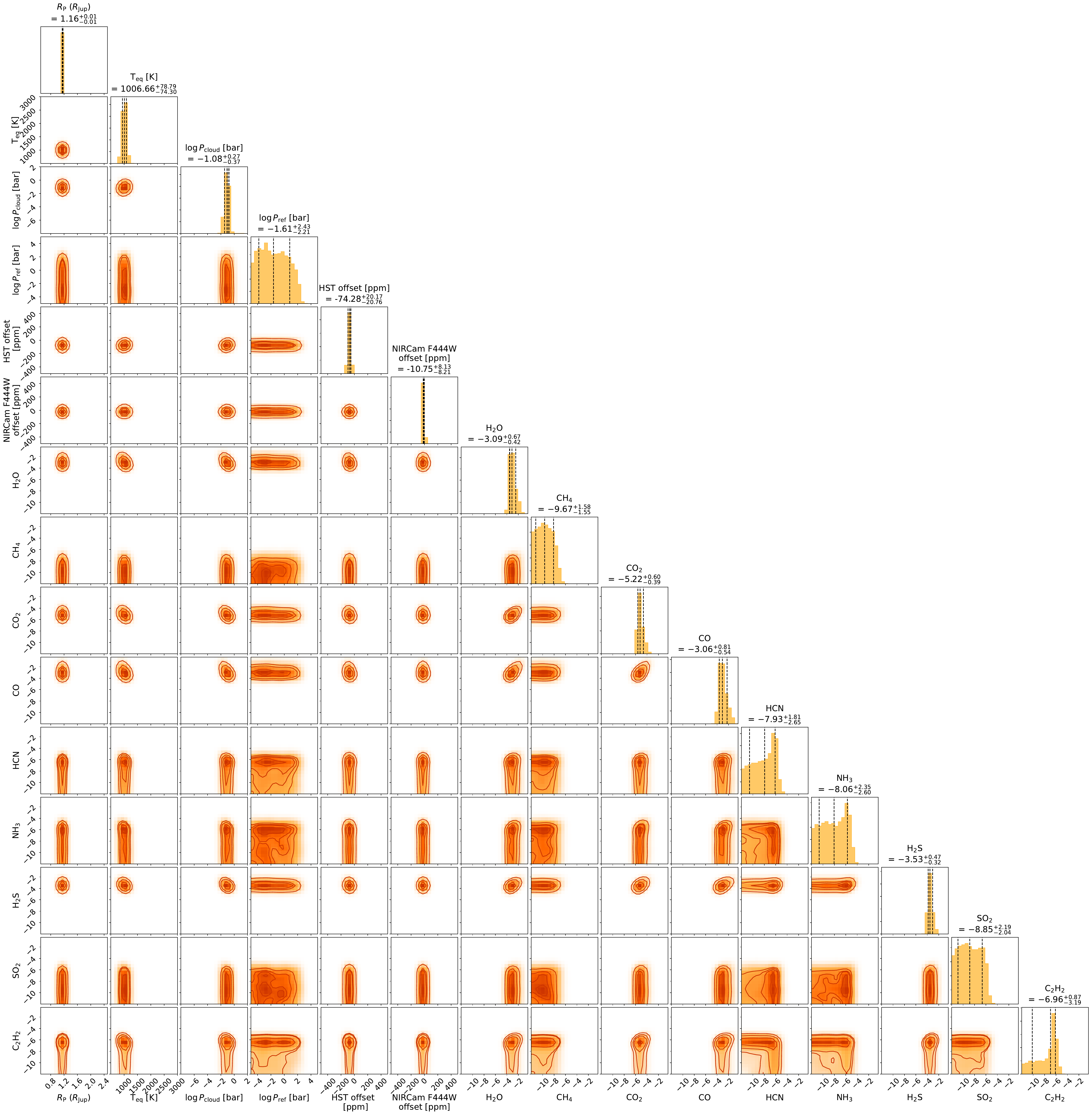}
  \put(45,70){\includegraphics[width=0.55\textwidth]{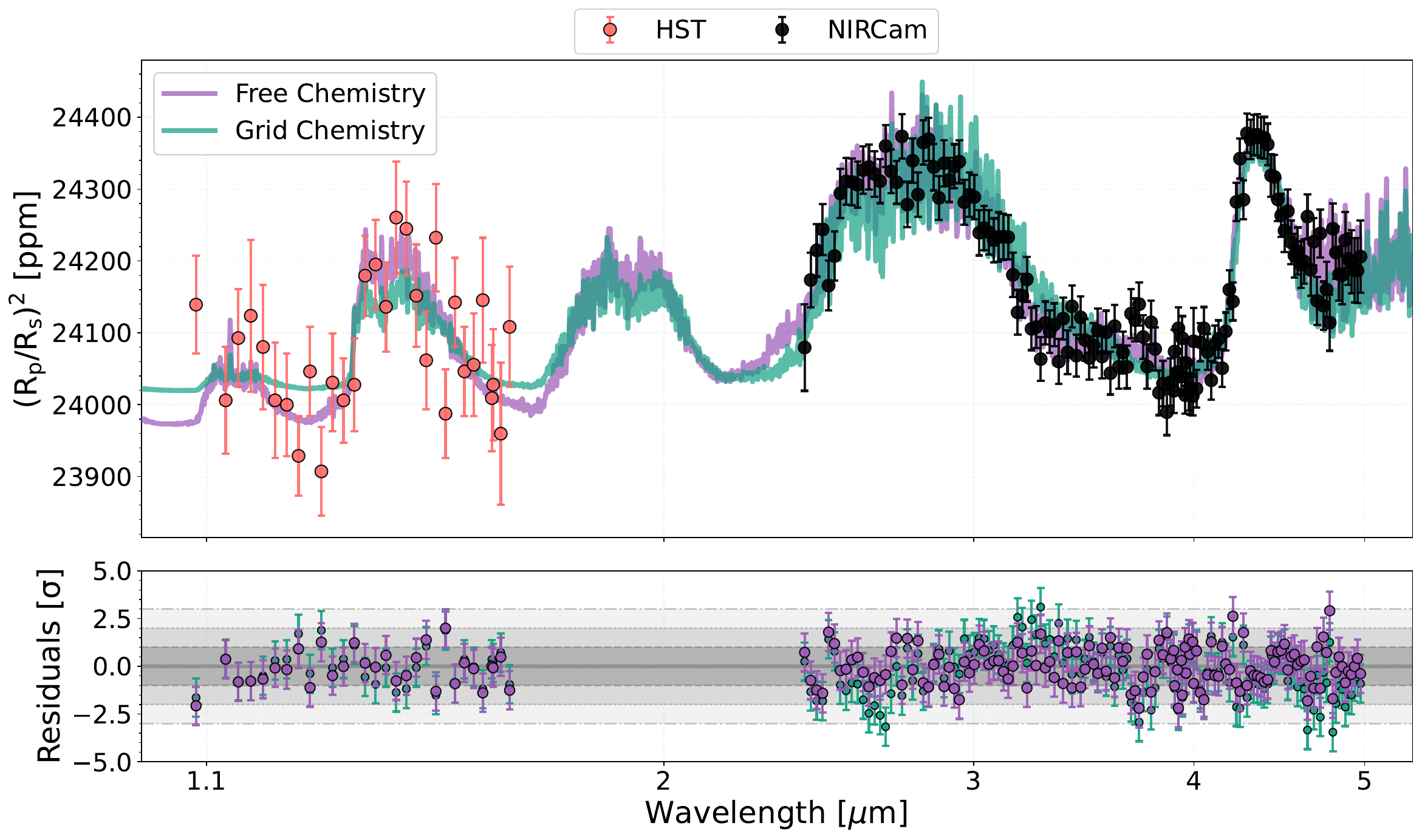}}
\end{overpic}
\caption{Similar to Figure \ref{fig:best_fit_corner_HAT-P-18b} for HD 189733 b.}
\end{figure*}

\renewcommand{\thefigure}{18}
\begin{figure*}
\centering
	\includegraphics[width=\textwidth]{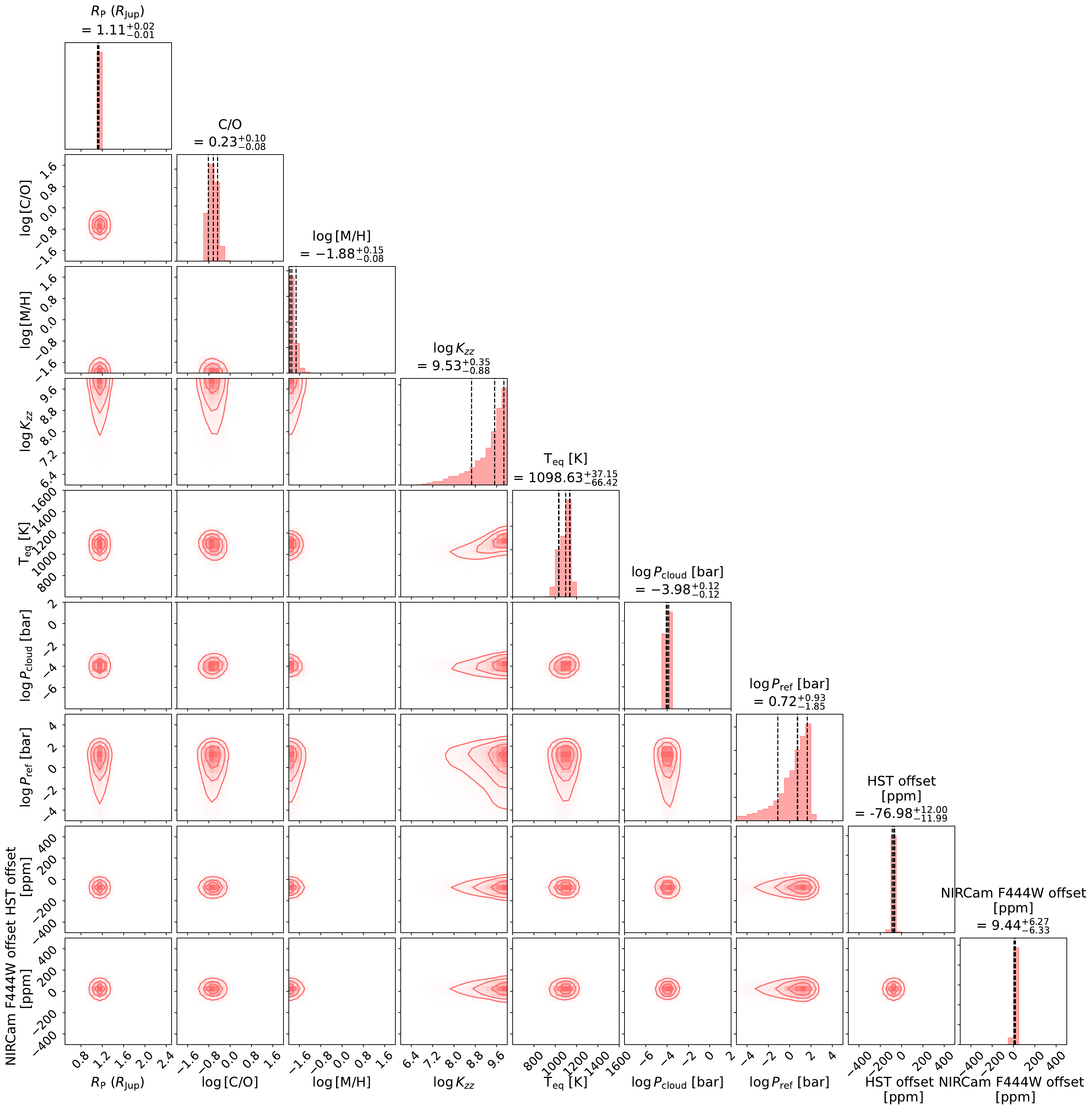}
  \caption{Similar to Figure \ref{fig:corner_hat-p-18b_vulcan} for HD 189733 b.}
   \label{fig:}
\end{figure*}

\renewcommand{\thefigure}{19}
\begin{figure*}
\centering
\begin{overpic}[width=\textwidth]{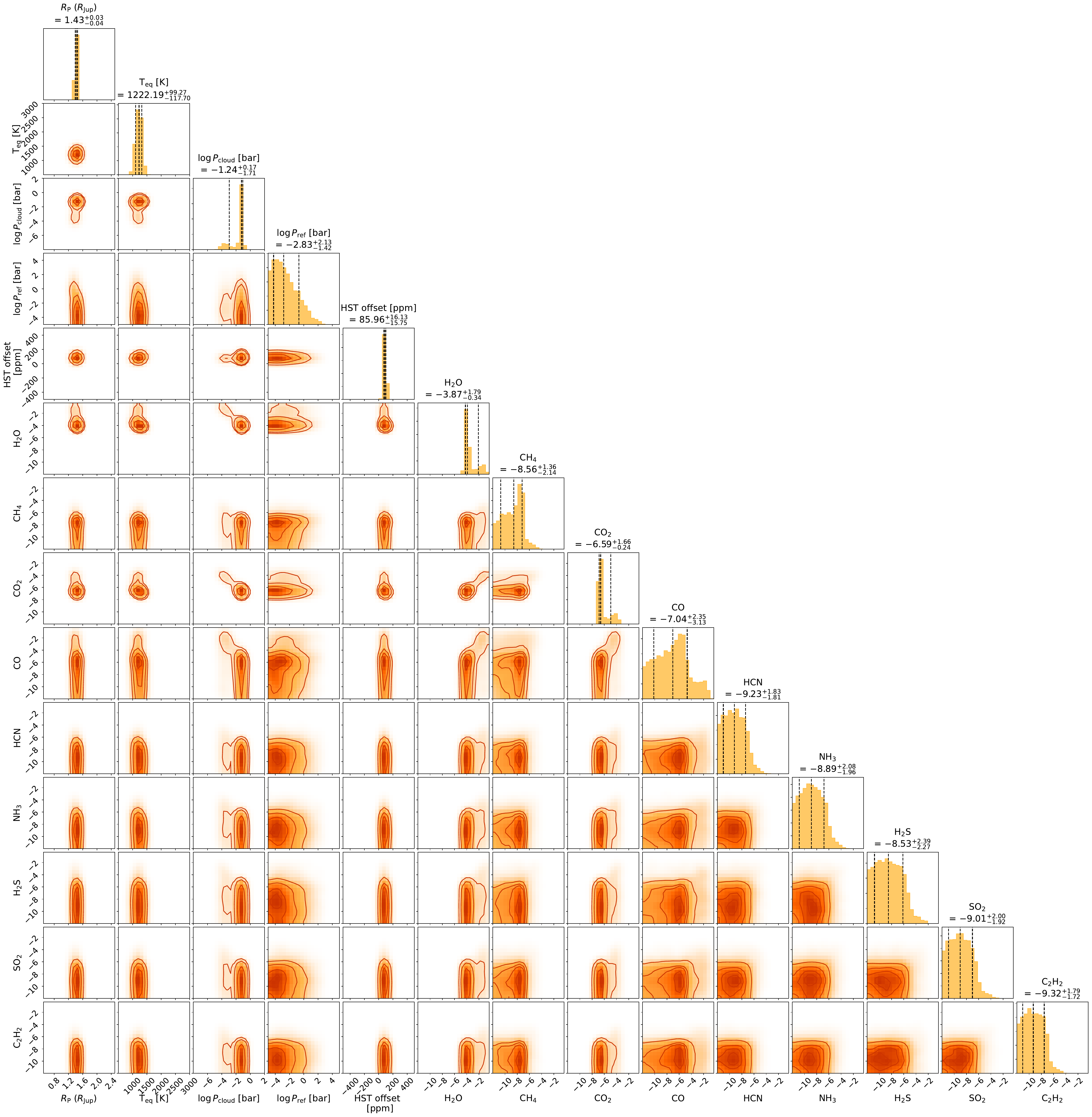}
  \put(45,70){\includegraphics[width=0.55\textwidth]{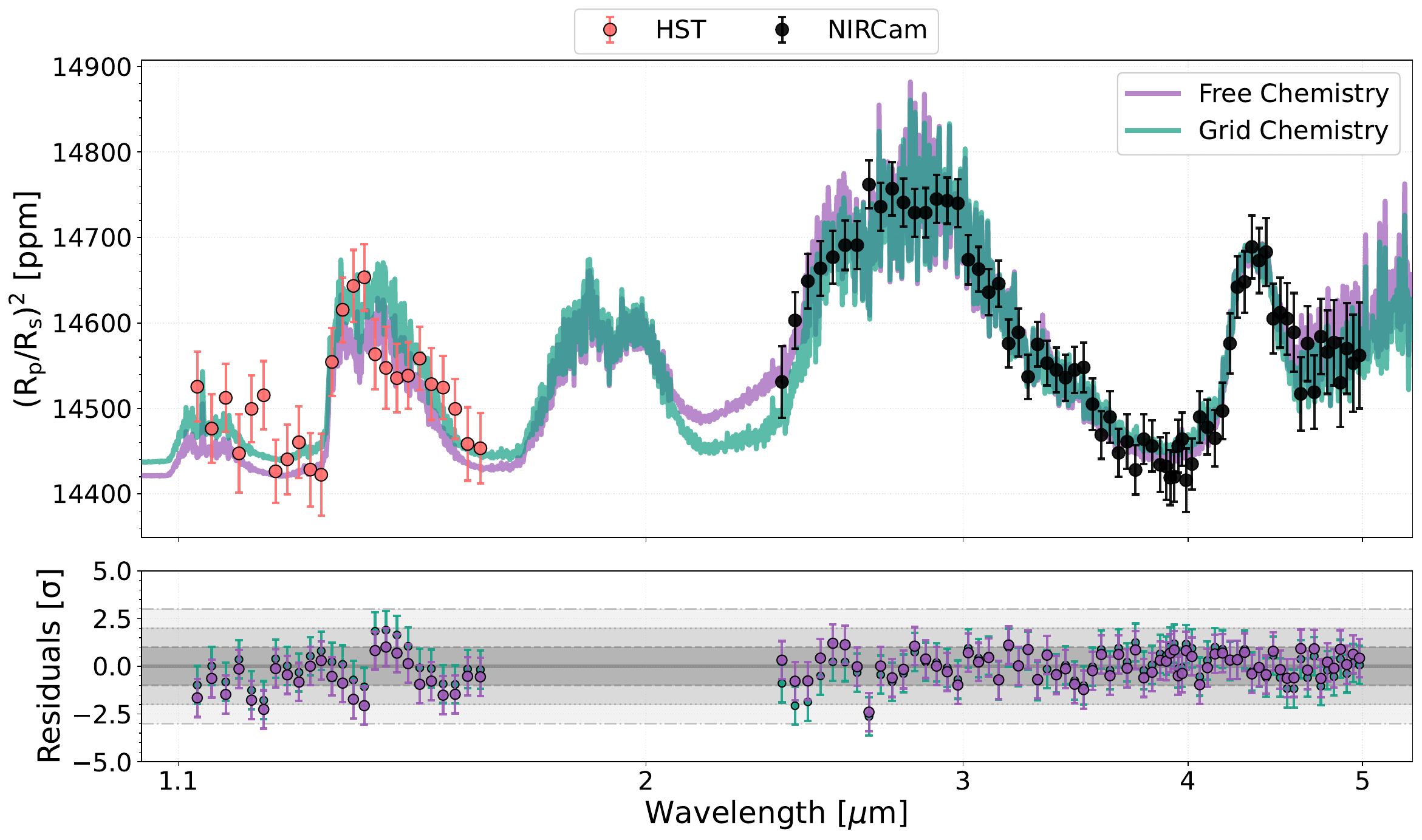}}
\end{overpic}
\caption{Similar to Figure \ref{fig:best_fit_corner_HAT-P-18b} for HD 209458 b.}
\end{figure*}

\renewcommand{\thefigure}{20}
\begin{figure*}
\centering
	\includegraphics[width=\textwidth]{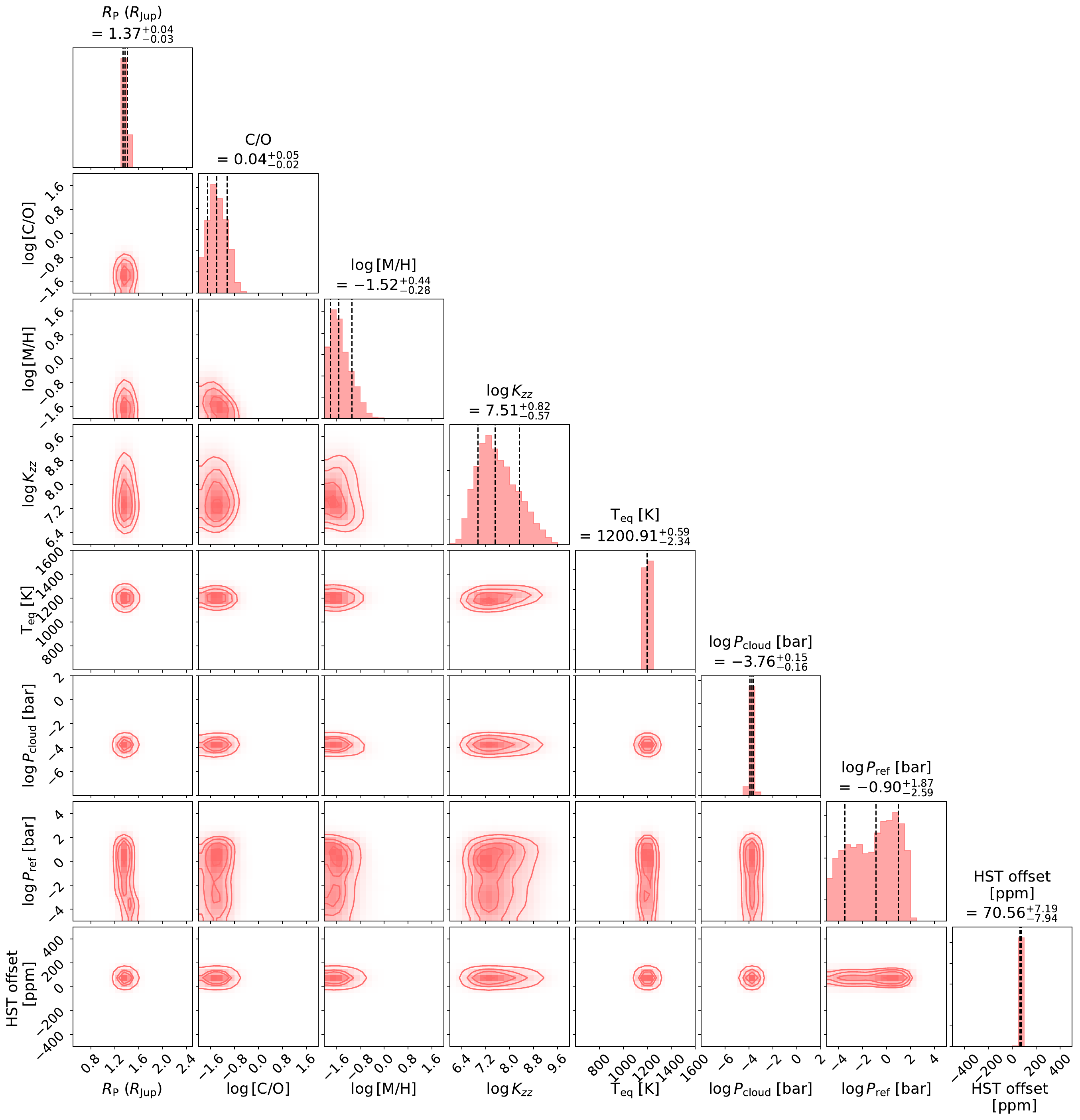}
  \caption{Similar to Figure \ref{fig:corner_hat-p-18b_vulcan} for HD 209458 b.}
   \label{fig:}
\end{figure*}

\renewcommand{\thefigure}{21}
\begin{figure*}
\centering
\begin{overpic}[width=\textwidth]{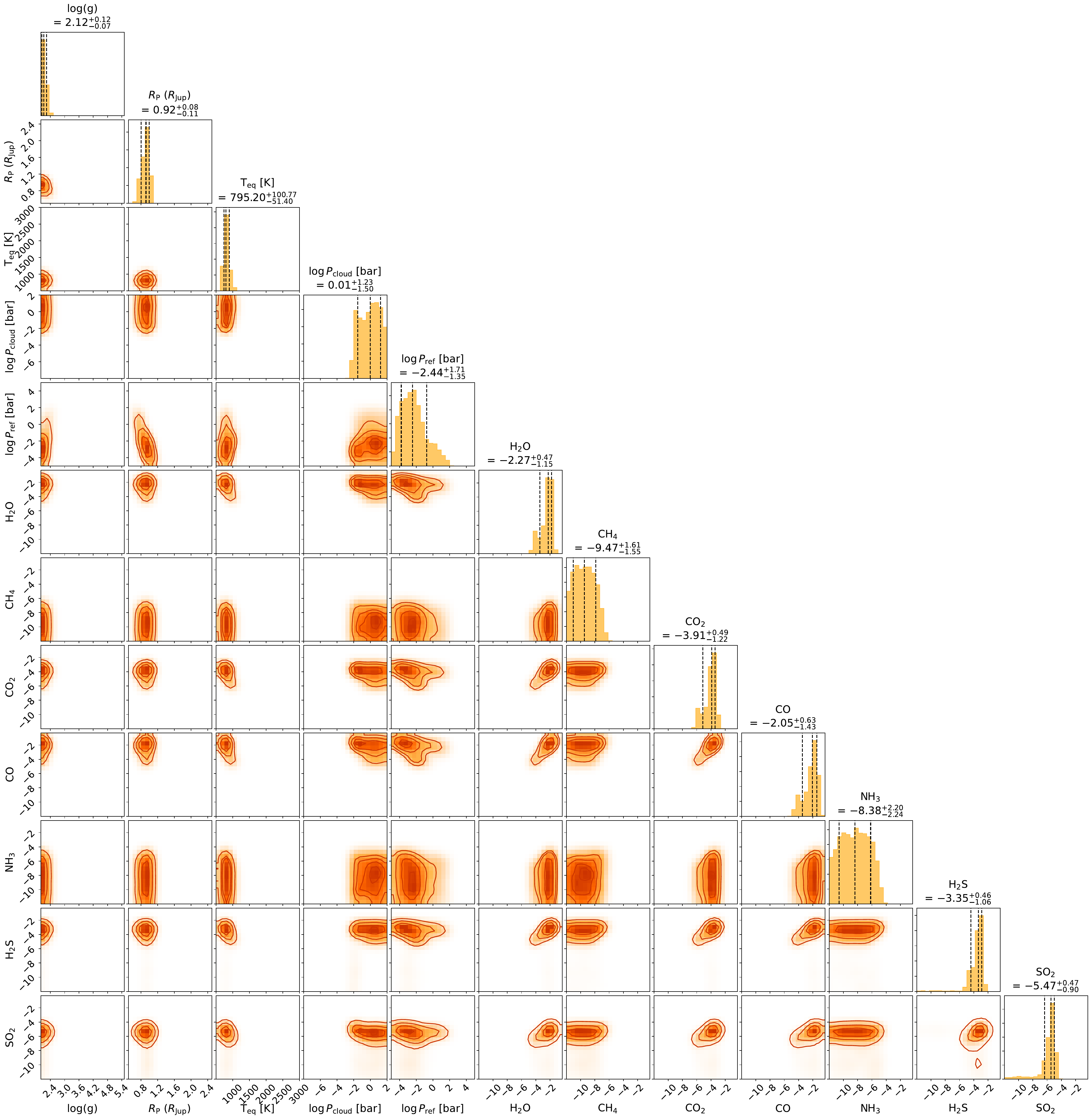}
  \put(45,70){\includegraphics[width=0.55\textwidth]{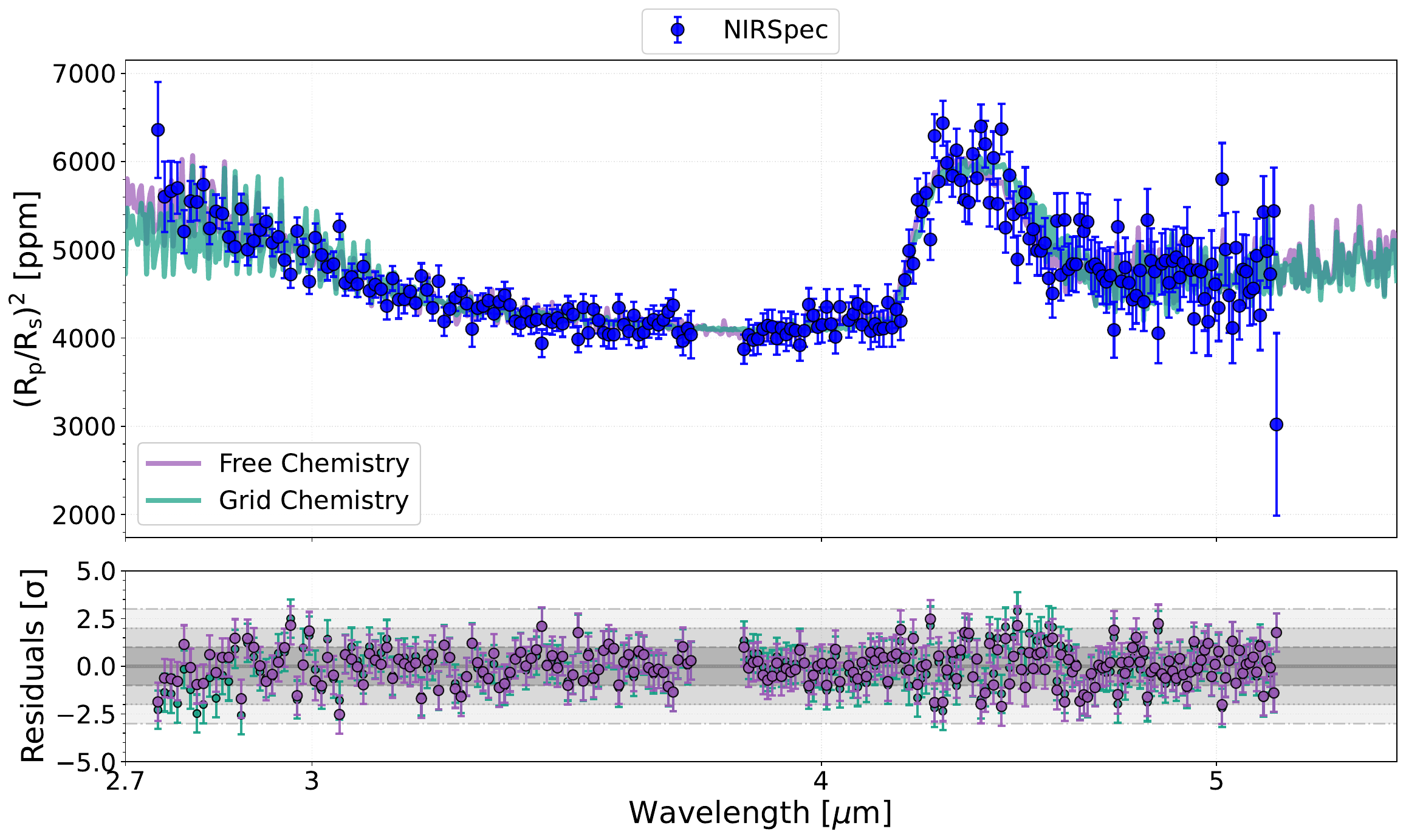}}
\end{overpic}
\caption{Similar to Figure \ref{fig:best_fit_corner_HAT-P-18b} for HIP 67522 b.}
\end{figure*}

\renewcommand{\thefigure}{22}
\begin{figure*}
\centering
	\includegraphics[width=\textwidth]{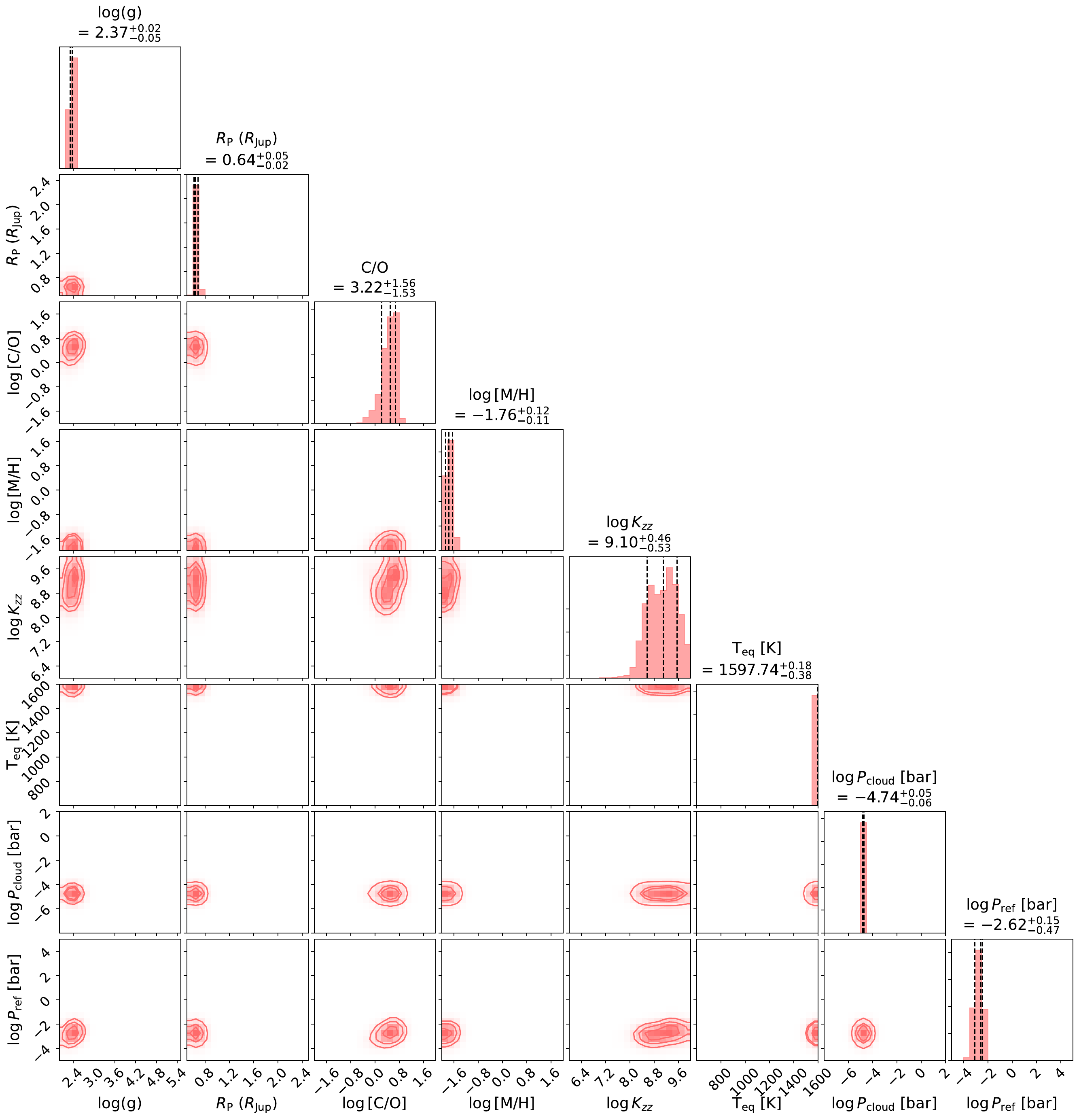}
  \caption{Similar to Figure \ref{fig:corner_hat-p-18b_vulcan} for HIP 67522 b.}
   \label{fig:}
\end{figure*}

\renewcommand{\thefigure}{23}
\begin{figure*}
\centering
\begin{overpic}[width=\textwidth]{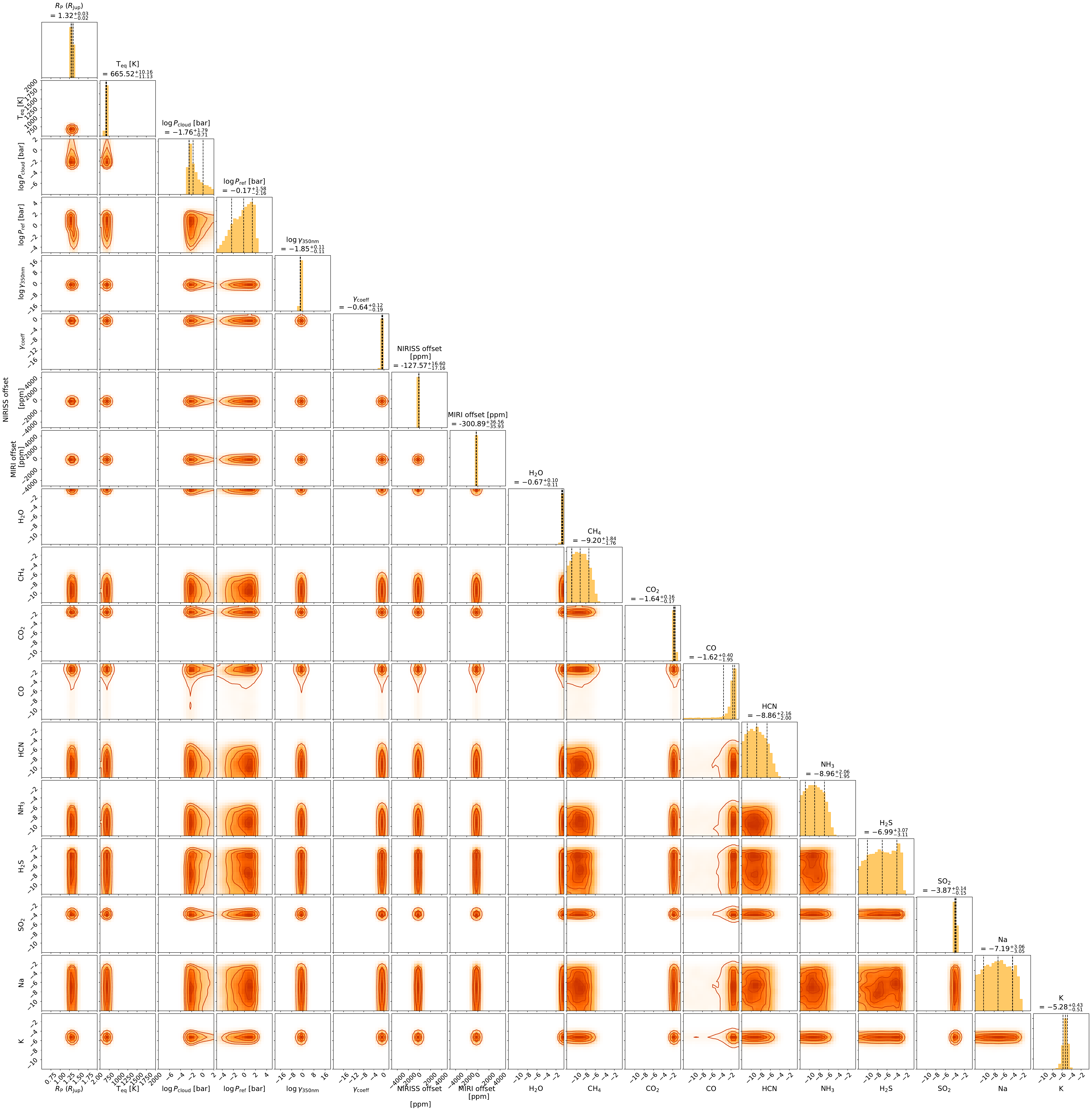}
  \put(45,70){\includegraphics[width=0.55\textwidth]{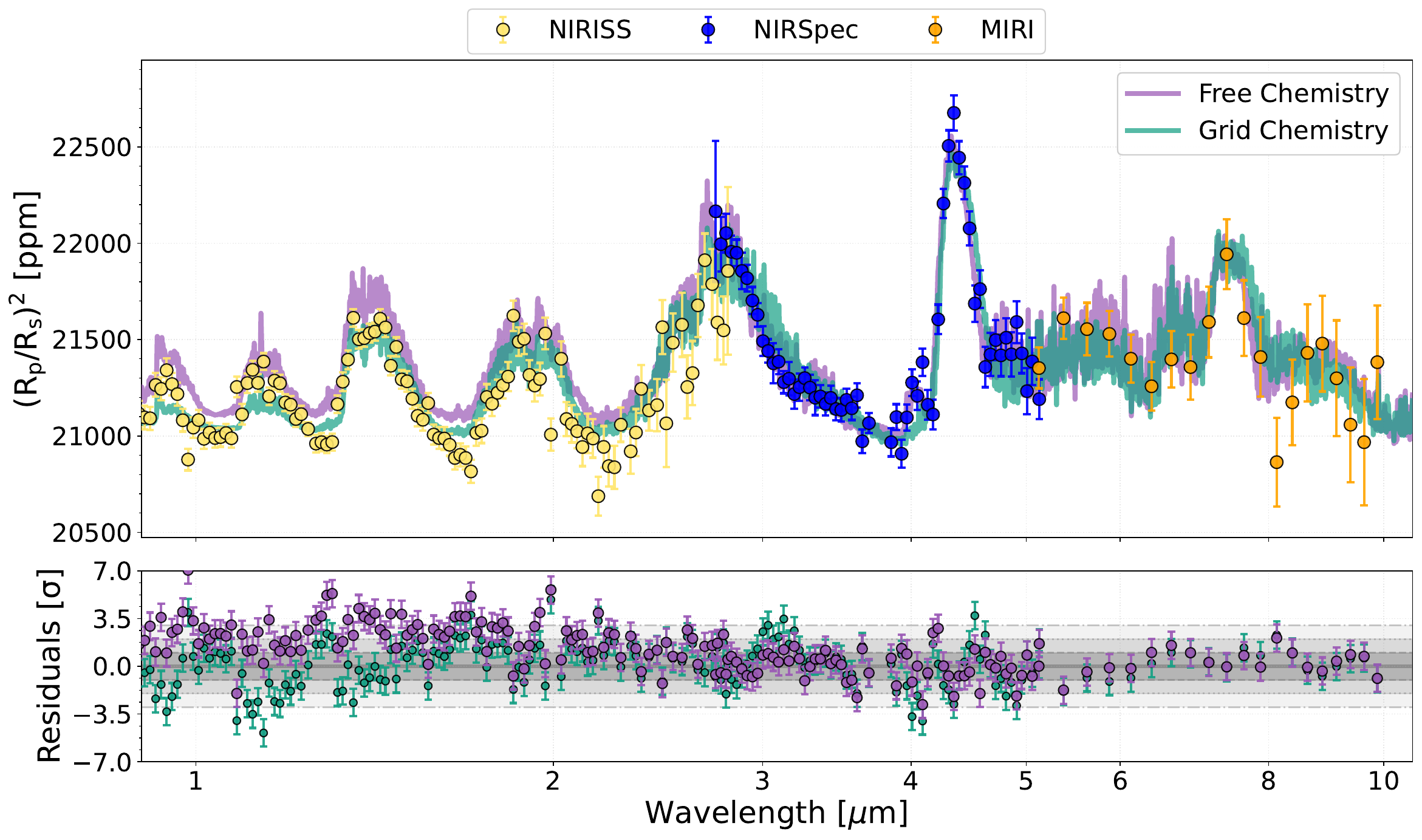}}
\end{overpic}
\caption{Similar to Figure \ref{fig:best_fit_corner_HAT-P-18b} for WASP-39 b.}
\end{figure*}

\renewcommand{\thefigure}{24}
\begin{figure*}
\centering
	\includegraphics[width=\textwidth]{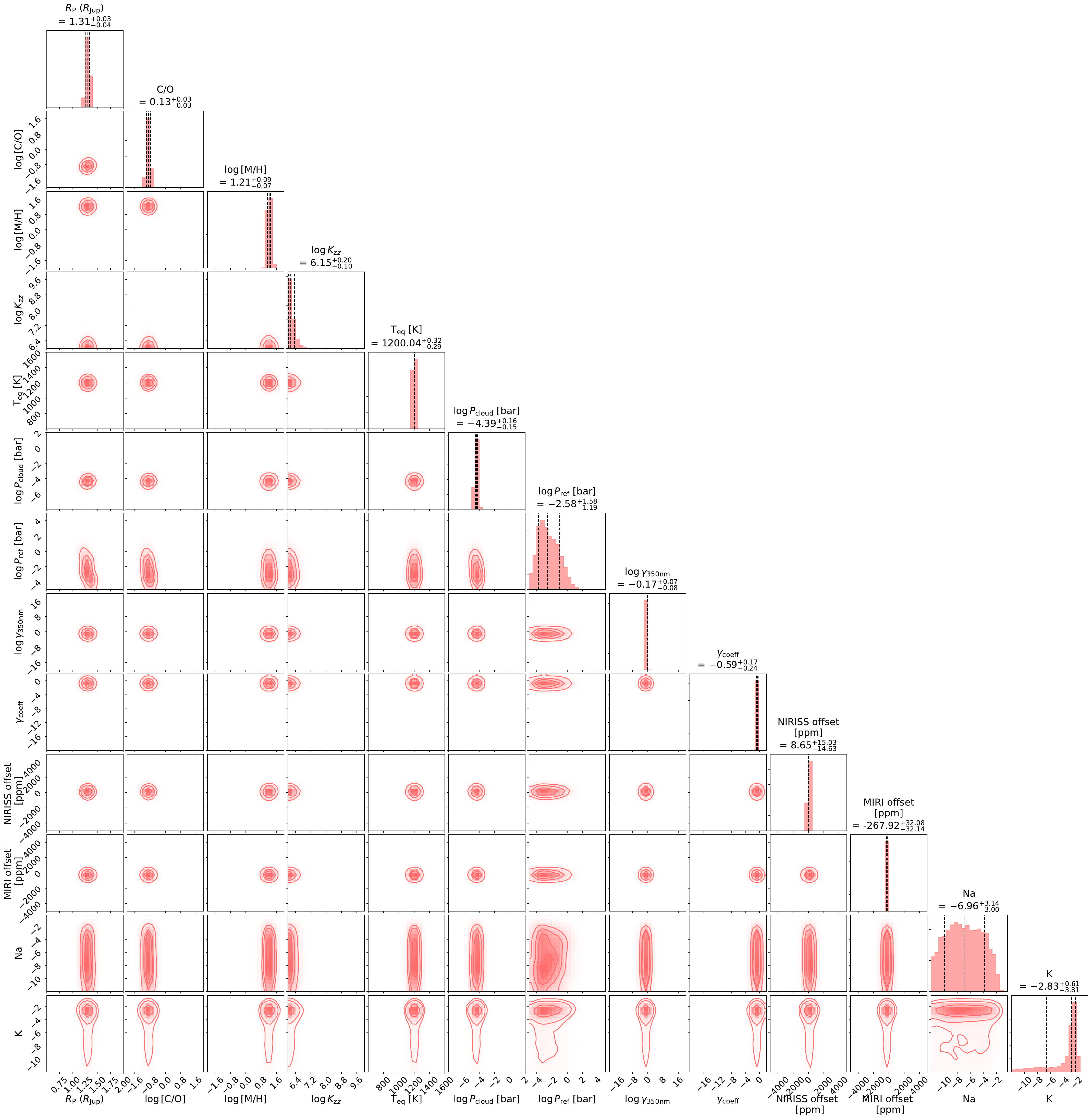}
  \caption{Similar to Figure \ref{fig:corner_hat-p-18b_vulcan} for WASP-39 b.}
   \label{fig:}
\end{figure*}

\renewcommand{\thefigure}{25}
\begin{figure*}
\centering
\begin{overpic}[width=\textwidth]{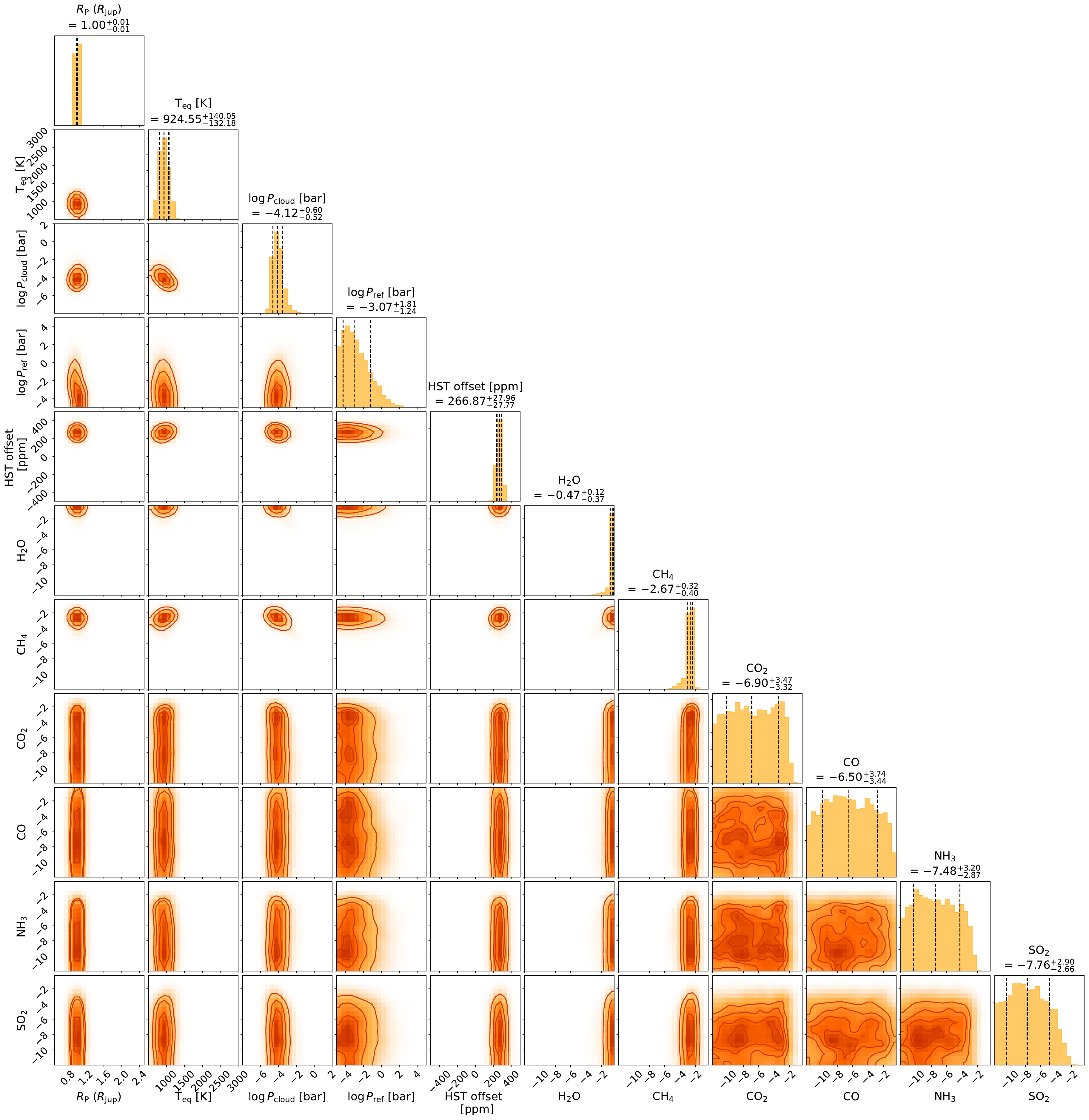}
  \put(45,70){\includegraphics[width=0.55\textwidth]{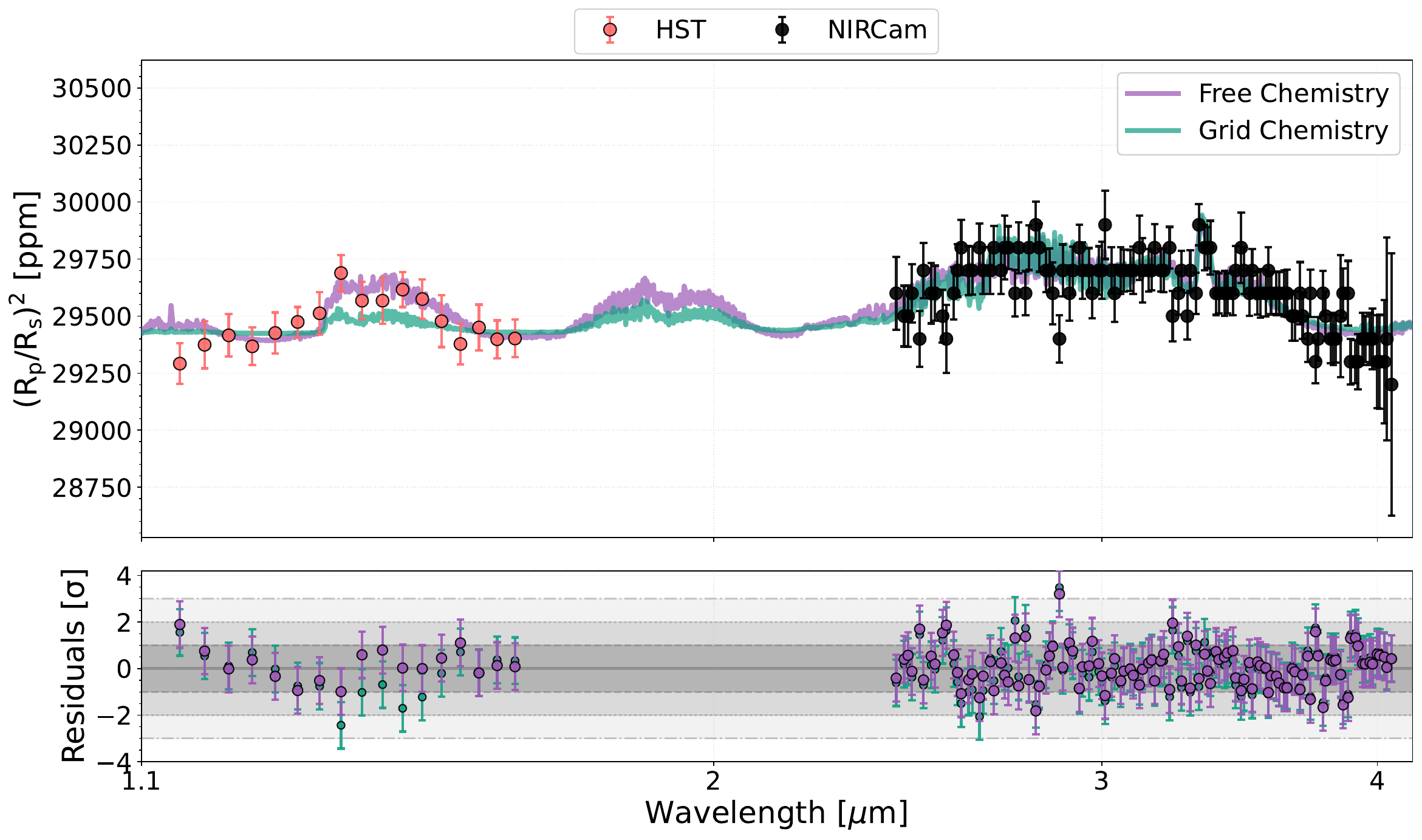}}
\end{overpic}
\caption{Similar to Figure \ref{fig:best_fit_corner_HAT-P-18b} for WASP-80 b.}
\end{figure*}

\renewcommand{\thefigure}{26}
\begin{figure*}
\centering
	\includegraphics[width=\textwidth]{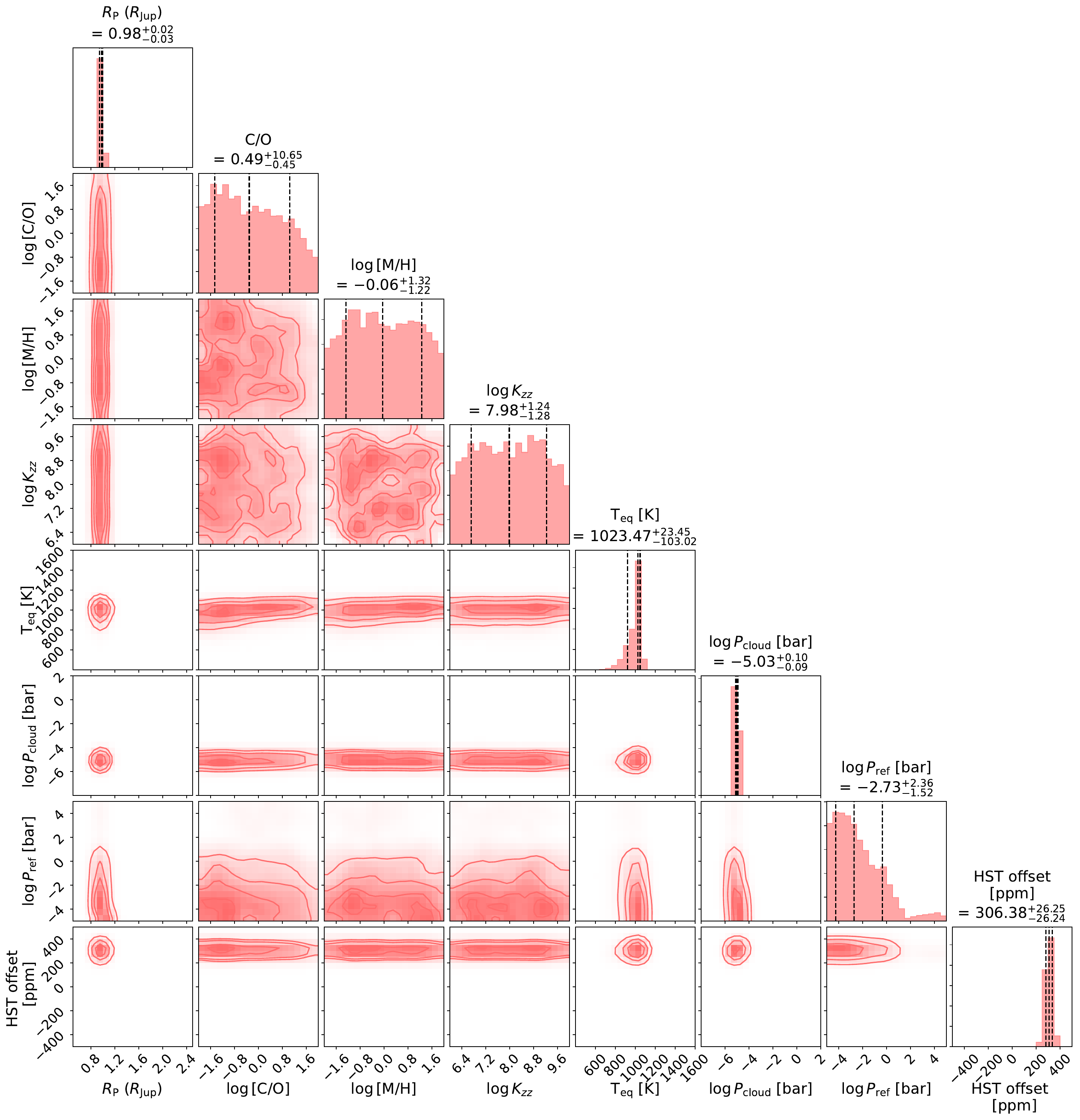}
  \caption{Similar to Figure \ref{fig:corner_hat-p-18b_vulcan} for WASP-80 b.}
   \label{fig:}
\end{figure*}

\renewcommand{\thefigure}{27}
\begin{figure*}
\centering
\begin{overpic}[width=\textwidth]{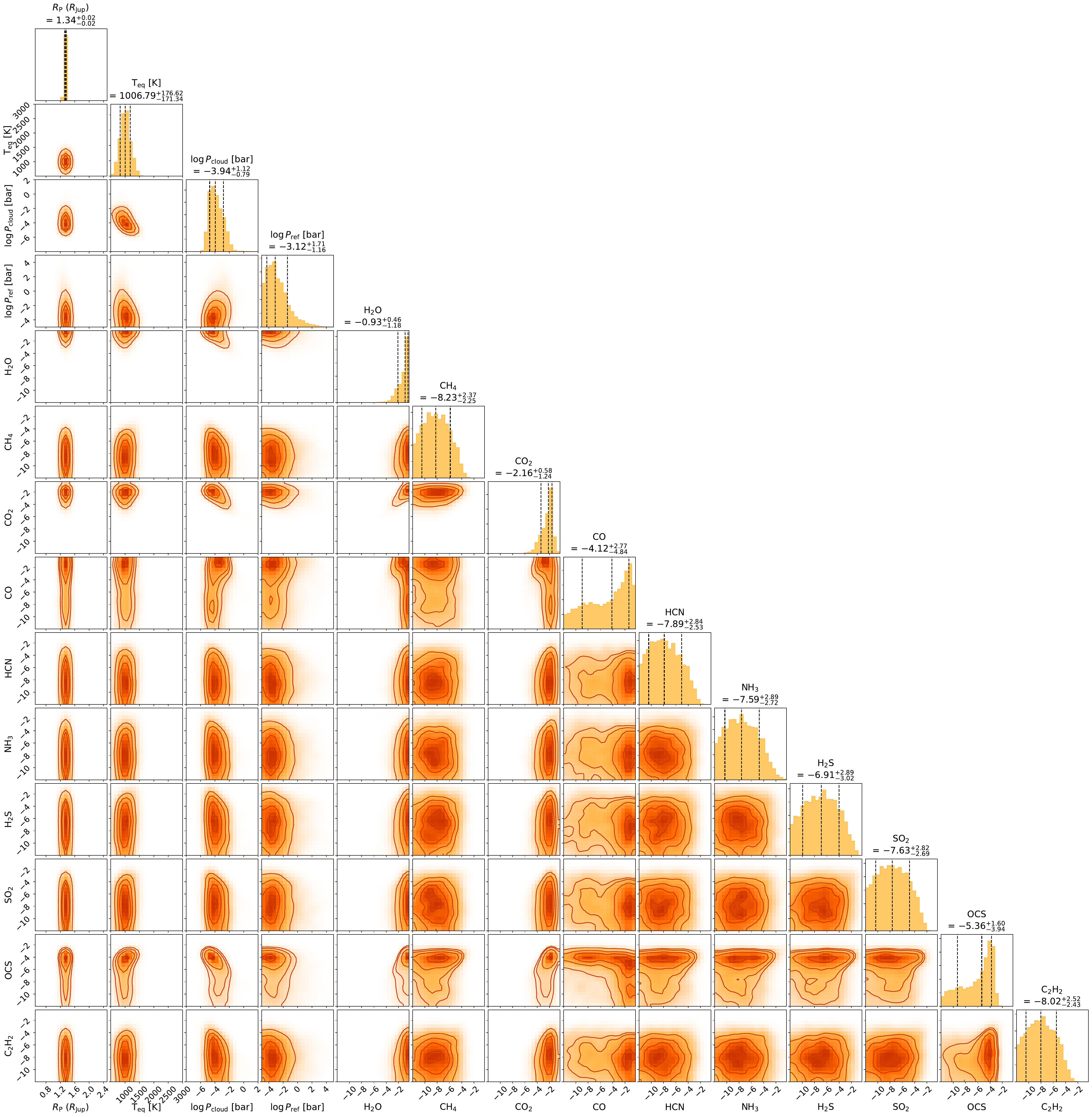}
  \put(45,70){\includegraphics[width=0.55\textwidth]{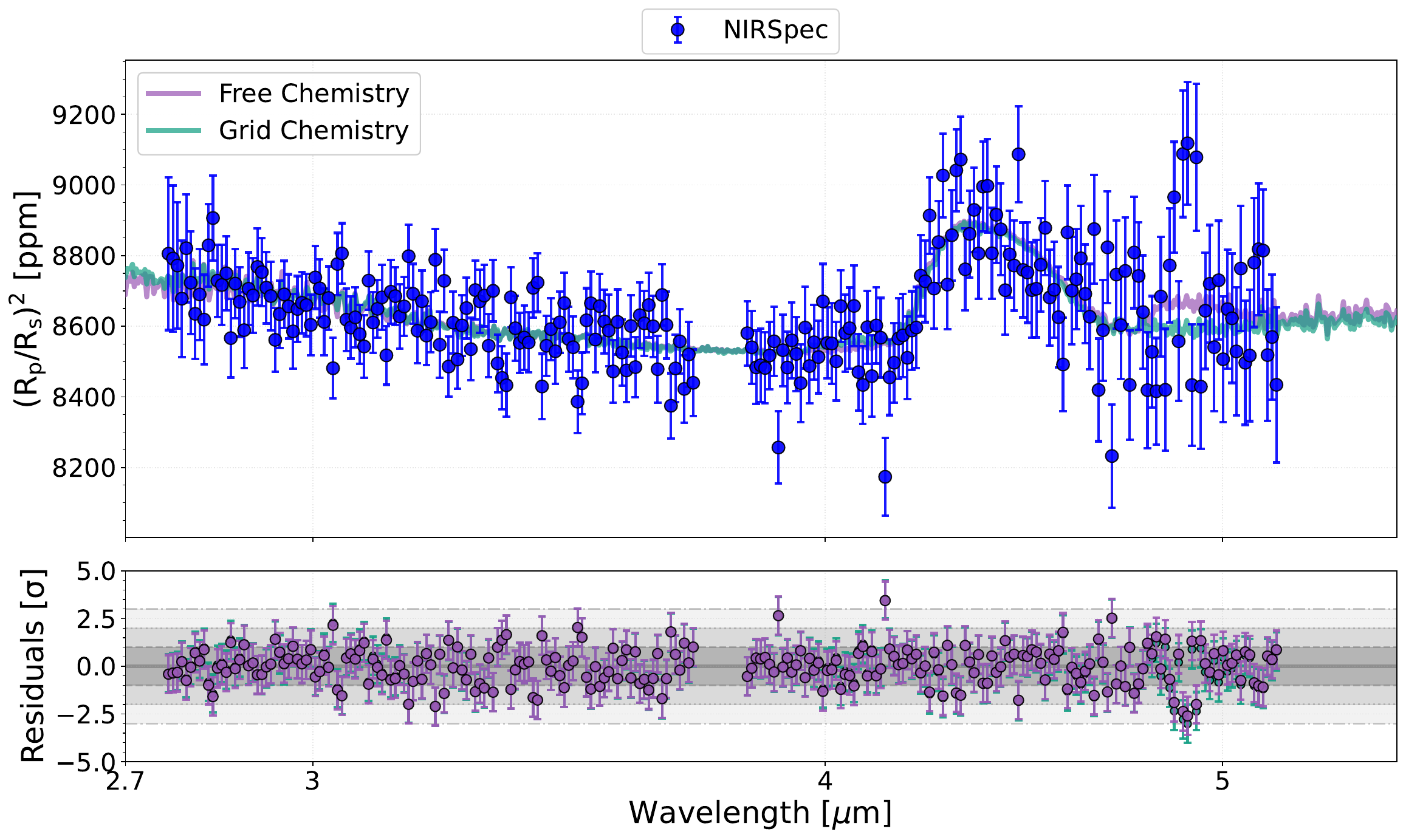}}
\end{overpic}
\caption{Similar to Figure \ref{fig:best_fit_corner_HAT-P-18b} for WASP-15 b.}
\end{figure*}

\renewcommand{\thefigure}{28}
\begin{figure*}
\centering
	\includegraphics[width=\textwidth]{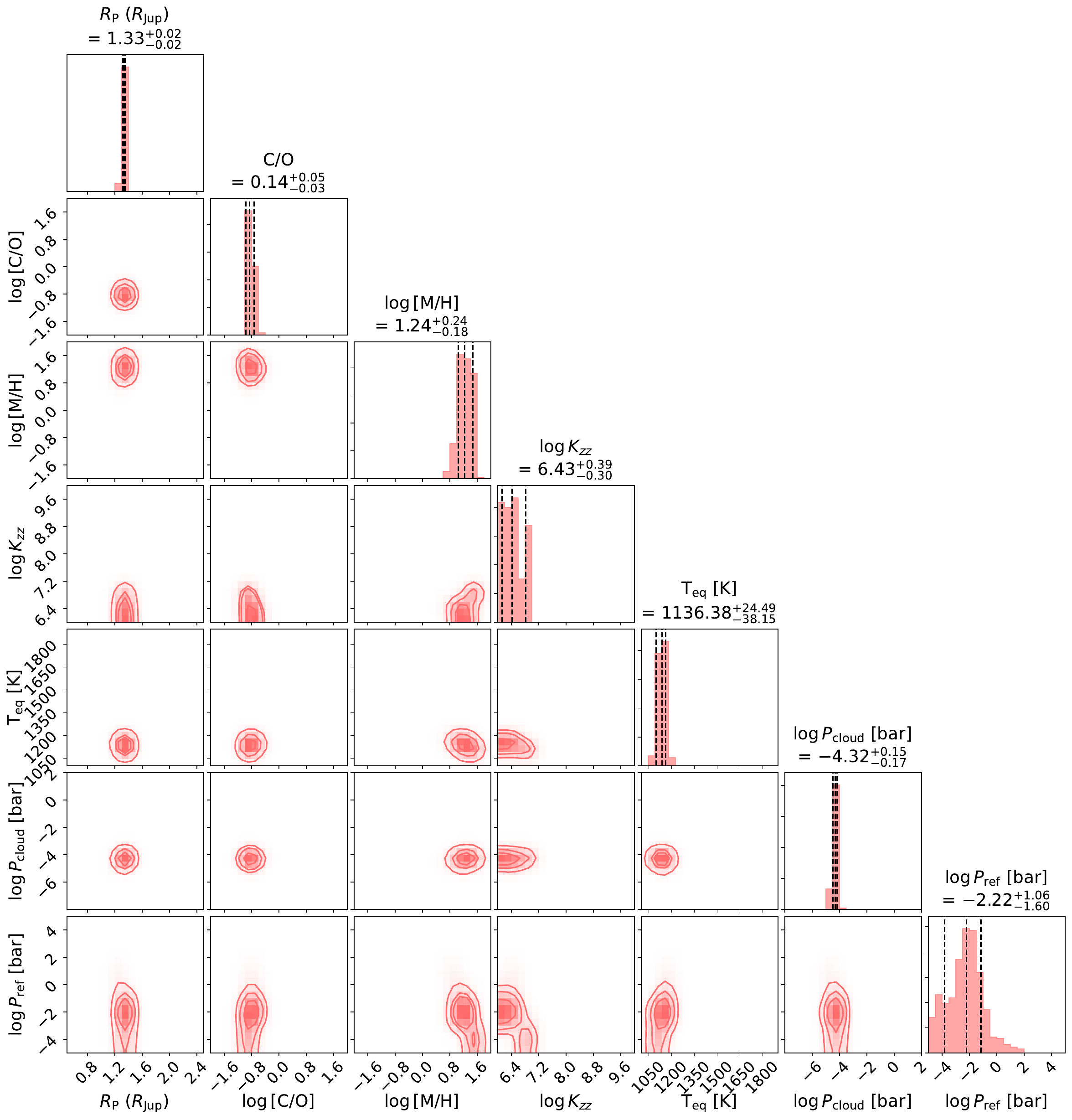}
  \caption{Similar to Figure \ref{fig:corner_hat-p-18b_vulcan} for WASP-15 b.}
   \label{fig:}
\end{figure*}

\renewcommand{\thefigure}{29}
\begin{figure*}
\centering
\begin{overpic}[width=\textwidth]{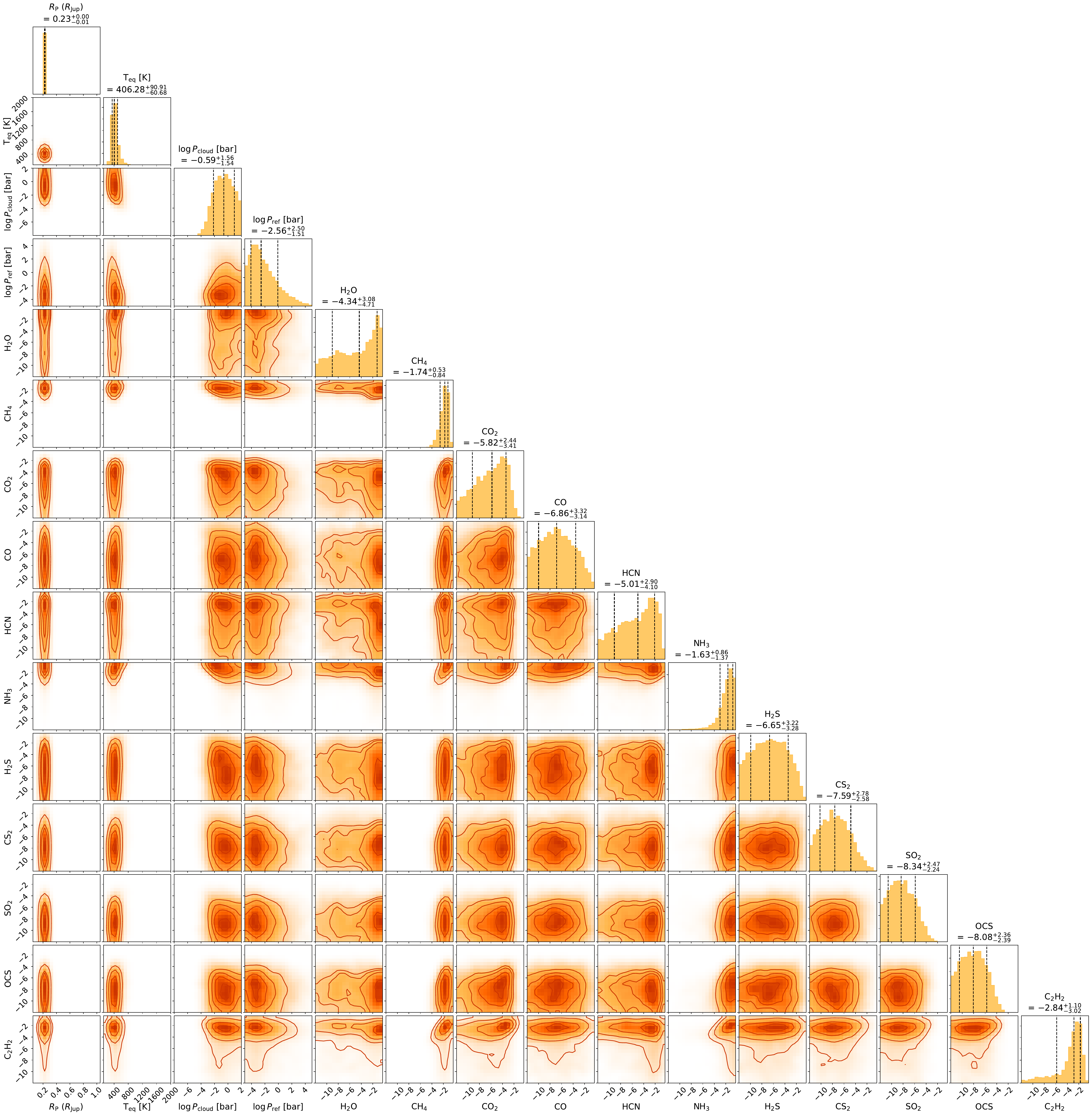}
  \put(45,70){\includegraphics[width=0.55\textwidth]{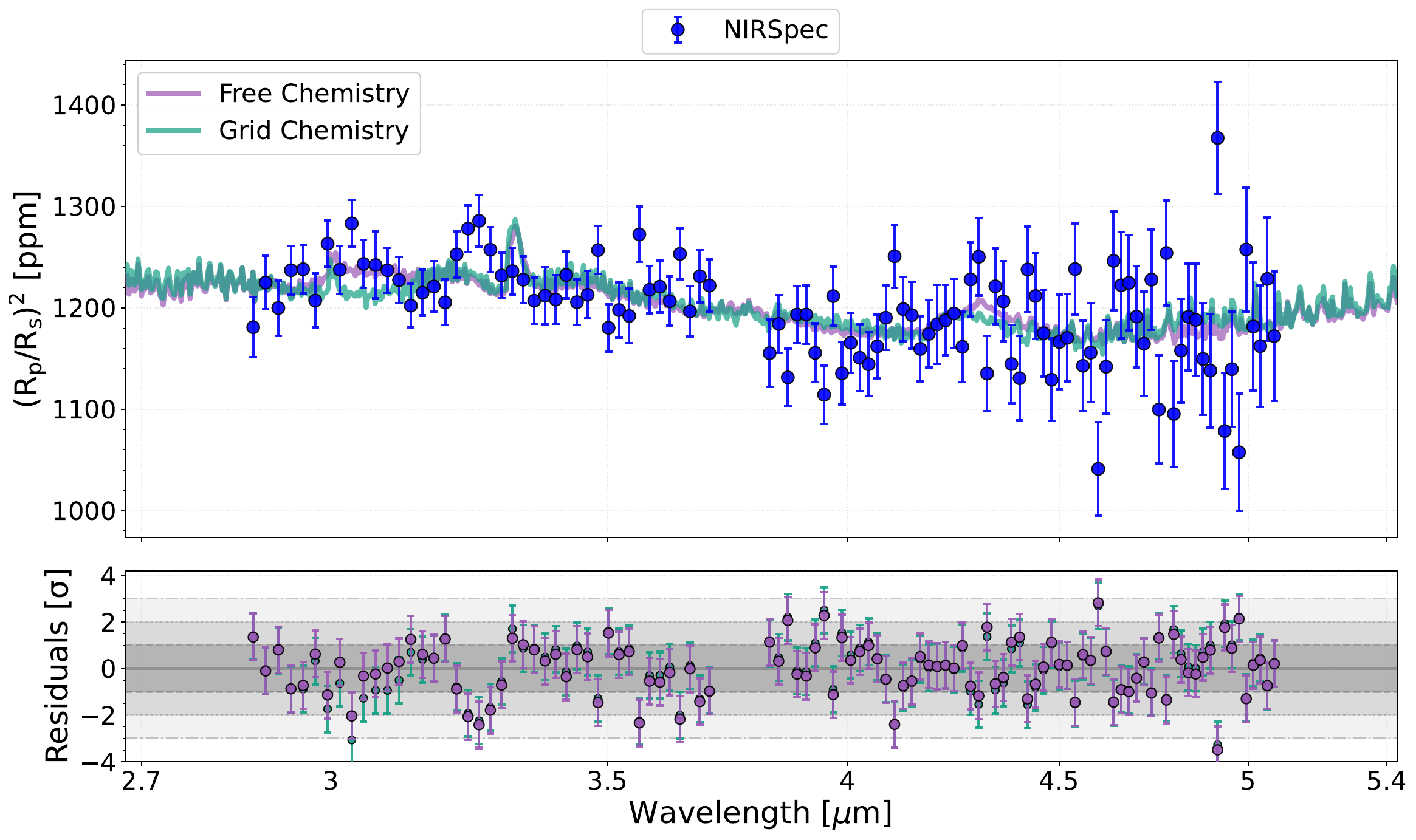}}
\end{overpic}
\caption{Similar to Figure \ref{fig:best_fit_corner_HAT-P-18b} for TOI-836 c.}
\end{figure*}

\renewcommand{\thefigure}{30}
\begin{figure*}
\centering
	\includegraphics[width=\textwidth]{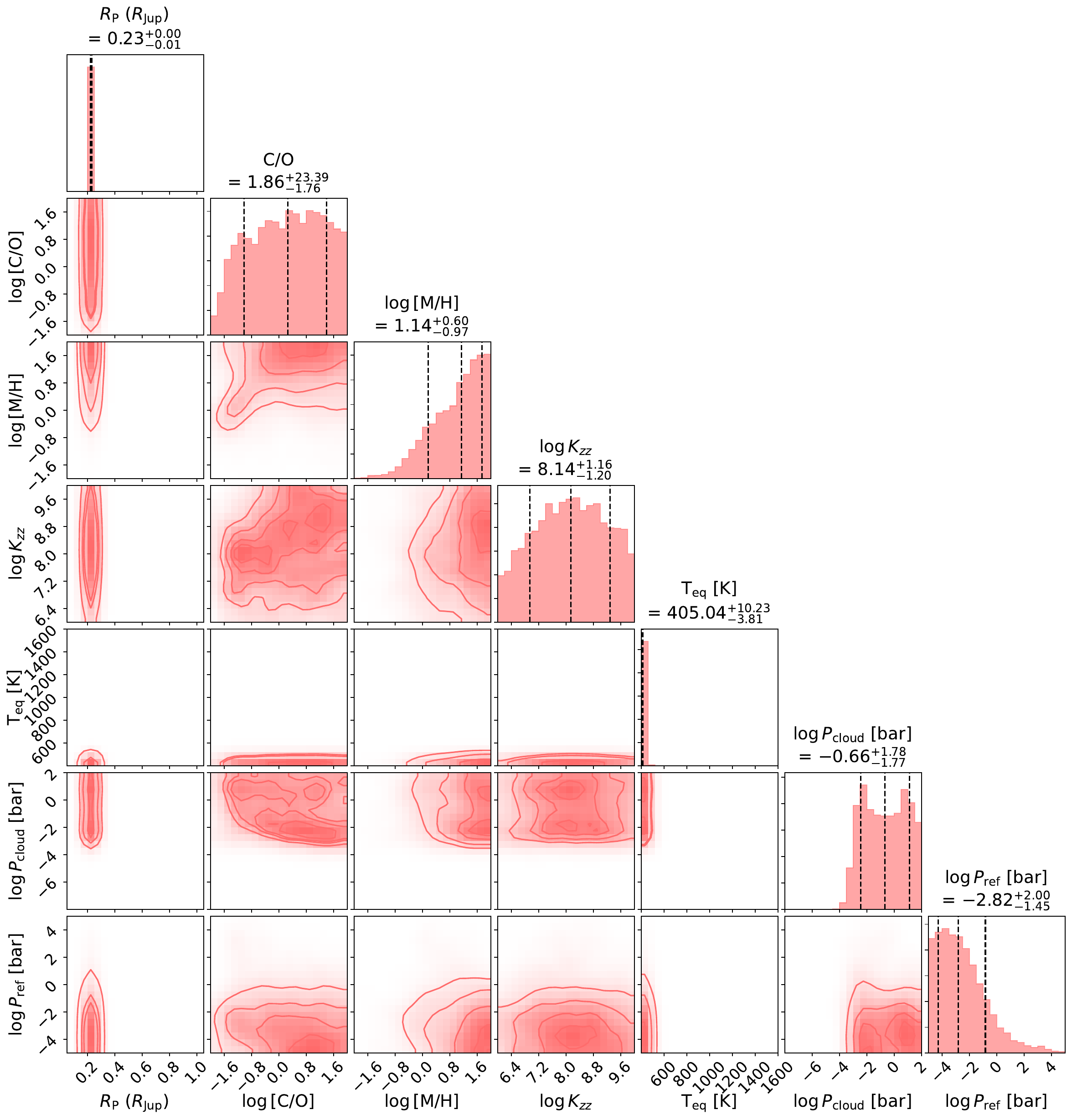}
  \caption{Similar to Figure \ref{fig:corner_hat-p-18b_vulcan} for TOI-836 c.}
   \label{fig:}
\end{figure*}

\renewcommand{\thefigure}{31}
\begin{figure*}
\centering
\begin{overpic}[width=\textwidth]{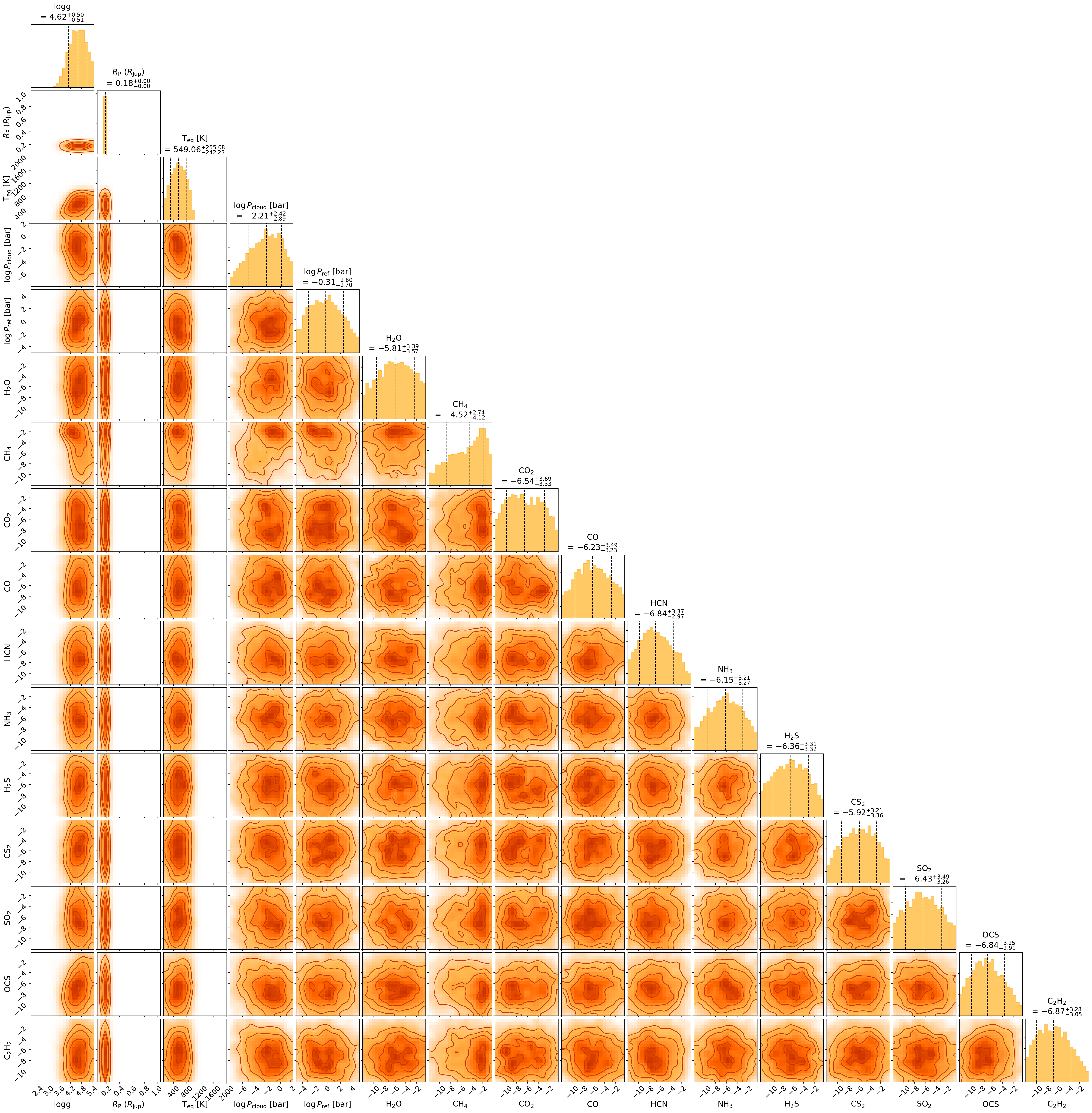}
  \put(45,70){\includegraphics[width=0.55\textwidth]{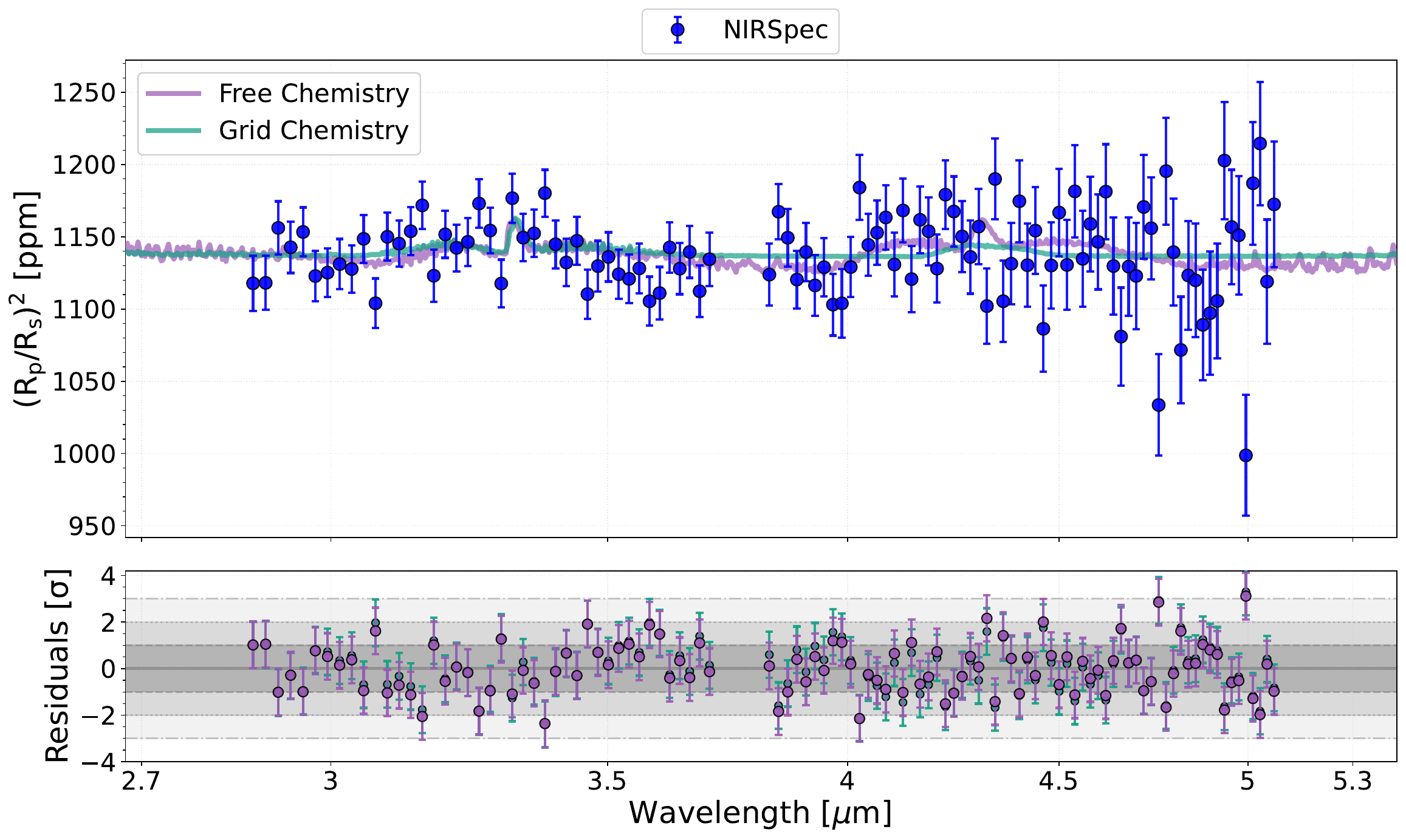}}
\end{overpic}
\caption{Similar to Figure \ref{fig:best_fit_corner_HAT-P-18b} for TOI-776 c.}
\end{figure*}

\renewcommand{\thefigure}{32}
\begin{figure*}
\centering
	\includegraphics[width=\textwidth]{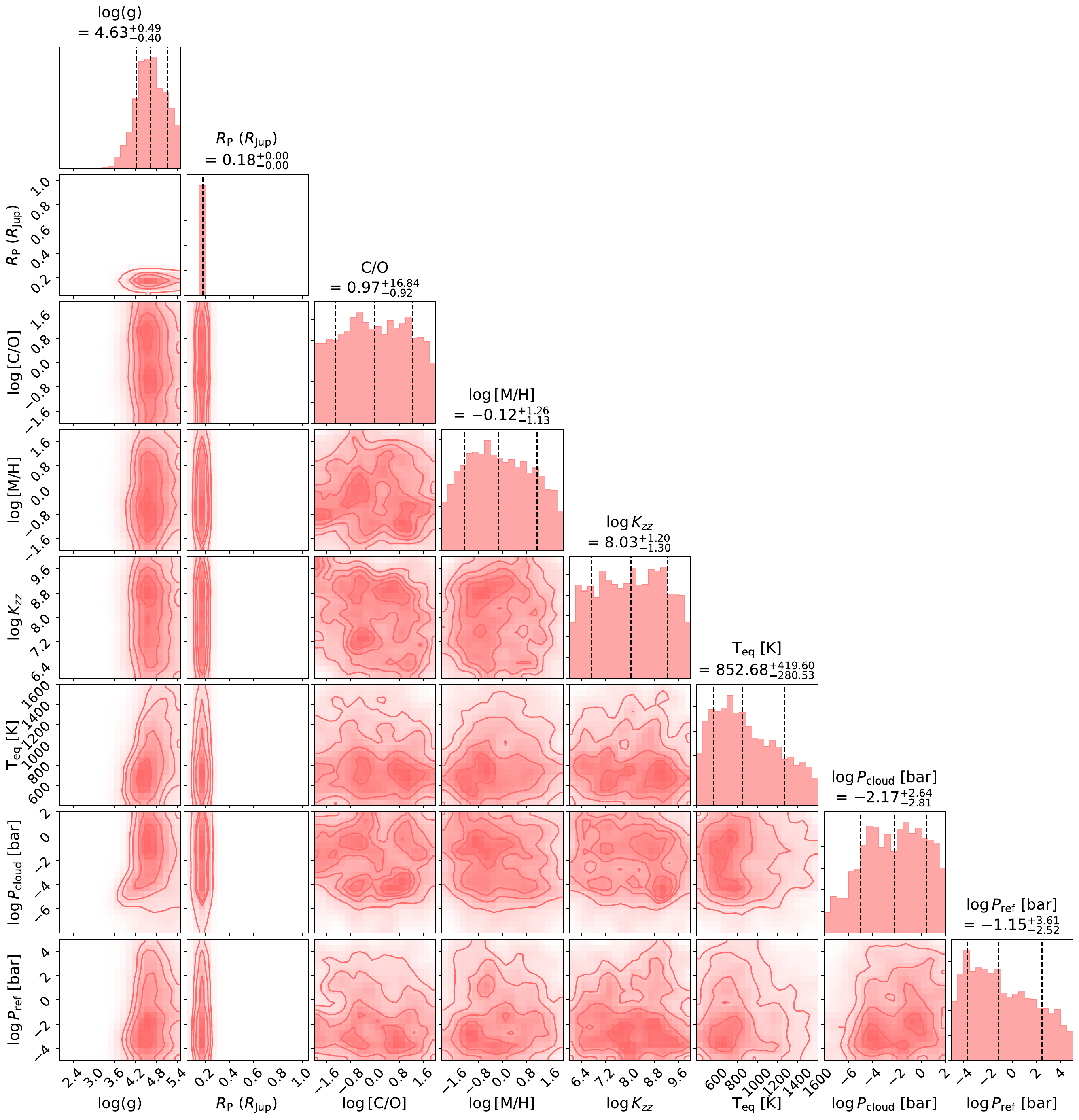}
  \caption{Similar to Figure \ref{fig:corner_hat-p-18b_vulcan} for TOI-776 c.}
   \label{fig:}
\end{figure*}

\renewcommand{\thefigure}{33}
\begin{figure*}
\centering
\begin{overpic}[width=\textwidth]{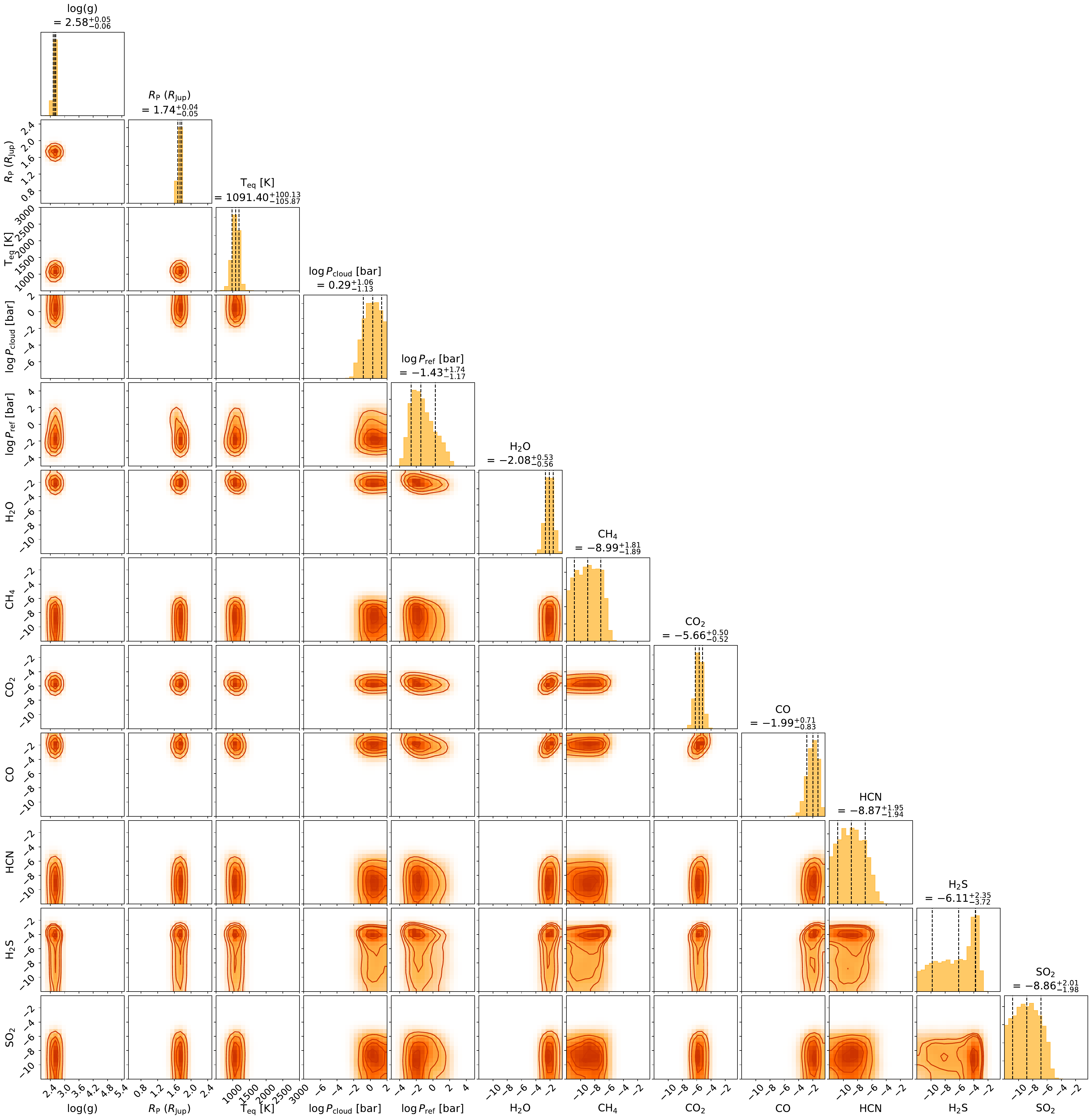}
  \put(45,70){\includegraphics[width=0.55\textwidth]{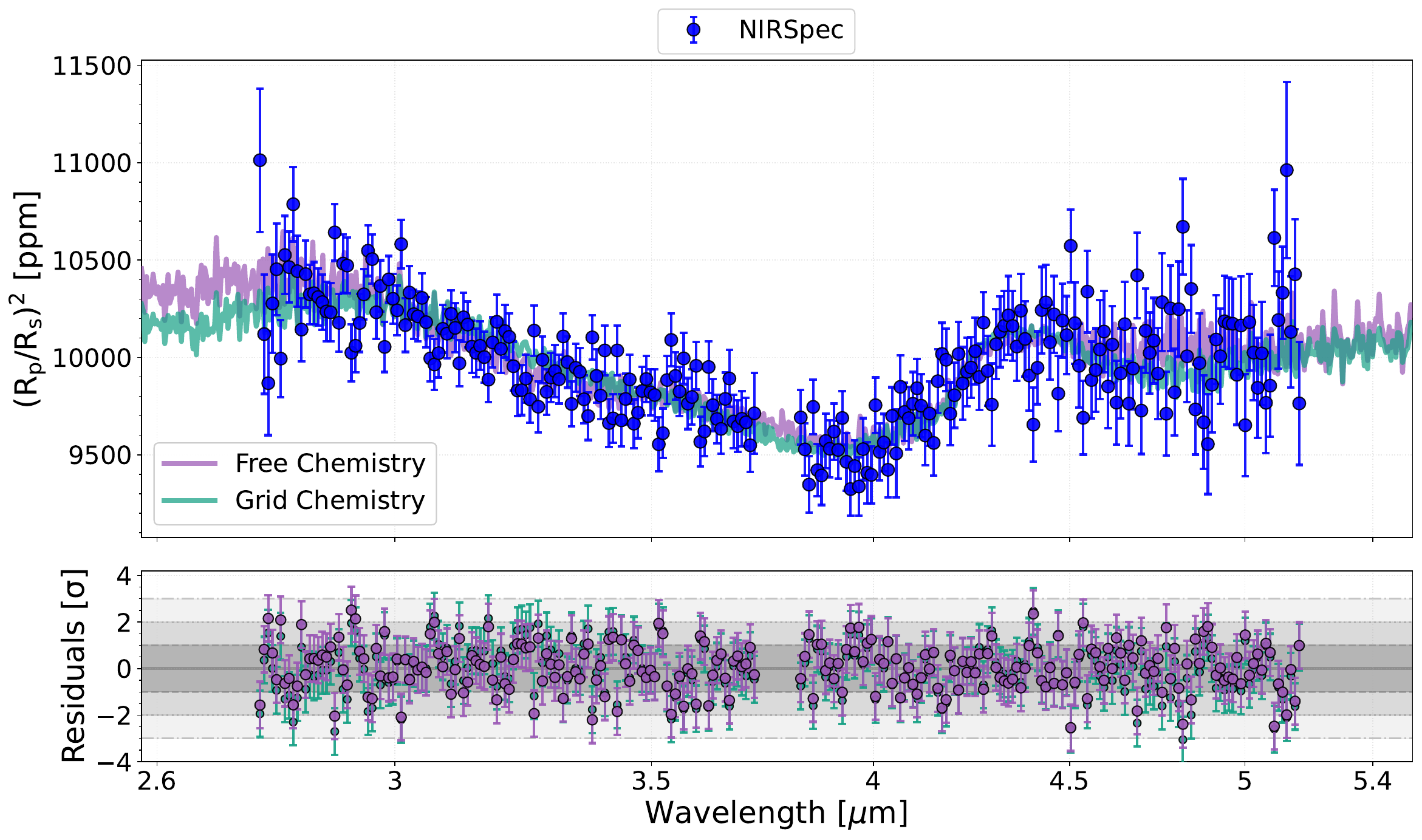}}
\end{overpic}
\caption{Similar to Figure \ref{fig:best_fit_corner_HAT-P-18b} for TrES-4 b.}
\end{figure*}

\renewcommand{\thefigure}{34}
\begin{figure*}
\centering
	\includegraphics[width=\textwidth]{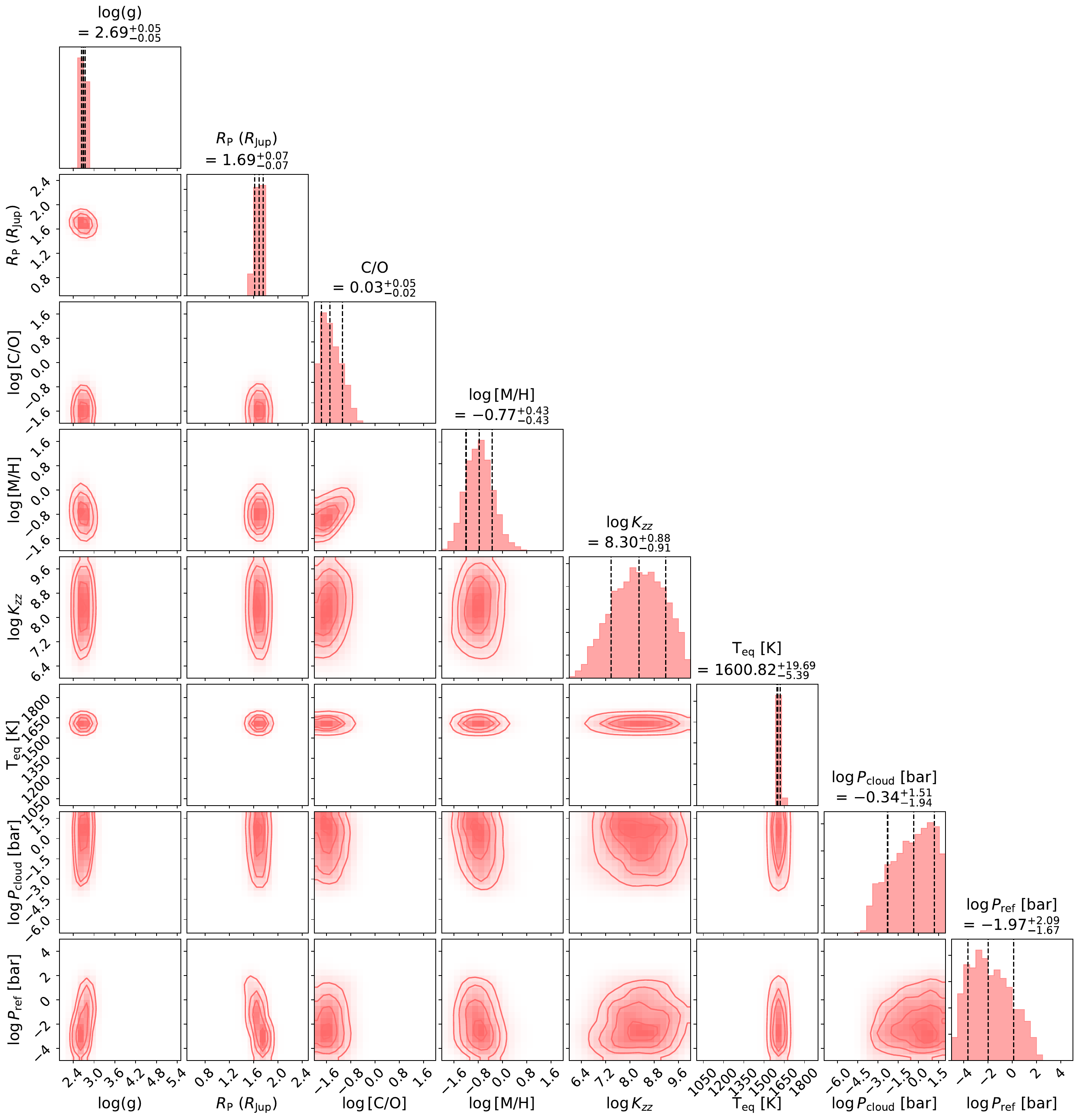}
  \caption{Similar to Figure \ref{fig:corner_hat-p-18b_vulcan} for TrES-4 b.}
   \label{fig:}
\end{figure*}


\renewcommand{\thefigure}{35}
\begin{figure*}
\centering
\begin{overpic}[width=\textwidth]{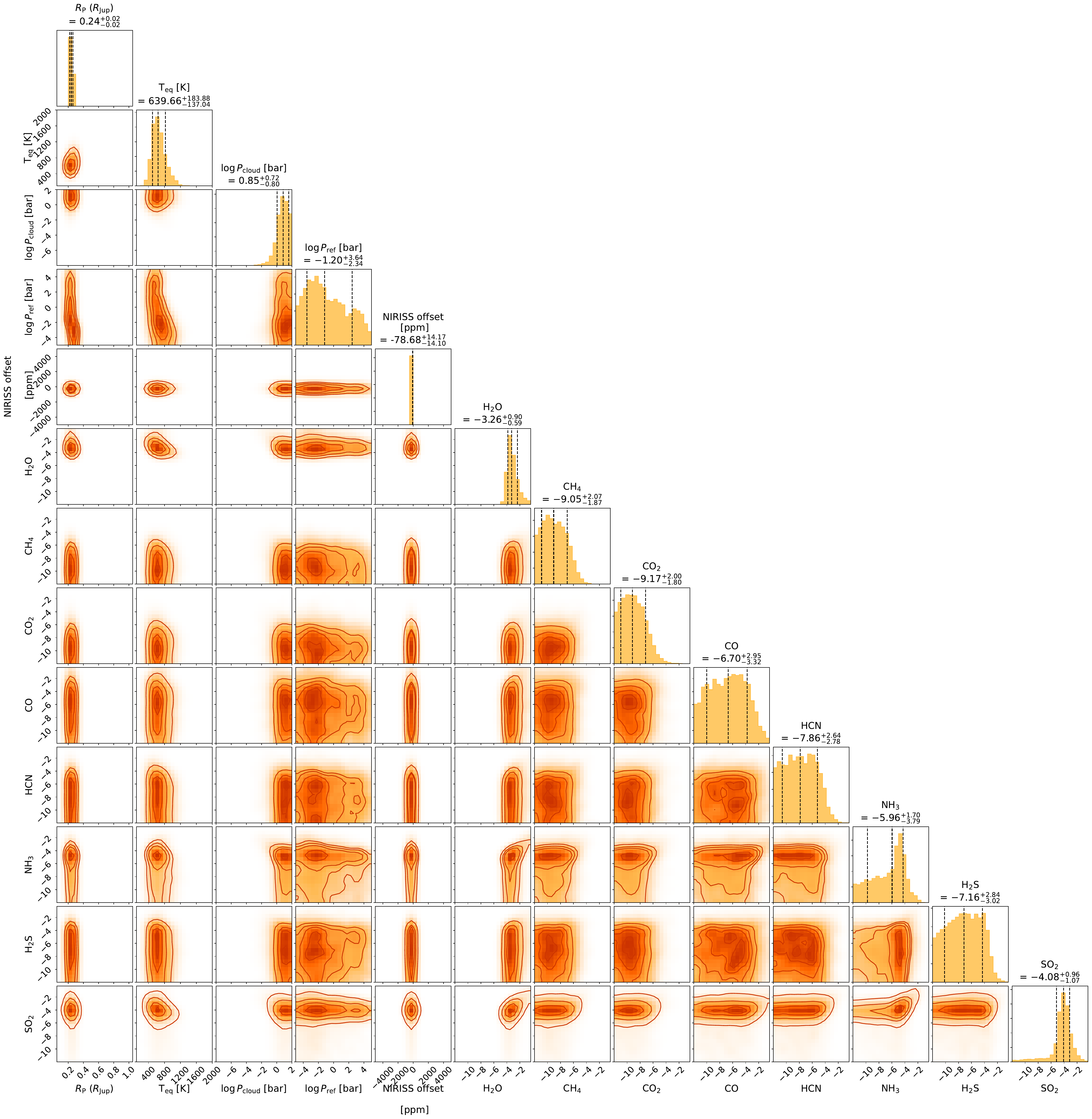}
  \put(45,70){\includegraphics[width=0.55\textwidth]{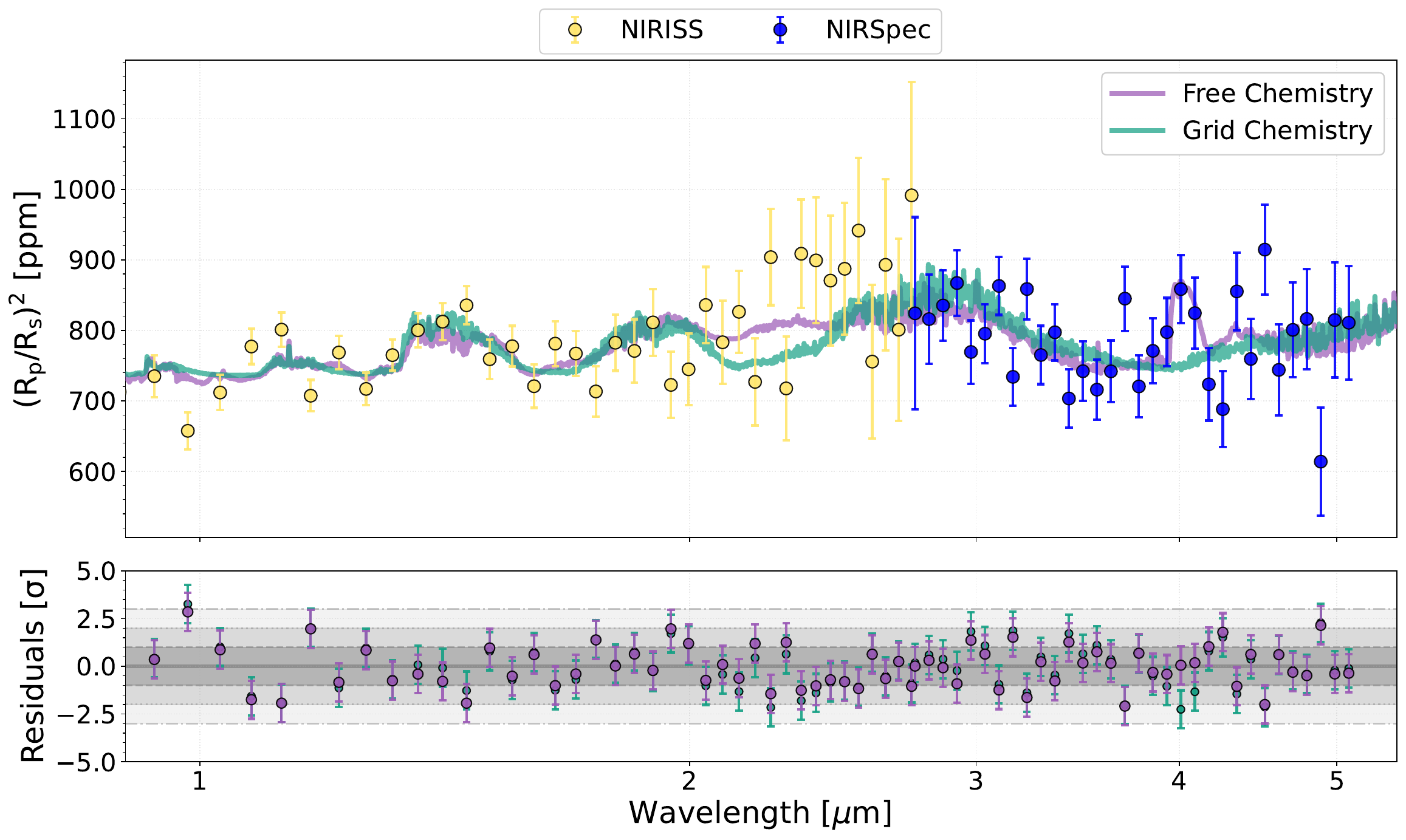}}
\end{overpic}
\caption{Similar to Figure \ref{fig:best_fit_corner_HAT-P-18b} for TOI-421 b.}
\end{figure*}

\renewcommand{\thefigure}{36}
\begin{figure*}
\centering
	\includegraphics[width=\textwidth]{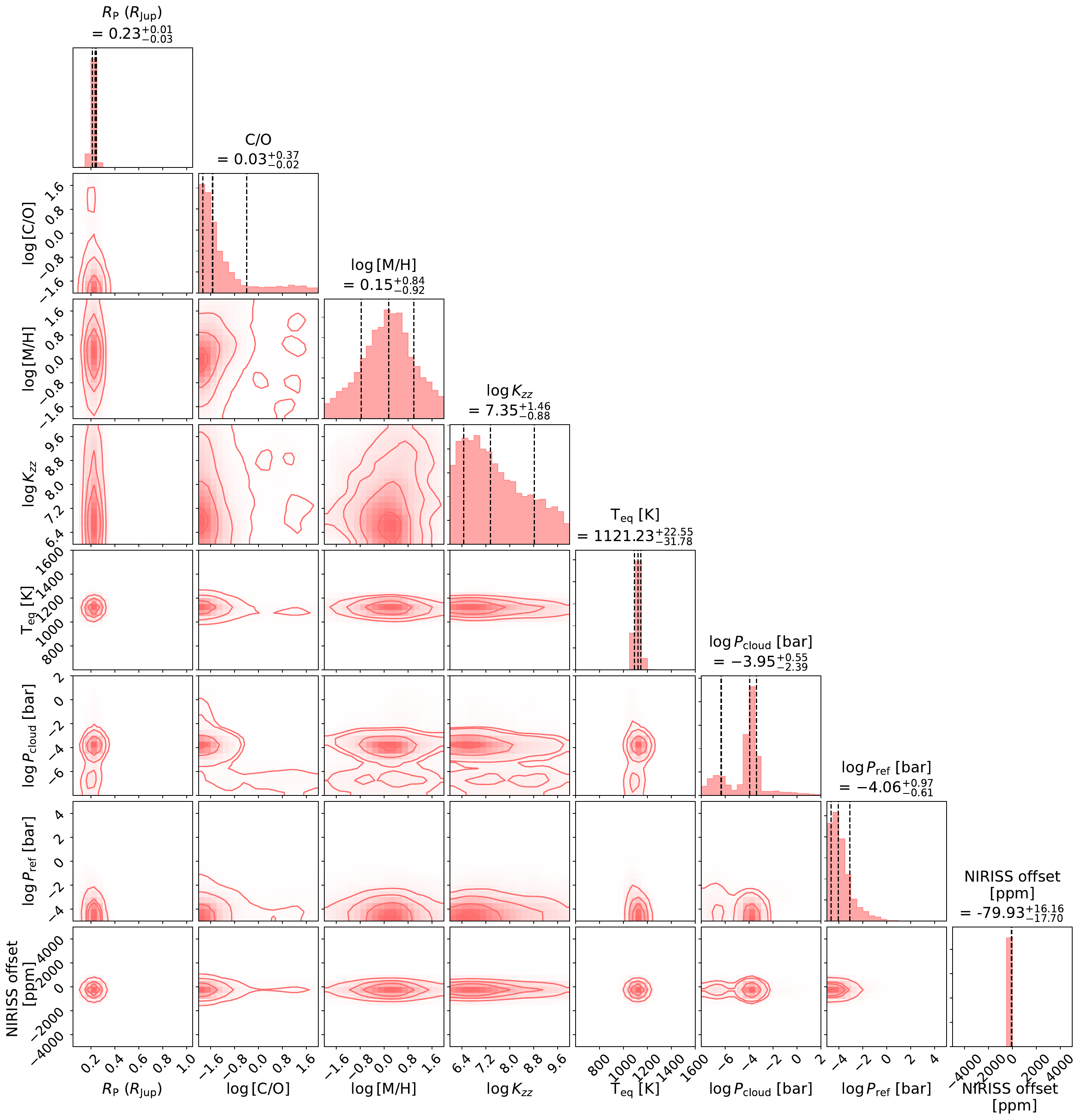}
  \caption{Similar to Figure \ref{fig:corner_hat-p-18b_vulcan} for TOI-421 b.}
   \label{fig:}
\end{figure*}
\renewcommand{\thefigure}{37}
\begin{figure*}
\centering
\begin{overpic}[width=\textwidth]{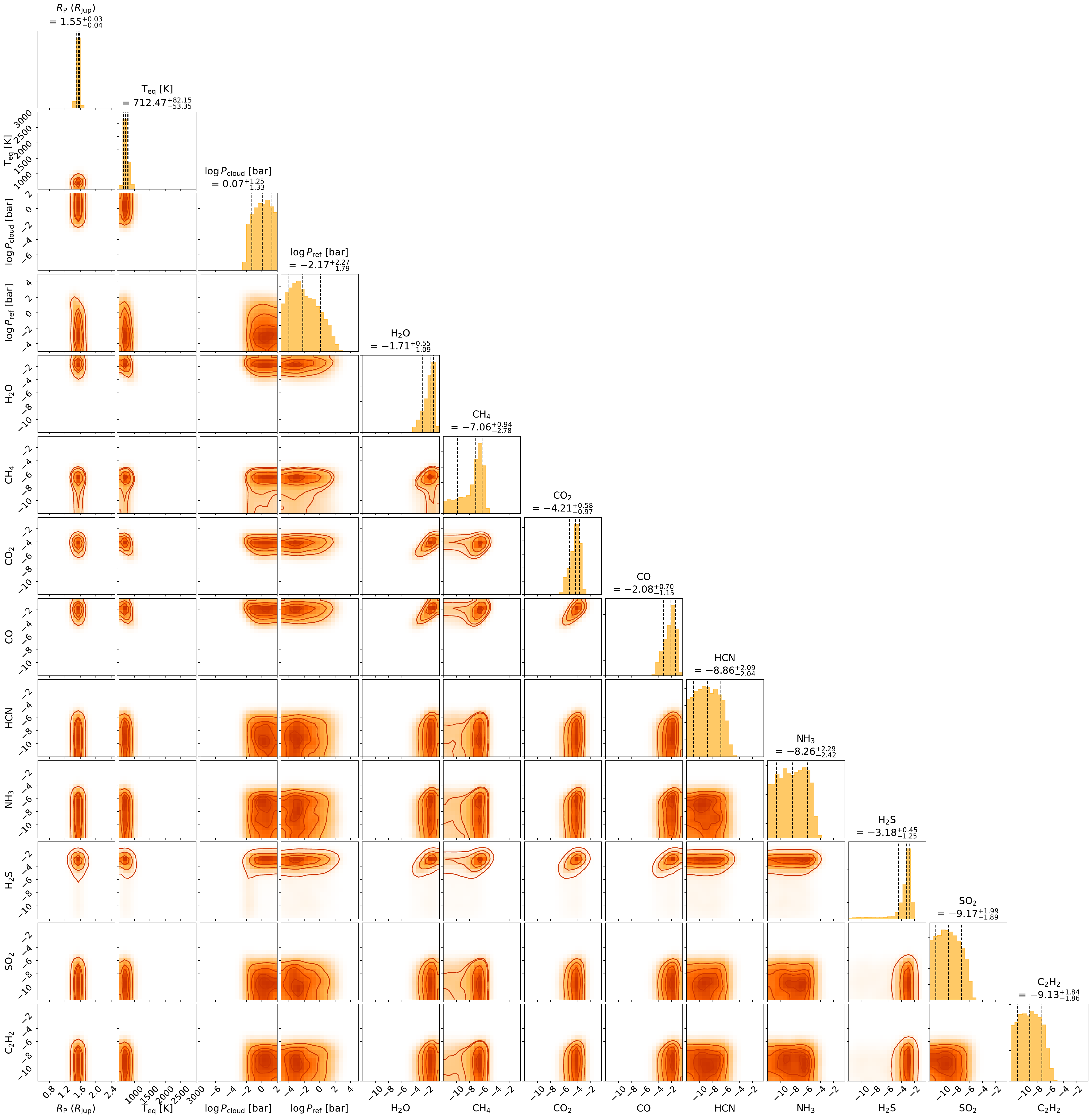}
  \put(45,70){\includegraphics[width=0.55\textwidth]{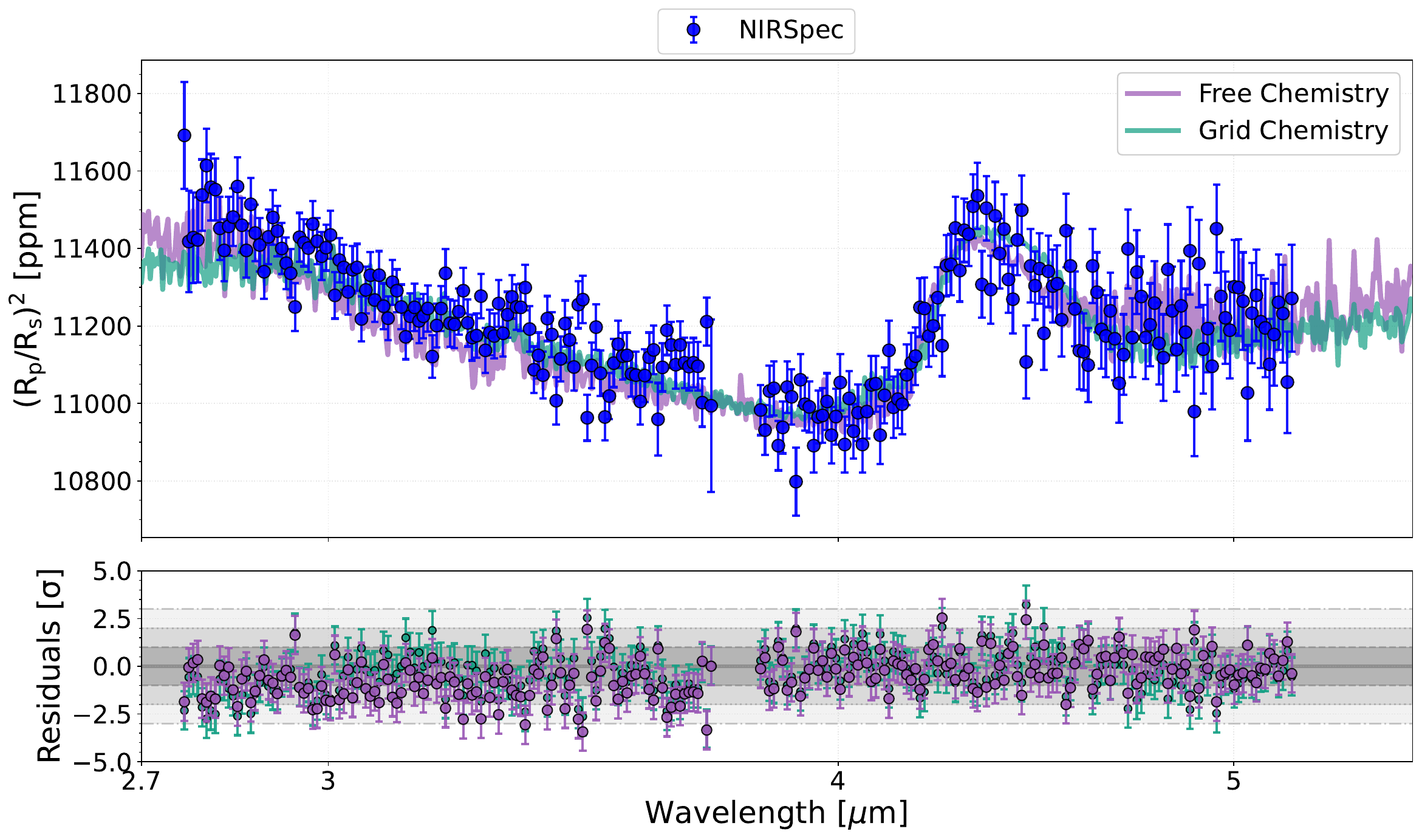}}
\end{overpic}
\caption{Similar to Figure \ref{fig:best_fit_corner_HAT-P-18b} for WASP-94 A b.}
\end{figure*}

\renewcommand{\thefigure}{38}
\begin{figure*}
\centering
	\includegraphics[width=\textwidth]{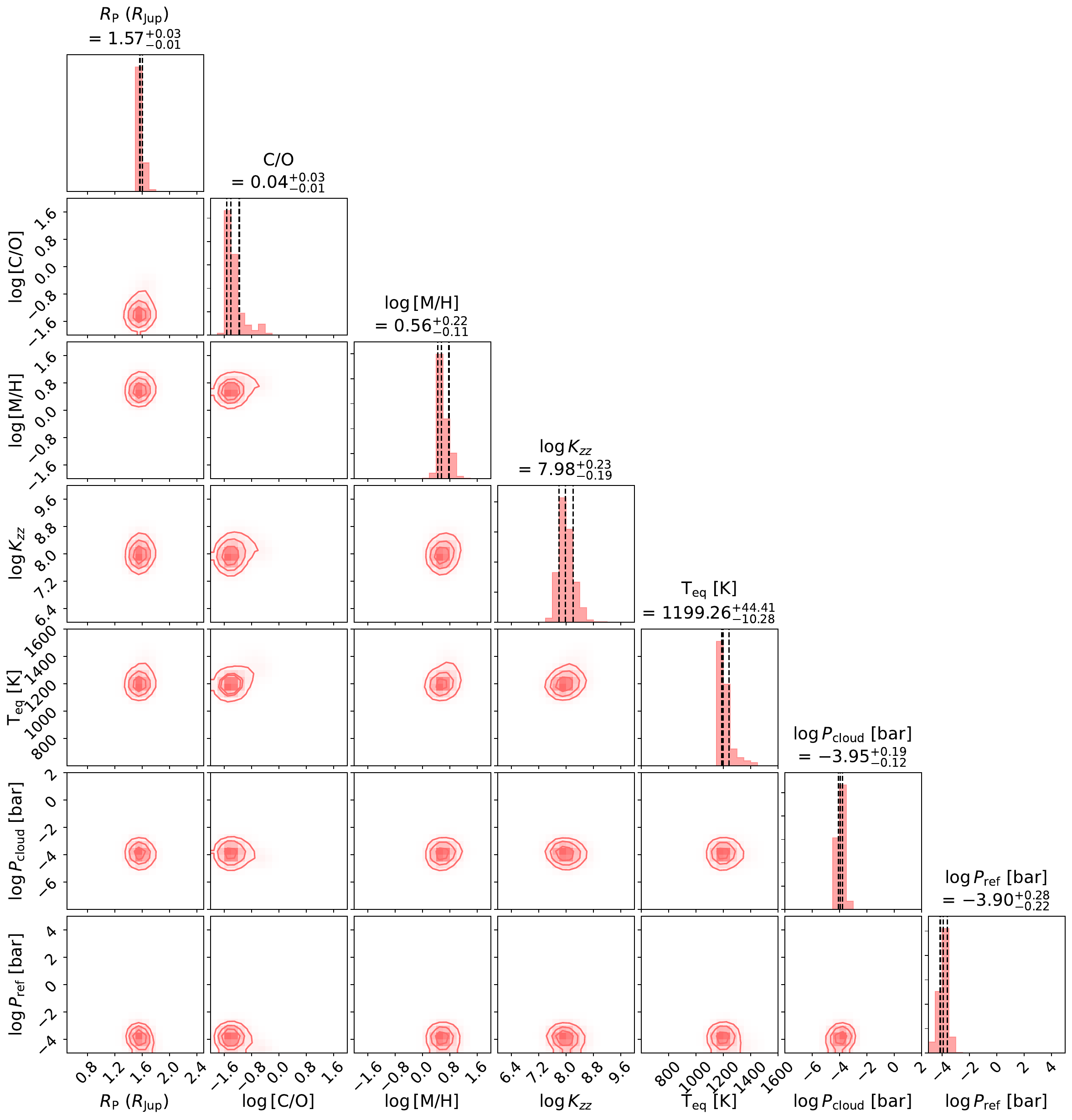}
  \caption{Similar to Figure \ref{fig:corner_hat-p-18b_vulcan} for WASP-94 A b.}
   \label{fig:}
\end{figure*}

\renewcommand{\thefigure}{39}
\begin{figure*}
\centering
\begin{overpic}[width=\textwidth]{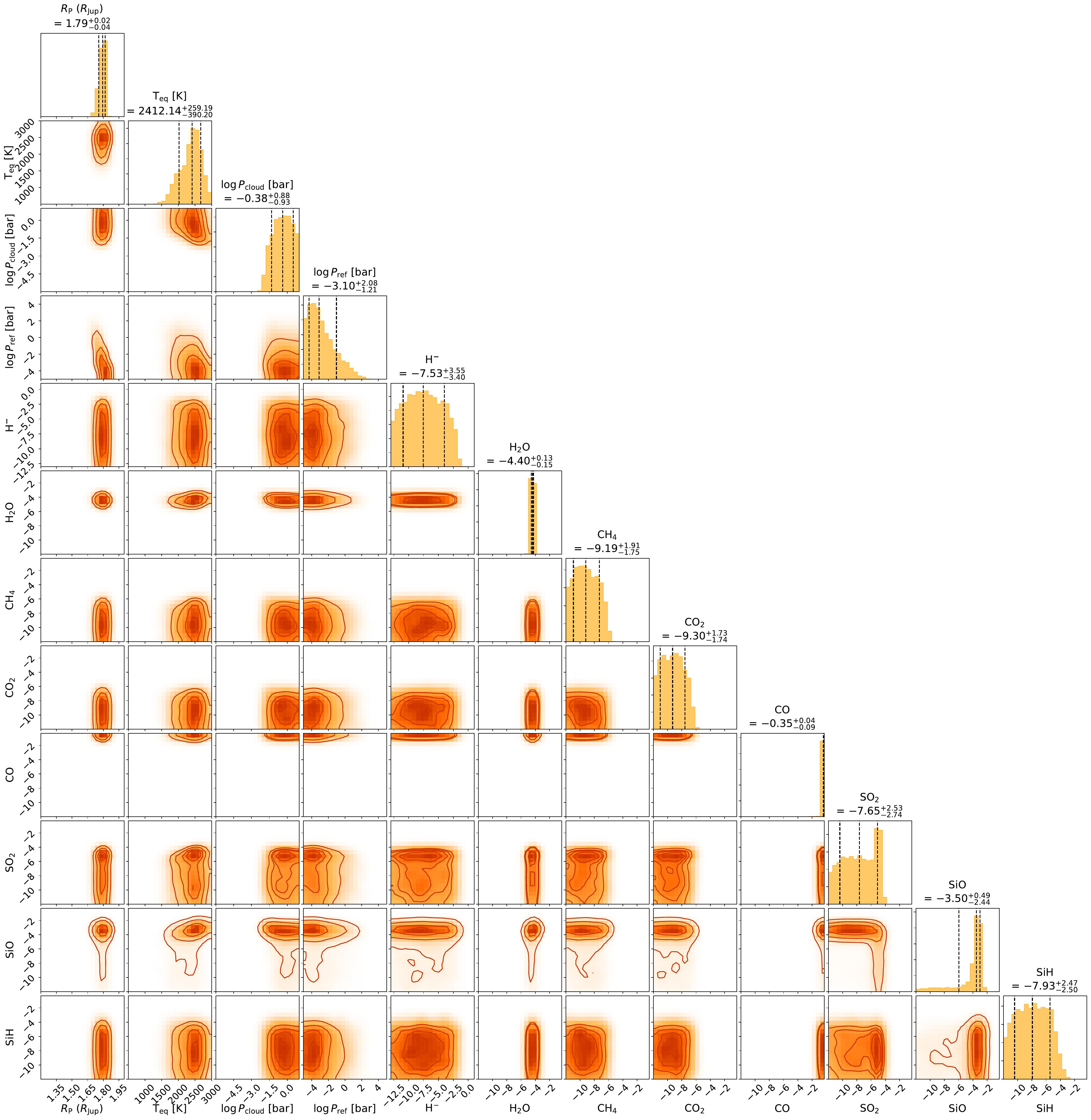}
  \put(45,70){\includegraphics[width=0.55\textwidth]{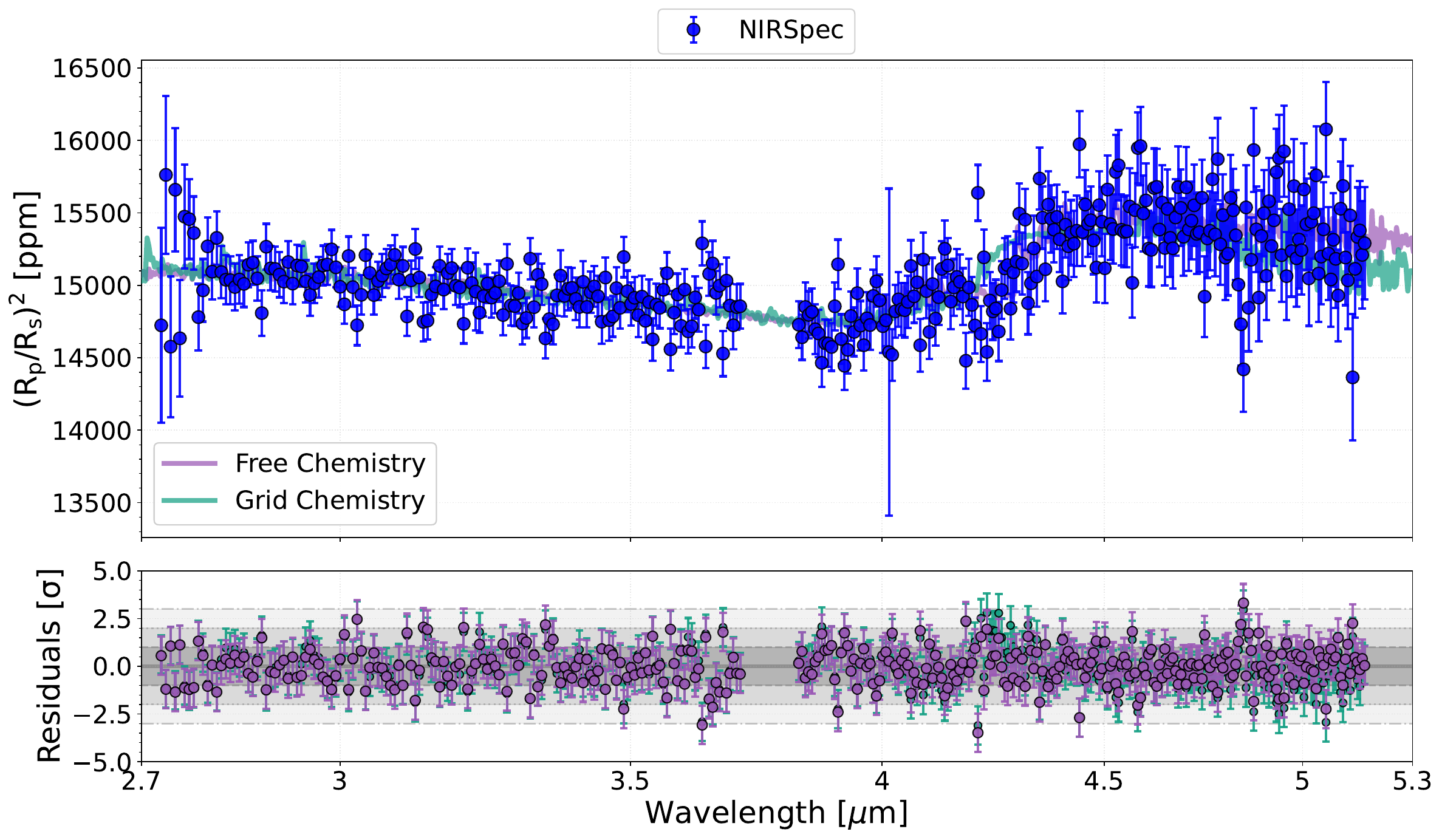}}
\end{overpic}
\caption{Similar to Figure \ref{fig:best_fit_corner_HAT-P-18b} for WASP-121 b.}
\end{figure*}

\renewcommand{\thefigure}{40}
\begin{figure*}
\centering
	\includegraphics[width=\textwidth]{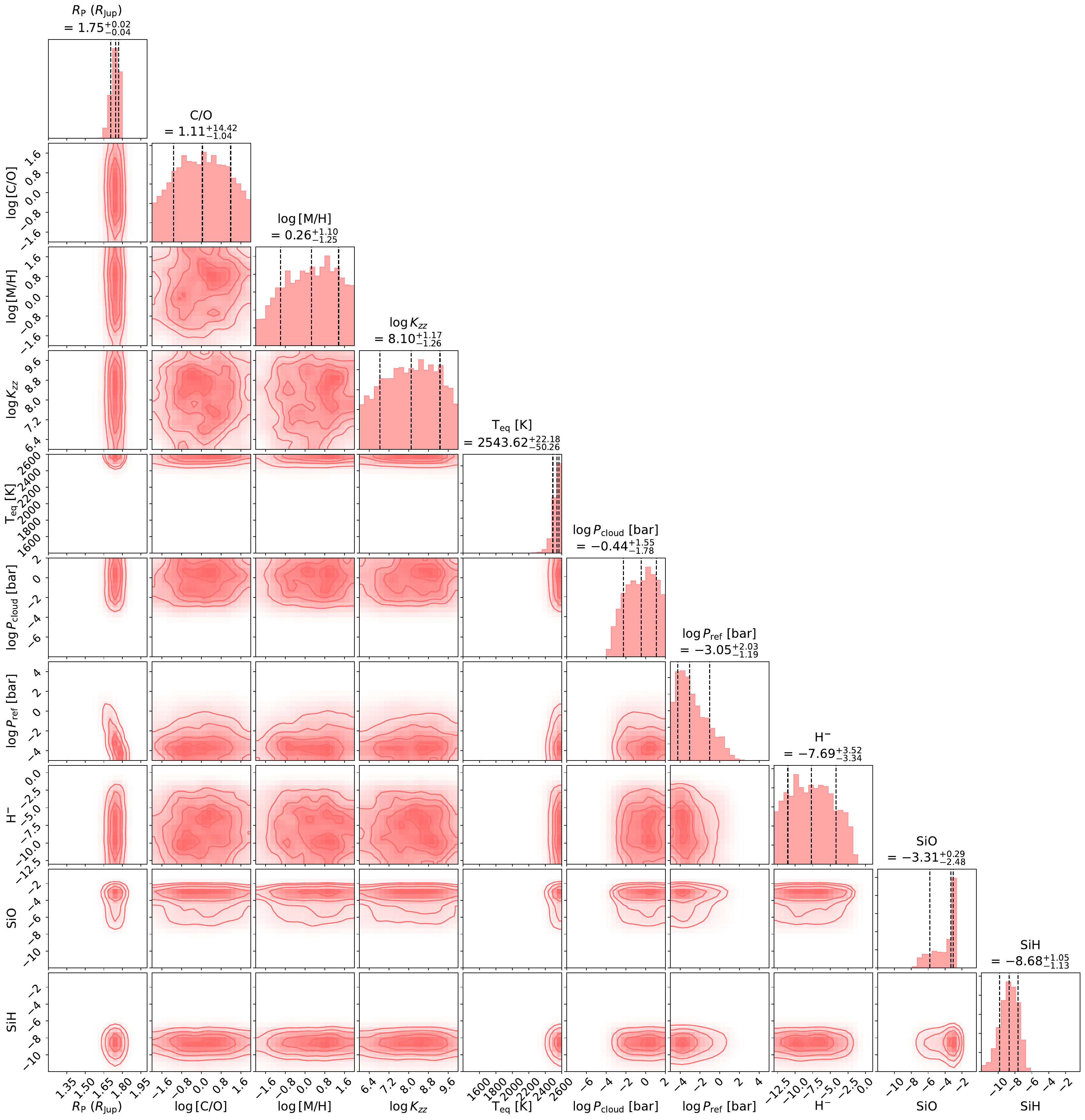}
  \caption{Similar to Figure \ref{fig:corner_hat-p-18b_vulcan} for WASP-121 b.}
   \label{fig:}
\end{figure*}

\renewcommand{\thefigure}{41}
\begin{figure*}
\centering
\begin{overpic}[width=\textwidth]{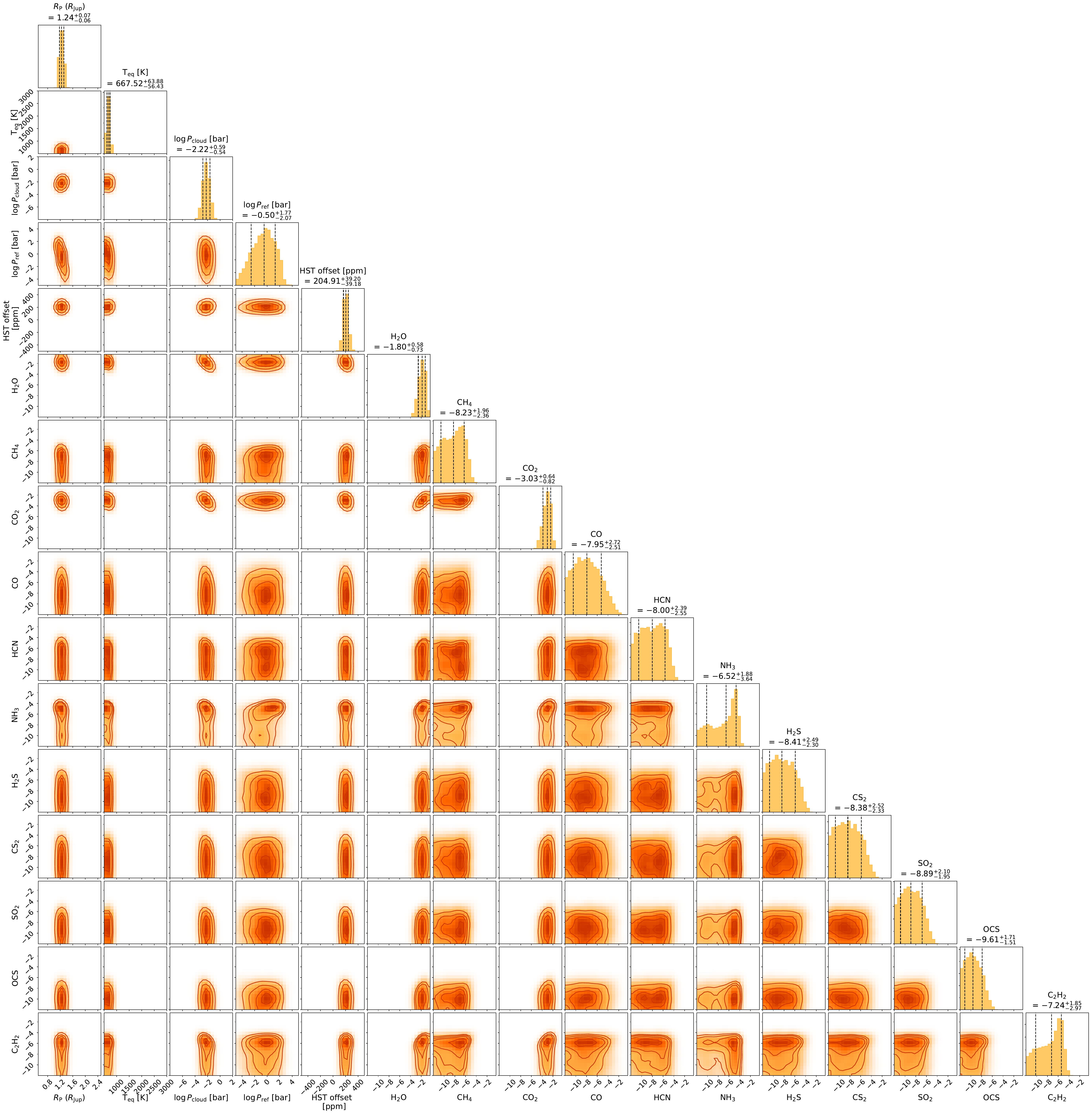}
  \put(45,70){\includegraphics[width=0.55\textwidth]{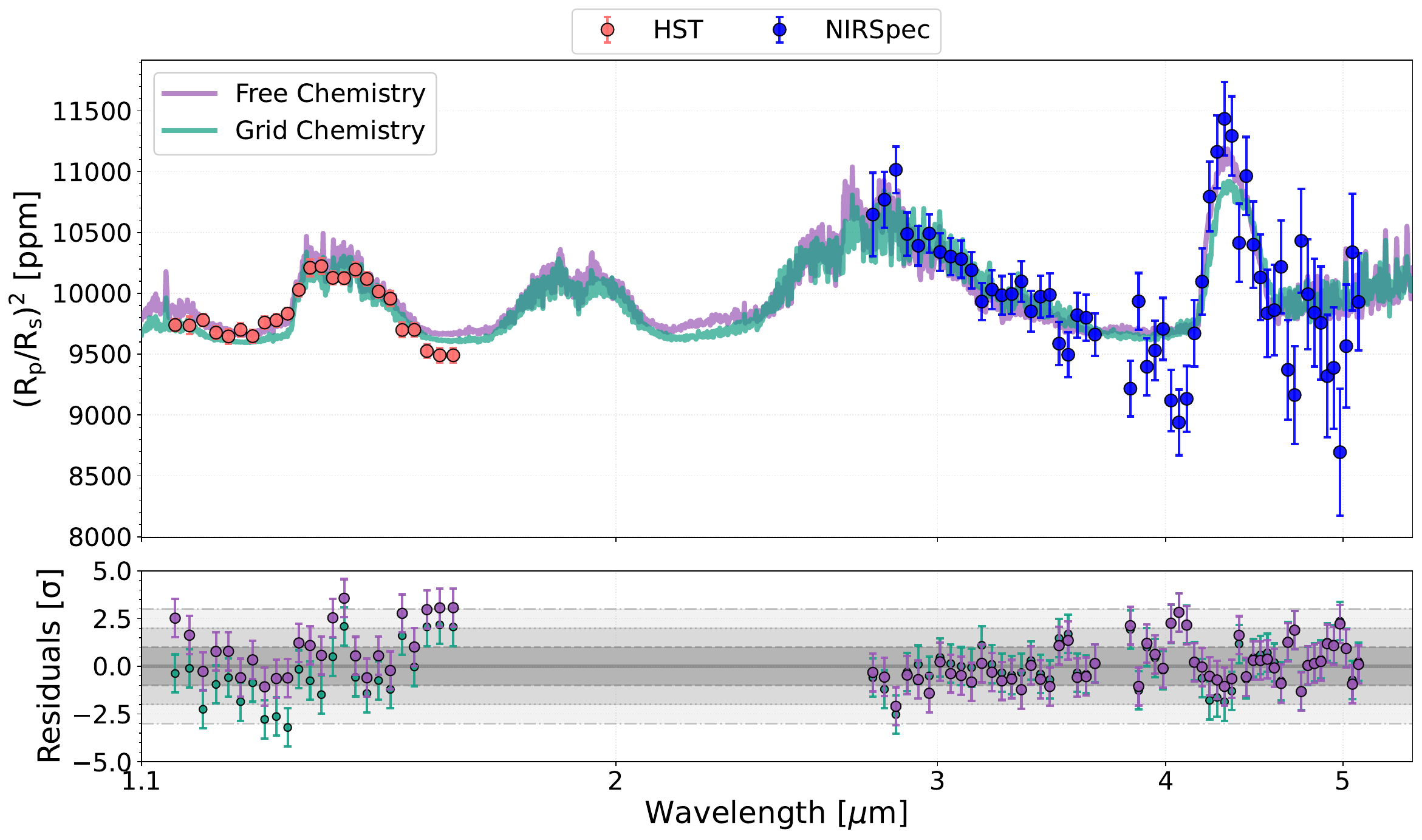}}
\end{overpic}
\caption{Similar to Figure \ref{fig:best_fit_corner_HAT-P-18b} for WASP-127 b.}
\end{figure*}

\renewcommand{\thefigure}{42}
\begin{figure*}
\centering
	\includegraphics[width=\textwidth]{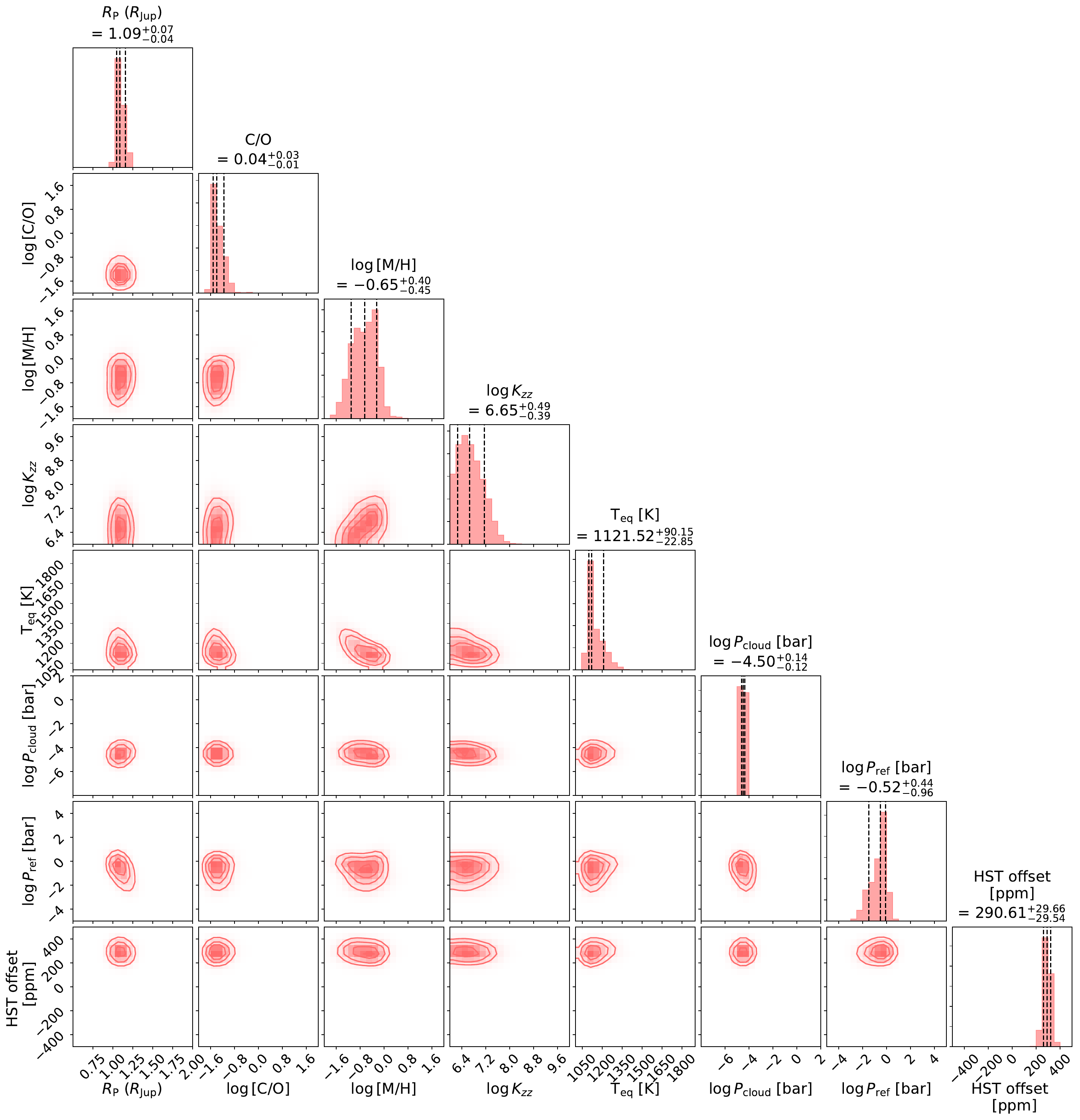}
  \caption{Similar to Figure \ref{fig:corner_hat-p-18b_vulcan} for WASP-127 b.}
   \label{fig:best_fit_corner_WASP-127b}
\end{figure*}
\renewcommand{\tablename}{Supplementary Table}
\renewcommand{\thetable}{3}
\begin{table}
\caption{Comparison of model fits for fixed-gravity and free-gravity retrievals. This table summarizes the reduced $\mathrm{\chi^2}$ values and Bayesian Evidences (ln Z) obtained for two versions of our free-chemistry retrievals: one in which the planetary surface gravity was held fixed, and another in which gravity was treated as a free parameter. The preferred model for each planet is determined using a combination of $\mathrm{\chi^2}$ ln Z, with higher Bayesian evidence given priority in cases where the two metrics disagree. For most planets, the fixed-gravity retrieval yields the stronger statistical support. However, for HIP 67522 b, TOI-776 c, and TrES-4 b, the free-gravity model provides the higher evidence and is therefore adopted as the final solution.}
\label{tab:gravity_comparison_overall}%
\begin{tabular}{@{}lccccc@{}}
\toprule
& \multicolumn{2}{@{}c@{}}{Free Gravity\footnotemark[1]} &     & \multicolumn{2}{@{}c@{}}{Fixed Gravity\footnotemark[2]} \\\cmidrule{2-3}\cmidrule{5-6}%
Planet & $\mathrm{\chi^2_{a}}$ & ln ($Z\mathrm{_a}$) &     & $\mathrm{\chi^2_{b}}$ &  ln ($Z\mathrm{_b}$)\\
\midrule
HAT-P-18 b  & 1.32 & 6884.9 $\pm$ 0.0 &     & 1.17 & 6960.2 $\pm$ 0.1 \\
\noalign{\smallskip}
WASP-17 b   & 2.15 & \textcolor{black}{1642.1 $\pm$ 0.1} &     & \textcolor{black}{2.11} & \textcolor{black}{1657.9 $\pm$ 0.0} \\
\noalign{\smallskip}
GJ 3470 b   & 0.93 & 1536.2 $\pm$ 0.0 &     & 0.93 & 1536.8 $\pm$ 0.0 \\
\noalign{\smallskip}
GJ 1214 b   & 1.31 & 758.1  $\pm$ 0.0 &     & 1.34 & 758.2  $\pm$ 0.9 \\
\noalign{\smallskip}
K2-18 b     & 1.38 & 2967.7 $\pm$ 0.0 &     & 1.35 & 2969.5 $\pm$ 0.1 \\
\noalign{\smallskip}
TOI-270 d   & 1.16 & 1582.6 $\pm$ 0.0 &     & 1.15 & 1584.3 $\pm$ 0.0 \\
\noalign{\smallskip}
HD 189733 b & 1.01 & 1447.0 $\pm$ 0.1 &     & 1.03 & 1454.3 $\pm$ 0.0 \\
\noalign{\smallskip}
HD 209458 b & 0.66 &  818.5 $\pm$ 0.0 &     & 0.64 &  820.8 $\pm$ 0.1 \\
\noalign{\smallskip}
HIP 67522 b & 0.93 & 1586.3 $\pm$ 0.0 &     & 1.55 & 1531.7 $\pm$ 0.1 \\
\noalign{\smallskip}
WASP-80 b   & 0.72 &  994.3 $\pm$ 0.0 &     & 0.73 &  994.4 $\pm$ 0.2 \\
\noalign{\smallskip}
WASP-39 b  & 1.65 & 1600.0 $\pm$ 0.0 &     & 1.68 &  1602.6 $\pm$ 0.0\\
\noalign{\smallskip}
WASP-107 b  & 1.64 & 1462.2  $\pm$ 0.1 &     & 1.60 & 1537.7 $\pm$ 0.6\\
\noalign{\smallskip}
WASP-15 b   & 0.96 & 1823.0  $\pm$ 0.2 &     & 0.95 & 1825.2 $\pm$ 0.5\\
\noalign{\smallskip}
TOI-836 c   & 1.42 & 903.8   $\pm$ 0.0 &     & 1.50 & 905.4  $\pm$ 0.4\\
\noalign{\smallskip}
TOI-776 c   & 1.41 & 953.8   $\pm$ 0.5 &     & 1.38 & 952.1  $\pm$ 0.2 \\
\noalign{\smallskip}
TrES-4 b    & 1.05 & 1731.4  $\pm$ 0.1 &     & 1.24 & 1716.2 $\pm$ 0.0\\
\noalign{\smallskip}
TOI-421 b   & 1.43 & 573.5   $\pm$ 0.0 &     & 1.44 & 576.3  $\pm$ 0.1 \\
\noalign{\smallskip}
WASP-94 A b & 0.97 & 1916.9  $\pm$ 0.0 &     & 0.97 & 1921.1 $\pm$ 0.0 \\
\noalign{\smallskip}
WASP-121 b  & 1.00 & \textcolor{black}{2455.5  $\pm$ 0.0} &     & \textcolor{black}{1.01} & 2456.9 $\pm$ 0.0\\
\noalign{\smallskip}
WASP-127 b  & 1.80 & 531.1   $\pm$ 0.0 &     & 1.75 &  537.5 $\pm$ 0.0 \\
\botrule
\end{tabular}
\footnotetext[1]{``Free-gravity” case: Gravity allowed to vary as a free parameter in the retrieval.}
\footnotetext[2]{``Fixed-gravity” case: Gravity held fixed to the value listed in Supplementary Tables \ref{tab:priors} and \ref{tab:priors_chem}.}
\end{table}
\renewcommand{\tablename}{Supplementary Table}
\renewcommand{\thetable}{4}
\begin{table}
\caption{Molecular and atomic opacity sources included in this study, along with the corresponding references for each line list. The table summarizes all species incorporated into the radiative-transfer calculations, including Rayleigh scattering species, and collision-induced absorption (CIA) species.}
\label{tab:linelist}%
\begin{tabular}{@{}lclc@{}}
\toprule
Chemical Species & Line list Refrence & Chemical Species &  Line list Refrence   \\
\midrule
$\mathrm{H_2O}$   & \citep{rothman2010hitemp} &  $\mathrm{CH_4}$  &  \citep{yurchenko2017hybrid}   \\
\noalign{\smallskip}
CO   & \citep{rothman2010hitemp} & $\mathrm{CO_2}$  & \citep{yurchenko2020exomol}     \\
\noalign{\smallskip}
$\mathrm{NH_3}$   & \citep{coles2019exomol} & $\mathrm{CO_2}$  & \citep{yurchenko2020exomol}     \\
\noalign{\smallskip}
$\mathrm{H_2S}$   & \citep{azzam2016exomol} & $\mathrm{SO_2}$  & \citep{underwood2016exomol}     \\
\noalign{\smallskip}
$\mathrm{C_2H_2}$   & \citep{chubb2020exomol} & $\mathrm{CS_2}$  & \citep{gordon2022hitran2020}     \\
\noalign{\smallskip}
 $\mathrm{SCO}$   & \citep{rothman2010hitemp, regalia2002line, muller2005cologne, sung2009line, toth2010line, golebiowski2014high,  gordon2017hitran2016} & $\mathrm{N_2O}$  & \citep{daumont2001line,gordon2017hitran2016}     \\
\noalign{\smallskip}
$\mathrm{AlO}$   & \citep{patrascu2015exomol} & $\mathrm{SiO}$  & \citep{yurchenko2022exomol}     \\ 
\noalign{\smallskip}
$\mathrm{SiH}$   & \citep{yurchenko2018exomol} & $\mathrm{Na}$  & \citep{allard2019new}     \\
\noalign{\smallskip}
$\mathrm{K}$   & \citep{burrows2003calculations} & $\mathrm{H_2}$ (Rayleigh Scattering)  & \citep{dalgarno1962rayleigh}     \\
\noalign{\smallskip}
$\mathrm{H_2}$–$\mathrm{H_2}$ and $\mathrm{H_2}$–He (CIA)  & \citep{borysow1988collison, borysow1989collision, borysow1989collision2, borysow2002collision, richard2012new} & $\mathrm{He}$ (Rayleigh Scattering)  & \citep{chan1965refractive}     \\
\noalign{\smallskip}
$\mathrm{H^{-}}$ (bound free absorption)  & \citep{2008oasp.book.....G} & $\mathrm{SiO_2}$ (condensate)  & \citep{kitzmann2018optical}     \\
\botrule
\end{tabular}
\end{table}

\renewcommand{\tablename}{Supplementary Table}
\renewcommand{\thetable}{5a}
\begin{sidewaystable*}
\rotatebox{0}{
           \begin{minipage}{\textwidth}
\centering
\caption{Bayesian priors used in the atmospheric retrievals. For each planet, the table lists the prior ranges for the free parameters and the values for the fixed parameters:  planetary radius (R$_p$) in the unit of Jupiter radius (R$_J$), equilibrium temperature ($T_{\mathrm{eq}}$) in Kelvin, log value of reference pressure ($P_{\mathrm{ref}}$) in bar, log value of cloud-top pressure ($P_{\mathrm{cloud}}$) in bar, dataset offset parameters in parts per million (ppm) (when multiple datasets are present), the log-normal particle distribution width ($\sigma_{\mathrm{lnorm}}$), log value of Eddy diffusion coefficient $K_{zz}$ (in cm$^2$/second), cloud sedimentation factor ($f_{\mathrm{sed}}$), Rayleigh-scattering slope (\textcolor{black}{$\log\gamma_{350\,\mathrm{nm}}$}), scattering amplitude (\textcolor{black}{$\gamma_{\mathrm{coeff}}$}), surface gravity ($\log g$) in cgs, and stellar radius ($R_{s}$) in the unit of Sun's radius ($R_{\odot}$). The stellar radius values are taken from Supplementary Table \ref{tab:vulcan_temp_grids}. For most planets, gravity was held fixed at the literature value due to the higher Bayesian evidence relative to models where gravity was treated as a free parameter (see Supplementary Table \ref{tab:gravity_comparison_overall}). For a few targets, HIP 67522 b, TOI-776 c, and TrES-4 b, treating gravity as a free parameter significantly improves the fit, and a prior range was therefore applied.}

\label{tab:priors}
\resizebox{\textwidth}{!}{
\begin{tabular}{lcccccccccccc}
\hline
\hline
\noalign{\smallskip}
Planet & $\mathrm{R_{p}}$ &  $\mathrm{T_{eq}}$ &  log$\mathrm{_{10}(P_{ref} [bar]})$ & log$\mathrm{_{10}(P_{cloud}[bar]})$& All Offsets & $\mathrm{\sigma_{lnorm}}$ & log$\mathrm{_(K_{zz}} [cm^2/s])$ & $\mathrm{f_{sed}}$ & \textcolor{black}{log ($\mathrm{\gamma_{350nm}}$)} & \textcolor{black}{$\mathrm{\gamma_{coeff}}$} & log(g) & $\mathrm{R_{s}}$\\
 &  ($\mathrm{R_J}$) & (K) & &  & (ppm) &  &  &  & & & (cgs) &  ($\mathrm{R_{\odot}}$)\\
\noalign{\smallskip}
\hline
\noalign{\smallskip}
  HAT-P-18 b    &  \begin{math}\mathcal{U}\end{math} [0.5, 2.5]  & \begin{math}\mathcal{U}\end{math}[300, 3000]  & \begin{math}\mathcal{U}\end{math}[-5, 5]  & \begin{math}\mathcal{U}\end{math}[-8, 2] & \begin{math}\mathcal{U}\end{math}[-500, 500] & -- &-- & --& \begin{math}\mathcal{U}\end{math}[-20, 20] & \begin{math}\mathcal{U}\end{math}[-20, 2] &  2.693  &   0.74 \\
  \noalign{\smallskip}
  WASP-17 b     & \begin{math}\mathcal{U}\end{math} [0.5, 2.5] & \begin{math}\mathcal{U}\end{math}[500, 3000]  & \begin{math}\mathcal{U}\end{math}[-5, 5]  & -- & \begin{math}\mathcal{U}\end{math}[-5000, 5000] & \begin{math}\mathcal{U}\end{math}[1.05, 3] & \begin{math}\mathcal{U}\end{math}[5, 13] & \begin{math}\mathcal{U}\end{math}[0.1, 10.1]& -- &--   & 2.7426 & 1.57 \\
\noalign{\smallskip}
GJ 3470 b     & \begin{math}\mathcal{U}\end{math}[0.2, 2.5] & \begin{math}\mathcal{U}\end{math}[500, 3000]  & \begin{math}\mathcal{U}\end{math}[-5, 5]  & \begin{math}\mathcal{U}\end{math}[-8, 2] & \begin{math}\mathcal{U}\end{math}[-500, 500] & -- & -- & -- & -- & --  & 2.8143 & 0.499  \\
\noalign{\smallskip}
GJ 1214 b     & \begin{math}\mathcal{U}\end{math} [0.05, 1.05] & \begin{math}\mathcal{U}\end{math}[100, 2000]  & \begin{math}\mathcal{U}\end{math}[-5, 5]  & \begin{math}\mathcal{U}\end{math}[-8, 2] & \begin{math}\mathcal{U}\end{math}[-5000, 5000] & -- & -- & -- & -- & --  & 3.0427 & 0.2162  \\
\noalign{\smallskip}
K2-18 b     & \begin{math}\mathcal{U}\end{math} [0.05, 1.05] & \begin{math}\mathcal{U}\end{math}[100, 2000]  & \begin{math}\mathcal{U}\end{math}[-5, 5]  & \begin{math}\mathcal{U}\end{math}[-8, 2] & \begin{math}\mathcal{U}\end{math}[-500, 500] & -- & -- & -- & \begin{math}\mathcal{U}\end{math}[-20, 20] & \begin{math}\mathcal{U}\end{math}[-20, 2]  & 3.192 & 0.468  \\
\noalign{\smallskip}
TOI-270 d     & \begin{math}\mathcal{U}\end{math} [0.05, 1.05] & \begin{math}\mathcal{U}\end{math}[100, 2000]  & \begin{math}\mathcal{U}\end{math}[-5, 5]  & \begin{math}\mathcal{U}\end{math}[-8, 2] & \begin{math}\mathcal{U}\end{math}[-500, 500] & -- & -- & -- & -- & --  & 3.013 & 0.374  \\
\noalign{\smallskip}
HD 189733 b     & \begin{math}\mathcal{U}\end{math} [0.5, 2.5] & \begin{math}\mathcal{U}\end{math}[500, 3000]  & \begin{math}\mathcal{U}\end{math}[-5, 5]  & \begin{math}\mathcal{U}\end{math}[-8, 2] & \begin{math}\mathcal{U}\end{math}[-500, 500] & -- & -- & -- & -- & --  & 3.3411 & 0.776  \\
\noalign{\smallskip}
HD 209458 b     & \begin{math}\mathcal{U}\end{math} [0.5, 2.5] & \begin{math}\mathcal{U}\end{math}[500, 3000]  & \begin{math}\mathcal{U}\end{math}[-5, 5]  & \begin{math}\mathcal{U}\end{math}[-8, 2] & \begin{math}\mathcal{U}\end{math}[-500, 500] & -- & -- & -- & -- & --  & 2.9715 & 1.186  \\
\noalign{\smallskip}
HIP 67522 b     & \begin{math}\mathcal{U}\end{math} [0.5, 2.5] & \begin{math}\mathcal{U}\end{math}[500, 3000]  & \begin{math}\mathcal{U}\end{math}[-5, 5]  & \begin{math}\mathcal{U}\end{math}[-8, 2] & -- & -- & -- & -- & -- & --  & \begin{math}\mathcal{U}\end{math}[2, 5.5] & 1.38  \\
\noalign{\smallskip}
WASP-39 b     & \begin{math}\mathcal{U}\end{math} [0.5, 2.5] & \begin{math}\mathcal{U}\end{math}[500, 3000]  & \begin{math}\mathcal{U}\end{math}[-5, 5]  & \begin{math}\mathcal{U}\end{math}[-8, 2] & \begin{math}\mathcal{U}\end{math}[-5000, 5000] & -- & -- & -- & \begin{math}\mathcal{U}\end{math}[-20, 20] & \begin{math}\mathcal{U}\end{math}[-20, 2]  & 2.63 & 0.94 \\
\noalign{\smallskip}
WASP-80 b     & \begin{math}\mathcal{U}\end{math} [0.5, 2.5] & \begin{math}\mathcal{U}\end{math}[500, 3000]  & \begin{math}\mathcal{U}\end{math}[-5, 5]  & \begin{math}\mathcal{U}\end{math}[-8, 2] & \begin{math}\mathcal{U}\end{math}[-500, 500] & -- & -- & -- & -- & --  & 3.12 & 0.636  \\
\noalign{\smallskip}
WASP-107 b     & \begin{math}\mathcal{U}\end{math} [0.5, 2.5] & \begin{math}\mathcal{U}\end{math}[400, 3000]  & \begin{math}\mathcal{U}\end{math}[-5, 5]  & \begin{math}\mathcal{U}\end{math}[-8, 2] & \begin{math}\mathcal{U}\end{math}[-500, 500] & -- & -- & -- & -- & --  & 2.43 & 0.67  \\
\noalign{\smallskip}
WASP-15 b     & \begin{math}\mathcal{U}\end{math} [0.5, 2.5] & \begin{math}\mathcal{U}\end{math}[500, 3000]  & \begin{math}\mathcal{U}\end{math}[-5, 5]  & \begin{math}\mathcal{U}\end{math}[-8, 2] & -- & -- & -- & -- & -- & --  & 2.82 & 1.50 \\
\noalign{\smallskip}
TOI-836 c     & \begin{math}\mathcal{U}\end{math} [0.05, 1.05] & \begin{math}\mathcal{U}\end{math}[100, 2000]  & \begin{math}\mathcal{U}\end{math}[-5, 5]  & \begin{math}\mathcal{U}\end{math}[-8, 2] & -- & -- & -- & -- & -- & --  & 3.14 & 0.665 \\
\noalign{\smallskip}
TOI-776 c     & \begin{math}\mathcal{U}\end{math} [0.05, 1.05] & \begin{math}\mathcal{U}\end{math}[100, 2000]  & \begin{math}\mathcal{U}\end{math}[-5, 5]  & \begin{math}\mathcal{U}\end{math}[-8, 2] & -- & -- & -- & -- & -- & --  & \begin{math}\mathcal{U}\end{math}[2, 5.5] & 0.547 \\
\noalign{\smallskip}
TrES-4 b      & \begin{math}\mathcal{U}\end{math} [0.5, 2.5] & \begin{math}\mathcal{U}\end{math}[500, 3000]  & \begin{math}\mathcal{U}\end{math}[-5, 5]  & \begin{math}\mathcal{U}\end{math}[-8, 2] & -- & -- & -- & -- & -- & --  & \begin{math}\mathcal{U}\end{math}[2, 5.5] & 2.008 \\
\noalign{\smallskip}
TOI-421 b      & \begin{math}\mathcal{U}\end{math} [0.05, 1.05] & \begin{math}\mathcal{U}\end{math}[100, 2000]  & \begin{math}\mathcal{U}\end{math}[-5, 5]  & \begin{math}\mathcal{U}\end{math}[-8, 2] & \begin{math}\mathcal{U}\end{math}[-5000, 5000] & -- & -- & -- & -- & --  & 2.974 & 0.866 \\
\noalign{\smallskip}
WASP-94 A b      & \begin{math}\mathcal{U}\end{math} [0.5, 2.5] & \begin{math}\mathcal{U}\end{math}[500, 3000]  & \begin{math}\mathcal{U}\end{math}[-5, 5]  & \begin{math}\mathcal{U}\end{math}[-8, 2] & -- & -- & -- & -- & -- & --  & 2.6959 & 1.48 \\
\noalign{\smallskip}
WASP-121 b      & \begin{math}\mathcal{U}\end{math}[0.5, 2.5] & \begin{math}\mathcal{U}\end{math}[500, 3000]  & \begin{math}\mathcal{U}\end{math}[-5, 5]  & \begin{math}\mathcal{U}\end{math}[-8, 2] & -- & -- & -- & -- & -- & --  & 2.97 & 1.458\\
\noalign{\smallskip}
WASP-127 b      & \begin{math}\mathcal{U}\end{math} [0.5, 2.5] & \begin{math}\mathcal{U}\end{math}[500, 3000]  & \begin{math}\mathcal{U}\end{math}[-5, 5]  & \begin{math}\mathcal{U}\end{math}[-8, 2] & \begin{math}\mathcal{U}\end{math}[-500, 500] & -- & -- & -- & -- & --  & 2.38 & 1.39 \\
\noalign{\smallskip}
\noalign{\smallskip}
\hline
\hline
\end{tabular}}
\end{minipage}}

\end{sidewaystable*}

\renewcommand{\tablename}{Supplementary Table}
\renewcommand{\thetable}{5b}
\begin{sidewaystable*}
\color{black}
\rotatebox{0}{
\begin{minipage}{\textwidth}

\centering
\caption{\textcolor{black}{Bayesian priors used in the chemically consistent disequilibrium retrievals. C/O and metallicity (M/H) are sampled over $\mathcal{U}[0.01,100]$. The eddy-diffusion coefficient used in the chemistry grid is sampled as $\log_{10}(K_{zz}/\mathrm{cm^2\,s^{-1}})=\mathcal{U}[6,10]$. Other parameters, including reference pressure, cloud parameters, offsets, gravity, and stellar radius, follow the same final setup as in the free-chemistry retrievals.}}
\label{tab:priors_chem}
\scriptsize
\resizebox{\textwidth}{!}{
\begin{tabular}{l*{14}{c}}
\hline
\hline
\noalign{\smallskip}
Planet & $\mathrm{R_p}$ & C/O & $\log_{10}(\mathrm{M/H})$ & $T_{\rm eq}$ & $\log_{10}(K_{zz})$ & $\log_{10}(P_{\rm ref}[\mathrm{bar}])$ & $\log_{10}(P_{\rm cloud}[\mathrm{bar}])$ & All Offsets & $\sigma_{\rm lnorm}$ & $f_{\rm sed}$ & $\log (\mathrm{\gamma_{350nm}}$) & $\mathrm{\gamma_{coeff}}$ & $\log g$ & $R_s$ \\
& ($R_J$) & & & (K) & (cgs) & & & (ppm) & & & & & (cgs) & ($R_\odot$) \\
\noalign{\smallskip}
\hline
\noalign{\smallskip}

HAT-P-18 b & $\mathcal{U}[0.5,2.5]$ & $\mathcal{U}[0.01,100]$ & $\mathcal{U}[0.01,100]$ & $\mathcal{U}[600,1600]$ & $\mathcal{U}[6,10]$ & $\mathcal{U}[-5,5]$ & $\mathcal{U}[-8,2]$ & $\mathcal{U}[-500,500]$ & -- & -- & $\mathcal{U}[-20,20]$ & $\mathcal{U}[-20,2]$ & 2.693 & 0.74 \\
\noalign{\smallskip}
WASP-17 b & $\mathcal{U}[0.5,2.5]$ & $\mathcal{U}[0.01,100]$ & $\mathcal{U}[0.01,100]$ & $\mathcal{U}[1000,1900]$ & $\mathcal{U}[6,10]$ & $\mathcal{U}[-5,5]$ & -- & $\mathcal{U}[-5000,5000]$ & $\mathcal{U}[1.05,3]$ & $\mathcal{U}[0.1,10.1]$ & -- & -- & 2.7426 & 1.57 \\
\noalign{\smallskip}
GJ 3470 b & $\mathcal{U}[0.2,2.5]$ & $\mathcal{U}[0.01,100]$ & $\mathcal{U}[0.01,100]$ & $\mathcal{U}[400,1600]$ & $\mathcal{U}[6,10]$ & $\mathcal{U}[-5,5]$ & $\mathcal{U}[-8,2]$ & $\mathcal{U}[-500,500]$ & -- & -- & -- & -- & 2.8143 & 0.499 \\
\noalign{\smallskip}
GJ 1214 b & $\mathcal{U}[0.05,1.05]$ & $\mathcal{U}[0.01,100]$ & $\mathcal{U}[0.01,100]$ & $\mathcal{U}[400,1600]$ & $\mathcal{U}[6,10]$ & $\mathcal{U}[-5,5]$ & $\mathcal{U}[-8,2]$ & $\mathcal{U}[-5000,5000]$ & -- & -- & -- & -- & 3.0427 & 0.2162 \\
\noalign{\smallskip}
K2-18 b & $\mathcal{U}[0.05,1.05]$ & $\mathcal{U}[0.01,100]$ & $\mathcal{U}[0.01,100]$ & $\mathcal{U}[400,1600]$ & $\mathcal{U}[6,10]$ & $\mathcal{U}[-5,5]$ & $\mathcal{U}[-8,2]$ & $\mathcal{U}[-500,500]$ & -- & -- & $\mathcal{U}[-20,20]$ & $\mathcal{U}[-20,2]$ & 3.192 & 0.468 \\
\noalign{\smallskip}
TOI-270 d & $\mathcal{U}[0.05,1.05]$ & $\mathcal{U}[0.01,100]$ & $\mathcal{U}[0.01,100]$ & $\mathcal{U}[400,1600]$ & $\mathcal{U}[6,10]$ & $\mathcal{U}[-5,5]$ & $\mathcal{U}[-8,2]$ & $\mathcal{U}[-500,500]$ & -- & -- & -- & -- & 3.013 & 0.374 \\
\noalign{\smallskip}
HD 189733 b & $\mathcal{U}[0.5,2.5]$ & $\mathcal{U}[0.01,100]$ & $\mathcal{U}[0.01,100]$ & $\mathcal{U}[600,1600]$ & $\mathcal{U}[6,10]$ & $\mathcal{U}[-5,5]$ & $\mathcal{U}[-8,2]$ & $\mathcal{U}[-500,500]$ & -- & -- & -- & -- & 3.3411 & 0.776 \\
\noalign{\smallskip}
HD 209458 b & $\mathcal{U}[0.5,2.5]$ & $\mathcal{U}[0.01,100]$ & $\mathcal{U}[0.01,100]$ & $\mathcal{U}[600,1600]$ & $\mathcal{U}[6,10]$ & $\mathcal{U}[-5,5]$ & $\mathcal{U}[-8,2]$ & $\mathcal{U}[-500,500]$ & -- & -- & -- & -- & 2.9715 & 1.186 \\
\noalign{\smallskip}
HIP 67522 b & $\mathcal{U}[0.5,2.5]$ & $\mathcal{U}[0.01,100]$ & $\mathcal{U}[0.01,100]$ & $\mathcal{U}[600,1600]$ & $\mathcal{U}[6,10]$ & $\mathcal{U}[-5,5]$ & $\mathcal{U}[-8,2]$ & -- & -- & -- & -- & -- & $\mathcal{U}[2,5.5]$ & 1.38 \\
\noalign{\smallskip}
WASP-39 b & $\mathcal{U}[0.5,2.5]$ & $\mathcal{U}[0.01,100]$ & $\mathcal{U}[0.01,100]$ & $\mathcal{U}[600,1600]$ & $\mathcal{U}[6,10]$ & $\mathcal{U}[-5,5]$ & $\mathcal{U}[-8,2]$ & $\mathcal{U}[-5000,5000]$ & -- & -- & $\mathcal{U}[-20,20]$ & $\mathcal{U}[-20,2]$ & 2.63 & 0.94 \\
\noalign{\smallskip}
WASP-80 b & $\mathcal{U}[0.5,2.5]$ & $\mathcal{U}[0.01,100]$ & $\mathcal{U}[0.01,100]$ & $\mathcal{U}[400,1600]$ & $\mathcal{U}[6,10]$ & $\mathcal{U}[-5,5]$ & $\mathcal{U}[-8,2]$ & $\mathcal{U}[-500,500]$ & -- & -- & -- & -- & 3.12 & 0.636 \\
\noalign{\smallskip}
WASP-107 b & $\mathcal{U}[0.5,2.5]$ & $\mathcal{U}[0.01,100]$ & $\mathcal{U}[0.01,100]$ & $\mathcal{U}[400,1600]$ & $\mathcal{U}[6,10]$ & $\mathcal{U}[-5,5]$ & $\mathcal{U}[-8,2]$ & $\mathcal{U}[-500,500]$ & -- & -- & -- & -- & 2.43 & 0.67 \\
\noalign{\smallskip}
WASP-15 b & $\mathcal{U}[0.5,2.5]$ & $\mathcal{U}[0.01,100]$ & $\mathcal{U}[0.01,100]$ & $\mathcal{U}[1000,1900]$ & $\mathcal{U}[6,10]$ & $\mathcal{U}[-5,5]$ & $\mathcal{U}[-8,2]$ & -- &  --- & -- & -- & -- & 2.82 & 1.50 \\
\noalign{\smallskip}
TOI-836 c & $\mathcal{U}[0.05,1.05]$ & $\mathcal{U}[0.01,100]$ & $\mathcal{U}[0.01,100]$ & $\mathcal{U}[400,1600]$ & $\mathcal{U}[6,10]$ & $\mathcal{U}[-5,5]$ & $\mathcal{U}[-8,2]$ & -- & -- & -- & -- & -- & 3.14 & 0.665 \\
\noalign{\smallskip}
TOI-776 c & $\mathcal{U}[0.05,1.05]$ & $\mathcal{U}[0.01,100]$ & $\mathcal{U}[0.01,100]$ & $\mathcal{U}[400,1600]$ & $\mathcal{U}[6,10]$ & $\mathcal{U}[-5,5]$ & $\mathcal{U}[-8,2]$ & -- & -- & -- & -- & -- & $\mathcal{U}[2,5.5]$ & 0.547 \\
\noalign{\smallskip}
TrES-4 b & $\mathcal{U}[0.5,2.5]$ & $\mathcal{U}[0.01,100]$ & $\mathcal{U}[0.01,100]$ & $\mathcal{U}[1000,1900]$ & $\mathcal{U}[6,10]$ & $\mathcal{U}[-5,5]$ & $\mathcal{U}[-8,2]$ & -- & -- & -- & -- & -- & $\mathcal{U}[2,5.5]$ & 2.008 \\
\noalign{\smallskip}
TOI-421 b & $\mathcal{U}[0.05,1.05]$ & $\mathcal{U}[0.01,100]$ & $\mathcal{U}[0.01,100]$ & $\mathcal{U}[600,1600]$ & $\mathcal{U}[6,10]$ & $\mathcal{U}[-5,5]$ & $\mathcal{U}[-8,2]$ & $\mathcal{U}[-5000,5000]$ & -- & -- & -- & -- & 2.974 & 0.866 \\
\noalign{\smallskip}
WASP-94 A b & $\mathcal{U}[0.5,2.5]$ & $\mathcal{U}[0.01,100]$ & $\mathcal{U}[0.01,100]$ & $\mathcal{U}[600,1600]$ & $\mathcal{U}[6,10]$ & $\mathcal{U}[-5,5]$ & $\mathcal{U}[-8,2]$ & -- & -- & -- & -- & -- & 2.6959 & 1.48 \\
\noalign{\smallskip}
WASP-121 b & $\mathcal{U}[0.5,2.5]$ & $\mathcal{U}[0.01,100]$ & $\mathcal{U}[0.01,100]$ & $\mathcal{U}[1400,2600]$ & $\mathcal{U}[6,10]$ & $\mathcal{U}[-5,5]$ & $\mathcal{U}[-8,2]$ & -- & -- & -- & -- & -- & 2.97 & 1.458 \\
\noalign{\smallskip}
WASP-127 b & $\mathcal{U}[0.5,2.5]$ & $\mathcal{U}[0.01,100]$ & $\mathcal{U}[0.01,100]$ & $\mathcal{U}[400,1600]$ & $\mathcal{U}[6,10]$ & $\mathcal{U}[-5,5]$ & $\mathcal{U}[-8,2]$ & $\mathcal{U}[-500,500]$ & -- & -- & -- & -- & 2.38 & 1.39 \\

\noalign{\smallskip}
\hline
\hline
\end{tabular}}
\end{minipage}}
\end{sidewaystable*}

\renewcommand{\tablename}{Supplementary Table}
\renewcommand{\thetable}{6}
\begin{sidewaystable*}
\rotatebox{0}{
           \begin{minipage}{\textwidth}
\centering
\caption{
List of molecular and atomic species included in the free and grid-chemistry retrievals for each planet in the sample. A ``Yes" indicates that the corresponding species was considered in the retrieval, with its mass-fraction assigned by a log-uniform prior between 10$^{-12}$ and 10$^{-0.3}$.  For the condensate species $\mathrm{SiO_2}$ and the $\mathrm{H^{-}}$ ion, we used log-uniform priors of [10$^{-11}$, 10$^0$] and [10$^{-14.6}$, 10$^{0.4}$], respectively. 
The species list reflects the dominant absorbers expected in the 0.6–12 $\mathrm{\mu}$m wavelength range covered by JWST and HST and follow the species commonly used in the literature for the respective planets.}
\label{tab:molecule_priors}
\resizebox{\textwidth}{!}{
\begin{tabular}{lccccccccccccccccccc}
        \hline
        \hline
        \noalign{\smallskip}
       Name & $\mathrm{H_2O}$ & $\mathrm{CH_4}$ & $\mathrm{CO}$ & $\mathrm{CO_2}$ & $\mathrm{NH_3}$ & $\mathrm{HCN}$ & $\mathrm{H_2S}$ & $\mathrm{SO_2}$ & $\mathrm{C_2H_2}$ & $\mathrm{CS_2}$ & $\mathrm{SCO}$ & $\mathrm{N_2O}$   & $\mathrm{AlO}$ & $\mathrm{SiO}$ & $\mathrm{SiO_2}$ (c) & $\mathrm{H^{-}}$ & $\mathrm{SiH}$ & $\mathrm{Na}$ & $\mathrm{K}$\\
        \noalign{\smallskip}
        \hline
        \noalign{\smallskip}
        HAT-P-18 b & Yes & Yes & Yes & Yes & Yes & Yes & Yes & Yes & -- & -- & -- & -- & -- & -- & -- & -- & -- & -- & --\\ 
        \noalign{\smallskip}
         \noalign{\smallskip}
        WASP-17 b & Yes & Yes & Yes & Yes & -- & -- & Yes & -- & -- & Yes & -- & -- & Yes  & -- & Yes & Yes & -- & Yes & Yes\\ 
        \noalign{\smallskip}
         \noalign{\smallskip}
          GJ 3470 b & Yes & Yes & Yes & Yes & Yes & Yes & -- & Yes & -- & -- & -- & -- & --  & -- & -- & -- & -- & -- & --\\ 
           \noalign{\smallskip}
         \noalign{\smallskip}
          GJ 1214 b & Yes & Yes & Yes & Yes & Yes & Yes & Yes & Yes & Yes & Yes & Yes & -- & --  & -- & -- & -- & -- & -- & --\\ 
          \noalign{\smallskip}
         \noalign{\smallskip}
          K2-18 b & Yes & Yes & Yes & Yes & Yes & Yes & -- & -- & -- & Yes & Yes & Yes & --  & -- & -- & -- & -- & -- & --\\ 
          \noalign{\smallskip}
         \noalign{\smallskip}
          TOI-270 d & Yes & Yes & Yes & Yes & Yes & Yes & Yes & Yes & Yes & Yes & Yes & Yes & --  & -- & -- & -- & -- & -- & --\\ 
           \noalign{\smallskip}
         \noalign{\smallskip}
          HD 189733 b & Yes & Yes & Yes & Yes & Yes & Yes & Yes & Yes & Yes & -- & -- & -- & --  & -- & -- & -- & -- & -- & --\\ 
           \noalign{\smallskip}
         \noalign{\smallskip}
          HD 209458 b & Yes & Yes & Yes & Yes & Yes & Yes & Yes & Yes & Yes & -- & -- & -- & --  & -- & -- & -- & -- & -- & --\\ 
           \noalign{\smallskip}
         \noalign{\smallskip}
          HIP 67522 b & Yes & Yes & Yes & Yes & Yes & -- & Yes & Yes & -- & -- & -- & -- & --  & -- & -- & -- & -- & -- & --\\ 
           \noalign{\smallskip}
         \noalign{\smallskip}
         WASP-39 b & Yes & Yes & Yes & Yes & Yes & Yes & Yes & Yes & -- & -- & -- & -- & --  & -- & -- & -- & -- & Yes & Yes\\ 
         \noalign{\smallskip}
         \noalign{\smallskip}
          WASP-80 b & Yes & Yes & Yes & Yes & Yes & -- & -- & Yes & -- & -- & -- & -- & --  & -- & -- & -- & -- & -- & --\\ 
          \noalign{\smallskip}
         \noalign{\smallskip}
          WASP-107 b & Yes & Yes & Yes & Yes & Yes & -- & Yes & Yes & -- & -- & -- & -- & --  & -- & -- & -- & -- & -- & --\\ 
          \noalign{\smallskip}
         \noalign{\smallskip}
          WASP-15 b & Yes & Yes & Yes & Yes & Yes & Yes & Yes & Yes & Yes & -- & Yes & -- & --  & -- & -- & -- & -- & -- & --\\ 
          \noalign{\smallskip}
         \noalign{\smallskip}
          TOI-836 c & Yes & Yes & Yes & Yes & Yes & Yes & Yes & Yes & Yes & Yes & Yes & -- & -- & -- & -- & -- & -- & -- & --\\ 
           \noalign{\smallskip}
         \noalign{\smallskip}
          TOI-776 c & Yes & Yes & Yes & Yes & Yes & Yes & Yes & Yes & Yes & Yes & Yes & -- & --  & -- & -- & -- & -- & -- & --\\ 
          \noalign{\smallskip}
         \noalign{\smallskip}
          TrES-4 b & Yes & Yes & Yes & Yes & -- & Yes & Yes & Yes & -- & -- & -- & -- & --  & -- & -- & -- & -- & -- & --\\ 
           \noalign{\smallskip}
         \noalign{\smallskip}
          TOI-421 b & Yes & Yes & Yes & Yes &Yes & Yes & Yes & Yes & -- & -- & -- & -- & -- & -- & -- & -- & -- & -- & --\\ 
          \noalign{\smallskip}
         \noalign{\smallskip}
          WASP-94 A b & Yes & Yes & Yes & Yes & Yes & -- & Yes & Yes & Yes & -- & -- & -- & --  & -- & -- & -- & -- & -- & --\\ 
          \noalign{\smallskip}
         \noalign{\smallskip}
          WASP-121 b & Yes & \textcolor{black}{Yes} & Yes & Yes & -- & -- & -- & Yes & -- & -- & -- & -- & --  & Yes & -- & Yes & Yes & -- & --\\ 
           \noalign{\smallskip}
         \noalign{\smallskip}
          WASP-127 b & Yes & Yes & Yes & Yes & Yes & Yes & Yes & Yes & Yes & Yes & Yes & -- & --  & -- & -- & -- & -- & -- & --\\ 
        
        \noalign{\smallskip}
        \hline
        \hline
\end{tabular}}
\end{minipage}}
\end{sidewaystable*}

\renewcommand{\tablename}{Supplementary Table}
\renewcommand{\thetable}{7}
\begin{table}
\caption{Stellar and planetary parameters used to generate the chemical-kinetics grids using VULCAN. For each system, the table lists the planet radius (in the unit of Jupiter radius), surface gravity, stellar radius (in the unit of Sun radius), star–planet separation (a), and the set of isothermal temperatures used in the VULCAN simulations.}
\label{tab:vulcan_temp_grids}%
\begin{tabular}{@{}lcccccccc@{}}
\toprule
Planet & Planet Radius\footnotemark[1] &  log(g)\footnotemark[2] & Star Radius & Star Radius & a [AU]  & Ref.  & Isothermal Temperature \\
Planet & [$\mathrm{R_J}$] &  cgs &  [$\mathrm{R_{\odot}}$] & Ref. &   &  & Grid for VULCAN (K)\\
\midrule
HAT-P-18 b  & 0.995 & 2.6930 & 0.74   & \cite{stassun2019revised}   & 0.055  &  \cite{bonomo2017gaps} &   [600, 800, 1200, 1600] \\
\noalign{\smallskip}
WASP-17 b   & 1.870 & 2.7426 & 1.57    & \cite{stassun2019revised}   & 0.051  & \cite{bonomo2017gaps}  &   [1000, 1300, 1600, 1900] \\
\noalign{\smallskip}
GJ 3470 b   & 0.408 & 2.8143 & 0.499    & \cite{stassun2019revised}   & 0.035  & \cite{2016MNRAS.463.2574A}  &   [400, 800, 1200, 1600] \\
\noalign{\smallskip}
GJ 1214 b   & 0.244 & 3.0427 & 0.2162    & \cite{2024ApJ...963L..37M}   & 0.015  & \cite{2024ApJ...963L..37M}  &   [400, 800, 1200, 1600] \\
\noalign{\smallskip}
K2-18 b     & 0.211 & 3.1920 &  0.468   & \cite{howard2025planet}   & 0.143  & \cite{2018AJ....155..257S}  &   [400, 800, 1200, 1600] \\
\noalign{\smallskip}
TOI-270 d   & 0.178 & 3.0130 &  0.374   & \cite{stassun2019revised}   & 0.072  & \cite{van2021masses}  &   [400, 800, 1200, 1600] \\
\noalign{\smallskip}
HD 189733 b & 1.119 & 3.3411 & 0.776    & \cite{stassun2019revised}   & 0.031  & \cite{2019PASP..131k5003A}  &   [600, 800, 1200, 1600] \\
\noalign{\smallskip}
HD 209458 b & 1.400 & 2.9715 & 1.186    & \cite{stassun2019revised}   & 0.047  & \cite{salz2015high}  &   [600, 800, 1200, 1600] \\
\noalign{\smallskip}
HIP 67522 b & 0.891 & 2.8466 & 1.38    & \cite{2024ApJ...973L..30B}   & 0.075  & \cite{2024ApJ...973L..30B}  &   [600, 800, 1200, 1600] \\
\noalign{\smallskip}
WASP-80 b   & 0.999 & 3.12 & 0.636    & \cite{stassun2019revised}   & 0.034  & \cite{2015MNRAS.450.2279T}  &   [400, 800, 1200, 1600] \\
\noalign{\smallskip}
WASP-39 b  & 1.279 & 2.6292 & 1.013    & \cite{stassun2019revised}   & 0.048  & \cite{mancini2018gaps}  &   [600, 800, 1200, 1600] \\
\noalign{\smallskip}
WASP-107 b  & 0.924 & 2.43 & 0.67    &\cite{howard2025planet}    & 0.055  & \cite{2017MNRAS.469.1622M}  &   [400, 800, 1200, 1600] \\
\noalign{\smallskip}
WASP-15 b   & 1.408 & 2.82 & 1.50    & \cite{2013MNRAS.434.1300S}   & 0.052  & \cite{2013MNRAS.434.1300S}  &   [1000, 1300, 1600, 1900] \\
\noalign{\smallskip}
TOI-836 c   & 0.231 & 3.14 & 0.665    & \cite{2023MNRAS.520.3649H}   & 0.075  & \cite{2023MNRAS.520.3649H}  &   [400, 800, 1200, 1600] \\
\noalign{\smallskip}
TOI-776 c   & 0.183 & 3.2078 & 0.547    &\cite{fridlund2024planets}    & 0.10  & \cite{fridlund2024planets}  &   [400, 800, 1200, 1600] \\
\noalign{\smallskip}
TrES-4 b    & 1.706 & 2.8727 & 2.008    &\cite{stassun2019revised}    & 0.050  & \cite{turner2016ground}  &   [1000, 1300, 1600, 1900] \\
\noalign{\smallskip}
TOI-421 b   & 0.236 & 2.9740 &  0.866   & \cite{krenn2024characterisation}   & 0.055  &\cite{krenn2024characterisation}   &   [600, 800, 1200, 1600] \\
\noalign{\smallskip}
WASP-94 A b & 1.720 & 2.6959 & 1.48    & \cite{2017AJ....153..136S}   & 0.055  & \cite{neveu2014wasp}  &   [600, 800, 1200, 1600]\\
\noalign{\smallskip}
WASP-121 b  & 1.865 & 2.97 & 1.458    & \cite{2016MNRAS.458.4025D}   & 0.025  & \cite{2016MNRAS.458.4025D}   &   [1400, 1800, 2200, 2600]\\
\noalign{\smallskip}
WASP-127 b  & 1.370 & 2.38 &  1.39   &\cite{lam2017dense}    & 0.052  & \cite{lam2017dense}  &   [400, 800, 1200, 1600]\\
\botrule
\end{tabular}
\footnotetext[1]{For planet radius, the values are taken from the same references mentioned in Table \ref{tab:exoplanet_properties}.}
\footnotetext[2]{Gravity values are taken from the \href{https://exo.mast.stsci.edu/}{Exo.MAST} database.}
\end{table}

\clearpage
\renewcommand{\tablename}{Supplementary Table}
\renewcommand{\thetable}{8}
\begin{table*}
\centering
\caption{Bayesian priors and retrieved posterior constraints from the equilibrium-chemistry retrievals for TOI-270 d and WASP-15 b. The table lists the prior ranges for the retrieved parameters, the adopted values for fixed system parameters, and the posterior median values with 1$\sigma$ uncertainties. Retrieved parameters include the planetary radius ($R_{\rm p}$) in units of Jupiter radius ($R_{\rm J}$), equilibrium temperature ($T_{\rm eq}$) in Kelvin, the log value of the cloud-top pressure ($P_{\rm cloud}$) in bar, the log value of the reference pressure ($P_{\rm ref}$) in bar, the log value of the quench pressure ($P_{\rm quench}$) in bar, atmospheric metallicity parameterized as $\mathrm{[M/H]}$, C/O, internal temperature ($T_{\rm int}$), the infrared opacity $\log_{10}(\kappa_{\rm IR})$, the Guillot parametric temperature-profile parameter $\gamma$, and dataset offset parameters where applicable. The stellar radius ($R_\star$) and surface gravity ($\log g$) were held fixed to the adopted system values.}
\label{tab:eq_validation_priors}
\scriptsize
\resizebox{\textwidth}{!}{
\begin{tabular}{@{}lcccc@{}}
\toprule
Parameter 
& \multicolumn{2}{c}{TOI-270 d} 
& \multicolumn{2}{c}{WASP-15 b} \\
\cmidrule(lr){2-3}
\cmidrule(lr){4-5}
& Priors & Retrieved value 
& Priors & Retrieved value \\
\midrule
\noalign{\smallskip}
\multicolumn{5}{l}{1. Fixed system parameters} \\
\noalign{\smallskip}
\noalign{\smallskip}
$R_\star$ ($R_\odot$) 
& 0.38 &  
& 1.477 &  \\

$\log g$ (cgs) 
& 3.013 &  
& 2.82 &  \\

\midrule
\noalign{\smallskip}
\multicolumn{5}{l}{2. Retrieved atmospheric free parameters} \\
\noalign{\smallskip}
\noalign{\smallskip}
$R_{\rm p}$ ($R_{\rm J}$) 
& $\mathcal{U}[0.05,1.05]$ 
& 0.20$\mathrm{^{\text{+} 0.00 }_{- 0.01}}$ 
& $\mathcal{U}[0.5,2.0]$ 
& 1.35$\mathrm{^{\text{+}0.02 }_{-0.02 }}$ \\
\noalign{\smallskip}

$T_{\rm eq}$ (K) 
& $\mathcal{U}[100,1000]$ 
& 379.52$\mathrm{^{\text{+}80.41 }_{- 167.27}}$ 
& $\mathcal{U}[400,2400]$ 
& 1009.97$\mathrm{^{\text{+}224.46 }_{-194.65 }}$ \\
\noalign{\smallskip}

$\log_{10}(P_{\rm cloud}[\mathrm{bar}])$ 
& $\mathcal{U}[-8,2]$ 
& -3.85$\mathrm{^{\text{+} 0.36}_{- 0.43}}$ 
& $\mathcal{U}[-8,2]$ 
& -3.47$\mathrm{^{\text{+}3.44 }_{-0.82 }}$ \\
\noalign{\smallskip}

$\log_{10}(P_{\rm ref}[\mathrm{bar}])$ 
& $\mathcal{U}[-5,5]$ 
& -3.11$\mathrm{^{\text{+}1.53 }_{-1.19 }}$ 
& $\mathcal{U}[-5,5]$ 
& -2.86$\mathrm{^{\text{+}1.75 }_{-1.31 }}$ \\
\noalign{\smallskip}

$\log_{10}(P_{\rm quench}[\mathrm{bar}])$ 
& $\mathcal{U}[-6,3]$ 
& -1.54$\mathrm{^{\text{+} 2.71}_{-2.60}}$ 
& $\mathcal{U}[-6,3]$ 
& -3.72$\mathrm{^{\text{+}1.83 }_{-1.42 }}$ \\
\noalign{\smallskip}

$\log_{10}\mathrm{[M/H]}$ 
& $\mathcal{U}[-2,2]$ 
& 1.73$\mathrm{^{\text{+}0.18 }_{-0.28}}$ 
& $\mathcal{U}[-2,2]$ 
& 1.66$\mathrm{^{\text{+}0.22 }_{-0.31 }}$ \\
\noalign{\smallskip}

C/O 
& $\mathcal{U}[0.01,2.0]$ 
& 0.26$\mathrm{^{\text{+}0.28 }_{-0.14 }}$ 
& $\mathcal{U}[0.01,2.0]$ 
& 0.27$\mathrm{^{\text{+}0.22 }_{-0.13 }}$ \\
\noalign{\smallskip}

$T_{\rm int}$ (K) 
& $\mathcal{U}[200,600]$ 
& 485.57$\mathrm{^{\text{+}71.99 }_{-171.91 }}$ 
& $\mathcal{U}[200,600]$ 
& 422.12$\mathrm{^{\text{+}112.95 }_{-130.32 }}$ \\
\noalign{\smallskip}

$\log_{10}(\kappa_{\rm IR})$ 
& $\mathcal{U}[-7,-1]$ 
& -4.38$\mathrm{^{\text{+}1.85}_{-1.67 }}$ 
& $\mathcal{U}[-7,-1]$ 
& -3.88$\mathrm{^{\text{+}1.82 }_{-1.93 }}$ \\
\noalign{\smallskip}

$\gamma$ 
& $\mathcal{U}[0.4,2.4]$ 
& 1.33$\mathrm{^{\text{+}0.67 }_{-0.61 }}$ 
& $\mathcal{U}[0.4,2.4]$ 
& 1.32$\mathrm{^{\text{+}0.67}_{- 0.56}}$ \\
\noalign{\smallskip}

Dataset offset 
& $\mathcal{U}(-500,500)$ 
& 90.40$\mathrm{^{\text{+}13.16 }_{-13.06 }}$ 
& -- 
& -- \\

\bottomrule
\end{tabular}}
\vspace{0.6em}

\begin{minipage}{0.96\textwidth}
\footnotesize
\textit{Note:} Retrieved values are reported as posterior medians with 1$\sigma$ uncertainties. The molecular opacity species included through the equilibrium-chemistry calculation are $\mathrm{H_2O}$, $\mathrm{CH_4}$, $\mathrm{CO_2}$, CO, HCN, $\mathrm{NH_3}$, $\mathrm{H_2S}$, and $\mathrm{C_2H_2}$ for both TOI-270 d and WASP-15 b.
\end{minipage}
\end{table*}

\renewcommand{\tablename}{Supplementary Table}
\renewcommand{\thetable}{9}

\begin{table*}[htbp]
\centering

\caption{Comparison of Bayesian evidence estimates obtained with MultiNest sampling efficiencies of 0.8 and 0.3 for five representative planets. Both the free-chemistry and VULCAN disequilibrium-grid retrievals were repeated with all other retrieval parameters kept unchanged. The difference is defined as $\Delta\ln Z=\ln Z_{0.3}-\ln Z_{0.8}$.}

\label{tab:efficiency_lnz_comparison}

\small
\begin{tabular}{@{}llccc@{}}
\toprule
Planet & Retrieval framework &
$\ln Z$ (0.8) &
$\ln Z$ (0.3) &
$\Delta\ln Z$ \\
\midrule

GJ 3470 b
& Free chemistry
& $1536.8\pm0.0$
& $1537.1\pm0.2$
& $+0.3$ \\

& VULCAN grid
& $1537.9\pm0.1$
& $1538.2\pm0.0$
& $+0.3$ \\

\midrule

WASP-80 b
& Free chemistry
& $994.4\pm0.2$
& $994.4\pm0.0$
& $0.0$ \\

& VULCAN grid
& $990.6\pm0.0$
& $990.9\pm0.3$
& $+0.3$ \\

\midrule

HD 209458 b
& Free chemistry
& $820.8\pm0.1$
& $820.9\pm0.0$
& $+0.1$ \\

& VULCAN grid
& $816.4\pm0.1$
& $817.1\pm0.0$
& $+0.7$ \\

\midrule

WASP-15 b
& Free chemistry
& $1825.2\pm0.5$
& $1824.4\pm0.1$
& $-0.8$ \\

& VULCAN grid
& $1825.8\pm0.1$
& $1822.7\pm0.0$
& $-3.1$ \\

\midrule

WASP-121 b
& Free chemistry
& $2459.6\pm0.1$
& $2459.6\pm0.1$
& $0.0$ \\

& VULCAN grid
& $2435.5\pm0.0$
& $2435.7\pm0.0$
& $+0.2$ \\

\bottomrule
\end{tabular}

\end{table*}

\renewcommand{\figurename}{Supplementary Fig.}
\renewcommand{\thefigure}{43}
\clearpage
\begin{figure*}
\centering
\begin{overpic}[width=\textwidth]{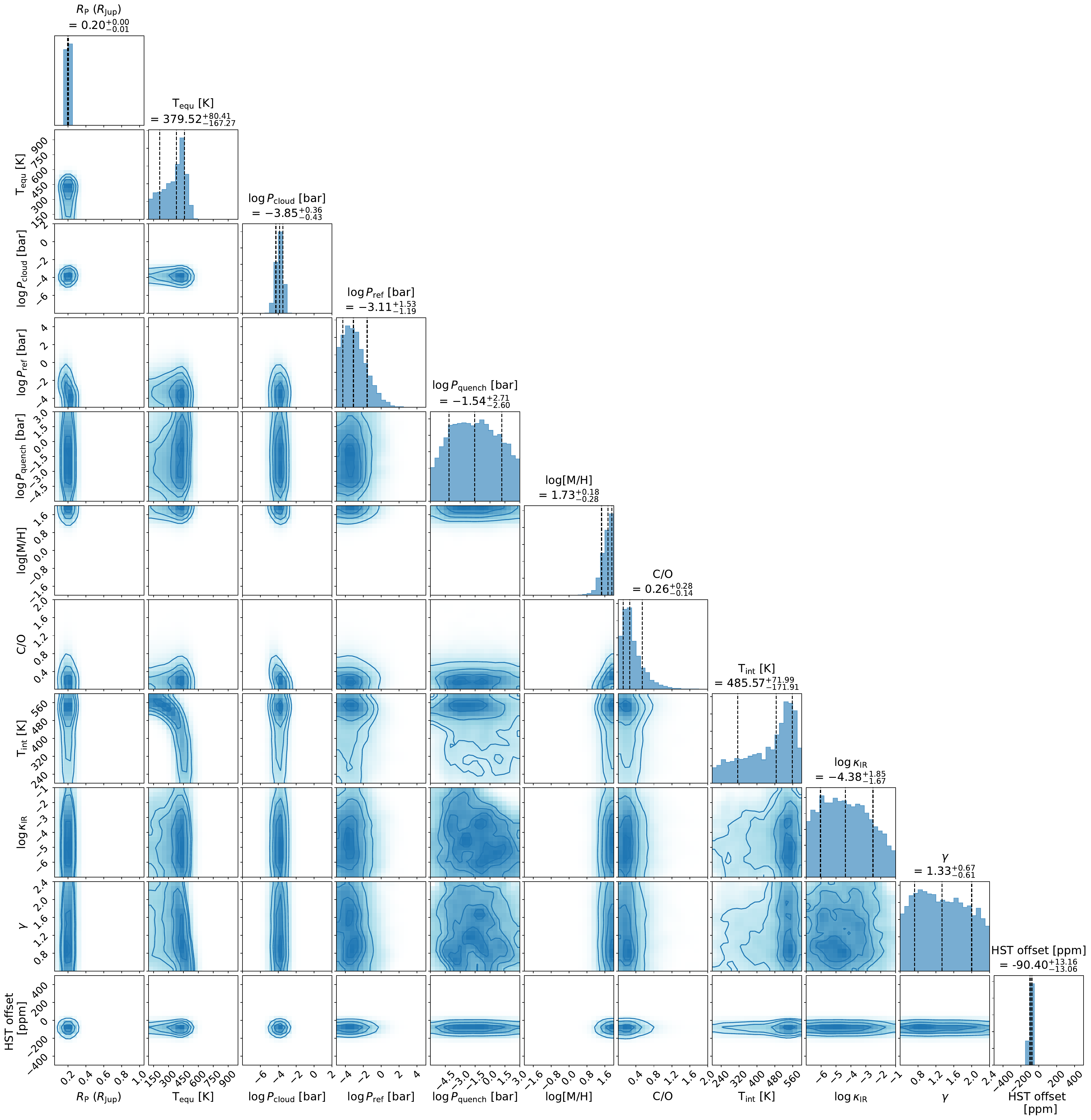}
  \put(45,70){\includegraphics[width=0.55\textwidth]{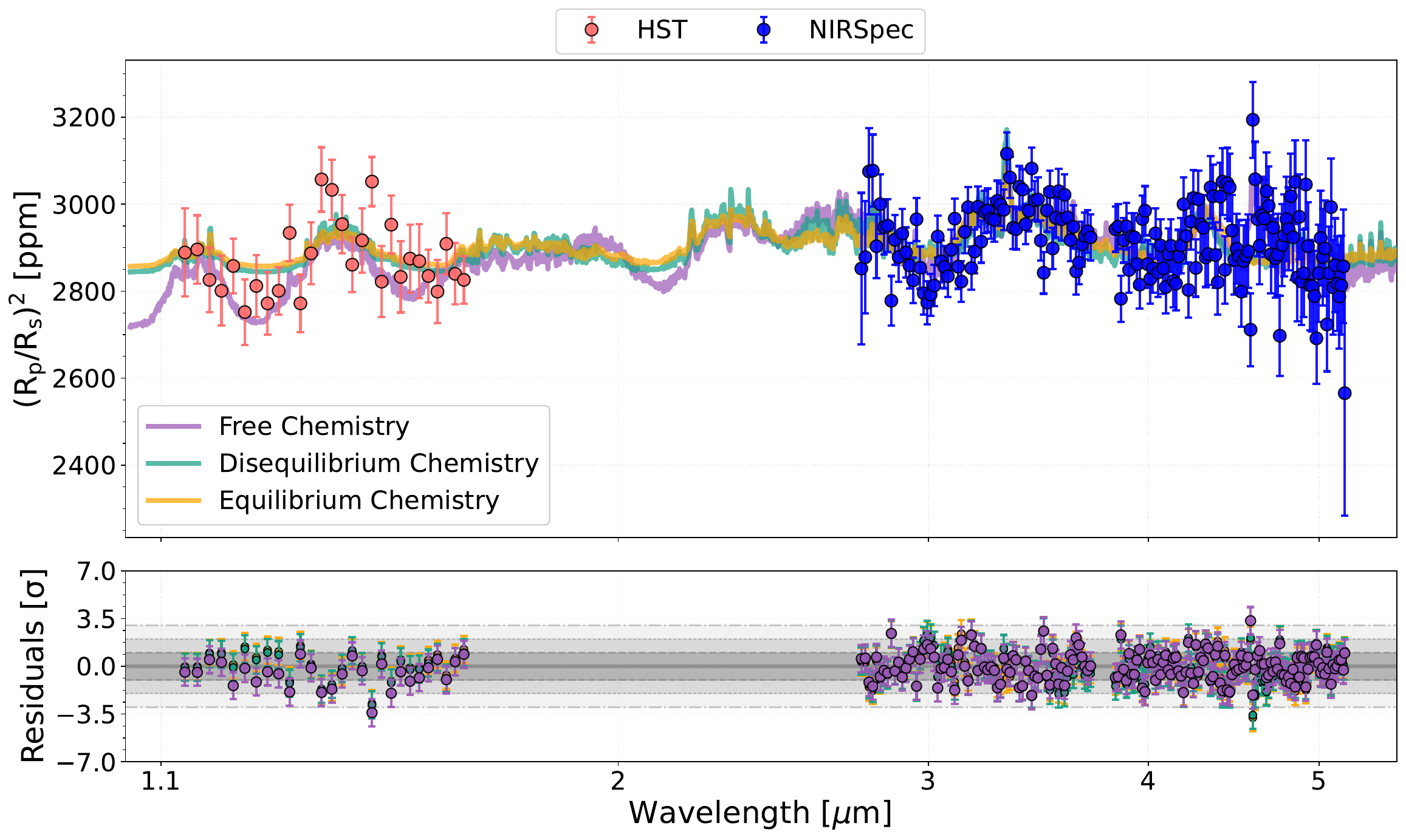}}
\end{overpic}
\caption{Equilibrium chemistry grid retrieval for TOI-270 d. The inset shows the observed spectrum compared with the best-fit models from the free-chemistry retrieval, the disequilibrium chemistry-grid retrieval, and the equilibrium chemistry-grid retrieval. The corner plot shows the posterior distribution from the equilibrium-grid retrieval.}
\label{fig:TOI270d_eq_validation}
\end{figure*}

\renewcommand{\figurename}{Supplementary Fig.}
\renewcommand{\thefigure}{44}
\begin{figure*}
\centering
\begin{overpic}[width=\textwidth]{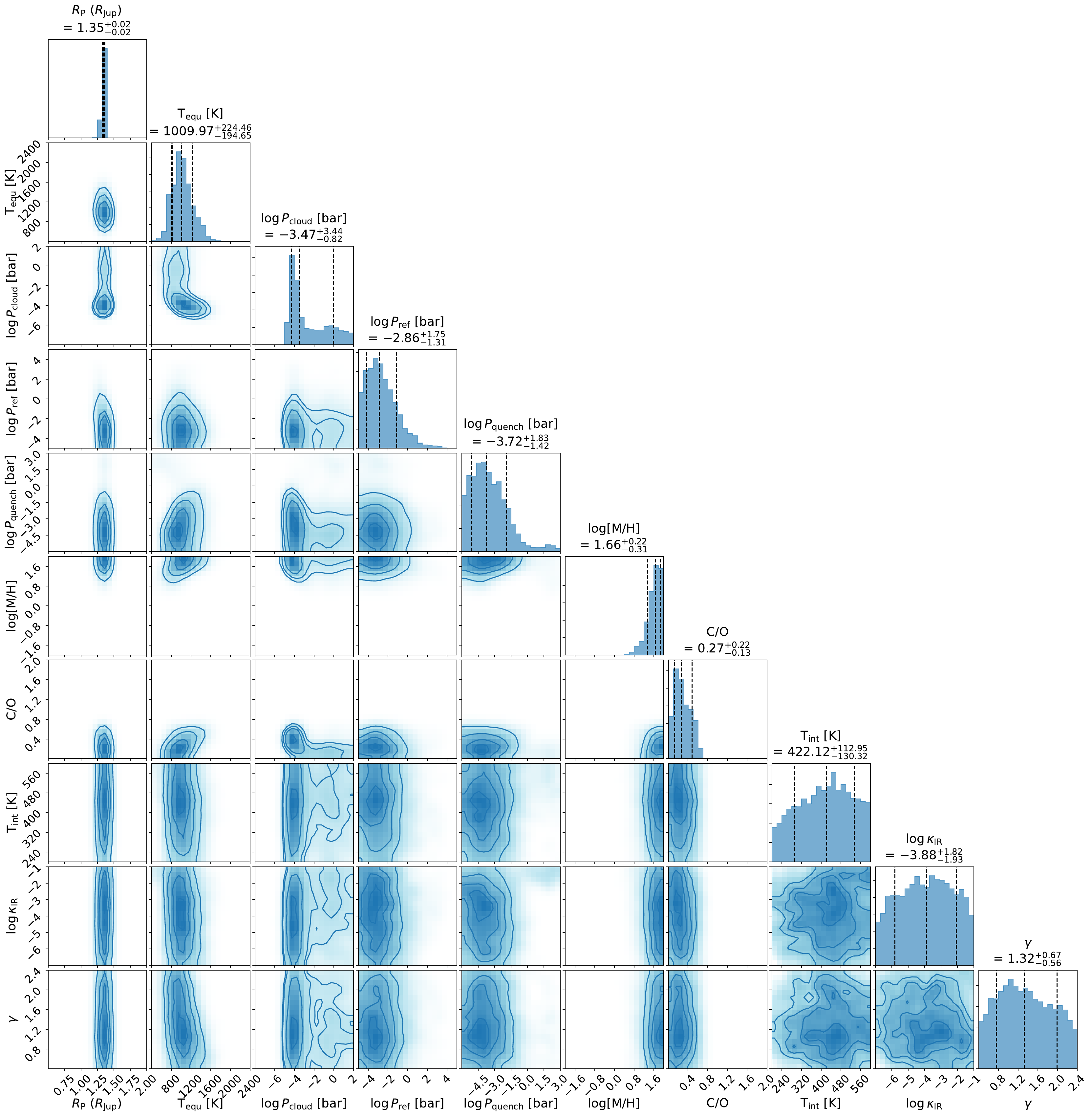}
  \put(45,70){\includegraphics[width=0.55\textwidth]{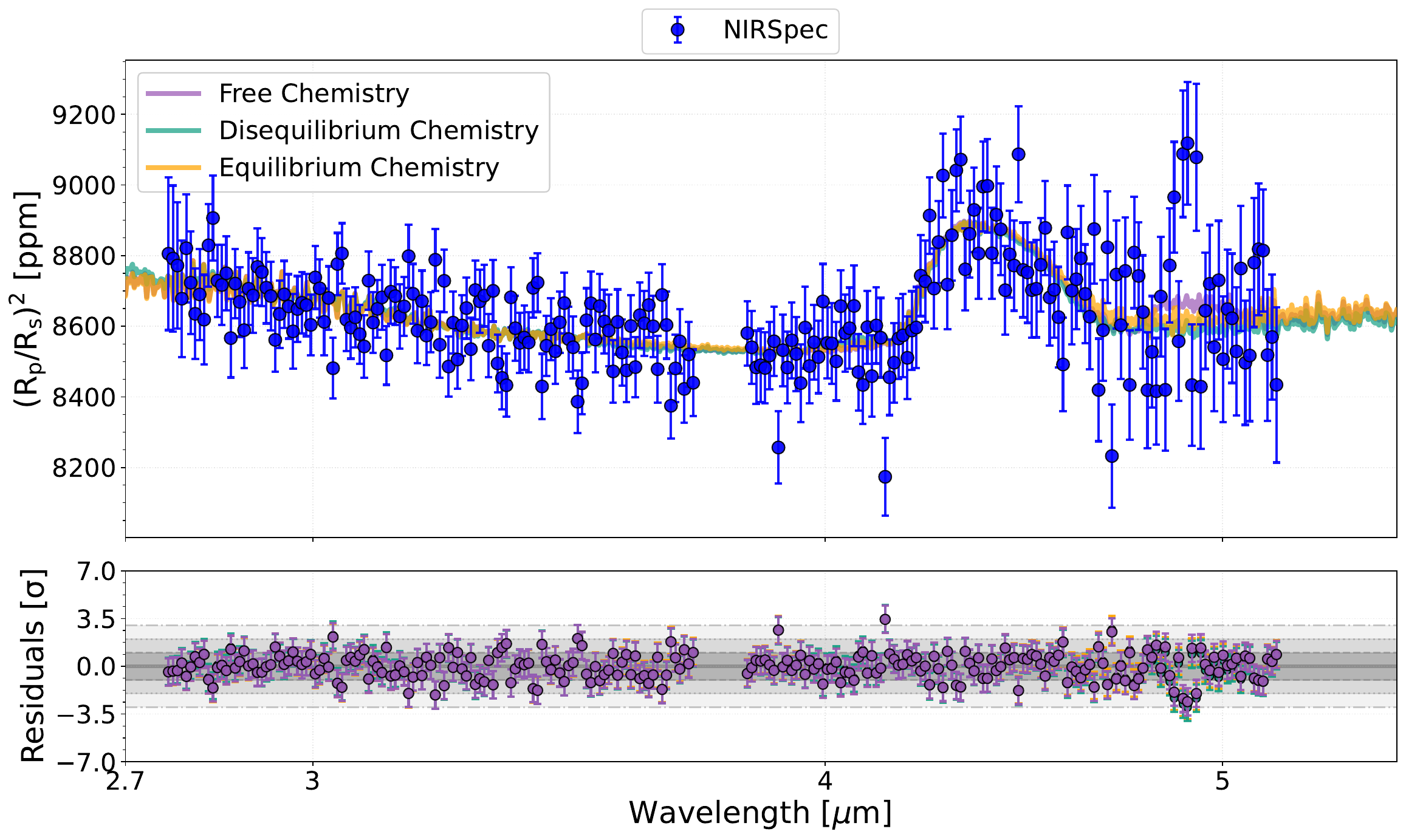}}
\end{overpic}
\caption{Equilibrium chemistry grid retrieval for WASP-15 b, similar to Supplementary Figure \ref{fig:TOI270d_eq_validation}.}
\label{fig:WASP15b_eq_validation}
\end{figure*}

\end{document}